\documentclass[twocolumn,twocolappendix]{aastex701}

\usepackage{amsmath}
\usepackage{makecell}
\usepackage{xspace}

\newcommand{\teff}{$T_{\mathrm{eff}}$}

\newcommand{\logg}{log~$g$}
\newcommand{\feh}{[Fe/H]}

\def\code#1{\texttt{#1}}

\newcommand{\degree}{$^{\circ}$}
\newcommand{\hii}{\hbox{{\rm H}\kern 0.1em{\sc ii}{\rm }}}

\newcommand{\todo}{\ifmmode \text{\color{red}\Huge{\(\bullet\)}} \else {\color{red}{\Huge$\bullet$}}\fi}
\newcommand{\tido}{\ifmmode {{\color{red}\bullet}} \else {\color{red}$\bullet$}\fi}

\newcommand{\vonex}{`\texttt{1.0.0}' cross-match}
\newcommand{\lvmvis}{\texttt{LVMvis}}

\def\ero{{eROSITA}\xspace}

\def\srgero{{SRG/eROSITA}\xspace}

\submitjournal{ApJS}

\graphicspath{{./}{figures/}}
\usepackage{multirow}

\begin{document}

\title{The Twentieth Data Release of the Sloan Digital Sky Survey: First All-Sky BOSS Spectra, eROSITA-SDSS-V Mapper Coordinated Observations, and a Preview of the Local Volume Mapper
}

\shorttitle{SDSS-V DR20}

\correspondingauthor{Emily Griffith} 
\email{spokesperson@sdss.org} 

\author{SDSS Collaboration}
\affiliation{Sloan Digital Sky Survey}
\email{spokesperson@sdss.org}

\author[0000-0001-8341-3940]{Mojgan Aghakhanloo}
\affiliation{Department of Astronomy, University of Virginia, Charlottesville, VA 22904-4325, USA}
\email{mvy4at@virginia.edu}

\author{David Aguilar}
\affiliation{Department of Physics and Astronomy, California State University, San Bernardino, San Bernardino, CA 92407, USA}
\email{}

\author[0000-0003-1908-8463]{James Aird}
\affiliation{Institute for Astronomy, University of Edinburgh, Royal Observatory, Edinburgh EH9 3HJ, UK}
\email{james.aird@ed.ac.uk}

\author[0009-0000-0733-2479]{Andr\'{e}s Almeida}
\affiliation{Department of Astronomy, University of Virginia, Charlottesville, VA 22904-4325, USA}
\email{tac6na@virginia.edu}

\author[0009-0006-0837-4287]{Bella Abigail Sanabria Alonso}
\affiliation{Universidad Nacional Aut\'{o}noma de M\'{e}xico. Instituto de Astronom\'{i}a. A.P. 70-264, 04510, Ciudad de M\'{e}xico, M\'{e}xico}
\email{bsanabria@astro.unam.mx}

\author[0000-0002-2962-1391]{Hillary Diane Andales}
\affiliation{Department of Astronomy and Astrophysics, University of Chicago, Chicago, IL 60637, USA}
\email{handales@uchicago.edu}

\author[0000-0002-6404-9562]{Scott F. Anderson}
\affiliation{Department of Astronomy, University of Washington, Box 351580, Seattle, WA 98195, USA}
\email{sfander@uw.edu}

\author[0000-0002-6270-8624]{Stefan Arseneau}
\affiliation{Department of Astronomy \& Institute for Astrophysical Research, Boston University, 725 Commonwealth Ave., Boston, MA 02215, USA}
\email{arseneau@bu.edu}

\author[0000-0001-5179-980X]{Consuelo Gonz\'{a}lez \'{A}vila}
\affiliation{Las Campanas Observatory, Ra\'{u}l Bitr\'{a}n 1200, La Serena, Chile}
\email{consuelo@carnegiescience.edu}

% ...

\begin{abstract}

This paper presents the twentieth data release (DR20) from the Sloan Digital Sky Survey, the third data release of its fifth generation (SDSS-V). SDSS-V is a panoptic spectroscopy survey that is mapping the stars, gas, and galaxies through three scientific programs: the Milky Way Mapper (MWM), the Local Volume Mapper (LVM), and the Black Hole Mapper (BHM). DR20 presents the first optical (BOSS) SDSS-V spectra from southern hemisphere for the MWM and BHM surveys; new optical MWM and BHM data from the northern hemisphere are also available, for a total over 3 million spectra of 1.5 million stars and half a million galaxies and quasars, with galactic and extragalactic x-ray targets coordinate with eROSITA DR2. DR20 includes integral field spectroscopy maps from LVM of six targets and 169 tiles, spanning Galactic HII regions, planetary nebulae, and nearby galaxies. Additionally, eighteen value added catalogs are also released with DR20, based on SDSS-V MWM and BHM data, and we present a new LVM visualization tool including an RGB HiPS map as a value added product.

\end{abstract}


\section{Introduction} \label{sec:intro}

Since the Early Data Release (EDR) in June 2001 \citep{Stoughton_2002_sdssEDR}, the Sloan Digital Sky Survey \citep[SDSS;][]{York_2000_sdss} has been mapping the cosmos and regularly publicly releasing data. This paper presents the twentieth data release (DR20) of SDSS, the third data release of SDSS-V.

The quarter century of SDSS data releases have spanned five generations of the survey and numerous science programs. In the first generation, SDSS-I \citep[EDR, DR1-DR5;][]{York_2000_sdss} conducted a photometric survey in the $ugriz$ band passes of over $8000\,\text{deg}^2$ of the sky as well as a low-resolution ($R\sim 1800$) optical spectroscopic survey of stars, galaxies, and quasars with the 2.5m Sloan Foundation Telescope at Apache Point Observatory \citep[APO;][]{Gunn_2006_sloantelescope}. In SDSS-II \citep[DR6-DR7;][]{sdss_dr7}, the legacy imaging survey was completed, along with the SDSS-II supernova survey \citep{Frieman_2008_sdss2supernovae} and the Sloan Extension for Galactic Understanding and Exploration Survey \citep[SEGUE-1;][]{Yanny_09_SEGUE}, which captured spectra of over 200,000 stars in the Milky Way (MW).

The third generation, SDSS-III \citep[DR8-DR12;][]{Eisenstein_11_sdss3overview}, brought major instrumentation updates to the Sloan Foundation Telescope, including the integration of new fiber-fed, multi-object spectrographs (MOS). The upgraded hardware enabled the Baryon Oscillation Spectroscopic Survey \citep[BOSS;][]{Dawson2013_boss}, to map galaxy clustering, and SEGUE-2 \citep{Rockosi_2022_segue2}, to further study old stellar populations, with upgraded low-resolution optical spectrographs \citep{Smee_2013_bossspectrographs}. SDSS-III also conducted the Multi-object APO Radial Velocity Exoplanets Large-area Survey \citep[MARVELS;][]{Ge_2008_marvels} with a $R\sim 10,000$ optical spectrograph, as well as the Apache Point Galactic Evolution Experiment \citep[APOGEE;][]{Majewski_2017_apogeeoverview} with a moderate resolution ($R\sim22,500$) near-infrared spectrograph \citep{Wilson_2019_apogeespectrographs} designed to study red giant branch (RGB) stars in the MW.

The survey's fourth generation saw expansion to the southern hemisphere, adding an APOGEE spectrograph clone to the du Pont 2.5m Telescope at Las Campanas Observatory \citep[LCO;][]{Bowen_1973_duPontTelescope}. With these twin spectrographs, SDSS-IV \citep[DR13-17;][]{Blanton_2017_sdss4} conducted the APOGEE-2 survey of RGB stars at APO and LCO \citep{Zasowski_2017_apogee2targeting}, expanding coverage to the inner Galactic disk. At APO, a new optical, integral field unit (IFU) studied the stellar and gas properties of low-redshift galaxies in the Mapping Nearby Galaxies at APO survey \citep[MANGA;][]{Bundy_2015_manga} while the low-resolution, optical spectrographs completed the extended BOSS survey \citep[eBOSS;][]{Dawson_2016_eboss}.

The current, fifth generation (SDSS-V) began observations in 2020 \citep{kollmeier2026}. With major hardware upgrades, including a robotic focal plane system \citep[FPS;][]{Pogge_2020_fps} that integrated BOSS and APOGEE spectrographs onto both the Sloan Foundation Telescope and du Pont Telescope, as well as a new wide-field IFU instrumentation system built at LCO \citep[LVM-I;][]{Herbst2024}, SDSS-V set out to panoptically map the night sky. 

SDSS-V is comprised of three ``mapper'' surveys: the Black Hole Mapper (BHM; Section~\ref{sec:bhm}), a spectroscopic survey of quasars probing the growth and evolution of black holes; the Milky Way Mapper (MWM; Section~\ref{sec:mwm}), a stellar spectroscopic survey of our Galaxy; and the Local Volume Mapper (LVM; Section~\ref{sec:lvm}), an optical IFU survey of the Milky Way and nearby galaxies illuminating interstellar medium (ISM) and stellar feedback physics. The BHM and MWM surveys leverage a partnership with the eROSITA X-ray survey, enabling coordinated multi-wavelength observations of high-energy extra-galactic and stellar sources (Sections~\ref{sec:bhm_spiders} and ~\ref{sec:mwm_erosita}).

In this paper describing the twentieth release of SDSS, we present the first SDSS-V spectra from the southern hemisphere, including over 3 million BOSS optical spectra of 0.5M extragalactic and 1.5M stellar sources, as well as 169 IFU tiles covering 6 targets from the LVM survey. An overview of the scope of DR20 is given in Section~\ref{sec:scope} and information on how to access the SDSS-V data can be found in Section~\ref{sec:data_access}. Section~\ref{sec:pipelines} describes the data reduction and data analysis pipelines used for all mappers. Sections~\ref{sec:bhm}, \ref{sec:mwm}, and~\ref{sec:lvm} detail the data released by BHM, MWM, and LVM in DR20, respectively. Section~\ref{sec:ToO} summarizes the new Target of Opportunity observing mode and Section~\ref{sec:targeting} discusses the release of the survey targeting database. Value added catalogs associated with the BHM and MWM are described in Section~\ref{sec:vacs}. Finally, we summarize DR20 in Section~\ref{sec:summary} and provide an outlook on future SDSS-V data releases.

\section{Scope of DR20}\label{sec:scope}

DR20 for the first time releases SDSS-V stellar and extra-galactic spectra from the southern hemisphere observed from LCO. All optical spectra observed through MJD 60708 (corresponding to 2 February 2025) from both LCO and APO are included in this data release. For BHM, this corresponds to a release of over 500k extra-galactic sources. DR20 is the first large data release for the BHM SPIDERS program (Section \ref{sec:bhm_spiders}), which coincides with the eROSITA-DE second public data release\footnote{\url{https://erosita.mpe.mpg.de/dr2/}}. For MWM, 1.5M stellar targets are released. In DR20, both programs have increased their sample sizes by a factor of four or more. All previously observed SDSS-V optical spectra released in DR19 have been re-reduced with a new tagged version of the BOSS pipeline (see Section \ref{sec:boss_pipeline}) and are also included in DR20. Likewise, all optical stellar spectra have been processed with new versions of the \textsc{Astra} stellar analysis pipelines (Section \ref{sec:astra}). No new infrared data have been included in DR20, but all previously observed spectra and \textsc{Astra} stellar parameters and abundances have been added to DR20 ---  BOSS related files are updated in $v\_astra=0.8.1$ as released in DR20, whereas for users interested in APOGEE data, please look at $v\_astra=0.6.0$, which was previously released in DR19. New in DR20 is the inclusion of spectra observed within the SDSS-V Targets of Opportunity program, which consists of one target and spectrum, summarized in Section \ref{sec:ToO}. 

Building on their first publicly released tile in DR19, LVM is now releasing 6 targets with a total of 169 tiles, summarized in Table {\ref{tab:lvm_dr20_targets}}. These data include newly reduced spectra for the Helix Nebula, released in DR19. In addition to spectra, LVM is also releasing data products from their data analysis pipeline (DAP), which include maps of emission line fluxes and kinematics (see Section \ref{sec:lvm_data_product} for an overview). With DR20, we release a new spectral visualization tool, \texttt{LVMvis}, to inspect and interact with the LVM data (see Section~\ref{sec:lvmvis}).

As in DR19, we also release updated targeting information in DR20, described in Section~\ref{sec:targeting}, and value added catalogs (VACs) based on BHM and MWM spectra. In total, we are releasing 14 new VACs, updating or extending four VACs that were previously released in DR19, and releasing one new value added product (VAP). Section \ref{sec:vacs} provides a complete overview of these catalogs.

All data products and data access routes are described on our website, \url{www.sdss.org/dr20}.

\section{Accessing the Data}\label{sec:data_access}

\subsection{The Science Archive Server}\label{sec:SAS}

To support the immense scale of SDSS-V, which spans millions of targets across both hemispheres, the Science Archive Server (SAS) has evolved into a formidable cyberinfrastructure system. This petabyte-scale environment features a dedicated 1000-core computational cluster and specialized database servers. These resources are custom-built to manage the full lifecycle of survey operations, from automated data reduction and analysis pipelines to advanced visualization tools.  The SAS relies on the Center for High Performance Computing (CHPC) at the University of Utah, whose reliability is critical for the timely distribution of SDSS-V data. The sheer volume of processing is significant: In the last 12 months, the SAS computational cluster received more than 14,500 job submissions using over 6 million core hours.

DR20 represents a massive operational expansion of the foundations laid in DR19. While DR19 introduced the first major spectroscopic datasets for the MWM and BHM, as well as IFU data from the LVM, DR20 scales these efforts by including optical data from the Southern hemisphere. The SAS serves as a collaborative nexus, hosting pipeline outputs across development and production tags—known internally as {\tt sdsswork}—and the data products used to generate value-added catalogs (VACs). It provides the data to load the CAS with catalog-level tables and mirrors the flat-files accessible within SciServer Compute, as discussed in Section \ref{sec:CAS}. For the creation of DR20, nearly 100 astronomers from over 50 international institutes were granted University of Utah affiliate accounts, providing access to the CHPC interactive nodes on the SAS.

The SAS is equipped with five specialized virtual machines (VM) leased from the CHPC, which are dedicated to providing the infrastructure for collaboration web servers, astra pipeline automation, and both public and proprietary data access to the SAS.  New for SDSS-V is a dedicated VM for data visualization web application development, with details described in Section \ref{sec:zora}. Furthermore, a dedicated data/software team maintains the suite of access options that have made SDSS a standard in the field, such as {\tt sdss-access},\footnote{\url{https://sdss-access.readthedocs.io/}} driving scientific discovery while training the next generation of researchers.

\subsection{The Catalog Archive Server and SciServer}\label{sec:CAS}

The Catalog Archive Server (CAS) has been fully integrated into the SciServer science platform \citep{sciserver} as of DR17, and as a result offers a versatile and performance-optimized collaborative data-driven environment for SDSS users to perform their complex analyses and workflows server-side without having to download large amounts of data. 

SciServer brings along the classic SkyServer and CasJobs applications with significant upgrades to integrate them fully within SciServer. One can still use SkyServer in anonymous mode, but to save your results server-side then one can log in to SciServer and also use CasJobs for more features and workbench functionality.

While there is no new functionality in SkyServer for DR20, there are internal changes to render/serve the new data that is part of this release. The existing tools in SkyServer that are specific to SDSS data, as well as more generic tools like SQLxMatch \citep{sciserver_xmatch} that provide a cross-matching capability between SDSS and other data sets that are colocated with the SDSS data, continue to function as before and have been updated where applicable to handle newly added data.

A new addition to CasJobs for DR20 is the eROSITA database, which consists of source catalogs from the x-ray eROSITA mission in the eFEDS area \citep{Brunner2022}, as well as for the all-sky eRASS1 \citep{Merloni2024}, in both hard and soft bands, and includes counterpart tables in both cases \citep{Salvato2022}.

SciServer Compute is a Docker/VM environment that is the heart of the science platform. It consists of an interactive Jupyter Lab interface within a containerized software environment with access to large data volumes, including the entire SDSS SAS archive. This system provides an especially powerful and versatile analysis capability that can be used in both interactive and batch mode, thus complex workflows encoded in Jupyter notebooks can be submitted and run in batch mode. Python libraries for using SkyServer and Casjobs within Jupyter Notebooks are included with the default SDSS Docker image, which also includes dedicated SDSS libraries such as \texttt{sdss-access} and \texttt{semaphore}. \footnote{\url{ https://semaphore.readthedocs.io}}

There are many Value-Added Catalogs (see Section \ref{sec:vacs}) in DR20, containing specialized or derived data based on the core SDSS data set, and most of these are also available as tables in the CAS (accessible via SkyServer, CasJobs, and SciServer Compute), as in previous releases.

SciServer Compute strives to be an open analysis environment, with public SDSS Jupyter Notebook tutorials that provide users with a comprehensive set of examples on how to analyze and visualize SDSS data (see Section \ref{sec:tutorials}).

\subsection{Zora and Valis}\label{sec:zora}
Zora and Valis are the primary user interfaces for accessing and exploring new data in SDSS-V.  They provide ways of searching for SDSS data, exploring observed target metadata, and visualizing or accessing spectral data.  Zora is the primary front-end User Interface (UI), while Valis is the programmatic Application Programming Interface (API) python back-end. Users can opt to use the browser-based Zora or the Valis API directly for programmatic access. 

The core features of Zora and Valis remain the same as with DR19, with support for accessing SDSS-V data from the BHM and MWM surveys. For DR20, Zora and Valis have been updated to serve the corresponding SDSS data delivered for DR20. New features released in DR20: i.) the access and display of Astra pipeline-specific derived parameters; ii.) the interactive display of pipeline model spectra where available for given targets; and iii.) a random target selector on the Zora home page to quickly access to a random interesting SDSS-V target.

\subsection{LVM Visualization Tool: \lvmvis} \label{sec:lvmvis}


\begin{figure*}
      \centering
      \includegraphics[width=\linewidth]{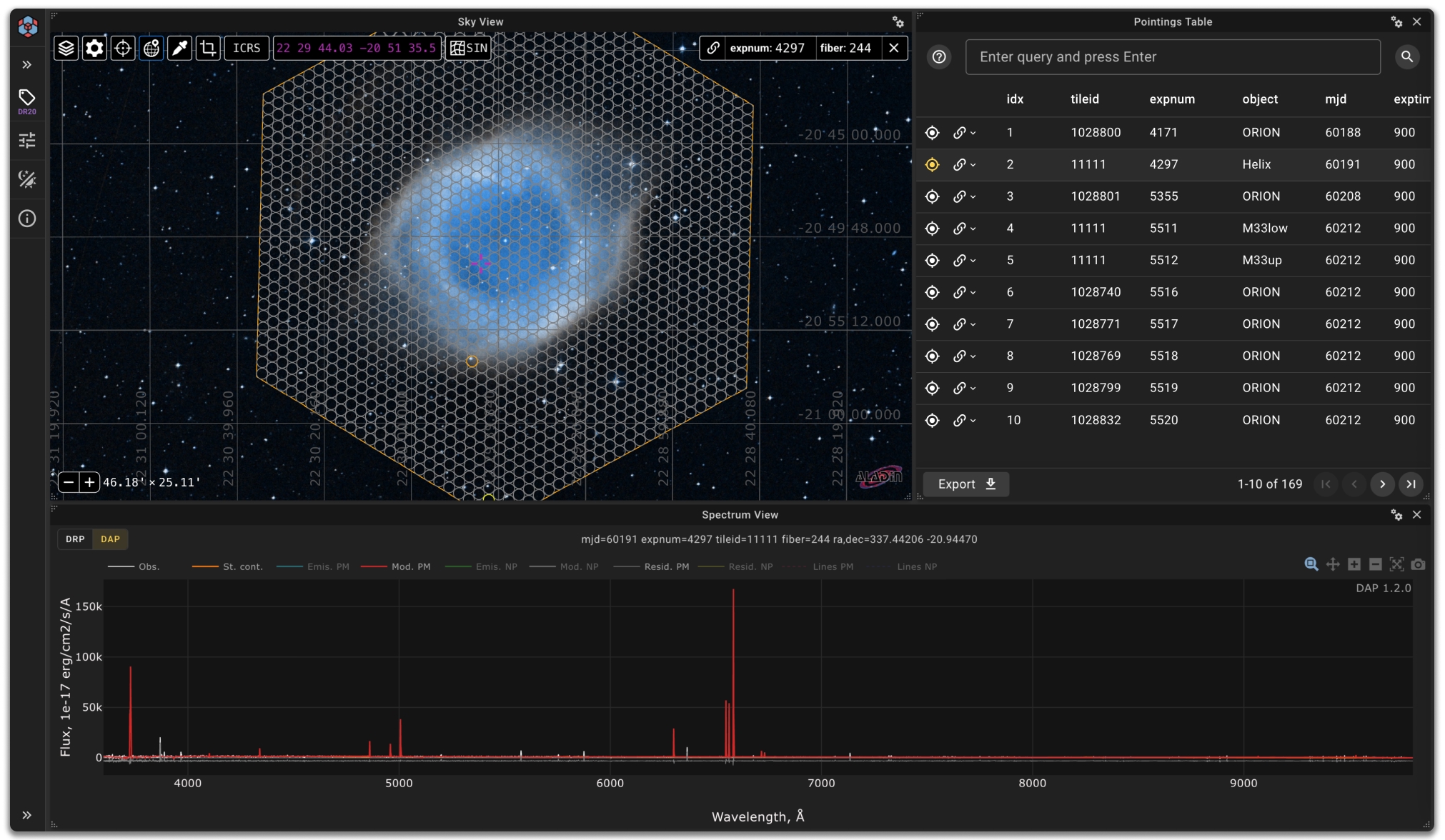}
       \caption{Screenshot of \lvmvis\ showing an observation of the Helix nebula included in the DR20 data release (expnum=4297). The top-left panel shows the sky view with the Digital Sky Survey (R+B) image as background and the hexagonal LVM science IFU footprint overlaid. The top-right panel lists available LVM exposures with their metadata; selections mirrored interactively between the table and the sky view. The bottom panel shows the DRP and DAP outputs for fiber=240, including the observed flux-calibrated spectrum, stellar continuum model, emission-line model, and residuals, with several prominent ISM lines clearly detected across the full 3600--9800\,\AA\ wavelength range. The tool is publicly accessible at https://dr20.sdss.org/lvmvis/.}
      \label{fig:lvmvis}
\end{figure*}

\lvmvis\ is an interactive, browser-based visualization tool developed to facilitate exploration of LVM data products (see~Figure~\ref{fig:lvmvis}).
It provides direct access to both the LVM outputs of the data reduction and data analysis (that will be described in Sections \ref{sec:lvm_drp} and \ref{sec:lvmdap}), helping users to understand what observations are available and to assess data quality before downloading or further analysis. 

The interface combines a sky view based on Aladin Lite \citep{baumann2022, aladinlite2024}, a table of LVM pointings, and an interactive spectral viewer in an adjustable layout.
Users can select pointings through linked sky and table views, filter records in the table with column-based AND/OR conditions, inspect fiber locations, and view reduced spectra with available best-fit models for individual fibers.

Using Aladin Lite as the sky-view component allows \lvmvis\ to show wide-field LVM pointings against HiPS maps \citep{2015A&A...578A.114F} from other surveys.\footnote{Additional HiPS maps are available from the HiPS aggregator: \url{https://aladin.cds.unistra.fr/hips/list}}
In addition to visualizing the spectra released in DR20, \lvmvis\ also presents a color RGB HiPS map constructed from LVM measurements of $H\alpha$, [O{\sc iii}]$\lambda5007$, and [S{\sc ii}]$\lambda\lambda6717,31$ collected before MJD~61086.
This map uses emission-line flux cuts scaled with an asinh stretching function to enhance the visibility of structures across both faint and very bright regions and is intended as a preview of the expected quality, scale, and sky coverage of the LVM survey.
Future releases will include HiPS maps in FITS format based on LVM DAP products, enabling quantitative measurements of physical parameters; their construction and integration into \lvmvis\ will be described in Katkov et al. (in prep.).

The tool is publicly accessible at \url{https://dr20.sdss.org/lvmvis/}. Details of how to use it are given on the \lvmvis\ webpage: \url{https://www.sdss.org/dr20/data_access/lvmvis/}

\subsection{Tutorials}\label{sec:tutorials}

Beginning in DR19, SDSS released a number of tutorials using SDSS data products, primarily in a Jupyter Notebook format. DR19 included tutorials exploring a variety of data summary files as well as BOSS and APOGEE spectra, tutorials highlighting operational decisions such as target offsets and sdss\_id, a tutorial for LVM data, and a few tutorials for our value added catalogs (VACs). Where applicable, all of these tutorials have been updated to use DR20 data files.

DR19 and DR20 tutorials continue to be available on SciServer. \footnote{\url{apps.sciserver.org/compute/}}

\subsubsection{DR20 Improvements}\label{subsubsec:tutorial_DR20}

There is heterogeneity in the design of DR20 tutorials. While many used \texttt{sdss-access} to access files, some simply direct users to download data files. \texttt{sdss-access} is a robust framework that allows SDSS data to be retrieved from any computer, whether the data are already stored locally (as is the case with SciServer) or needs to be downloaded and cached. For DR20, all previous tutorials now use \texttt{sdss-access} in a way that is transparent to the end user. This approach is the recommended method for accessing SDSS data.

The SDSS website has also been updated to sort tutorials according to various criteria (e.g., MWM or BHM, optical or infrared, summary files or spectra, etc.), allowing users to easily find relevant tutorials.

\subsubsection{New for DR20}

MWM has extended tutorials focusing on stellar spectra to include model spectra from relevant pipelines, and added a new tutorial exploring best fit models for individual elements. Another new tutorial builds a machine learning model using data from ASPCAP and \cite{Pinsonneault25} to predict ages for all ASPCAP stars.

BHM has updated tutorials to use the latest BOSS pipeline results, and developed a new tutorial exploring various spectra pixel masks available from the BOSS pipeline.

LVM has developed five new tutorials. There are tutorials exploring LVM spectra and some pipeline features, a tutorial creating mosaic images from LVM data products, and an extensive tutorial exploring the Helix Nebula, effectively recreating much of the content from \cite{HelixDR19o}.

Four tutorials highlighting VACs are provided. The BOSS visual inspection VAC (\ref{vac:inspect_boss}), Fermi Blazar VAC (\ref{vac:fermi_redshifts}), Open Cluster (OCCAM) VAC (\ref{vac:occam}), and CLAM stellar parameters VAC (\ref{vac:clam_parameters}) all provide tutorials demonstrating usage of their catalogs.

\subsection{User Support and Help Desk}\label{sec:helpdesk}
To assist users in navigating the various data access interfaces, pipelines, and data products, the SDSS maintains a centralized user support system. Users can submit inquiries, report technical issues, or request clarification on datasets by contacting the SDSS Help Desk at \url{helpdesk@sdss.org}.

\section{Pipelines}\label{sec:pipelines}

\subsection{BOSS Data Reduction Pipeline}\label{sec:boss_pipeline}
The optical BOSS data released in DR20 are processed with version v6\_2\_1 of the BOSS pipeline software \texttt{idlspec2d} (\citealt{Bolton2012, Dawson_2013_boss}; Morrison et al. in prep). Updates made to one of the BOSS spectrographs when it was moved to the LCO Du Pont telescope combined with the associated observing strategy necessitated significant update to the pipeline from the DR19 v6\_1\_3 version of the pipeline.  All SDSS-V versions of \texttt{idlspec2d} are available for download from the SDSS GitHub, with the version used in DR20 and described here available at \url{https://github.com/sdss/idlspec2d/releases/tag/v6_2_1}. 

\begin{figure}
    \centering
    \includegraphics[width=0.9\linewidth]{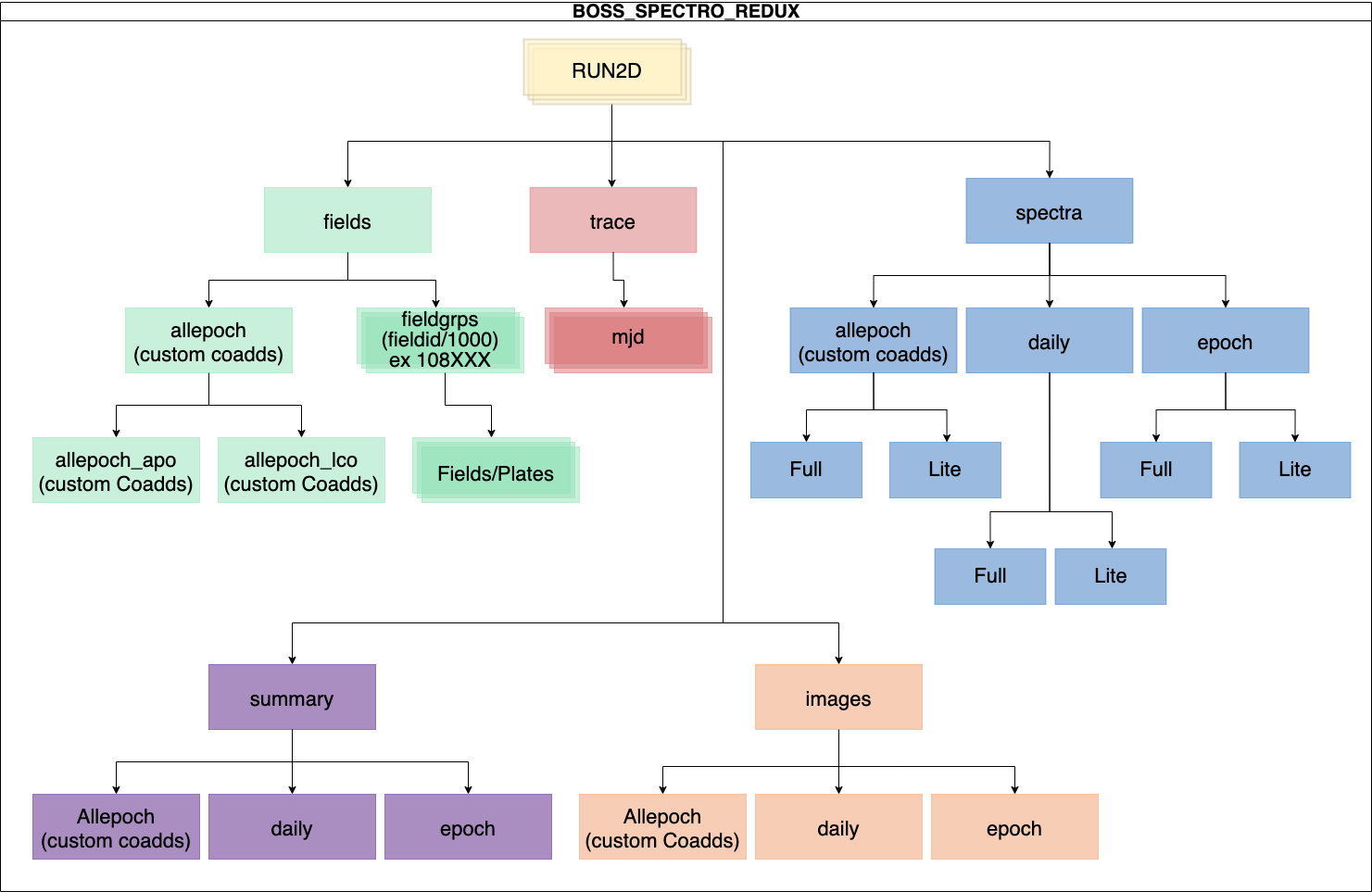}
    \caption{A schematic of the updated BOSS pipeline output file tree. It shows the updated hierarchical structure implemented in the DR20 version of the pipeline, with parent folders for the different product levels and coadd types produced by the BOSS DRP. A more detailed version of the diagram can be found in the pipeline’s \href{https://sdss-idlspec2d.readthedocs.io/en/v6_2_1/tree.html}{ReadTheDocs.}}
    \label{fig:BOSS_drp_tree}
\end{figure}

The pipeline updates included a number of small changes ranging from new quality assurance tools and metadata updates, to tweaks of parameters to improve calibration of LCO data (and FPS data in general). It also contains some other operational updates such as removal of depreciated code and cleanup of the code repository, added Read the Docs documentation\footnote{\url{https://sdss-idlspec2d.readthedocs.io/}} for running the pipeline, and reorganization of the output file tree structure. The file tree structure (outlined in \autoref{fig:BOSS_drp_tree}) was reorganized to assist in easier navigation of rapidly growing files produced by the pipeline associated with the large number of fields and products produced in SDSS-V. 

The most significant update involves a modification to the fiber tracing algorithm used in the pipeline. In SDSS-V (and pre-SDSS-V) plates, flats (and arcs) were taken with each field to calibrate and trace the spectra. However, in FPS operations, the number of flat field calibrations taken during the night was significantly reduced to just a few each night to minimize overheads. However, it was determined early on that this caused some issues with fiber tracing due to flexure in the BOSS instrument, especially at LCO. This was addressed by adding additional code that correlates the position of the arc lamp lines taken with each science field with those of an arc taken in association with the flats to build trace corrections that can be applied to the traces built from the flats as a first pass to the tracing of the science frames. This procedure is outlined in more detail by Morrison et al. in prep, which more fully describes the current pipeline.

\subsection{APOGEE Data Reduction Pipeline}\label{sec:apogee_drp}

The APOGEE Data Reduction Pipeline (DRP) converts the raw detector readouts from the APOGEE spectrographs into science-ready, wavelength-calibrated spectra suitable for subsequent stellar parameter and abundance analysis. The pipeline processes individual exposures through a sequence of calibration and reduction steps that include detector-level corrections, spectral extraction, wavelength calibration, sky subtraction, telluric correction, flux calibration, and the combination of dithered exposures into visit spectra. Multiple visits to the same target are subsequently aligned in radial velocity and co-added to produce the final high signal-to-noise stellar spectra. The DRP also derives precise radial velocities and provides a comprehensive set of quality-assurance metrics and data-quality flags for all observations. 

For SDSS-V, \citet{sdssVdr19} describes the updates to the APOGEE DRP to handle the data from the robotic fiber positioning system and Fabry-Perot interferometer calibrations and to improve other aspects of the pipeline. Because no new APOGEE data or reductions are released in DR20, the DR19 paper provides the up-to-date description of the APOGEE DRP for DR20. Further updates and improvements are being implemented in the DRP to be applied to the APOGEE data released in DR21.

\subsection{MWM Data Analysis Pipeline: \textsc{Astra}} \label{sec:astra}


\textsc{Astra} is a software package that manages and orchestrates the analysis of Milky Way Mapper spectra (Casey et al., in prep.). \textsc{Astra}
 ingests the reduced data products from the APOGEE and BOSS data 
reduction pipelines and collates various metadata (photometry, astrometry, and metrics related to observations and reductions). It then analyses those spectra through various pipelines (depending on the star and spectrum type), and
 produces science-ready data products that store stellar parameters, 
chemical abundances, line measurements, and other quantities. Because MWM has both BOSS and APOGEE spectra over a wide range of \teff, \textsc{Astra} is used to bring together analysis methods into a single software framework.

For DR20, there are two important notes. First, \textsc{Astra} was only run on the BOSS data. This means users interested in data analysis pipelines for APOGEE spectra should reference \cite{sdssVdr19} and its related data products. Second, many of the BOSS pipelines are unchanged for DR20. In the summary of the pipelines below, we will explicitly state if a BOSS data analysis pipeline has been changed or not for DR20. In Table~\ref{tab:astra_pipe}, we list the pipelines included in \textsc{Astra}, the spectra that they processed, whether the most recent data analysis is from DR19 or DR20, and if there are updates to the underlying data analysis pipeline code for DR20. Note that each pipeline was refactored at least slightly to work within the \textsc{Astra} framework.

\begin{deluxetable*}{llllll}
\label{tab:astra_pipe}
\tablecaption{List of Data Analysis Pipelines (DAPs) included in the \textsc{Astra} Framework. Each DAP includes a reference for the specific pipeline, the SDSS-V spectra is best applies to, the recommended use of the results, for which data release the DAP was most recently run, and if there have been updates to the DAP for DR20.}
\tablehead{\colhead{Pipeline} & \colhead{Ref.} & \colhead{Relevant Spectra} & \colhead{Recommended Use} & \colhead{DR} & \colhead{DR20 Pipe. Update}}
\startdata
APOGEENet (v2; v3) & S24  & all APOGEE spectra with \teff $> 20,000$K &  YSOs and stars & DR19 & \nodata  \\
\hline
ASPCAP & GP16 & all APOGEE co-added spectra & Stellar parameters \teff$<20,000$K; & DR19 & \nodata \\
& & & elemental abundances \teff$< 7000 $K & & \\
\hline
FERRE & AP23 & all APOGEE co-added spectra & & \nodata\\
&& (through ASPCAP) \\
\hline
The Payne & T19 & all APOGEE spectra &  & DR19 & \nodata \\
\hline
AstroNN & L19 & all APOGEE spectra & Stellar parameters and element abundances & DR19 & \nodata \\
\hline
AstroNNdist & ... & all APOGEE co-added spectra & Spectrophotometric distances & DR19 & \nodata \\
\hline
BOSSNet & S24 & BOSS spectra from MWM cartons &  Stellar parameters for OBAFGK stars & DR20 & No \\
\hline
MDwarfType & G26; C26 &  BOSS spectra from MWM cartons& Spectral Type from K5.0 to M8.5;  & DR20 & Yes \\
&  &  & Metallicity class from usdM to dM & & \\
\hline
SLAM & Q26 & BOSS spectra of dwarfs& Stellar parameters for M dwarfs  & DR20 & No \\
 & & & with [Fe/H] $> -0.6$ & & \\
\hline
SnowWhite & GF26 & BOSS Spectra of  White Dwarfs& Classification; Stellar parameters for DA WDs & DR20 & Yes \\
\hline 
corv & C23 & BOSS Spectra of White Dwarfs & Radial Velocities for DA WDs  & DR20 & No \\
\hline
LineForest & S24  & BOSS spectra from MWM cartons& Line Indices for H, Li, Ca, and other lines,   & DR20 & No \\
& &  & particularly for YSOs & & \\
\hline
CLAM & C26b & all APOGEE and BOSS spectra & Continuum normalization & DR20 & No \\
\hline
\enddata
\tablecomments{
References: S24 \citep{BOSSNET2024AJ....167..173S}, GP16 \citep{ASPCAP.2016AJ....151..144G}, AP23 \citep{Allende-Prieto2023}, T19 \citep{ThePayne2019ApJ...879...69T}, N15 \citep{Ness2015}, L19 \citep{AstroNN2019MNRAS.483.3255L}, C26 (Casey et al., in prep.), G26 \citep{Galligan2026}, Q26 \citep{slam2025}, GF26 (Gentile Fusillo et al., in prep.), C23 \citep{corv}, S24 \citep{LineForest2024AJ....167..125S}, C26b \citep{Casey2026}
}
\end{deluxetable*}

Below we provide a brief summary of these pipelines for the reader and, importantly, identify any changes that may have occurred to them between DR19 and DR20. For additional details on these pipelines, please reference \citet{sdssVdr19}. 
\begin{itemize}

    \item \textbf{BOSSNet:} BOSSNet is the primary general purpose pipeline to extract stellar parameters ($T_{\rm eff}$, $\log g$, [Fe/H]) for objects across the entire HR diagram, between $1,700<T_{\rm eff}<100,000$ K, and $0<\log g<10$. See \citet{BOSSNET2024AJ....167..173S} for full description. The code is unchanged for DR20.
    
    \item \textbf{MDwarfType:} The MDwarfType pipeline \citep[Casey et al., in prep.;][]{Galligan2026} compares spectra against a set of semi-empirical M-dwarf template spectra and estimates the spectral subtype from K5.0 to M8.5, evaluated to a half-subtype, and a “metallicity class” with a half-integer value from 0.5 to 12.5, from most metal-rich to most metal-poor. For DR20, the code has been changed to improve the normalization of the spectra. Specifically, both the observed and template spectra are now rectified using a second-degree polynomial.

    \item \textbf{SLAM:} Described in \cite{slam2025}, this pipeline trains the Stellar LAbel Machine (SLAM) on APOGEENet, ASPCAP and LAMOST parameters for M dwarfs in wide binaries with F, G, K primaries. The SLAM provides \teff, \logg, [Fe/H], and [$\alpha$/Fe] for M dwarfs with BOSS spectra for stars with [Fe/H] $> -0.6$. The code is unchanged for DR20.

    \item \textbf{{\tt corv}:} The {\tt corv} pipeline \citep{corv} is optimized for measuring RVs of Hydrogen rich (DA) white dwarfs by linearly interpolating between the Montreal models of DA–type white dwarfs to forward model white dwarf spectra. The code is unchanged for DR20.

    \item \textbf{SnowWhite:} SnowWhite (Gentile Fusillo, in prep.) is a pipeline specifically designed for analyzing BOSS spectra of white dwarfs. It consists of both a classifier that reports the type of white dwarf based on line indices and, for DA white dwarfs, a parameter estimator. Stellar parameters are based on model spectrum fits around hydrogen lines, using Gaia astrometry and photometry todifferentiate between potentially multimodal solutions. For DR20, the classifier has been retrained by adding an additional $\sim16,000$ white dwarfs from DR19. Additionally, the fitting for stellar parameters now depends on an expanded, more uniform grid that covers a wider range of parameter space. Also, the Gaia $B_P-R_P$ color is now being used to better differentiate between cold and hot white dwarfs.

    \item \textbf{LineForest:} The LineForest pipeline extracts equivalent widths of various lines in BOSS spectra, including all of the accessible H lines in Balmer and Paschen series, various Ca lines, He lines, Li I, and others. See \citet{LineForest2024AJ....167..125S} for full description. The code is unchanged for DR20.

    \item \textbf{CLAM:} The CLAM pipeline uses a constrained linear absorption model to simultaneously fit stellar absorption and continuum. This is used to  continuum-normalize all APOGEE spectra (co-adds and visits) and BOSS spectra (co-adds and visits). The method is similar to \citet{Casey2026}, and will be described in more detail in Casey et al. (in prep). In DR20, a variant of the CLAM that maps stellar labels to basis weights is introduced through a VAC, which is described in Medan et al. (in prep) and Section~\ref{vac:clam_parameters}.

\end{itemize}

 As stated above, no new APOGEE data is included in DR20. To work with the \textsc{Astra} parameters for APOGEE spectra, users should reference \citet{sdssVdr19} for a full description of these pipelines. We include a brief summary of the additional DR19 APOGEE spectra pipelines here for the reader.
 \begin{itemize}
     \item \textbf{APOGEENet:} APOGEENet \citep{Olney2020AJ....159..182O} is a convolutional neural network that has been updated with an expanded training set, adding hot stars to the sample of cooler dwarfs, red giants, and YSOs \citep{Sprague2022AJ....163..152S}. APOGEENet is the only APOGEE pipeline that incorporates YSOs into its training or synthetic spectra, where other pipelines (e.g.,~ASPCAP) likely give inaccurate parameters for this class of objects \citep[e.g.,][]{INSYNC2014ApJ...794..125C}.
     \item \textbf{AstroNN:} AstroNN \citep{AstroNN2019MNRAS.483.3255L} uses a deep-learning artificial neural network to determine stellar parameters (\teff, \logg, and metallicity) and 18 individual element abundances with associated uncertainty from a rest-frame resampled APOGEE spectrum. AstroNN was updated specifically for DR19 to be retrained on DR17 ASPCAP results \citep{Abdurrouf_2022}. AstroNN is complemented by AstroNNDist, a pipeline to estimate reddening and distances for APOGEE co-added spectra.
     \item \textbf{ThePayne:} ThePayne, \citep{ThePayne2019ApJ...879...69T}
 trained a neural net on synthetic spectra in the APOGEE wavelength 
region. We use the identical model to predict stellar parameters and 
abundances for 25 elements between 3000 K $<$ \teff $<$ 8000 K, $-1.5 <$ [Fe/H] $< +0.5 $ and $-1.0 <$ [X/Fe] $< +1.2$ from MWM APOGEE spectra. 
    \item \textbf{ASPCAP:}  \textsc{Astra} included a new version of ASPCAP for DR19 \citep[][Casey et al., in prep.]{ASPCAP.2016AJ....151..144G}. A description of the changes from DR17 can be found in \cite{sdssVdr19}.
    \item \textbf{FERRE:} ASPCAP is a sequence of recipes to execute FERRE \citep{Allende-Prieto2023}, which interpolates from multi-dimensional grids of stellar spectra. FERRE itself is unchanged in DR20.
 \end{itemize}

\subsection{LVM Data Reduction Pipeline}\label{sec:lvm_drp}
\label{sec:lvmdrp}
The LVM Data Reduction Pipeline
(\texttt{LVM-DRP};\footnote{\url{https://github.com/sdss/lvmdrp}}
Mej\'ia-Narv\'aez et al., in prep.) is the software system responsible
for transforming the raw detector output from the LVM instrument software into
fully calibrated, science-ready spectra.  It is built upon a collection
of routines derived from the \texttt{Py3D} package
\citep{husemann2013}, originally developed for the CALIFA survey
\citep{sanchez2006}, and has been re-implemented in \texttt{Python~3.11} to
meet the specific requirements of the LVM data format and observing
strategy. The LVM DR20 data products described in this paper were reduced
with version~\texttt{1.2.0} of the pipeline (released in November 2025).

Version~\texttt{1.2.0} implemented a series of important improvements since DR19 (version~\texttt{1.1.1}) that increased the accuracy of the spectral extraction and all downstream reduction stages, namely: the implementation of a physically motivated fiber profile model represented by the convolution of a rounded top-hat with a Gaussian profile to capture the instrument PSF; fiber flat fields to account for spectrograph shutter timing effects and; a template-matching flux calibration based on synthetic stellar spectra.

The \texttt{LVM-DRP} consists of two pipelines: a calibrations pipeline responsible for producing the calibration frames needed to reduce a science exposure; and a science pipeline which main task is to produce science-ready spectra. Both pipelines will be detailed in Mej\'ia-Narv\'aez et al., in prep. In the following we briefly describe the science pipeline.

\subsubsection{Pipeline overview and processing stages}
\label{sec:lvmdrp:stages}

The science pipeline reduces the raw data provided by the LVM Instrument (LVM-I)
described in detail in \citet{Konidaris2024} and \citet{Herbst2024},
and briefly outlined in Sec. \ref{sec:LVMstatus}. The instrument consists of
four telescopes, each equipped with a IFU: (i) a science one (T1, 1801 fibers), (ii) two sky-monitoring telescopes (T2 and T3, 59 and 60 fibers respectively) pointing toward the east and
west sky, and (iii) a fourth telescope (T4, 24 fibers) that simultaneously observes spectrophotometric standard stars. These IFUs feed a set of three DESI-like bench spectrographs. In turn, each spectrograph splits the
incoming beam into three channels via a set of dichroics: a blue channel
(3600--5800\,\AA), a red channel (5750--7570\,\AA), and a near infrared
channel (7520--9800\,\AA), together covering the full optical range at
a spectral resolving power of $R\sim4000$ at H$\alpha$.  This design allows
contemporaneous science, sky, and standard star observations within
every science exposure, all necessary for a proper reduction of the observed
frame.

The \texttt{LVM-DRP} processes each science exposure through the following
sequence of well-defined steps, propagating uncertainties and bad pixels at every
stage.

\begin{itemize}

\item \textbf{Pre-processing:}  Raw detector frames are overscan-trimmed and subtracted, and a static bad-pixel mask is applied.

\item \textbf{Detrending:}  The remaining bias structure left by the overscan is subtracted at this step, pixel-level flat fielding is applied, the data are converted to units of electrons and Poisson uncertainties are computed. Detector artifacts (e.g., cosmic rays) identified in the detrended frame are flagged and masked prior to further processing.

\item \textbf{Stray-light subtraction:}  A two-dimensional model of the scattered light field is constructed and subtracted from each detector frame, using inter-fiber block regions to constrain the smooth background. 

\item \textbf{Astrometric solution:}  On-sky fiber positions (right ascension and declination for each fiber in the system) are computed from guider-camera data and stored in the \texttt{SLITMAP} extension, enabling direct sky-plane reconstruction without further calibration.  Heliocentric velocity corrections are also added to the frame headers without being applied.

\item \textbf{Spectral extraction:}  Before the extraction, the pipeline calculates zero-point corrections due to thermal fiber shifts in the cross-dispersion direction. Then, individual fiber spectra are extracted using a profile-fitting algorithm that employs a physically motivated fiber profile model. The extraction integrates over detector pixels to minimize resampling noise and systematics. The extracted spectra from all three spectrographs are then row-stacked to produce a single multi-fiber frame per spectrograph channel.

\item \textbf{Wavelength calibration:}  For each fiber, pixel-to-wavelength mapping and line spread function (LSF) as a function of wavelength are added as FITS image extensions, both of which were derived beforehand by the calibrations pipeline from a series of arc-lamp exposures. The wavelength solution is subsequently refined for each science frame by matching the positions of sky emission lines, thereby correcting for intra-night thermal shifts of the spectrographs in the dispersion direction.

\item \textbf{Fiber-flat correction:}  The relative fiber-to-fiber throughput variations are measured by the calibrations pipeline using a sequence twilight sky flat exposures.  Starting from version~\texttt{1.2.0}, a dedicated algorithm accounts for shutter-timing effects that can introduce illumination discontinuities across the IFU, substantially improving the uniformity of the resulting flat-fielded frame.

\item \textbf{Wavelength resampling:}  Each fiber spectrum is resampled onto a common, linear wavelength grid with a sampling of $0.5$\,\AA\ per pixel, using a simple rebinning code that conserves flux and minimizes interpolation artifacts.

\item \textbf{Flux calibration:} Sensitivity curves are derived from observations of spectrophotometric standard stars obtained simultaneously with the science exposure via telescope~T4. Since version~\texttt{1.1.0}, additional flux-calibration constraints are obtained from field stars serendipitously falling within the IFU footprint.  The version~\texttt{1.2.0} adopts a template-matching approach for improved accuracy.  Calibrated spectra are delivered in physical units of erg\,s$^{-1}$\,cm$^{-2}$\,\AA$^{-1}$\,fiber$^{-1}$.

\item \textbf{Channel combination:}  The three overlapping spectral channels (blue, red, and infrared) are stitched together into a single, continuous spectrum covering the full wavelength range 3600--9800\,\AA\ for every
  fiber.

\item \textbf{Sky subtraction:}  The night-sky background is removed using data collected by the sky telescopes T2 and T3. For every science exposure, one sky telescope is directed at a relatively dark patch of the sky within $10^{\circ}$ separation from the science field. The other sky telescope is pointed at a dark region close to the Galactic Pole, to minimize ``contamination'' from the Milky Way. Sky spectra from both telescopes are averaged across fibers and combined into a single sky spectrum. Then, both line and continuum sky components are treated separately for both telescopes. A combined sky spectrum is constructed by combining the sky lines from the far sky field and the continuum contribution from the near sky field. Finally, this spectrum is subtracted from the flux-calibrated science spectra.
\end{itemize}

Each pipeline run yields a set of FITS files stored in
the Science Archive Server (SAS). The primary science product is the flux-calibrated sky-subtracted row-stacked spectrum
file \texttt{lvmSFrame-\textit{EXPNUM}.fits} (\texttt{EXPNUM} being the exposure number), where each row
corresponds to an individual fiber from either telescope. The format of this file
is described in Appendix \ref{appendix:LVM_DRP}.

\subsection{LVM Data Analysis Pipeline}
\label{sec:lvmdap}

The LVM Data Analysis Pipeline \citep[LVM-DAP;][]{lvmdap}\footnote{\url{https://github.com/sdss/lvmdap}} is designed to extract the physical properties of the ionized gas and stellar components from the reduced spectra produced by the \texttt{LVM-DRP}. It builds upon the analysis framework developed for previous integral field spectroscopy surveys, and in particular on the \texttt{pyFIT3D} fitting package \citep{pypipe3d}, adapting it to the specific characteristics of LVM data, and in particular, the wide range of spatial scales that are sampled.

The analysis is performed on a fiber-by-fiber basis, treating each spectrum independently (in the current implementation). The pipeline follows a multi-stage procedure that combines both parametric and non-parametric approaches to characterize the stellar continuum and the emission line properties. 

First, the non-linear parameters of the stellar component are estimated, including the systemic velocity, velocity dispersion, and dust attenuation. This is achieved through an iterative fitting procedure in which emission lines are masked (in the 1st iteration), and the observed spectrum modeled using a linear combination of a limited set of stellar templates. Then it is constructed a refined model of the stellar continuum through a linear combination of a much larger set of templates, which are shifted, broadened, and  attenuated by the non-linear parameters derived in the first step.

A particularly important difference of the LVM-DAP with respect to previous similar analyses in the literature \citep[e.g.][]{sanchez22} is the nature of the adopted stellar templates. They consist of a set of spectra that minimizes the degeneracies in the observational space (i.e., the spectra themselves) while retaining the maximum information in the physical space (\teff, $\log(g)$, [Fe/H], [$\alpha$/Fe]) characterized by a probability distribution function (PDF). We define this particular type of stellar templates, that can be used to model from one single star to a complete stellar population, as Resolved Stellar Population \citep[RSP;][]{lvmdap} in contrast to the well known Single/Synthetic Stellar Population concept \citep[SSP;][]{tinsley72,bc03,conroy13}. 

Once the stellar component is modeled, it is subtracted from the observed spectrum to produce a gas-pure spectrum. This residual spectrum is used to derive the properties of the ionized gas emission lines. The analysis proceeds in two main steps. First, a parametric fit is performed on the strongest emission lines to estimate their integrated fluxes, velocities, and velocity dispersions. This step uses a combination of pseudo-random exploration and non-linear optimization techniques implemented within \texttt{pyFIT3D}. In the current implementation each emission line is modeled with a single Gaussian function. The results of this initial fit are then refined through a second optimization stage that improves the accuracy and stability of the derived parameters. The emission line model could be removed from the original spectrum to refine the stellar population model, in a second iteration of the procedure described above, in which the fitted emission lines are not masked anymore.

In addition to this evaluation of the emission line properties using a parametric modeling, the pipeline also estimates the properties of the emission lines without assuming any particular model, based on a weight-moment analysis of a large set of emission lines. These measurements provide a complementary characterization of the emission line properties, particularly in cases where the line profiles deviate from simple Gaussian shapes. Beside that, as it is a much faster procedure, it allow us to explore a much larger number of emission lines. We will evaluate in the future if it is required to perform both analyses on the same set of emission lines.

The final data products of the LVM-DAP include, for each fiber, the properties of the stellar population (e.g., kinematics, dust attenuation, average stellar properties, like \teff), as well as a set of emission line properties, including fluxes, kinematics, and equivalent widths, and their corresponding errors. These outputs are organized in a single FITS file per exposure/frame with multiple extensions comprising a set of tables with the results of each stage of the analysis. The format was described in detail in \citet{lvmdap} and \citet{HelixDR19}. We include a brief summary in Appendix \ref{appendix:LVM_DAP}.

The LVM-DAP has a modular structure allowing for continuous improvements and extensions, including the incorporation of additional spectral diagnostics, modifications of one particular analysis (for instance the stellar kinematics) and the optimization of the analysis for large datasets (distributing the analysis between different cores or even computers). Its application to the LVM survey enables the systematic characterization of the physical conditions of the ionized interstellar medium across large areas of the sky. For the 
strong emission lines it has been proved to be very stable, validated with a large set of simulations \citep[][]{lvmdap} and external comparisons \citep{HelixDR19,Singh2026}, however there are several improvements that are still required, and they will be presented in future data releases.

\section{Black Hole Mapper}\label{sec:bhm}

Earlier introductions to the Black Hole Mapper (BHM) program in SDSS-V may be found in relevant sections of the recent SDSS-V overview paper \citep{kollmeier2026}, as well as in the SDSS-V data release papers for both DR18 \citep{almeida2023} and DR19 \citep{sdssVdr19}. 
As a high-level summary, BHM in SDSS-V emphasizes especially the study of quasars and other active galactic nuclei (AGN), as among Universe’s most luminous objects, powered by accretion onto supermassive black holes (SMBHs). BHM exploits – with order(s) of magnitude advances – two hallmark characteristics of quasars: their marked variability on a range of timescales, and prodigious luminosity extending to X-rays. By the conclusion of the SDSS-V survey, the ultimate core components of BHM (there are additional ancillary components) are aimed to provide:

\begin{itemize}
\item Repeat time-domain optical (BOSS) spectra of $\sim 10^{4.5}$ known quasars over a broad range of timespans from days to decades, sampling changes on light-travel, dynamical, thermal, etc. timescales,  in order to: (i) measure BH masses via the core Reverberation Mapping (RM; Section \ref{sec:bhm_rm}) from high-cadence and intensive repeat spectroscopy in selected fields; and (ii) from its broader sky coverage but more modest cadence repeat spectroscopy program termed AQMES (All-Quasar Multi-Epoch Spectroscopy; Section \ref{sec:bhm_aqmes}), study broad line region (BLR) dynamics, capture and discover accretion events such as those in changing look quasars; and understand the astrophysics of quasar accretion and outflows. Such time-domain approaches probe spatially unresolved size scales.

\item Optical (BOSS) follow-up spectroscopy of $\sim 10^{5.5}$ eROSITA \citep{Merloni2024} X-ray source counterparts, via the BHM core program SPIDERS (SPectroscopic IDentification of ERosita Sources; see Section \ref{sec:bhm_spiders})
spectra providing counterpart identifications and redshifts, to study the demographics, evolution, and astrophysics of X-ray sources. These counterparts are mainly quasars/AGN, but also include X-ray emitting galaxy clusters and stars.
\end{itemize}

DR20 includes the release of $\sim$1.1 million BOSS spectra for $\sim$0.5 million
distinct BHM objects (a 3-4$\times$  expansion vs. DR19, and a more than an order of magnitude expansion vs. DR18), collected using both the Sloan Foundation 2.5m Telescope at APO in New Mexico, and in DR20 also from the du Pont 2.5m Telescope at LCO in Chile. These DR20 data extend across nearly the full suite of BHM core programs as well as multiple ancillary programs, and emphasize both
northern/equatorial fields (from APO) and also the first SDSS-V/BHM public release
of BHM data in more-southern fields (from LCO).

\begin{figure}
      \centering
      \includegraphics[width=\linewidth]{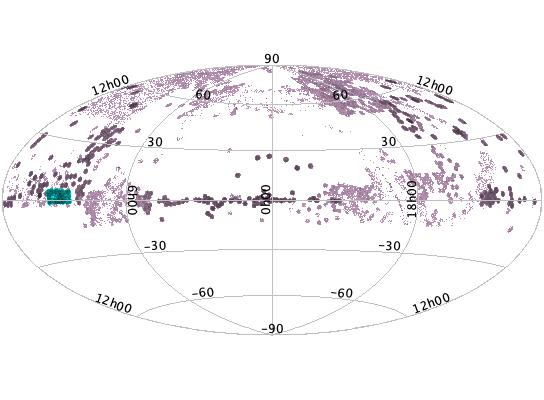}
      \vskip -1.1cm
        \includegraphics[width=\linewidth]{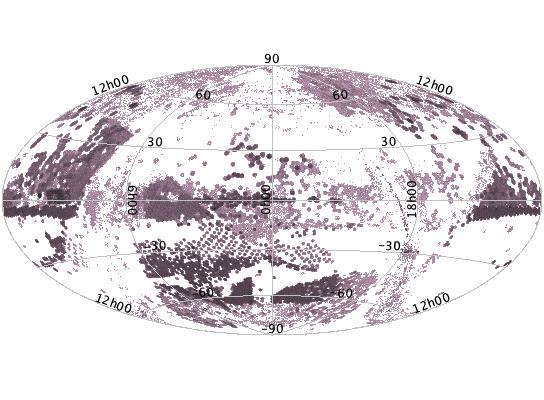}
       \caption{Expansion of BHM BOSS spectra and sky coverage from DR18/19 (upper panel) to DR20 (lower panel). Both panels show the distribution on the sky in (RA, Dec) of SDSS-V/BOSS optical science spectroscopic targets. For comparison, in the upper panel, the small and approximately rectangular equatorial region near RA=9h (blue-green) depicts the $\sim$24k spectra for $\sim$12k distinct science targets from the 140 sq. deg. eFEDS area, which comprised the initial main SDSS-V DR18 spectra release \citep{almeida2023}. This upper panel also highlights the dramatic expansion of BHM/BOSS data taken at APO included in DR19 (violet). The lower panel highlights the further expansion in BHM/BOSS science spectroscopy and sky coverage in DR20 (violet). In DR20, there are $\sim$1 million BHM/BOSS science released for $\sim$0.5 million distinct objects, taken from both APO and LCO, and with abundant spectra for BHM core science programs AQMES and RM (which are programs done mainly from APO), SPIDERS (done mainly from LCO), and as well as a variety of BHM-connected ancillary programs.
       }
      \label{fig:bhmdr20}
\end{figure}

Figure \ref{fig:bhmdr20} demonstrates how the BHM survey area, number of targets, and number of BOSS optical spectra taken have markedly expanded since the start of the SDSS-V Survey. The initial SDSS-V DR18 \citep{almeida2023} included $\sim$24k BOSS optical science spectra for $\sim$12k distinct targets, most from the small 140 deg$^2$ eFEDS \citep{Brunner2022} X-ray mini-survey region (see the small, nearly rectangular patch, near RA=9h and at a near-equatorial Dec in Figure \ref{fig:bhmdr20} upper panel, color-coded in blue-green).  The upper panel of Figure \ref{fig:bhmdr20} highlights the rapid expansion of BOSS data from DR18 to DR19 (latter highlighted in violet), with BOSS plate and FPS data from APO. This includes $\sim$380k science spectra for $\sim$115k distinct targets, many of which have multi-epoch observations, that were primarily selected for spectroscopic targeting by BHM. DR19 provided expanded emphasis for time-domain BHM spectra (e.g., from RM and AQMES, and ancillary programs) from APO in the North. The lower panel of Figure \ref{fig:bhmdr20} depicts (in violet) the expansion in BOSS optical science spectroscopy from BHM included in DR20, encompassing data spanning wide sky areas at both APO and LCO. This includes $\sim$1M (daily-coadd) spectra for $\sim$0.5M distinct BHM-led objects, with the most marked expansion from both DR18 and DR19 in the hundreds of thousands of BHM identification/redshift spectra for the SPIDERS program---an order of magnitude expansion over eFEDS from DR18 and the largest such homogeneous X-ray/optical sample yet.

\begin{table*}
\centering
\caption{BHM: Approximate Numbers of Science Targets Observed \& BOSS Spectra Released (DR18 - DR20)}
\label{tab:bhm_sum}
\begin{tabular}{l c c c}

\hline
\multirow{2}{*}{BHM Program} & DR18  & DR19  & DR20 \\
 & targets  (spectra)  & targets  (spectra)  & targets (spectra) \\
\hline
\multicolumn{4}{l}{SPIDERS/eFEDS (eROSITA sources)}\\
{\hspace{6pt}} SPIDERS AGN & {\it 9.0k (18k)} & {\it 20k (39k)} & {\it 184k (310k)} \\
{\hspace{6pt}} SPIDERS clusters & {\it 3.2k (6.1k)} & {\it 4.6k (8.7k)} & {\it 28k (46k)} \\
{\hspace{6pt}} SPIDERS commissioning  & & {\it 0.4k (1.4k)} & {\it 1.0k (2.5k)} \\
SPIDERS/eFEDS (sub-) total & 12k   (24k) & 25k   (49k) & 213k   (358k)\\
\hline
\multicolumn{4}{l}{AQMES (Quasar monitoring)}\\
{\hspace{6pt}} AQMES-Wide core & & {\it 6.8k (19k)} & {\it 13k (31k)} \\
{\hspace{6pt}} AQMES-Medium core & & {\it 2.3k (16k)} & {\it 2.5k (25k)} \\
{\hspace{6pt}} AQMES other & & {\it 32k (100k)} & {\it 47k (142k)} \\
AQMES (sub-) total &  & 41k   (135k) & 63k   (198k)\\
\hline
\multicolumn{4}{l}{RM (Reverberation Mapping)}\\

{\hspace{6pt}} RM intensively monitored \texttt{}& & {\it 1.1k (79k)} & {\it 1.7k (152k)} \\
{\hspace{6pt}} RM other \texttt{}& & {\it 1.1k (31k)} & {\it 1.7k (32k)} \\
RM (sub-) total &  & 2.2k  (110k) & 3.4k   (184k)\\
\hline
\multicolumn{4}{l}{BHM other (ancillary, related)}\\
{\hspace{6pt}} CSC (Chandra Source Catalog) \texttt{} & & {\it 6.8k (19k)} & {\it 16k (37k)} \\
{\hspace{6pt}} Gaia-unWISE AGN, \& bright galaxies \texttt{} & & {\it 33k (59k)} & {\it 165k (279k)} \\
{\hspace{6pt}} Open fiber (extragalactic) \texttt{} & & {\it 7.1k (7.8k)} & {\it 45k (47k)} \\
BHM other (sub-) total &  & 47k  (86k) & 226k   (363k)\\
\hline
BHM TOTAL & 12k (24k) & 115k (380k) & 0.50M (1.1M)\\
\hline
\end{tabular}

\vspace{0.2cm}\
\noindent{\footnotesize
{\it Note:} Tabulated numbers of BHM targets in this table highlight one simple accounting approach according to the 'FIRSTCARTON' parameter, which identifies the specific SDSS-V targeting carton making the primary assignment of the spectroscopic fiber. The number of spectra tabulated here also refers specifically to 'daily' coadds, i.e., combined spectral exposures from the same night.
}
\end{table*}

Approximate, fiducial numbers of unique BHM targets observed with BOSS, and associated numbers of BOSS spectra (often multi-epoch, but here tabulated instead according to 'daily' coadds) included in SDSS-V data releases are provided in Table \ref{tab:bhm_sum}, according to their BHM primary carton ('FIRSTCARTON') targeting selection category.\footnote{For simple accounting in Table \ref{tab:bhm_sum}, we also list BHM-affiliated BOSS spectroscopic follow-up of X-ray sources from the Chandra Source Catalog (CSC), the Gaia-unWISE (GUA) AGN survey, a bright galaxy survey set derived from Legacy Survey data, and a suite of mainly-extragalactic SDSS-V 'Open fiber' programs; e.g., see Appendix \ref{sec:appendix_bhm} herein, or \citet{kollmeier2026} and references therein.}
Additional details highlighting each of the core BHM science programs---SPIDERS, AQMES, and RM--- emphasizing DR20 updates and results are provided next in the following three subsections.

\subsection{BHM: Scientific Programs}\label{sec:bhm_programs}

\subsubsection{SPIDERS: The SPectroscopic IDentification of ERosita Sources}
\label{sec:bhm_spiders}
The SPIDERS program in SDSS-V is the culmination of almost a decade of SDSS efforts in designing and implementing large systematic spectroscopic follow-up programs to characterize very wide-area X-ray surveys. Within SDSS-IV, the first generation of the SPIDERS survey was conceived in order to exploit the synergy between the SRG/eROSITA all-sky survey (set to deliver tens of X-ray sources per square degree in need of spectroscopic identification) and BOSS optical spectroscopic surveys capable of identifying hundreds of optical sources per square degree, down to a magnitude limit sufficient to classify typical eROSITA sources \citep[see e.g.][]{Merloni2012}. Delays in the launch of SRG prevented the full scope of the program to be realized during SDSS-IV. Instead, a pilot program based on ROSAT all-sky survey targets was implemented \citep{Dwelly2017,Comparat2020,Clerc2020}. With the launch of SRG in July 2019, and the start of scientific operations of the eROSITA telescope \citep{Predehl2021}, the conditions for the full implementation of the SPIDERS programmatic goals were finally met.

The SPIDERS program in BHM strives to achieve a highly complete follow-up
of X-ray sources (mainly AGN, but also including clusters of galaxies, X-ray binaries and other
X-ray active stellar targets) detected by eROSITA in its all-sky (or wide area) surveys. eROSITA targets have been split into BHM-led (mainly extragalactic: $|b| > 15^{\circ}$) and MWM-led (mainly galactic) cartons, based
on the best possible classification obtained at the time of targeting. Within BHM, the `core’ goal
of the eROSITA spectroscopic follow-up program is to obtain BOSS spectra in much of `German' hemisphere \citep[Western Galactic, see][]{Merloni2024}, which exceeds $10^4$ deg$^2$ of extragalactic sky, mainly accessible in the South from LCO, but also includes up to about $\sim$3000 deg$^2$ of high-latitude area accessible from APO. 

The targeting of eROSITA-selected sources has evolved during the SDSS-V survey. Initial wide area SPIDERS cartons \citep[described by ][]{almeida2023,sdssVdr19} were based on early reductions of eRASS1 data \citep[official reductions of eRASS1 data are presented by][]{Merloni2024,Bulbul2024}. These have now been superseded by new SPIDERS cartons based on early reductions of deeper eROSITA survey data.
Given the depth of the eROSITA X-ray catalogs, and the capabilities of SDSS 2.5 meters telescopes, the AGN component of SPIDERS primarily makes use of X-ray sources detected from eROSITA's first 18 months of survey operation (termed `eRASS:3' because it contains data from three passes over the whole sky), which were completed in June 2021. The latest reduction of the eRASS:3 dataset, and the corresponding X-ray catalogs are described and released in Ramos-Ceja et al. (submitted) as part of the second public data release of eROSITA-DE (DR2). The component of SPIDERS which follows up candidate galaxy clusters now exploits a bespoke reduction of the eRASS:4 (24 month) dataset.
 
Science-specific cross-identification strategies were followed for point-like X-ray sources \citep[mainly AGN and stars, see][]{Salvato2025} and extended ones \citep[clusters of galaxies, see][]{Kluge2024}. Section~\ref{sec:erosita_cat} describes the updated parent catalogs of eROSITA sources and their optical counterparts which we have used to inform SPIDERS target selection. All updates to SPIDERS target cartons since SDSS DR19 are described in Section~\ref{sec:bhm_targeting}. 
A more complete description of the design and status of the SPIDERS program will be presented by Merloni et al. (in prep.).

\subsubsection{AQMES: All-Quasar MultiEpoch Spectroscopy}
\label{sec:bhm_aqmes}

The primary goal of the Black Hole Mapper All-Quasar Multi-Epoch Spectroscopy (AQMES) program is to quantify quasar spectroscopic variability on time scales ranging from months to more than a decade. To that end it will ultimately combine new spectroscopic observations of $\sim 10^4$ relatively bright quasars, with $i \leq 19.1$ AB mag, with archival spectra of the same quasars taken in SDSS-I--IV. The vast majority of the targets have absolute magnitudes of $-22 > M_i > -29$ and redshifts $z\lesssim 3.5$.
The AQMES program is presented in detail in a dedicated paper (Eracleous et al., in prep.).

To put the temporal baseline in the proper perspective, we note that it is of the same order as the dynamical time of the outer broad-line region (BLR),
$t_{\rm dyn} \equiv (R^3 / GM_{\rm BH})^{1/2} \!\!
= 8.7~ R_{\rm 17}^{3/2}\, M_8^{-1/2}\;$years,
where $R=10^{17} R_{17}\;{\rm cm}$ is the distance from the central black hole and $M_{\rm BH}=10^8\,M_8\; {\rm M_\odot}$ is the mass of the central supermassive black hole. For reference, the corresponding light crossing time is $R/c = 39\;R_{17}\;$light-days and the radial distance, expressed in terms of the gravitational radius ($R_{\rm g}=GM_{\rm BH}/c^2$), is $R/R_{\rm g}=6750\;R_{17}\,M_8^{-1}$. With its cadence and by employing previous SDSS spectra, the AQMES survey can probe changes in the emission and absorption lines that are associated with dynamical changes in the corresponding gas.

AQMES has two broad scientific goals. First, it is an exploratory study of spectroscopic variability, aimed at identifying patterns on time scales on the order of the dynamical time of the BLR. Second, it systematically gathers spectra that are suitable for constructing statistical measures of spectroscopic variability that can be compared with physical models of the dynamics of the emission- or absorption-line gas. At the same time, AQMES has the potential of uncovering new variability modes or identifying large samples of quasars undergoing rare modes of variability (e.g., ``changing-look'' quasars), allowing us to better understand their nature.

To achieve its scientific goals AQMES relies on a two-tier observing strategy. In the AQMES-Medium tier, the survey aims to carry out 10 new observations of each of $\sim$2000
previously observed quasars distributed in
35-40 fields. In the AQMES-Wide tier the survey aims to carry out at least one new observation of each of approximately
$\sim$20,000 quasars distributed in 
400-500 fields; approximately 14\% of these fields are planned to be observed one more time and approximately 3\% will be observed two more times. Thus, at the end of the survey and after the new spectra are combined with archival spectra, it will be possible to probe spectroscopic variability over time intervals between approximately 60 and 6,000 days. 

Within DR20, the AQMES data include spectra for $\sim$15k unique quasars from the core AQMES samples mentioned above, adding at least one new epoch to their archival SDSS-I--IV spectroscopy.  This includes $>$1200 AQMES-Medium quasars that will have achieved (at typical per pixel $S/N>5$) at least eight new SDSS-V spectral epochs by DR20, and are advancing toward their planned minimum of 10 total new epochs by the close of the survey; and $>$12,000 AQMES-Wide quasars that will have been observed in at least one new SDSS-V epoch by DR20, with about half of these AQMES-Wide quasars observed in at least two new SDSS-V epochs by DR20.
We note that the SDSS-V DR20 includes repeated spectroscopy of many other, previously known quasars and AGN, which are however not part of the highly complete, core AQMES pool of targets. 
Together, these data have already led to the identification of many new extremely variable and/or changing-look quasars \cite[e.g.,][]{Zeltyn_etal_2022,Zeltyn_etal_2024}, with additional statistical analyses of the DR20 content of AQMES forthcoming (Zeltyn et al., in prep.). The multi-epoch spectroscopy was also used to test the AGN and host decomposition methodology employed for SPIDERS \citep{Aydar2026}.

\subsubsection{RM: Reverberation Mapping}
\label{sec:bhm_rm}

The Black Hole Mapper Reverberation Mapping (BHM-RM) program
is securing multi-epoch spectro-photometric observations of $\sim$1600 broad-line
quasars spanning a wide range of redshifts. The primary goal of this program is to measure black hole masses ($M_{\rm BH}$) in $\gtrsim$1000 quasars using the technique of reverberation mapping (RM). 
RM works as follows: Rapid variations near the supermassive black hole (SMBH) drive
reverberations that propagate outward at the speed of light, probing the surrounding accretion flow.
The observed ultraviolet and optical continuum and photo-ionized emission-line variations are correlated, with time delays between the signals interpreted as light travel times, allowing us to measure the relative sizes of the reverberating regions. Continuum lags increasing with wavelength probe the accretion disk temperature, which decreases with radial distance from the SMBH. Similarly, emission-line time lags relative to continuum variations measure the characteristic size of the photo-ionized emission-line gas in the broad line-emitting region (BLR) at larger radii. 

Once the distance to the BLR is determined via RM, one can combine measurements of the BLR line widths with sizes from time lags to obtain virial estimates for the black hole mass ($M_{\rm BH}=f\,R_{\rm BLR}\,\Delta V^2/G$, where $R_{\rm BLR}$ is the characteristic radius of the BLR measured via RM, $\Delta V$ is the velocity width of the broad emission line, and $f$ is a dimensionless factor accounting for the geometry and orientation of the BLR). RM measurements $R_{\rm BLR}$ across a wide range of redshifts require high-cadence, long-baseline spectroscopic monitoring, as the expected time delay scales with luminosity, and time dilation due to higher redshifts very quickly increases the observed time lags. 

\begin{figure}
    \centering
    \includegraphics[trim=15mm 5mm 45mm 5mm,clip,width=\linewidth]{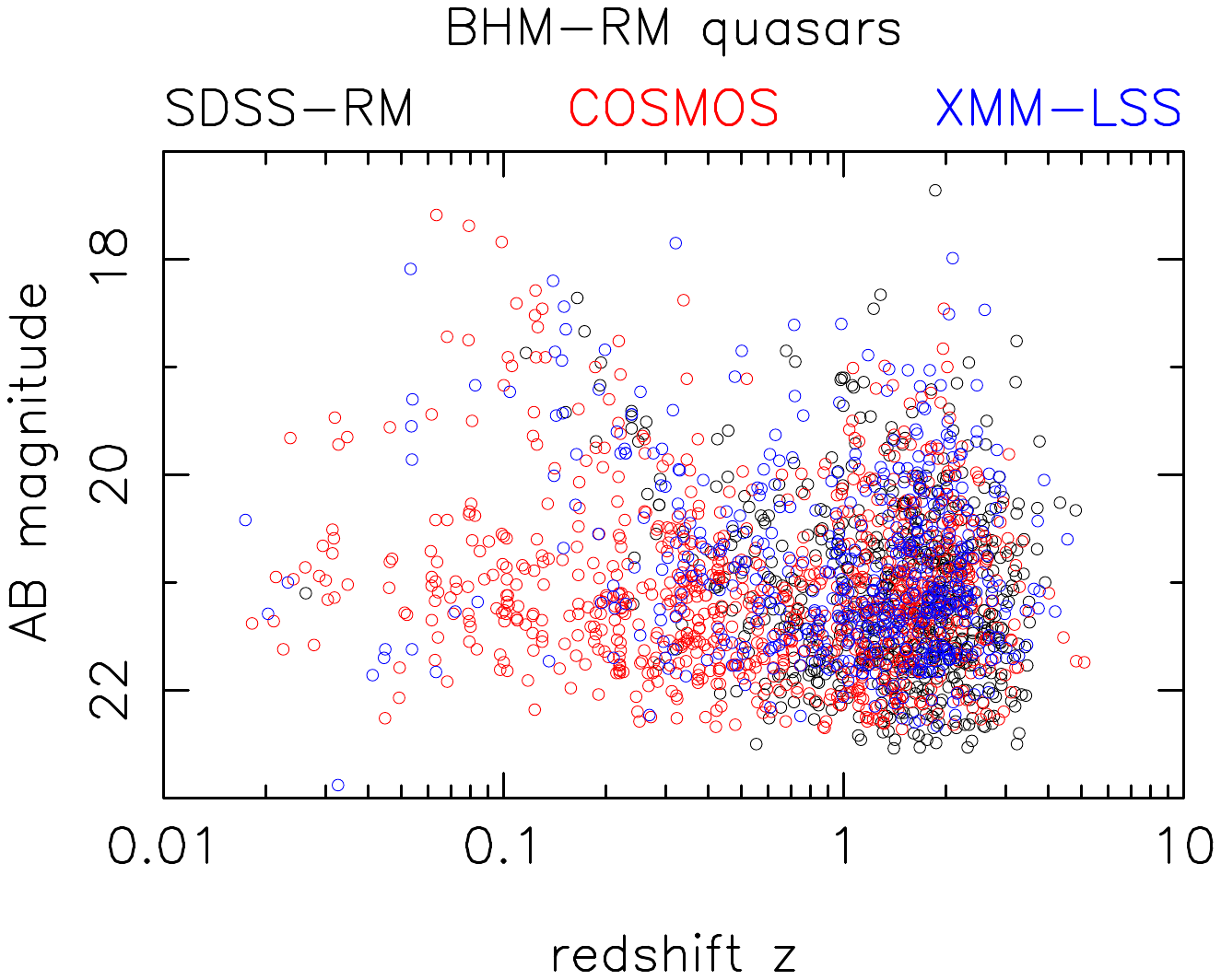}
    \caption{Redshift and $i$-band AB magnitude distribution of 
    BHM-RM quasars.}
    \label{fig:bhmq}
\end{figure}

\begin{figure}
    \centering
    \includegraphics[trim=45mm 10mm 60mm 15mm,clip,width=\linewidth]{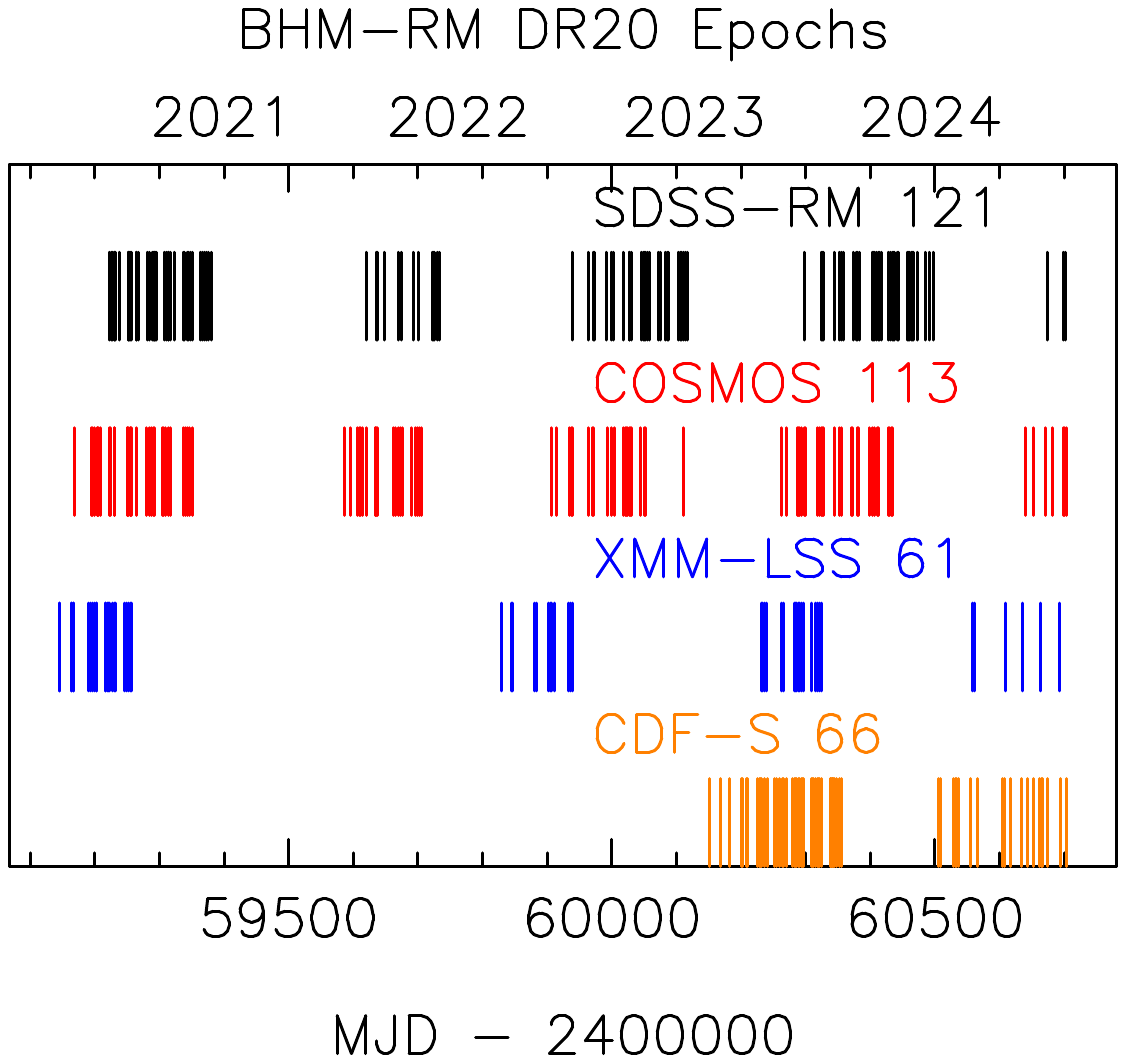}
    \caption{Time sampling and number of epochs for the
    BHM-RM quasar monitoring spectra released in DR20.
    }
    \label{fig:bhrm_epochs}
\end{figure}

The BHM-RM BOSS spectroscopy included in DR20 covers broad-line quasars in three primary BHM-RM fields observed from APO: SDSS-RM, COSMOS, and XMM-LSS, and in the CDF-S field observed from LCO. BHM-RM is monitoring up to $\sim 380$ quasars in each field, with remaining BOSS fibers allocated for calibrations. Figure~\ref{fig:bhmq} shows the redshift and magnitude distribution of the BHM-RM quasar sample. The BHM-RM spectra are taken as a series of 15-minute exposures, calibrated with standard star and sky background spectra, and co-added to achieve a uniform signal-to-noise ratio per epoch.
The co-added spectra are usually secured within a single night, but may be spread over a few nights when needed.

For DR20, the v$6\_2\_1$-reduced spectra are released, including RM observations up through MJD\,60708.
Figure~\ref{fig:bhrm_epochs} shows the number and
time distribution of observations for each BHM-RM field.
These BHM-RM spectra extend the 7-year spectroscopic baseline 
covered by the SDSS-RM precursor study of 849 broad-line quasars in the SDSS-RM field
\citep{Shen2015,Shen2024ApJS..272...26S}, 
which yielded time lags and several hundred
black hole mass estimates based on reverberation lags for
H$\alpha$ and H$\beta$ at low redshifts \citep{Grier2017ApJ...851...21G},
for Mg\,{\sc II} at intermediate redshifts \citep{Homayouni2020ApJ...901...55H},
and for C\,{\sc IV} at higher redshifts \citep{Grier2019ApJ...887...38G}.
Scaling relations anchored by these and similar RM-based mass estimates 
\citep{DallaBonta2020ApJ...903..112D, DallaBonta2025A&A...696A..48D} then provide a basis for single-epoch black hole mass estimates needed for larger samples and at higher redshifts to study black hole growth and demographics over cosmic time.

The BHM-RM quasars are somewhat brighter than those covered by the SDSS-RM pilot program, due to the shift from 1000 to 500 fibers for the robotic FPS system.  The SDSS-IV spectrographs, using fiber plug plates, achieved a typical spectro-photometric accuracy of 5\%. In comparison, the early SDSS-V FPS spectra had larger spectro-photometric errors, sometimes 10-15\%, 
due to smaller diameter optical fibers and less accurate and repeatable placement by the robotic fiber positioners compared with plug plates. The spectro-photometric accuracy can be improved for low-redshift quasars with strong narrow emission lines, such as the narrow [O\,{\sc III}] lines near H$\beta$, available as internal calibrators. At higher redshifts where suitable narrow emission lines are typically not present, comparison of synthetic photometry with light curves from photometric imaging monitoring programs \citep{Zhuang_etal_2024} should yield improvements.

RM results from the SDSS-V monitoring include measurements of unique variability in the SDSS-RM quasar RM\,160 (\citealt{Fries_etal_2023, Fries_etal_2024}), including velocity-resolved RM measurements of the H$\beta$ and H$\alpha$ broad emission lines. As this is one of our brighter sources, the data were also of sufficient quality to carry out dynamical modeling of the BLR (\citealt{Stone_etal_2024}). We also made the first RM measurements in the [Ne\,\textsc{v}]\,$\lambda$3427 coronal emission line in one of the quasars in the COSMOS monitoring field (\citealt{Smith2025ApJ...995..185S}). It takes time to build up light curves of sufficient length to measure reverberation lags in our luminous quasars, which often have expected observed lags on the order of a year or more. Current efforts to measure RM lags with the first years of BHM-RM monitoring are underway (Robinson et al. 2026, in prep; Stone et al. 2026, in prep; Davis et al. 2026, in prep) and will include measurements in all four fields covered by DR20. In addition to RM, BHM-RM monitoring has been used to investigate broad absorption-line variability in BHM-RM quasars, leading to the discovery of an accelerating quasar outflow in the quasar SDSS-RM 613 (\citealt{Wheatley_etal_2024}).

\subsection{BHM: Targeting Updates}\label{sec:bhm_targeting}

This data release includes a number of new and revised BHM target cartons, which form part of the MOS targeting system. Perhaps the most significant updates are within the SPIDERS program. Here we have moved from basing our X-ray samples on the first 6 months of eROSITA sky survey data (eRASS1), to using X-ray source catalogs derived from 18 months of survey data (eRASS:3), or in some specific cases on even deeper X-ray data (eROSITA galaxy clusters, South Ecliptic Pole region). The increased X-ray survey depth results in a higher sky density of SPIDERS targets. In parallel, improvements to the main supporting optical/IR dataset (LegacySurvey DR10 catalog, see section \ref{sec:legacy_survey_dr10}) increases the size and uniformity of the extragalactic footprint available to the SPIDERS project, particularly in the South. The updated parent samples from which these new SPIDERS target cartons are drawn are described in Section \ref{sec:erosita_cat}. 

The primary carton which enables the wide-area SPIDERS AGN program (\texttt{bhm\_spiders\_agn\_lsdr10}) is based on the combination of eRASS:3 (18 months of eROSITA survey) and LegacySurvey DR10. The counterparts identification follows the approach of \citet{Salvato2025}, but this new carton 
contains more than four times as many targets as its predecessor (\texttt{bhm\_spiders\_agn\_lsdr8}).
The wide area coverage of LegacySurvey DR10 means that we were able to discontinue the previous SPIDERS AGN cartons 
that relied on the Pan-STARRS1, SkyMapper and SuperCosmos surveys, leading to a much more uniform sample. However, we have continued to use the Gaia catalog (in this case Gaia DR3) to provide supplementary optical coverage outside the LegacySurvey footprint. The resulting carton (\texttt{bhm\_spiders\_agn\_gaiadr3}) expands the brighter end of the SPIDERS AGN sample down to low Galactic latitudes.
We have further exploited the eROSITA dataset to identify a small number of additional cartons that expand the scope of the SPIDERS AGN program. The \texttt{bhm\_spiders\_agn\_hard} carton contains high-confidence hard X-ray sources selected from the 2.3--5\,keV band (eRASS:1 depth). The \texttt{bhm\_spiders\_agn\_sep} carton is based on a highly tuned reduction of the deepest eROSITA survey region surrounding the South Ecliptic Pole (SEP), where the high X-ray source density leads to some source confusion. The \texttt{bhm\_spiders\_agn\_tda} carton
contains a sample of eROSITA sources that exhibited significant X-ray variability across several independent seasons of observation. 

The SPIDERS Clusters program is implemented via the new \texttt{bhm\_spiders\_clusters\_lsdr10} carton, which is based on eRASS:4 X-ray data (24 months of eROSITA sky survey). Cluster member galaxies are identified via the `eROMAPPER' algorithm, as described in \citet{Kluge2024}. The inclusion of $i$-band information in the LegacySurvey DR10 catalog has improved the ability to identify spatially clustered red sequence galaxies. The \texttt{bhm\_spiders\_clusters\_lsdr10} carton contains more than five times as many target galaxies as its predecessor (\texttt{bhm\_spiders\_clusters\_lsdr8}).

In addition, we have made incremental improvements to many BHM target cartons, as new data sets and better informed selection criteria have become available. 
We have updated the RM target cartons to take account of spectroscopy that has clarified the true nature of a number of candidate QSOs.  Confirmed QSOs have been promoted to the \texttt{bhm\_rm\_known\_spec} carton, whereas a small number of targets have been excluded from further consideration (e.g. where spectroscopy has revealed them to be stars or to be compact blue star-forming galaxies). 
In addition, we introduced a new RM carton \texttt{bhm\_rm\_xrayqso} which includes a small number of candidate QSOs that were identified via X-ray detection and SED modeling \citep{Ni2021}.

The BHM cartons which target Chandra X-ray sources have been updated to take account of a new release of The Chandra Source Catalog \citep[CSC v2.1, ][]{Evans_2024}. The CSC\,v2.1 release includes additional X-ray data and improved reduction methods. The process whereby CSC\,v2.1 X-ray sources are matched to optical counterparts has been updated to include the LegacySurvey\,DR10 and Gaia\,DR3 catalogs, in addition to the PanSTARRS1 and 2MASS catalogs that were used in earlier iterations (see Section \ref{sec:csc_v2p1}).

For AQMES, previous cartons have been carried over to use the new \vonex, and we have added new cartons \texttt{bhm\_aqmes\_wide1}, \texttt{bhm\_aqmes\_wide1\_faint} which request a single epoch of SDSS-V spectroscopy for QSOs ($16< i_{\mathrm{psf}} < 19.1$ and $19.1< i_{\mathrm{psf}} < 21$ respectively) taken from SDSS DR16Q \citep{Lyke2020}. This new carton allows \texttt{robostrategy} to consider a shallower time allocation (one epoch vs two epochs) in AQMES `wide' sky areas, giving it greater flexibility to balance the demands for, and availability of, dark time across LSTs. 

As described in Section \ref{sec:targeting}, we made substantial changes to the survey allocation plan (allocation of survey time to different parts of the sky), including a reduction of the typical exposure time per field over much of the high Galactic latitude sky. In order to match the per target time-requests (cadences) to the available time resources (per-field allocations), we have created a set of parallel BHM cartons containing the same set of targets, but having shallower cadence requests. These new cartons are given the suffix `\texttt{\_d3}', reflecting that the targets can nominally be satisfied if assigned a fiber in three designs (in contrast to the four designs previously typically requested by the faintest targets). The target priorities in these cartons are organized so that, if and when a field allocation permits it, we prefer to assign a fiber to a target for all four designs, but with a fall-back option of three designs otherwise.  For targets receiving the new shallower cadences, the main impact will be to reduce the fraction of observed sources from which we are able to successfully extract a redshift, particularly at the faint end. However, this is offset by the much larger sky area that we expect to cover by surveying wider and shallower.  

We have modified the selection criteria for the \texttt{bhm\_gua\_*} cartons to use apparent Gaia magnitudes, rather than the de-reddened magnitudes that were erroneously used in earlier iterations. This has mainly had the effect of moving some targets from the \texttt{bhm\_gua\_bright} carton to the \texttt{bhm\_gua\_dark} carton.

During 2024, members of the SDSS-V team submitted a number of additional openfiber cartons which focus on extragalactic science themes. These programs seek to obtain obtain optical spectroscopy for i) several large samples of AGN/QSOs identified via their photometric/astrometric/variability properties in the optical, infrared, X-ray and/or radio, ii) samples of carefully selected low redshift galaxies that are expected to e.g. have a higher than average rate tidal disruption events, or to have characteristics reminiscent of Little Red Dots, and iii) a sample of blazar candidates from the literature. 

Please see Appendix \ref{sec:appendix_bhm} for a detailed description of all new BHM (and extragalactic openfiber) target cartons released as part of DR20.

\section{Milky Way Mapper}\label{sec:mwm}

The Milky Way Mapper (MWM) is one of the flagship surveys of SDSS-V, designed to deliver a comprehensive, multi-dimensional view of the stellar populations of the Milky Way. By combining wide-area, multi-epoch spectroscopy with Gaia astrometry, MWM provides precise radial velocities, stellar parameters, and chemical abundances for millions of stars across the full sky. The survey spans a broad range of Galactic environments---from the inner disk to the distant halo---and targets stars covering a wide range of masses, ages, and evolutionary stages. This approach enables a coherent reconstruction of the structure, formation history, and dynamical evolution of the Galaxy.

A defining feature of MWM is its systematic and homogeneous coverage of the sky, coupled with a strategy that leverages both bright-time and dark-time observing. The program utilizes the APOGEE and BOSS spectrographs to efficiently sample complementary stellar populations, with APOGEE focusing on brighter targets and infrared observations, and BOSS extending to fainter magnitudes in the optical. The survey design emphasizes flexibility and multiplexing, allowing multiple science goals to be pursued simultaneously while maintaining well-defined selection functions. In addition, the multi-epoch nature of the observations enables the characterization of time-variable phenomena, including stellar multiplicity and dynamical evolution.

MWM is organized into a set of interlocking sub-programs, each optimized to address specific aspects of Galactic structure and stellar astrophysics. These include surveys of young stellar populations, evolved stars, and halo tracers, among others. While each component employs tailored target selection strategies and observational requirements, they are unified by a common goal: to map the chemo-dynamical properties of the Milky Way with unprecedented scale and fidelity. 

\begin{figure}
    \centering
    \includegraphics[width=1\linewidth]{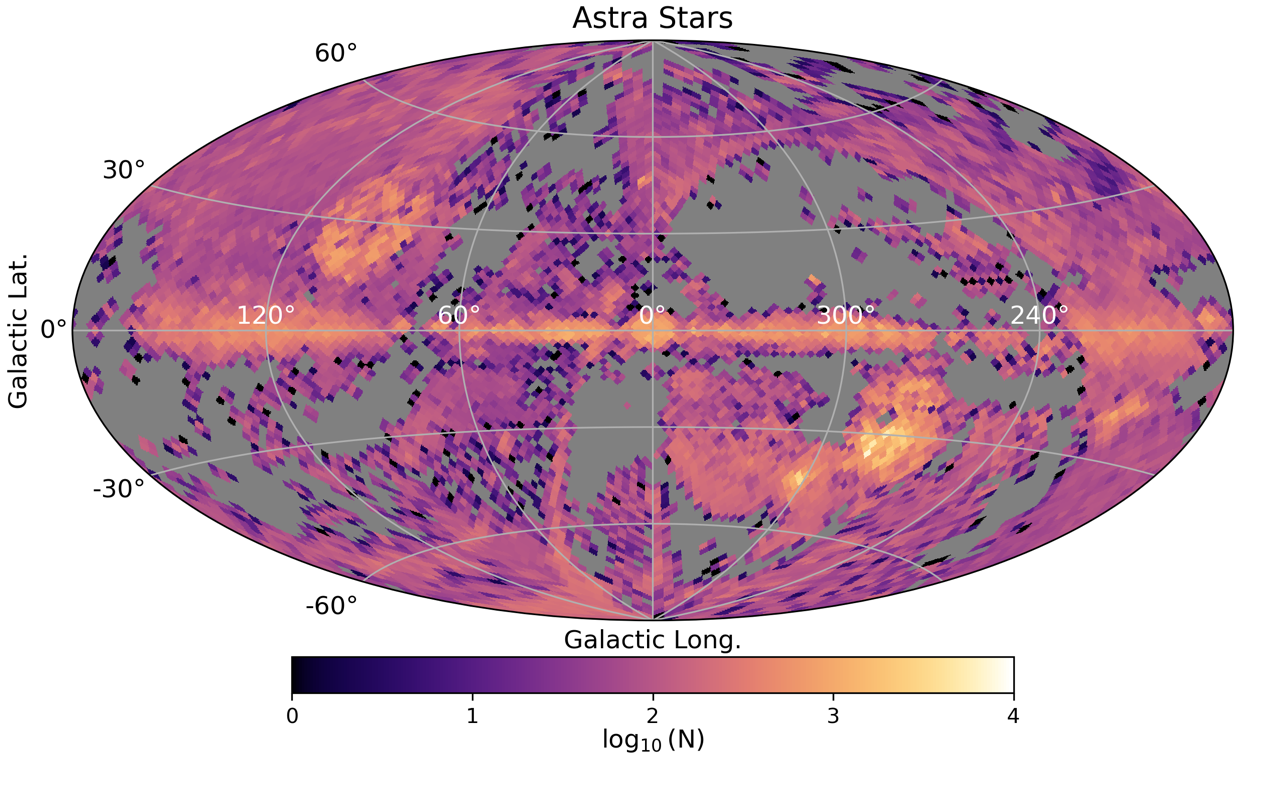}
    \caption{Sky coverage of MWM DR20 targets observed by BOSS. }
    \label{fig:mwm_all_sky}
\end{figure}

\begin{figure}
    \centering
    \includegraphics[width=1\linewidth]{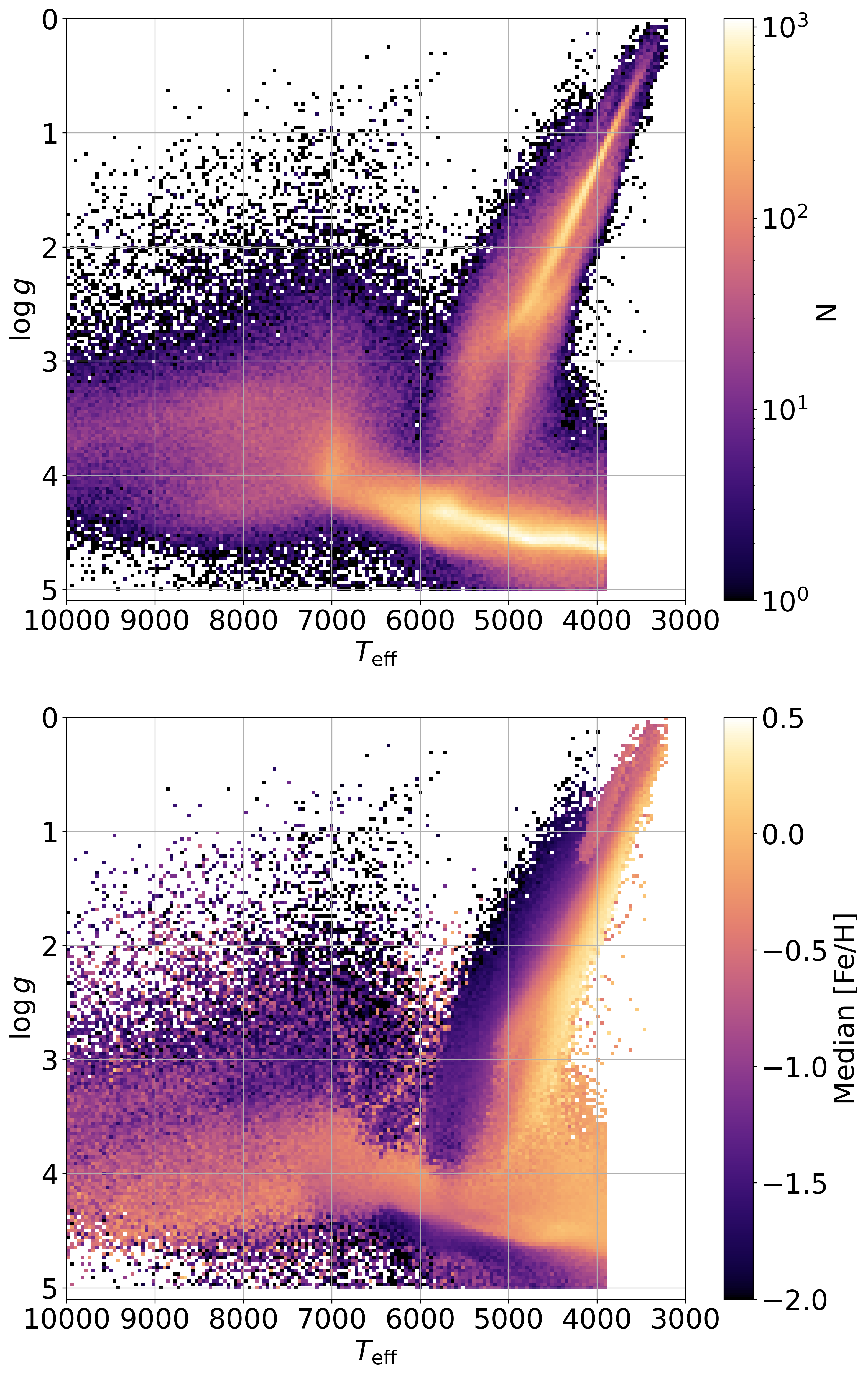}
    \caption{Kiel Diagram of MWM DR20 targets observed by BOSS. Stars with \texttt{flag\_bad=False}, \texttt{flag\_warn=False}, and \texttt{result\_flags=0} in BossNet have been removed. Top: Kiel diagram colored by the density of DR20 visits. Bottom: Kiel diagram colored by the median [Fe/H] from BossNet. }
    \label{fig:mwm_HR}
\end{figure}

DR20 includes the release of $\sim2.2$ million optical BOSS MWM spectra for $\sim1.2$ million stellar targets observed at both APO and LCO, spanning a significant portion of the sky (Figure~\ref{fig:mwm_all_sky}) and HR diagram (Figure~\ref{fig:mwm_HR}). DR20 represents a factor of $4-5$ increase in optical spectra over DR19. The new data span many MWM programs and are summarized in Table~\ref{tab:mwm_sum}. The following sections describe sub-programs that largely rely on BOSS optical spectra, highlighting their individual contributions to the broader scientific objectives of the MWM.

\begin{table*}
\centering
\caption{MWM: Numbers of Stars Observed and Spectra Released (DR19-DR20). For program descriptions, see \url{https://www.sdss.org/dr20/mwm/programs/}. In these statistics, a source can be targeted by multiple mappers and/or programs, causing it to appear in multiple rows in this table. Standard stars are not included in these counts, reducing the total number of targets from 1.5M to 1.2M.}
\label{tab:mwm_sum}
\begin{tabular}{l l c c}
\hline
Description & Mapper/Program & DR19 targets (spectra) & DR20 targets (spectra) \\
\hline
All MWM & mwm & 200k (460k) & 1.2M (2.2M) \\
Open Fiber & open & 210k (363k) & 745k (1.2M) \\
Binary Systems & mwm\_bin & 6k (14k) & 54k (84k) \\
Compact Binaries & mwm\_cb & 27k (84k) & 88k (243k) \\
Dust & mwm\_dust & 68 (81) & 562 (1k) \\
eROSITA Sources & mwm\_erosita & 4k (8k) & 52k (84k) \\
Filler & mwm\_filler & 21k (50k) & 214k (330k) \\
Galactic Genesis & mwm\_galactic, mwm\_gg & 5k (11k) & 44k (74k) \\
Halo & mwm\_halo & 38k (74k) & 251k (380k) \\
Science Validation & mwm\_legacy, mwm\_monitor, mwm\_validation & 16k (27k) & 69k (101k) \\
Magellanic Clouds & mwm\_magcloud & 0 (0) & 95k (133k) \\
OB Stars & mwm\_ob, mwm\_tess\_ob, mwm\_tessob & 38k (102k) & 175k (432k) \\
Planet Hosts & mwm\_planet & 4k (10k) & 19k (38k) \\
Binary Systems & mwm\_rv & 5k (10k) & 32k (45k) \\
Solar Neighborhood Census & mwm\_snc & 71k (131k) & 284k (483k) \\
Asteroseismic Red Giants & mwm\_tessrgb & 3k (8k) & 42k (80k) \\
White Dwarfs & mwm\_wd & 10k (38k) & 38k (109k) \\
Young Stellar Objects & mwm\_yso & 17k (60k) & 67k (191k) \\

\hline
\end{tabular}
\end{table*}

\subsection{MWM: Scientific Programs}\label{sec:mwm_programs}

\subsubsection{Young Galaxy}\label{sec:mwm_yso}

The Milky Way Mapper Young Galaxy program is designed to provide a comprehensive, all-sky census of young stellar populations, enabling detailed studies of star formation, early stellar evolution, and the structure of the Galactic disk. Young stars, defined here as objects younger than 100 Myr, span a wide range of masses, evolutionary stages, and observational characteristics, including diverse spectral energy distributions, strong variability, and a mixture of absorption and emission features. This diversity necessitates multiple complementary selection strategies. The program is therefore divided into two primary components: the Young Stellar Object (YSO) program and the OB-type (OB) program, which together probe the full mass spectrum and environments of recently formed stars.

The YSO component focuses on low- and intermediate-mass stars (FGKM types) in the pre-main-sequence phase, including objects embedded in or recently emerged from star-forming regions. These stars span a broad range of evolutionary stages, from protostars to disk-bearing and diskless pre-main-sequence stars, reflecting the complex and time-dependent processes of accretion, disk evolution, and magnetic activity. Because contraction timescales for low-mass stars can approach 100 Myr, the YSO program samples populations across a wide range of ages and environments, including clusters, associations, and dispersed populations. Many targets exhibit significant photometric and spectroscopic variability, driven largely by magnetic activity and accretion processes. This component enables detailed characterization of the physical and dynamical properties of young stellar populations on small spatial scales, including the structure, age distribution, and internal kinematics of star-forming regions. The full description of the program is presented in \citet{kounkel2023}. No changes to the targeting has been done since the previous data release. However, since DR19, the number of targets observed by BOSS has increased by $\sim200$\%---both due to the inclusion of the data from LCO and due to a more complete coverage of the Galactic plane from APO.

The OB component targets massive stars, aiming to construct a uniform and unbiased census of OB-type stars across the Galaxy, independent of their association with known star-forming regions. These stars serve as key tracers of recent star formation and Galactic structure. Recent work has shown that the spatial distribution of young massive stars reveals features of the Milky Way that differ from those inferred from older stellar populations, dust, Cepheids, or masers. By combining homogeneous spectroscopic observations from SDSS-V with Gaia astrometry, this program probes the large-scale structure and kinematics of the young Galactic disk, including the morphology of spiral arms. The multi-epoch nature of the observations further enables robust characterization of stellar multiplicity, providing constraints on binary fractions, orbital parameters, and the role of binarity in the evolution of massive stars and their remnants.

Together, the YSO and OB components address a set of interconnected questions central to the baryon cycle and star formation in galaxies. The program investigates the spatial and temporal structure of star-forming regions, including the degree of coevality and the impact of feedback from massive stars on the surrounding interstellar medium. It examines how the properties of young stars---such as their mass distribution, multiplicity, and angular momentum---depend on the conditions of their natal environments, including the structure and dynamics of molecular clouds. The survey also probes the role of dynamical interactions in shaping stellar populations, from small-N systems to larger clusters, and explores how angular momentum is partitioned between stellar rotation and multiplicity. By providing large, homogeneous samples of young stars with well-characterized properties, the program enables stringent tests of stellar structure and evolution models, including comparisons with asteroseismic and atmospheric observables. Finally, by mapping the distribution and kinematics of young stars across the sky, the Young Galaxy program delivers new constraints on the structure and dynamical state of the Milky Way disk as traced by its youngest stellar populations.

\subsubsection{Halo}\label{sec:mwm_halo}

The Milky Way Mapper Halo program is designed to provide a comprehensive dynamical and chemical characterization of the Galactic halo across a wide range of spatial scales, stellar populations, and metallicities. It comprises three complementary components---Distant Halo, Local Halo, and Metal-Poor Halo---that together probe extreme regimes of halo structure, formation history, and accretion processes. The Halo program started as a collection of open fiber programs in DR19 (targetdb version 0.5 and earlier), and due to early success was converted to part of the main survey (targetdb version 1.0 and later). DR20 is thus the first major data release of targets from the halo program.

The Distant Halo component aims to map, for the first time in a single survey, the full 4$\pi$ extent of the outer Galactic halo beyond distances of 20~kpc. Target selection is dominated by luminous tracers, primarily K-type red giants identified through WISE and Gaia XP photometry, supplemented by blue horizontal branch stars and RR Lyrae variables. The survey is designed to achieve as uniform a sky coverage as possible, with a requirement to observe over 10,000 red giant stars beyond 30 kpc. 

The Local Halo component focuses on the chemo-dynamical structure of halo stars within approximately 1 kpc of the Sun, extending down to low-mass main-sequence stars that will not get radial velocities in Gaia. Targets are selected primarily from Gaia DR3 based on high tangential velocities (v\_tan $>$ 150~km~s$^{-1}$, with prioritization above 200 km~s$^{-1}$), ensuring a clean halo-dominated sample. The program will obtain spectra for hundreds of thousands local halo stars.

The Metal-Poor Halo component targets the rare, chemically primitive population of the halo, including stars originating from accreted dwarf galaxies and those at the lowest metallicities. Candidate selection relies primarily on Gaia XP and WISE-based metallicity estimates, focusing on stars with [Fe/H] $< -2$, and is complemented by open-fiber opportunities drawing on surveys such as SkyMapper and SAGES. The program is designed to assemble a sample of at least 10,000 such stars to enable transformative constraints on early chemical evolution. 

Observations are carried out primarily with the BOSS spectrographs, though brighter targets (G $<$ 13) in the Local and Metal-Poor Halo samples are observed with APOGEE. Additional use of dark-time fiber offsets maintains coverage of intermediate-magnitude targets (13 $<$ G $<$ 16) with BOSS. Together, these strategies ensure broad and efficient sampling of the halo across distance, metallicity, and stellar type.
Note that since the halo program was added at a later time, astra's default BOSS pipeline (BOSSNet) generally does not produce good stellar parameters for halo stars. Two VACs, MINESweeper \citep{Chandra2026} and BOSS-CLAM \citet{medan2026clam}, should be used for halo stars instead.

\subsubsection{Solar Neighborhood Census}\label{sec:mwm_snc}


Studies of the Solar Neighborhood are crucial for an understanding of stellar populations across the full stellar mass range. Critically, a census of the Solar Neighborhood 
allows for the unique observation of the lowest mass M dwarfs, and even 
brown dwarfs, which are too faint to observe at larger volumes. 
Observations of low-mass M dwarfs are critical for studies of stellar 
populations, as they account for three out of every four stars \citep{Henry2006, RECONS}. Such observations in the Solar Neighborhood facilitate the understanding of star formation and its history at the lowest masses, and how it connects to more luminous stellar populations observed throughout the Galaxy. Additionally, with data from SDSS-V we can probe how these populations change with e.g.~metallicity, age and Galactic population.

To accomplish the above science goals, the Solar Neighborhood Census (SNC) comprises two complementary components. The first is the 100 pc sample, which is volume-limited based on data from Gaia. Within the 100 pc sample, brighter stars will be observed with APOGEE and the low-luminosity sources with BOSS. As the 100 pc sample is dominated by K and M dwarfs, to better sample higher mass stars a larger volume is needed. The second component is the extension sample, which selects stars $> 100$ pc that are more luminous than a linear cut in $M_G$ vs.~distance. By down-sampling this selection by a factor of 20, the resulting extension sample when combined with the 100 pc sample is roughly uniform over a wide range in luminosity, providing sufficient sampling across a wide mass range. Similar to the 100 pc sample, brighter stars are observed with APOGEE and fainter with BOSS.

There are two important changes from DR19 to DR20. First, the target selection has changed, most significantly for the extension sample. For the extension sample, previously the targets were limited to 250 pc and $M_G < 6$. For parts of DR20 the sample now is anything $> 100$ pc, more luminous than a linear cut in $M_G$ vs.~distance and fainter than an apparent magnitude cut. For the 100 pc sample, additional astrometric cuts have been applied towards the Galactic center and Magellanic Clouds to remove spurious sources. More details on these changes can be found in Section \ref{sec:mwm_targeting}.

Second, as the survey has progressed, we have developed a better understanding of the survey plan and speed. With this, we have better predictions of the completeness for the 100 pc sample. Primarily, the expectation for the number of two-epoch observations we will achieve by the end of the survey has reduced to $\sim90,000$ for APOGEE and $\sim47,000$ for BOSS. In addition, we have added one-epoch observations with BOSS, with an expectation of $\sim85,000$ targets by the end of the survey. Despite this, the dataset still completes a rich description of the Solar Neighborhood and we expect $>60\%$ of stars in e.g., the Gaia Catalog of Nearby Stars \citep[GCNS;][]{GCNS} to have a spectrum by the end of the survey. Within the Solar Neighborhood Census working group we are developing a framework for characterizing the selection function of the SDSS-V SNC relative to the GCNS, along with a forward modeling method to infer the properties of stellar subpopulations \citep{medan2026}. This will facilitate analyses of the SDSS-V SNC data in a statistically robust manner, despite its incompleteness.

\subsubsection{White Dwarfs}\label{sec:mwm_wd}


White dwarfs are uniquely powerful tracers of stellar and Galactic evolution that have until recently been difficult to study systematically because they are so faint and difficult to distinguish from other populations of stars. Before 2019, most of the white dwarfs known were serendipitously observed with spectroscopy from earlier iterations of SDSS, as white dwarfs sit in a similar color space as quasars (e.g., \citealt{Kleinman_2013}).

The advent of Gaia has changed things dramatically. Its precise parallaxes and photometry have enabled the construction of the first large, homogeneous, all-sky sample of white dwarf candidates, increasing the known population by roughly an order of magnitude and finally making large-scale spectroscopic follow-up practical \citep{GentileFusillo2019,Gentile_2021}. SDSS-V is well positioned to capitalize on this opportunity by obtaining BOSS spectroscopy for an extremely large Gaia-selected sample, transforming a photometric census with considerable extra information.

The scientific motivation for this effort is broad. White dwarfs retain a fossil record of intermediate-mass stellar evolution, and their mass distribution, luminosity function, atmospheric composition, and magnetic properties provide constraints on post-main-sequence mass loss, cooling physics, convection, and the star formation history of the Milky Way. A large, homogeneous spectroscopic sample also enables population-level studies of hydrogen- and helium-dominated atmospheres, the origin and incidence of strong magnetic fields, and the detection of debris and metal pollution from remnant planetary systems. In this sense, SDSS-V observations of white dwarfs are not merely catalog-building; they provide the empirical foundation for testing models of white dwarf structure and evolution across a wide range of cooling ages.

White dwarfs are a core carton of the Milky Way Mapper, and their relatively low space density (fewer than 10 objects down to G~$<$~20\,mag per square degree across the whole sky) makes surveys like SDSS-V incredibly efficient for collecting many new spectra. Before SDSS-V, roughly 42,000 white dwarfs had been observed spectroscopically up to DR17. In DR19, more than 16,000 unique white dwarfs were observed. In DR20, more than 49,000 unique white dwarfs have been observed, more than 80\% of which have not been observed in a previous SDSS data release. Crucially, more than 40\% of the newly observed white dwarfs are located in the southern hemisphere, where relatively few large spectroscopic surveys have to-date been undertaken. In DR20, the white dwarf targeting selection used improved astrometric selection from Gaia DR3 \citep{Gentile_2021}, as well as an expanded region of the Gaia color-magnitude diagram with an emphasis on the white dwarf selection function \citep{Rix2021}.

By targeting tens of thousands of Gaia-selected white dwarfs, including repeated observations when fields are revisited, SDSS-V is creating a spectroscopic resource large enough to study both the dominant white dwarf population as well as unique subclasses of rare stellar remnants.

\subsubsection{Galactic eROSITA Sources}\label{sec:mwm_erosita}
The main goal of the various MWM survey components targeting eROSITA is to boost our understanding of the demographics of Galactic X-ray sources. The necessary pre-requisite is a rigorous optical follow-up program as envisioned by the MWM sub-survey targeting Galactic \ero sources. This is for the simple reason that \ero may detect the interesting objects, but their nature remains uncovered as long as no optical follow-up is performed. Galactic point-like \ero sources described here are either accretion-powered or are coronal emitters. Synergies between \srgero and SDSS-LVM exist for X-ray extended sources like the general ISM and SNRs and are described elsewhere. 

{\it Compact accreting binaries:} Among the binary accreting sources, the cataclysmic variables (CVs) and their subtypes play a specially important role for the SDSS/\ero synergy. The objects harbor white dwarfs that accrete from a low-mass companion via Roche-lobe overflow. While the SDSS project has a long and successful legacy uncovering such systems many questions still remain unanswered in CV research, mainly because past CV discoveries happened by chance and not through systematic searches \citep[see e.g. the large compilations of CVs presented by ][and references therein]{szkody+11, inight+23}. Open questions regarding their space density \citep{schwope_flux_limited_samples}, the fraction of very low mass-transferring systems, so called period bouncers \citep{inight+23b}, population demographics between magnetic (mCVs) and non-magnetic systems including their orbital period distribution \citep{schreiber_period_gap}, the effects of magnetic braking and other phenomena on the angular momentum loss in these systems \citep{magnetic_braking_aml}, the origin of the WD magnetic field in mCVs \citep{schreiber_magnetic_WDs}, and finally also the contribution of the mCVs to the Galactic Ridge X-ray Emission \citep{GRXE}. Both volume-limited \citep{pala, inight_300pc} and flux limited \citep[see for example][]{schwope_flux_limited_samples} population studies of CVs have been done in the past, as this is one of the most robust ways to attempt to answer these above-mentioned questions. However, all of these studies had their limitations, as the volume-limited studies suffered from low-number statistics, while the flux-limited studies were biased to the brightest systems.

To this end, we aim to identify and create a complete inventory of all the CVs in the Western Galactic hemisphere (that which is available to the German \ero  Consortium). A first look into the eROSITA CV X-ray sky revealed the feasibility of creating a 500\,pc volume-limited sample of CVs, much larger than existing ones \citep{schwope+24b}. This will be used to perform population studies and address the mentioned questions regarding CVs and their evolution. First results are reported in \cite{schwope+24a}, Brink et al.~(2026, submitted.), and Hernandez-Diaz et al.~(2026, submitted). The number of targets that were defined, the number of observed spectra and the number of CVs are reported further below in the description of the corresponding VAC (see Sect.~\ref{vac:erosita_cvs}) 

{\it Coronal X-ray sources:} 
Stellar X-ray emission is found in virtually all regions of the Hertzsprung-Russell Diagram (HRD), for the SDSS/\ero follow-up program late-type stars play a dominant role. Cool late-type stars generate X-rays from magnetic activity that is powered by a stellar dynamo; in these stars the energy output in X-rays strongly declines (by roughly four orders of magnitude) with increasing age, i.e. $\log L_X/L_{\rm bol} \simeq -3\, \dots -7$. For an overview on stellar X-ray properties see e.g.~the review article by \cite{guedel_naze09}. 

The \ero/SDSS stellar X-ray survey will determine the combined X-ray and optical activity properties of diverse stellar populations by obtaining volume complete samples in the solar neighborhood and flux limited sample in the local Milky Way. It will study the X-ray emission and its evolution in relation to other stellar parameters like mass, age, and rotation \citep[see e.g.][and references therein]{fritzewski+25}. And it will do so for the latest spectral types to confine the activity boundary when the stellar photospheres effectively become neutral and a strong decrease of activity indicators is expected \citep{stelzer+22}. The most widely used observational diagnostics of magnetic activity for late-type stars is H$\alpha$ emission from a chromosphere. 

The \ero/Gaia/SDSS synergy will reveal the needed input to address the mentioned questions. To this end, \cite{freund+24} have developed a probabilistic method to match \ero X-ray sources with Gaia stellar counterparts. The resulting target input catalog has 222,406 entries of which 47,979 were observed and are published in DR20. The input catalog forms the largest yet available candidate catalog with a proven low contamination fraction \citep{schwope+24a} to study the combined X-ray and optical activity properties of stars all along the main sequence. 

\subsection{MWM: Targeting Updates}\label{sec:mwm_targeting}

SDSS-V continues to use a suite of target selection algorithms to define the survey sample, resulting in versioned ``targeting generation''. DR20 contains observations from the latest `v1.0' targeting generation (Section \ref{sec:targeting}).
Many MWM cartons were impacted by updates to the underlying input catalogs, most notably the change from Gaia DR2 to Gaia DR3. Additionally, some cartons adjusted their target selection criteria to better meet the needs of the respective scientific programs. In Appendix \ref{sec:appendix_mwm}, we discuss the kind of changes that have been made to the cartons in v1.0. In that appendix, we discuss three example cartons mwm\_galactic\_core\_dist\_apogee (\ref{ex1_mwm_gg}),  mwm\_bin\_rv\_short\_rgb\_apogee (\ref{ex2_mwm_rv_short}), and mwm\_wd\_gaia\_boss (\ref{ex3_mwm_wd}). The first two cartons are examples of changes to improve overall selection of their desired targets, and the final carton is entirely new in DR20.

The new and updated cartons were generally a result of three different factors. The first factor is the addition of new classes of targets being added or being promoted from open fiber programs. For example there were number of cartons added like manual\_mwm\_planet and manual\_mwm\_rv\_validation that contained custom lists of stars as versus an algorithmic target selection. The second major factor were improvements to the selection algorithms such as with mwm\_galactic\_core\_dist\_apogee and mwm\_bin\_rv\_short\_rgb\_apogee see Appendix \ref{sec:appendix_mwm} for more details. The final factor was responding to the results of the robostrategy (see section \ref{sec:robostrategy}) runs. As the observing information was updated with new robostrategy runs, it became clear that sample sizes were going to be smaller than originally planned. One way to mitigate the drop in the number of targets in cartons with multiple epochs was to use a nested strategy for cadences. Looking at Table \ref{tab:mwm_cartons} many cartons have the same name, but with different numbers appended to them. These are duplicate cartons, that are nested such that if the optimal number of epochs could not be achieved, we could still fulfill the needs of the program with an adequate number of epochs. For example mwm\_bin\_rv\_short\_mdwarf\_apogee\_18epoch, also has a 12 and 08 epoch version. The idea in that case is that 18 epochs is ideal for fully characterizing the orbits of M dwarf companions, but 12 epochs (less ideal), or a minimum of 8 epochs would still be enough to give us a reasonable orbit.

A few more minor changes are noteworthy for the seamless use of  DR20 products compared to DR19. First,  MWM standardized the carton naming scheme in v1.0  to explicitly  identify whether the BOSS or APOGEE instrument was being used. Previously, this was only done for science programs that utilized both instruments for their science goals.  
The use of the Semaphore package,\footnote{\url{https://github.com/sdss/semaphore/tree/0.2.4}} first released with DR19 \citep{sdssVdr19}, is invaluable for easily working with the targeting flags, particularly with the ``alt\_program'' and ``alt\_carton'' metadata for selecting targets with the same underlying selection functions but different names. Useful examples are provided in the DR20 tutorial.\footnote{ \url{https://www.sdss.org/dr20/tutorials/python/?id=SDSS-V_targeting_flags} } targets in v1.0 were able to take advantage of the ``offsetting'' technique, where the wing of a target's point spread function (PSF) is placed in the fiber instead of the PSF peak \citep{Medan2025}. This new operational ability effectively increased the bright limit of observed targets in BOSS cartons, such as those in the OB Star science program.

\section{Local Volume Mapper}\label{sec:lvm}

The Local Volume Mapper (LVM; \citealt{Drory2024,Blanc2024}) was designed to provide a comprehensive spectroscopic view of the ionized interstellar medium (ISM) across the Milky Way, the Magellanic Clouds, and a representative sample of nearby galaxies. Its primary scientific objective is to characterize stellar feedback at the energy injection scale, connecting the physical processes operating within individual H\,{\sc ii} regions to the large-scale structure and evolution of galaxies. By mapping how energy, momentum, and chemically enriched material are injected and redistributed within the ISM, the LVM aims to constrain the mechanisms that regulate star formation across a wide range of environments.

The LVM achieves these goals through wide-field optical integral field spectroscopy, enabling spatially resolved measurements of the physical conditions of the ionized gas, including electron density, temperature, ionization state, and chemical abundances. This approach allows the survey to link individual feedback sources, such as massive stars and supernova remnants, with the surrounding ISM, tracing the interplay between stellar populations and gas across scales ranging from sub-parsec structures in the Milky Way to kiloparsec scales in external galaxies. The resulting dataset constitutes a unique, homogeneous spectral atlas of ionized gas in the Local Volume.

The survey is enabled by the dedicated LVM Instrument \citep[LVM-I][]{Konidaris2024,Herbst2024}, a new facility developed as part of SDSS-V that combines a wide-field fiber-fed integral field unit with a system of multiple small telescopes feeding a set of spectrographs covering the full optical wavelength range (3600--9800\,\AA) at a resolving power of $R\sim4000$. Each observation corresponds to a tile with a $\sim0.5^{\circ}$ diameter field of view, sampled by 1801 science fibers of $35.3''$ diameter. The tiling strategy is designed to create contiguous mosaics over the main survey regions, including the Galactic plane and the Magellanic Clouds, with overlapping rows of fibers between adjacent tiles to ensure accurate cross-calibration and uniform spectrophotometric quality. In addition, for particular regions in the sky, like the Small and Large Magellanic Clouds, a nine-pointing dithering scheme was implemented, in order to 
ensure full sampling  of the field of view and provide a slight increase of the spatial resolution.

\begin{figure}
      \centering
      \includegraphics[width=\linewidth]{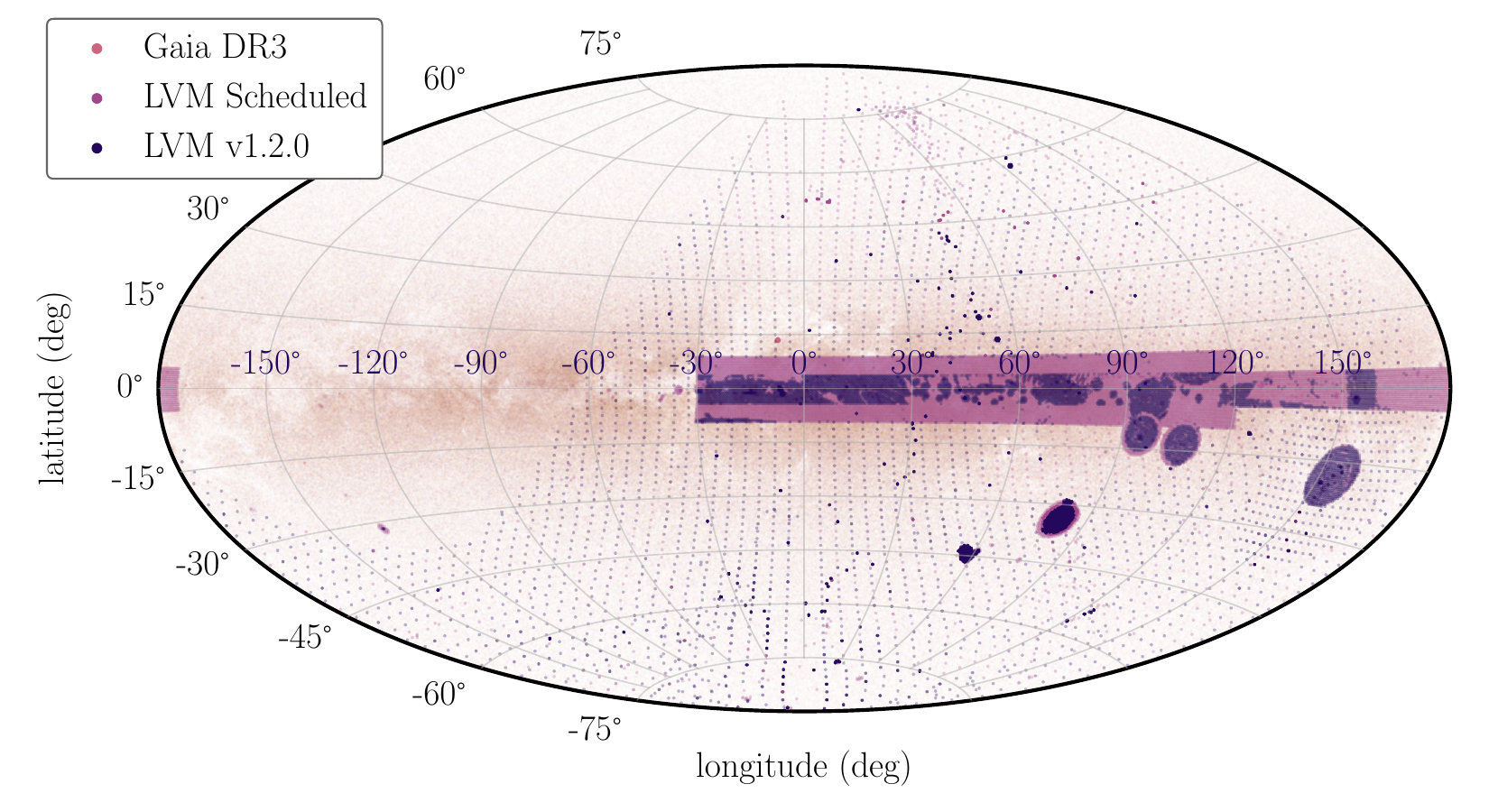}
       \caption{LVM: Comparison between the scheduled and observed tiles. Figure shows the distribution in the sky of the LVM full scheduled tiles (orange) and currently observed ones, up to November 2025 (blue), in comparison with the distribution of GAIA DR3 stars \citep{gaiadr3}, that clearly trace the location of the MW disk.}
      \label{fig:lvm_status}
\end{figure}

\subsection{LVM: Current status}
\label{sec:LVMstatus}

Since the early commissioning phase described in previous releases \citep{sdssVdr19}, the LVM survey has progressed significantly, moving into large-scale operations and accumulating several thousand observed tiles across its main targets. So far we have taken $\sim$17k exposures comprising $\sim$30M science spectra, observed up to November 2025, as illustrated in Figure \ref{fig:lvm_status}. As described in \citet{Drory2024}, the survey is targeting a set of regions that cover the required ranges of astrophysical environments to achieve its goals. Those regions include the Milky Way disk (MW), the Magellanic Clouds (MCs), selected nearby Galactic regions such as Orion (ORI) and the Gum Nebula (GUM), targeted fields such as THOR \citep{Bihr2015}, a sparse Full Sky component aimed to sample the MW Halo in a discrete way, and a sample of Local Volume (LV) galaxies.

The survey presents a heterogeneous level of completion across these components, as expected due to the implementation of a strategy in which certain regions had higher priorities. Thus, early science regions, such as ORI and THOR, are essentially complete, while GUM has recently reached full completion. The MCs are at an advanced stage, with $\sim 75\%$ of the planned observations completed, and the Full Sky component has reached $\sim 60\%$. The MW program, which represents the largest footprint of the survey, is currently at $\sim 35\%$ completion, reflecting its intrinsic ambitious scope. In contrast, the LV component remains in an early phase, with only a $\sim 25\%$ completion. Despite this apparent drawback, it is important to recall that the LV component is the one that is less affected by the lack of completeness: with $\sim$90 individual galaxies observed, it already comprises a very useful sample. 

All observations are processed using a dedicated DRP (Mejía-Narváez et al. in prep.), which produces fully calibrated row-stacked spectra (RSS) preserving the native sampling of the instrument, as described in Section \ref{sec:lvmdrp}. In addition, a Data Analysis Pipeline (DAP; \citealt{lvmdap}) has been developed to derive the observational and physical properties of both the stellar and ionized gas components (Section \ref{sec:lvmdap}). Unlike previous releases \citep{sdssVdr19}, in which only reduced spectra were provided as a preview of the data format, the current data release includes both DRP and DAP products, as we will describe below, enabling direct scientific exploitation of the dataset.

\begin{table*}
\centering
\caption{Summary of LVM DR20 targets. }
\begin{tabular}{l l l c c c}
\hline
NGC Name & Alternative Name & type & RA (deg) & Dec (deg) & Appr. Diameter \\
\hline
NGC\,2024 & Flame Nebula / Horsehead Nebula & HII complex & 05:41:40.6 & $-$02:39:58 & 5.5\degree \\
NGC\,2237 & Caldwell 49 / Rosette Nebula & HII region & 06:31:49.7 & +04:50:13 & 5.5\degree \\
NGC\,6514 & M20 / Trifid Nebula & HII region & 18:02:40.1 & $-$22:56:17 & 0.5\degree \\
NGC\,7293 & Helix Nebula & Planetary Nebula & 22:29:38.6 & $-$20:50:13 & 0.5\degree \\
\hline
NGC\,598 & M33 / Triangulum Galaxy & Local Volume Galaxy & 01:33:50.9 & +30:32:56 & 0.7\degree \\
NGC\,4945 & Caldwell 83 & Local Volume Galaxy & 13:05:27.6 & $-$49:28:30 & 1.5\degree \\
\hline
\end{tabular}
\label{tab:lvm_dr20_targets}
\end{table*}

\subsection{LVM: Science results}
\label{sec:LVMsci}

The scientific capabilities of the LVM are already demonstrated by a growing body of early science results that exploit its unique combination of wide-field coverage and spatially resolved spectroscopy. Detailed studies of individual H\,{\sc ii} regions, such as the Lagoon Nebula \citep{Singh2026}, the Trifid Nebula \citep{Sattler2026}, and the Rosette Nebula \citep{VillaDurango2025,Hilder2026}, have revealed the complex internal structure of ionized gas, including spatial variations in temperature, density, and chemical abundances at scales down to the pc scales. These analyses have provided new insights into long-standing problems such as the discrepancy between abundances inferred from collisionally excited lines compared to metal recombination lines and the mixing of chemically enriched material within nebulae. Observations of supernova remnants, such as RCW~86 \citep{Sarbadhicary2026}, demonstrate the ability of LVM to resolve shock structures and probe the microphysics of collisionless shocks in the interstellar medium. In addition, studies of massive stars in low-metallicity environments \citep{GonzalezTora2025} highlight the sensitivity of LVM to detect weak emission features, enabling the characterization of ionizing sources such as Wolf--Rayet stars and their contribution to the ionization budget. The combination of these results, together with the public release of high-quality datasets and analysis products \citep{lvmdap,HelixDR19}, illustrates the transformative potential of LVM to address key questions on stellar feedback, ISM structure, and chemical evolution across a wide range of Galactic and extragalactic environments. 

\subsection{LVM: Released Targets}
\label{sec:LVMrel}

\begin{figure*}
      \centering
      \includegraphics[width=\linewidth]{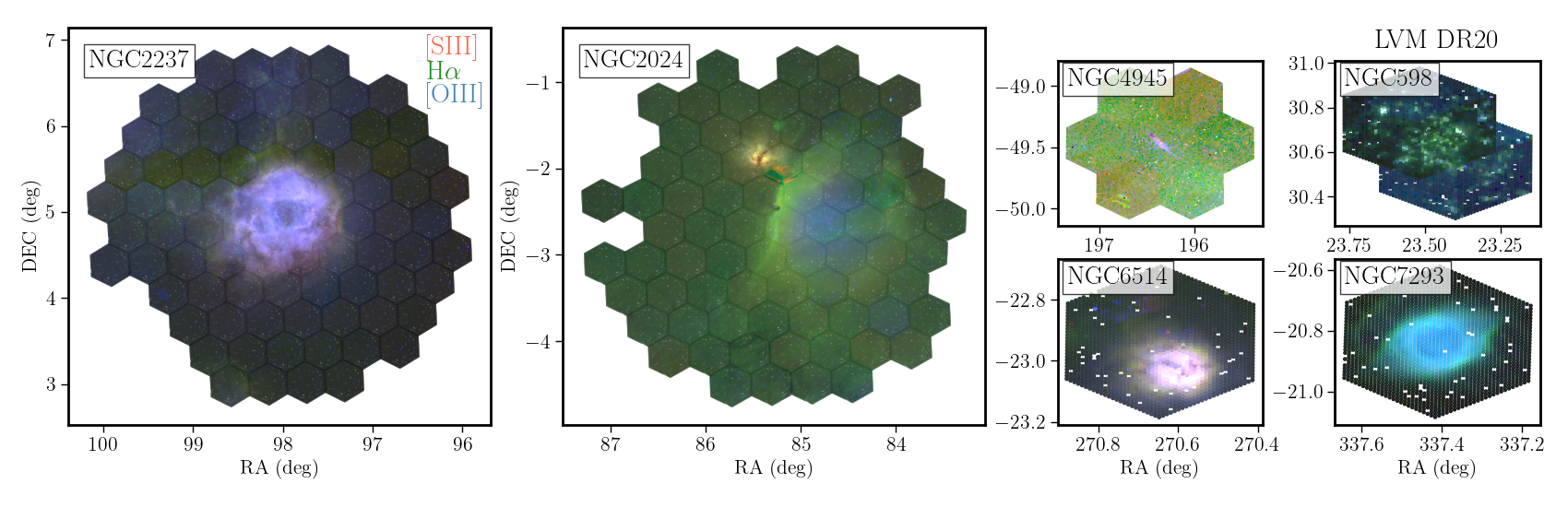}
       \caption{RGB emission-line maps of the six LVM targets included in DR20, constructed from the [OIII]$\lambda$5007 (blue), H$\alpha$\ (green), and
        [SIII]$\lambda$9069 (red) flux maps derived spaxel-by-spaxel by the LVM-DAP. The two large panels show the extended Galactic \hii\ region mosaics: the Rosette Nebula (NGC\,2237, left), and the Flame Nebula and Horsehead Nebula region (NGC\,2024, center) and . The four smaller panels show the remaining targets: the starburst galaxy NGC\,4945 (upper centre-right), the Triangulum Galaxy M\,33/NGC\,598 (upper right), the Trifid Nebula (NGC\,6514/M\,20, lower centre-right), and the Helix Nebula (NGC\,7293, lower right). The diverse range of environments represented, from Galactic \hii\ regions and a planetary nebula to nearby starburst and spiral galaxies, illustrates the        wide range science cases addressed by the LVM survey. Note that for the extragalactic targets (NGC\,4945 and NGC\,598) the maps include a contribution from foreground Milky Way emission, which has not been subtracted as described in the text.}
      \label{fig:lvm_DR20}
\end{figure*}

The LVM DR20 sample includes six representative targets spanning three Galactic HII regions at different distances, a planetary nebula, and two nearby galaxies. Their central coordinates and observing setups are listed in Table~\ref{tab:lvm_dr20_targets}, while Figure \ref{fig:lvm_DR20} illustrates the spatial coverage of the sampled regions. Together, these targets were selected to showcase the diversity of the LVM science cases, outlined in Section \ref{sec:LVMsci}, from spatially resolved ionization fronts and abundance diagnostics in Galactic nebulae to feedback-driven structures and galaxy-scale environments in nearby systems.

\paragraph{NGC\,2024 / Flame Nebula and Horsehead Nebula region.}
This field forms part of the Orion complex and includes the Flame Nebula (NGC~2024), IC~434, and the Horsehead Nebula at the interface with the Orion~B molecular cloud  \citep{Kreckel2024}. The LVM Orion observations demonstrated the ability of the survey to map the ionization structure of extended Galactic star-forming regions over very large areas, resolving the contrast between bright ionized nebulae, dark molecular structures, and photo-dissociation fronts. In particular, the Flame Nebula appears as a compact, dense \ion{H}{2} region with strong Balmer and sulfur emission, while the Horsehead is seen in projection as a dark inclusion against the bright IC~434 ionization front. These data highlight the power of LVM to trace the coupling between ionized gas, dust, and molecular material, as well as the escape of ionizing radiation into the surrounding diffuse medium \citep{Kreckel2024}.

\paragraph{NGC\,2237 / Rosette Nebula.}
The Rosette Nebula is a well-known Galactic \ion{H}{2} region ionized by the OB stars of the young cluster NGC~2244 and embedded within the remnant of a giant molecular cloud. The LVM observations cover the full optical extent of the nebula and reveal the spatial relationship between the evacuated central cavity, the surrounding ionized shell, the molecular cloud, and the dusty interfaces with ongoing star formation. The first LVM analysis of this target emphasized its morphological structure through the distribution of H$\alpha$, H$\beta$, [O\,{\sc iii}], [N\,{\sc ii}], and [S\,{\sc ii}] emission, together with ancillary CO and infrared data, showing that the Rosette provides an archetypal example of a feedback-shaped \ion{H}{2} region and an ideal laboratory for studying the interaction between massive stars and the surrounding ISM. \citep{VillaDurango2025,Hilder2026}

\paragraph{NGC\,6514 / M\,20 / Trifid Nebula.}
The Trifid Nebula is a compact and nearly spherical Galactic \ion{H}{2} region, predominantly ionized by the central O star HD~164492A, and therefore represents a benchmark case for the study of a relatively simple Str\"omgren sphere. The LVM observations spatially resolve its ionized gas at sub-parsec scales and have been used to derive maps of extinction, electron density, electron temperature, and ionic abundances. These analyses show a clear radial decrease in density, but comparatively small temperature fluctuations and only weak abundance gradients, indicating that M\,20 behaves as a relatively homogeneous nebula at the scales sampled by LVM. Its combination of simple geometry, ongoing star formation, and detectable faint auroral lines makes it a particularly valuable calibrator for ionization and abundance diagnostics \citep{Sattler2026}.

\paragraph{NGC\,7293 / Helix Nebula.}
The Helix Nebula is one of the nearest and best-studied planetary nebulae, and its large angular extent makes it a particularly well matched target for the LVM field of view \citep{HelixDR19}. The LVM observations provide spatially resolved maps of the principal ionic species across most of the nebula in a single pointing, revealing the classical ionization stratification from the compact He$^{++}$ core to the bright [O\,{\sc iii}] ring and the extended low-ionization envelope. The LVM-DAP analysis further shows that numerous faint auroral and diagnostic lines can be detected across the field, enabling measurements of weak-line morphology, radial trends, and ionized-gas kinematics. This target therefore serves both as a science showcase and as a demonstration of the diagnostic power of the LVM data products on extended nearby nebulae. As data for this target were already delivered in DR19, for DR20 we just provide an updated version of the data set.

\paragraph{NGC\,598 / M\,33.}

M\,33 is the nearest massive late-type galaxy in the northern Local Group and provides an important bridge between Galactic resolved nebulae and galaxy-wide external systems. Together with M\,31, it is one of the best laboratories to explore the evolution of stellar populations \citep{Smercina2023,Lazzarini2022}, chemical enrichment processes \citep{Rogers2022,Magrini2007}, and feedback processes \citep{Corbelli2025}. Its distance \citep[$D \approx 840$\,kpc;][]{Breuval2023} and angular size are well placed to combine the analysis of resolved, partially resolved, and integrated stellar populations, in combination with multiwavelength datasets covering from X-ray to atomic and molecular gas \citep[e.g.][]{Verley2009,Tabatabaei2022,Sarbadhicary2025}. This makes M\,33 a particularly compelling LVM target for studying feedback, the spatial relation between ionized gas and star-forming structures, and the connection between
cloud-scale environments and galaxy-scale evolution \citep{Corbelli2025,Koplitz2023}.
Within DR20, M\,33 extends the delivered sample beyond Galactic nebulae into the regime of spatially resolved nearby galaxies, linking LVM with the goals of the proposed LVM-II project within the After Sloan-5 proposal (K. Hawkins et al., in prep.)

\paragraph{NGC\,4945.}
NGC~4945 is a nearby ($D \approx 3.8$\,Mpc), nearly edge-on starburst galaxy hosting
both intense central star formation \citep{Emig2020} and a Seyfert~2 nucleus
\citep{Iwasawa1993}, making it one of the nearest composite AGN--starburst systems
\citep{Heckman1990,PorrazBarrera2024}. NGC~4945 is scientifically well suited for LVM because its large angular size and the presence of a large galactic wind that ionizes the ISM \citep{Heckman1990,Venturi2017,Mingozzi2019} make it an excellent laboratory for linking optical ionized-gas diagnostics with the broader multiphase outflow traced in X-rays \citep{PorrazBarrera2024} and at millimetre wavelengths \citep{Bolatto2021}. In the context of DR20, it illustrates the capability of LVM to target nearby galaxies with resolved feedback-driven extraplanar emission.

\subsubsection{Data Products}\label{sec:lvm_data_product}

The LVM DR20 data products comprise (i) the outputs of the DRP, and (ii) the products generated by the DAP. Together, these provide both calibrated spectroscopic data and higher-level measurements of the physical conditions of the ionized interstellar medium. See Appendix~\ref{appendix:LVM_DRP} and Appendix~\ref{appendix:LVM_DAP} for details on the released DRP and DAP files, respectively. The current data release represents the first substantial set of LVM observations, but we must acknowledge that it only represents $\sim$1\% of the full dataset currently acquired and analyzed by the LVM survey (e.g., Section \ref{sec:LVMstatus}). 

Figure~\ref{fig:lvm_DR20} provides a simultaneous view of the current
progress of the LVM survey  (already shown in Fig.~\ref{fig:lvm_status}) and of the quality of the delivered data products released in DR20. The RGB emission-line maps used to construct both the all-sky mosaic and the individual target insets are derived entirely from the LVM-DAP pipeline products described in Section~\ref{sec:lvmdap}, illustrating the quality of both the DRP and DAP data products. It is important to note that for extragalactic targets such as NGC\,4945 and NGC\,598, the foreground Milky Way component is preserved in the DRP files, as it is throughout the full LVM dataset. This component includes both foreground stars and diffuse emission-line flux from the MW itself. In most integral field spectroscopic observations of extragalactic targets, this contribution is sufficiently faint to be removed as part of the standard sky subtraction procedure \citep[e.g.][]{sanchez16,Law2016}. However,
since the MW ISM is itself a primary science target of the LVM survey, this component is intentionally not subtracted. As a consequence, the delivered DAP data products for these fields reflect the combined extragalactic and Galactic signal, as is evident in the RGB composite shown in Figure~\ref{fig:lvm_DR20} for NGC\,4945, where the diffuse MW foreground emission is clearly visible. 

\begin{figure}
      \centering
      \includegraphics[width=0.5\linewidth]{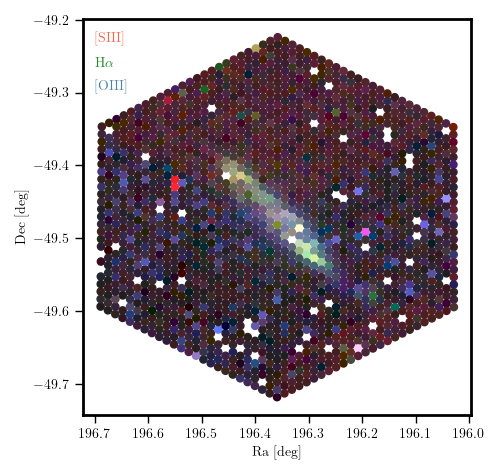}\includegraphics[width=0.5\linewidth]{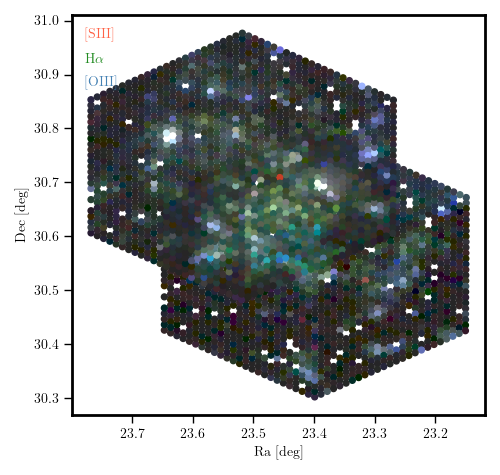}
       \caption{RGB emission-line maps of the two Local Volume (extragalactic) targets included in DR20 (NGC4945 and NGC598), constructed in a similar way as Fig. \ref{fig:lvm_DR20}, once subtracted the contribution of the Milky Way, as indicated in the text.}
      \label{fig:lvm_DR20_LV}
\end{figure}

To address these particular targets (LV), that in general would correspond to $\sim$1-2\% of the tiles foreseen for the LVM, we have developed a  MW foreground subtraction procedure. This procedure involves (i) a masking of the extragalactic target, (ii) a modeling of the MW emission line contribution based on the DAP analysis described before (using the non-masked regions and implementing an interpolation scheme), (iii) a detection and subtraction of the foreground stars based on a cross-matching with the GAIA DR3 catalog \citep{gaiadr3}, and (iv) a full re-analysis based on the standard DAP fine tuning the redshift range to match that of the considered extragalactic targets. This analysis is still experimental. However, we prefer to deliver its results to allow the community to evaluate the effect of the presence of the MW and explore the intrinsic properties of the distributed extragalactic targets. Figure \ref{fig:lvm_DR20_LV} illustrates those differences, showing the same figure as Fig. \ref{fig:lvm_DR20} for the two extragalactic targets (NGC4945 and NGC598), once subtracted the contribution of the MW. The corresponding data products are distributed at same directories as the standard analysis for the considered exposure using the following name convention: \texttt{LV\_dap-rsp108-sn20-EXPNUM.fits}. 

\subsection{LVM: Quality Control}
\label{sec:lvm_qc}

All DRP and DAP outputs for the DR20 tiles were subject to a dedicated quality control (QC) process carried out by a team of LVM collaboration members. The primary goals of this process were to identify problematic exposures, measure quantitative quality metrics for all delivered tiles, and flag or remove data products that do not meet the minimum quality requirements for public release so far. A visual inspection of all the DR20 exposures and DAP products by QC team members was a basis for excluding a small number of tiles from the original DR20 release candidate list because of failures in the DAP
emission-line fitting for a significant fraction of fibers, or particularly strong anomalies in
the reduced spectra due to calibration or sky subtraction issues. A subsequent quantitative quality assurance, relying on some of the quality metrics and visual inspection, summarized below resulted in the quality flags provided for each exposure in \texttt{qcall-\textit{drpver}-\textit{dapver}.fits} file in the form of a bitmask, enabling users to apply additional filtering appropriate to their science requirements. This file is distributed in the following link: \url{https://dr20.sdss.org/sas/dr20/spectro/lvm/analysis/1.2.0/1.2.0.251218/}. We strongly recommend consulting these flags before performing any quantitative analysis of the delivered data products.

Several metrics, characterizing the observation conditions, quality of reduced spectra and DAP products, were evaluated for each DR20 exposure. Several of them, as well as the results of the qualitative visual inspection of the spectra and quality of DAP fit, were converted to the QA flag in provided bitmask. Each metric occupies two bits encoding one of the following values: 00: \texttt{good}, 01: \texttt{warning}, 10: \texttt{major issue}, 11: \texttt{not evaluated}. Typically, \texttt{warning} refer to just a warning about a slight deviation of the particular metrics from the accepted range of \texttt{good} values defined by 1$\sigma$ observational uncertainties, and \texttt{major issue} refer to a clearly identifiable artifacts in the data or results of their analysis. We note that even exposures having \texttt{major issue} flags are ready for scientific analysis, unless any of the known issues listed in the Section~\ref{sec:lvm_qc:known_issues} is critical for it. 
We further note that in DR20, we release the prototype version of the LVM QA flags, that will likely be revised in the future data releases, in particular by implementing all the metrics listed here and removing any dependence on visual inspection. 

Metrics including presence of the bright stars in the field, sky brightness, relative flux calibration between different branches, absolute flux calibration accuracy, wavelength calibration and flat-fielding accuracy, sky subtraction quality, and DAP fitting results consistency were explored and converted to LVM QA flags in the DR20.

The other metrics characterizing the reduced spectra that were explored, but not resulted in dedicated QA flags in its current version, include derived LSF accuracy and consistency of the H$\alpha$ brightness distribution between the adjacent tiles. 
The fiber-to-fiber variations of the LSF and flat-fielding quality were characterized from the widths and normalized flux of night-sky emission lines. Although we found systematics in wavelength offsets between different group of fibers for some exposures, they are typically within 1/10th of the spectral pixel size and should not affect most of the science applications. Several exposures exhibit systematic differences in flat-fielding quality, currently reflected in visual inspection and corresponding quantitative QA flags. Similarly, some tiles, typically -- in the faint regions, exhibit noticeable differences in H$\alpha$ brightness with respect to the adjacent tiles, as currently reflected only in visual inspection QA flags. 

The quality of the DAP products was evaluated by exploring the emission line fitting residuals, consistency of the emission line flux ratios with theoretical expectations, consistency between the parametric and non-parametric results, and behavior of the reduced $\chi^2$ as a diagnostic of the overall quality of the DAP modeling for each fiber and exposure. In DR20 bitmask file, these metrics reflected by the DAP QA flags relying on both visual inspection quantitative estimates. 

The quantitative analysis of the ratio of the parametric non-parametric (\texttt{NP}) to parametric (\texttt{PEK}) emission-line fluxes
provided by DAP was examined as a function of line flux, equivalent width, and signal-to-noise ratio across all DR20 tiles. The two methods yield mutually consistent results with median \texttt{NP}/\texttt{PEK} ratios of unity and interquartile ranges well below 10\%, for emission lines with parametric S/N\,$\gtrsim$\,25, integrated flux $\gtrsim 3\times10^{-14}$\,erg\,s$^{-1}$\,cm$^{-2}$,
or equivalent width $\gtrsim$\,100\,\AA. At lower flux levels or
signal-to-noise ratios the scatter between the two methods increases
substantially, as expected for measurements approaching the noise floor. These
thresholds provide practical guidance for users selecting between the parametric
and non-parametric measurements for their science applications.

\subsubsection{Summary of known issues in DR20}
\label{sec:lvm_qc:known_issues}
From our QC measurements and visual inspections we revealed several issues present across multiple number of exposures in the current data release. Here we describe typical known issues. Their presence and prominence in each exposure is reflected by quality flags.

The most common issue is a jump in absolute flux between the different spectral branches, particularly prominent between the blue and red spectral channels. This issue affects the stellar continuum spectra and is taken into account correctly by DAP. The user should be careful when applying their own analysis codes to the reduced spectra. 

Another common issue present in many exposures is the over- or under-subtracted brightest night-sky emission lines OI $\lambda$ 5577, 6300, 6364\AA. In particular, this significantly affects all measurements of the [OI]$\lambda$6300, 6364\AA\ nebular emission lines for Galactic sources, for which redshift is not sufficient to separate them from the airglow lines. These lines should not be used for the analysis without additional bespoke procedures, such as the local background sky subtraction. 

Small differences in absolute flux are sometimes seen between the adjacent tiles outside the bright regions. These deviations from the median flux distribution are likely resulted from not ideal night-sky continuum subtraction and often correlate with the observation conditions. 

Several exposures exhibit systematic differences in fluxes between fibers from different spectrographs, which result in a prominent 3-sector pattern within the hexagonal LVM field. Such differences are likely resulted from imperfect flat-field correction for some exposures, and are typically present only in faint areas of DR20 tiles. One exception is the Flame nebula in Orion, hosting one of the brightest stars in the sky leading to the saturation across many fibers in one of the spectrographs (the nebula itself is not affected). 

For a few tiles, corresponding to relatively low surface-brightness areas, we noticed a non-physical line ratio of [NII] $\lambda$6584\AA/$\lambda$6548\AA\ emission lines significantly greater than 3. This issue seems to be resulted from suppression of [NII]$\lambda$6548\AA\ and does not affect other lines. This line ratio is fixed by atomic physics, and thus this issue should not affect any analysis relying on the brighter [NII]$\lambda$6584\AA\ line. Nevertheless, we warn the reader about this issue present in a small number of DR20 tiles. 

Finally, we note also that the telluric line correction is not applied in the current DRP version, and therefore the telluric absorption bands can be visible in red part of the stellar spectra of all released exposures. The telluric correction is being currently tested and will be implemented in the future versions.   

\section{Targets of Opportunity}\label{sec:ToO}

SDSS-V has introduced a new Target of Opportunity (ToO) mode of operation to obtain spectra of time-sensitive targets. In its current implementation, ToO targeting tags along with the main survey plan and replaces individual fibers in predefined survey fields from target lists that can be updated daily. This approach balances pre-planned survey priorities with new opportunities identified in photometric surveys and is ideal for observing transients selected uniformly over large sky area at relatively high density. This is because, unlike traditional ToO triggers, individual targets are not guaranteed and they cannot repoint the telescope to an unplanned field. This mode takes advantage of the flexibility of the FPS robotic positioning system and is in operation at both APO and LCO with the BOSS and APOGEE spectrographs. 

The SDSS ToO observing mode supports independent programs for target selection. To date, there are two ToO programs, one of which has supplied targets for spectra released in DR20. The relevant program, called \texttt{TNS}, selects targets based on alerts reported to the Transient Name Server (TNS) and targets them while they remain brighter than $g$ or $r<20.5$~AB~mag. Cuts are made to exclude targets with existing classification spectra or indications of regular AGN variability (characterized by a magnitude change of less than 1.5~AB~mag/year, but note that this does not completely remove AGN from targeting). One reduced spectrum is being released from this program, classified as a $z=1.04$ AGN. A second spectrum was taken on MJD 60454 as a test of the ToO targeting pipeline.

\section{Targeting}\label{sec:targeting}

In this section we describe major updates to the SDSS-V catalog and target databases, and changes to MOS target selection that were made between the targeting described in DR19 \citep{sdssVdr19}, and the last night of FPS operations that are included in this data release (MJD 60708, \texttt{robostrategy} plan \texttt{theta-2-boss-only-2}). 

For the initial iterations of SDSS-V MOS targeting (see DR18, DR19), the underlying cross-match of parent catalogs was anchored to the TESS Input Catalog \citep[][TIC v8]{Stassun_2019_TIC_v8}. The TIC is itself a compilation, for which the primary reference catalog is Gaia DR2.
Following the public releases of Gaia DR3 \citep{gaiadr3} and DESI LegacySurvey DR10 \citep{Dey_2019_DESIsurveys}, we carried out a major update to our underlying cross-match of parent catalogs. 
In this section we describe the newly added parent catalogs and the process of creating the \vonex. 

\subsection{New parent catalogs}\label{sec:catalogdb}

\subsubsection{LegacySurvey DR10}\label{sec:legacy_survey_dr10}
The LegacySurvey imaging catalogs \citep{Dey_2019_DESIsurveys,Zenteno2025} provide an important resource for many SDSS-V programs which feature targets beyond the reach of Gaia, including optically faint targets and moderately resolved galaxies. The LegacySurvey DR10 update\footnote{\url{https://www.legacysurvey.org/dr10}} provides significant additional sky area relative to previous releases, as well as new $i$-band imaging (in addition to $grz$ and \textit{WISE} bands). LegacySurvey DR10 includes sky areas observed with the DECam instrument, so is limited to Southern and Equatorial skies. We have supplemented this coverage with the Northern part of the LegacySurvey DR9 catalog (mostly Dec$>$+32.375\,deg, derived from The Beijing-Arizona Sky Survey and Mayall z-band Legacy Survey). The DR9 catalog has a slightly different native data model, and so we have transformed this DR10 format (padded with null values where necessary) to create a uniform merged database table (named \texttt{legacy\_survey\_dr10} in \texttt{catalogdb}). To cleanly manage the small sky overlap between the DR10 and DR9 catalogs, we have assigned objects a \texttt{survey\_primary} flag, which separates the surveys along the Dec$=$+32.375\,deg parallel within the North Galactic Cap. We note that (due to schedule pressures) the version of LegacySurvey DR10 catalog loaded into the SDSS catalog database is a pre-release version (Aug\,2022) which differs slightly (over a few tens of 0.25$\times$0.25\,deg `bricks') from the official DR10 catalog that was released a few months later by the LegacySurvey team. 

\subsubsection{Deeper eROSITA catalogs}\label{sec:erosita_cat}
One of the most significant additions for BHM was an updated set of parent catalogs containing optical counterparts to X-ray sources detected in the eROSITA all sky survey (eRASS). These new catalogs (named \texttt{erosita\_superset\_v1\_*} in \texttt{catalogdb}) are based on deeper X-ray data (and/or updated reductions) than were previously available. This update includes wide area (hemisphere) catalogs of sources derived from eRASS:3 (18 month survey depth, M. Ramos-Ceja et al., submitted) provided by the eROSITA AGN, Compact Object and Stellar working groups, and a catalog provided by the eROSITA Galaxy Clusters working group that is derived from the eRASS:4 survey (24 month depth). We also ingest several smaller specialized eROSITA parent catalogs derived from (i) very deep eROSITA data near the South Ecliptic Pole, (ii) time variable X-ray sources selected by comparing X-ray measurements between independent (one pass every 6 months) eROSITA sky surveys, and (iii) sources detected in the eROSITA hard X-ray band \citep[2.3--5\,keV;][]{Waddell2026}. All eROSITA catalogs are limited to the sky hemisphere where MPE controls the data rights (approx. $180<l<360$\,deg). Optical counterparts for these X-ray sources were selected from either the LegacySurvey DR10 catalog (supplemented with LegacySurvey DR9 in the North, see above), or from the Gaia DR3 catalog. The derivation of optical counterparts to X-ray sources (a context dependent task), was carried out by the individual eROSITA working groups before catalogs were ingested into the SDSS-V database.

\subsubsection{Updated Chandra Source Catalog parent sample}\label{sec:csc_v2p1}
In order to take advantage of recent Chandra X-ray data and reduction algorithms, as well as improved 
supporting optical surveys, we have again updated the parent catalog from which the BHM CSC target cartons (\texttt{bhm\_csc*}) are drawn. 
Our starting point is a sample of 362\,310 CSC\,v2.1 X-ray sources extracted from the production database of 14 November 2022,
restricted to sources with positive flux in at least one X-ray band \citep{Evans_2024}.
To identify optical/infrared counterparts
suitable for SDSS-V spectroscopic follow-up, we used the NWAY
probabilistic cross-matching algorithm \citep{Salvato2018} to match X-ray sources
to Gaia DR3 \citep{gaiadr3}, LegacySurvey DR10 \citep{Dey_2019_DESIsurveys}, Pan-STARRS1 DR2 \citep{Chambers2016_ps1},
and 2MASS \citep{Skrutskie_2006}. To remain within the magnitude limits
of the SDSS-V Black Hole Mapper, we restricted Gaia counterparts to
$14 < G < 20$ (Vega) and Legacy Survey or Pan-STARRS1 counterparts to
$14 < {\rm mag} < 21.5$ (AB) in at least one of the $g$, $r$, or $z$
bands. We additionally retained 2MASS counterparts with $H \le 14$ (Vega) to
enable APOGEE follow-up. We find 188\,647 Chandra sources meeting these criteria. 

\subsubsection{Compilation of pre-FPS optical spectroscopy}\label{sec:prefps_zcomp}
In previous iterations of BHM target selection, the process of determining which potential targets had previously been observed with good quality SDSS optical spectroscopy had become quite cumbersome, involving complex database joins and filtering logic.  
In an effort to simplify this process we have created a new catalog table \texttt{dr20\_sdss\_dr19p\_speclite} which documents all SDSS multi-object optical spectroscopic observations carried out until June 2021 (i.e. up to the end of plug plate observations at APO, and before the start of FPS operations). This table combines information from i) the final 
optical spectroscopy catalog released at the end of SDSS-IV\footnote{supplemented with \texttt{PLATE-MJD} = 10658--58439 from DR16} \citep{Abdurrouf_2021_sdssDR17}, 
ii) the catalog of stacked spectra in the eFEDS field 
released as part of DR18 \citep{almeida2023}, and iii) a catalog based on early reductions of all year-1 SDSS-V optical plate-based observations, a dataset that was released later (via an updated pipeline) as part of DR19 \citep{sdssVdr19}.
In cases where a single astrophysical object had been observed multiple times, we selected the `best' spectrum per object via a heuristic algorithm based on SNR, redshift warnings and plate quality, finding 6\,103\,248 entries for 5\,197\,982 distinct astrophysical objects. This compilation catalog was connected to the central catalog cross-match table in a non-standard way, reflecting its most common use as a veto table rather than as a parent catalog from which targets could be selected. This results in the \texttt{dr20\_catalog\_from\_sdss\_dr19p\_speclite} table, in which we have exactly one entry per spectrum listed in \texttt{dr20\_sdss\_dr19p\_speclite}.

\subsection{Updated cross-match algorithm}\label{sec:crossmatch}

Those cartons which were carried over unmodified w.r.t. earlier versions, and were recreated via the new \vonex, have received new version numbers (plans).  In general, the metadata for those targets (coordinates, magnitudes etc) may have changed slightly w.r.t. earlier iterations. In a very small number of cases, targets will have slipped into or out of cartons, despite the underlying sample remaining the same. This is an inevitable result of the new crossmatch re-assessing which pairs of catalog entries constitute single astrophysical objects, and which should be considered as separate targets.

\subsection{Updated MOS targeting generations}\label{sec:targeting_generations}
A `targeting generation' describes a well-defined collection of target cartons (and their meta-data) which informed SDSS survey design at some point. In addition to the targeting generations described in earlier SDSS-V data releases \citep{sdssVdr19}, we newly release six targeting generations that have influenced the survey designs included in DR20. 
Targeting generation \texttt{v1.0.2} was the first to be based on the \vonex\ (i.e. target selection that was founded on Gaia DR3 and Legacy Survey DR10, see section \ref{sec:crossmatch}), and was used by the \texttt{eta-3}, \texttt{eta-4} and \texttt{eta-5} robostrategy runs. 
Targeting generations \texttt{v1.0.3} and \texttt{v1.0.4} were incremental updates used by the \texttt{eta-6}, \texttt{eta-7}, \texttt{eta-8}, and \texttt{eta-9}   \texttt{robostrategy} runs. 
Targeting generation \texttt{v1.0.5} was a moderate update in which, in order to better match the updated survey exposure time allocation, some selected BHM target cartons were re-implemented to allow a fallback to shorter exposure times.
Targeting generation \texttt{v1.0.6-boss-only} was briefly used for LCO observations during APOGEE-S maintenance in early 2025; it includes only cartons containing BOSS targets.
Targeting generation \texttt{dr20.manual} collates a few cartons that were used in manually created (i.e. outside \texttt{robostrategy}) FPS designs.

\subsection{Updated survey plans and assignments}\label{sec:robostrategy}
As described by \citet{sdssVdr19}, we use the software tool \texttt{robostrategy} \citep{Blanton2025} to plan the SDSS-V FPS MOS survey. \texttt{robostrategy} algorithmically allocates observing resources to \texttt{field}s on the sky, and makes assignments of fibers to targets within the \texttt{designs} associated with each of those \texttt{field}s. At any one point in time, an observatory will observe \texttt{design}s taken from a single \texttt{robostrategy} `\texttt{plan}', but we have updated these \texttt{plan}s as the survey has progressed. These updates have been mainly driven by i) updates to target selection, ii) re-evaluation of survey speed, and iii) algorithmic changes to target assignment logic. The FPS observations that are being released as part of DR20 have been been carried out at APO using the \texttt{zeta-0}, \texttt{zeta-3}, \texttt{eta-5}, \texttt{eta-6}, \texttt{eta-8}, \texttt{eta-9},  and \texttt{theta-1} \texttt{plan}s. 
At LCO we have used the \texttt{zeta-4}, \texttt{eta-5}, \texttt{eta-6}, \texttt{eta-7}, \texttt{eta-8}, \texttt{eta-9}, \texttt{theta-1} and \texttt{theta-2-boss-only-2} \texttt{plan}s. These \texttt{robostrategy} \texttt{plan}s, and the dates when they were introduced, are documented by \citet[][their section 6 and table 1]{Blanton2025}.

\section{Value-Added Catalogs and Products}\label{sec:vacs}

In this section, we describe the eighteen VACs and one VAP released as part of DR20. The VAP and fourteen of the VACs are new to DR20, while four VACs are updated (Sections \ref{vac:quasar_props} and \ref{vac:spiders_dl1}) or extended from DR19 to include BOSS stars for DR20 (Sections \ref{vac:minesweeper_params} and \ref{vac:occam}). All VACs can be found through the SDSS website at \url{https://www.sdss.org/dr20/data_access/value-added-catalogs/}. The VAP can be found at \url{https://dr20.sdss.org/sas/dr20/vap/lvm/hips/rgb_Halpha_OIII_SII/v1/}.

\newcommand{\sasenvbase}{https://dr20.sdss.org/sas/dr20/env}
\newcommand{\sasenvfull}[2][]{
  \if\relax\detokenize{#1}\relax \expandafter\href\expandafter{\sasenvbase/#2/}{#2}%
  \else\expandafter\href\expandafter{\sasenvbase/#2/#1/}{#2/#1}%
  \fi
}
\newcommand{\sasenv}[2][]{
  \if\relax\detokenize{#1}\relax \expandafter\href\expandafter{\sasenvbase/#2/}{#2}%
  \else\expandafter\href\expandafter{\sasenvbase/#2/#1/}{#2}%
  \fi
}

\begin{deluxetable}{lcccc}
\tablecaption{Summary of VACs and VAP released in DR20 \label{tab:vac_sum}}
\tabletypesize{\small}
\tablehead{
\colhead{Name} & \colhead{Data Product} & \colhead{CAS Tables} & \colhead{Sec.}
}
\startdata
Quasar Spectra Property Catalog  & \sasenv{BHM\_QSOPROP}  & \texttt{DR20Q\_prop}  & \ref{vac:quasar_props}\\
\shortstack[l]{eFEDS SPIDERS: Optical Spectral\\ Fitting of X-ray Selected AGN}  & \sasenv{SPIDERS\_AGN}  & \texttt{efeds\_spiders\_agn}  & \ref{vac:efeds_spiders}\\
\shortstack[l]{Spectral Identification and Properties\\ of eROSITA DL1 Sources}  & \sasenv{DL1\_SDSS\_EROSITA}  & \texttt{DL1\_eROSITA\_eRASS3}  & \ref{vac:spiders_dl1}\\
\shortstack[l]{Disentangling the Influence of the\\ Host Galaxy in Quasar Spectra}  & \sasenv[qms\_hg]{BHM\_QSO\_HOST}  & \texttt{qms\_hg\_index\_diagram}  & \ref{vac:host_galaxy_spectra}\\
\shortstack[l]{Updated Redshifts and Classifications\\ of Fermi-detected Sources}  & \sasenv[boss\_fermi\_redshifts]{BHM\_BLAZAR}  & \texttt{fermi\_blazar}  & \ref{vac:fermi_redshifts}\\
Visual Inspection of BOSS Spectra  & \sasenv{BHM\_VI}  & \texttt{boss\_vi\_results}  & \ref{vac:inspect_boss}\\
\shortstack[l]{Catalog of DA White Dwarf\\ Binary Candidates}  & \sasenv[da\_white\_dwarf\_binaries]{MWM\_WHITEDWARF}  & \shortstack[l]{\tt{DA\_DWD\_Candidates,} \\ \tt{DA\_DWD\_RVs}} & \ref{vac:wd_binary}\\
Catalog of eROSITA Observed CVs  & \sasenv[eROSITA\_CVs]{MWM\_WHITEDWARF}  & \texttt{eROSITA\_CVs}  & \ref{vac:erosita_cvs} \\
Gyrochronology Ages for MWM Stars  & \sasenv[gyro\_age]{MWM\_STELLAR\_AGE}  & \tt{gyro\_age\_dwarf}  & \ref{vac:gyrochronology_mwm}\\
\shortstack[l]{MINESweeper Stellar Parameters\\ for BOSS Halo Stars}  & \sasenv{MWM\_MINESWEEPER}  & \texttt{minesweeper}  & \ref{vac:minesweeper_params}\\
\shortstack[l]{NLTE Abundances and Corrections\\ for APOGEE RGB Stars}  & \sasenv[apogee\_rgb\_nlte\_abundances]{MWM\_RGB}  & \texttt{payne4GAIN\_summary}  & \ref{vac:nlte_apogee}\\
\shortstack[l]{Open Cluster Chemical Abundance\\ and Mapping Catalog}  & \sasenv{BOSS\_OCCAM}  & \shortstack[l]{\tt{boss\_occam\_cluster,} \\ \tt{boss\_occam\_member}}  & \ref{vac:occam}\\
\shortstack[l]{Pseudo-Continua for M Dwarf\\BOSS Spectra}  & \sasenv[mdwarf-contin]{MWM\_MDWARF}  & \tt{mdwarf\_contin}  & \ref{vac:mdwarf_continua}\\
\shortstack[l]{Sodium ISM Absorption in\\MWM BOSS}  & \sasenv[NaI\_ISM]{MWM\_ISM}  & \texttt{boss\_ISM\_NaI\_absorption}  & \ref{vac:sodium_ism_mwm}\\
\shortstack[l]{Spectral Subtypes, Morphology, and\\ H$\alpha$ Emission for K \& M dwarfs} & \sasenv[boss\_spectral\_activity]{MWM\_MDWARF}  & \texttt{mdwarf\_active\_params}  & \ref{vac:classify_kmdwarfs}\\
Stellar Orbits for SDSS-V DR20  & \sasenv{MWM\_ORBITS}  & \texttt{GravPot16}  & \ref{vac:galactic_orbital_elements}\\
\shortstack[l]{Stellar Parameters from BOSS\\ Spectra with The CLAM}  & \sasenv[boss\_clam]{MWM\_STELLAR\_PARAMS}  & \texttt{boss\_clam\_params}  & \ref{vac:clam_parameters}\\
\shortstack[l]{The Large-scale kinematics of\\ Young Stars in disk}  & \sasenv[ob-kin]{MWM\_YSO}  & \tt{yso\_ob\_kin}  & \ref{vac:yso_kinematics}\\
\shortstack[l]{LVM HiPS RGB Map} & \sasenv[rgb\_Halpha\_OIII\_SII]{LVM\_HIPS\_VAP}  & N.A. & \ref{vac:LVM_map}\\
\enddata
\tablecomments{
A complete listing of SDSS VACs is available at \url{https://www.sdss.org/dr20/data_access/value-added-catalogs/}.
The data volume used by each of the data products listed in the above table may be found at \url{https://www.sdss.org/dr20/data_access/volume/}
}
\end{deluxetable}

\subsection{BHM VACs}

\subsubsection{DR20 Quasar Spectra Property Catalog}\label{vac:quasar_props}

The large volume of quasars observed by the SDSS has significantly advanced our understanding of the quasar population and the growth of SMBHs \citep[e.g.,][]{York_2000_sdss, Schneider_etal_2010, Dawson2013_boss, Lyke2020}. Building on previous SDSS quasar VACs \citep[e.g.,][]{Shen_etal_2011, Wu&Shen2022}, we present uniform SDSS DR20 quasar spectral measurements in this section.
We utilize the epoch-coadded spectra to improve the spectral S/N and separate the sample into quasar-dominated and host-dominated AGN populations (see \S~\ref{vac:host_galaxy_spectra}). The parent sample is constructed from the allepoch catalog, from which we exclude objects with unreliable redshift estimates ($\rm ZWARNING >0$) and very low redshift ($z<0.01$). We define the primary quasar sample as objects with $z>0.94$, where the spectra are dominated by quasar emission. For the remaining low-redshift objects, we apply additional selection criteria based on morphology and color to identify quasar-dominated sources. Specifically, we select point-like sources using the $i$-band PSF information and adopt a color cut $g-r < 0.4 + z$ to isolate blue quasars. The final sample consists of 243,258 targets.

We measure quasar spectral properties using a uniform fitting procedure based on the publicly available code PyQSOFit \citep{PyQSOFit}, following previous SDSS VAC analyses \citep[e.g.,][]{Shen_etal_2011, Wu&Shen2022}. The spectral modeling includes a power-law continuum, a third-order polynomial continuum, Fe{\sc ii} emission templates, and multi-component emission-line fitting. We provide the emission-line measurements, including peak wavelength, line fluxes, luminosities, FWHM, equivalent widths (EW), and centroid wavelength. Bolometric luminosities are derived from monochromatic continuum luminosities using the corrections of \citet{Richards_etal_2006}. SMBH masses are estimated using single-epoch virial relations based on the broad H$\beta$, Mg\,{\sc ii}, and C\,{\sc iv} emission lines and their corresponding continuum luminosities \citep{Vestergaard&Peterson2006}. 

\subsubsection{eFEDS SPIDERS: Optical spectral fitting of X-ray selected AGN}
\label{vac:efeds_spiders}

The collaboration between eROSITA and SDSS allows the astronomical community interested in AGN to have access to optical spectra of an X-ray selected sample, which is more diverse than optically-selected samples in terms of obscuration and contribution from the host galaxy \citep[e.g.,][]{HickoxAlexander2018}.
In the context of the SPIDERS program (see Section \ref{sec:bhm_spiders}), the eROSITA Final Equatorial Depth Survey \citep[eFEDS;][]{Brunner2022, Salvato2022} was the first eROSITA sample to be publicly released, and the SDSS follow-up of these X-ray-detected sources has also been available since DR18 \citep{almeida2023}.
Therefore, the eFEDS sample has extended multiwavelength coverage and many associated works based on this rich dataset,\footnote{\href{https://erosita.mpe.mpg.de/edr/eROSITAObservations/Catalogues/}{https://erosita.mpe.mpg.de/edr/}} with the specific analysis of the SDSS optical spectra in connection to the eROSITA X-ray data provided by \citet{Aydar2025}.

For the eFEDS SPIDERS VAC, we provide optical spectral measurements 
for the $\sim$13k AGN spectra with reliable redshift estimates presented in \citet{Aydar2025}.
To fit this diverse sample, we divided the data into quasar-dominated sources and sources with a significant contribution from the host galaxy.
The quasar-dominated AGN are those with $z>1$ or the point-like blue objects (according to Legacy Survey, \citealt{Dey_2019_DESIsurveys}) at $z\leq1$, and they were fitted using \texttt{PyQSOFit} \citep{PyQSOFit, Wu&Shen2022}.
The AGN that require a host galaxy decomposition are the red or extended sources at $z\leq1$, and they were fitted first with \texttt{pPXF} \citep{Cappellari2023, Bernal2025}.
The detailed methodology is described in \citet{Aydar2026}.

This VAC contains measurements of the continuum (e.g., power-law featureless continuum index, Fe II pseudo-continuum FWHM), emission lines (e.g., flux, equivalent width, FWHM, central wavelength), and, when possible, information regarding the stellar population (e.g., stellar mass, stellar velocity dispersion, average age and metallicity). 
We also include public data from eROSITA \citep{Brunner2022, LiuT2022} and from photometric counterparts \citep{Salvato2022}.
The data model and full description of the VAC will be provided in Aydar et al. (in prep) and on the SDSS VAC website (see Section \ref{sec:vacs} for the link).

\subsubsection{Spectral Identification of eROSITA Sources DR20: Optical and X-ray properties (DL1)}
\label{vac:spiders_dl1}

Following the collaboration between eROSITA \citep{Predehl2021, Merloni2024} and SDSS-V in the SPectral IDentification of ERosita Sources project (SPIDERS, Merloni et al. in prep.), as already presented in the SDSS DR19 \citep{sdssVdr19}, we make available the Data Level 1 (DL1) of sources that were targeted due to their eROSITA detection in the X-rays and followed up with SDSS to provide optical spectra.
This provides a unique opportunity to study a large sample of point-like X-ray-emitting sources, with crucial information that optical spectra provide, such as redshift, object class (e.g., quasars, stars, and compact objects), and more precise sky positions.
The fact that the same instruments were used for the observations avoids issues arising from different calibrations and telescope technical properties, which are often encountered in multiwavelength samples.

The main difference between SPIDERS DL1 from DR19 and DR20 is that we now consider results from the eROSITA All-Sky Survey for the first three scans of eROSITA, the so-called eRASS:3, to be released in eROSITA DR2 (Ramos-Ceja et al., submitted), instead of only relying on the results from the first scan (eRASS1) that were released in eROSITA DR1 \citep{Merloni2024}.
For sources that do not have eRASS:3 information available, we provide the eRASS1 data.
Hence, although the information provided by the SDSS has a similar structure, the data from eROSITA are more complete, as they include more observations of the same sources.
We note, however, that time-domain information is not available at the DL1, and that extended sources (e.g., galaxy clusters and supernova remnants) are treated differently since the measured X-ray flux depends on a modeled radial profile and redshift and cannot be reliably extracted from the eROSITA pipeline-generated catalogs, and therefore are also not included in the DL1.

We provide information for SDSS \texttt{daily} (individual optical observations for each source) and \texttt{allepoch} (stacked optical observations of the same object) observations separately.
The \texttt{daily} catalog contains 478\,255 spectra from 271\,510 sources, compared to the $\sim15$k spectra from DR19.
Regarding the \texttt{allepoch} catalog, we provide 263\,310 spectra, compared to less than 30k spectra from DR19.
This significant difference in the number of SPIDERS sources between DR19 and DR20 is related to LCO observations in the Southern Hemisphere, where most of the German-eROSITA targets are located.

For each X-ray point-like source detected with eROSITA and followed up by SDSS-V, we provide:

\begin{itemize}
    \item From SDSS: field, MJD, catalog ID, run2D, target index, object type, target flag, fiber position coordinates, redshift (with error and warning), median S/N, class, and best fit stellar template and properties (spectral type, effective temperature, and surface gravity from Astra).
    \item From eROSITA: unique X-ray source identifier (\texttt{DETUID}), flux in the full eROSITA band and error, MJD, morphological classification, source detection likelihood, X-ray position estimate, and positional error.
\end{itemize}

When available, we also provide public photometric data to complement the multiwavelength information, derived from Gaia DR3 \citep{gaiadr3}, unWISE \citep{unwise}, GALEX \citep{galex}, and 2MASS \citep{Skrutskie_2006}.

More details on the methodology for constructing eROSITA-SDSS Data Level 1 will be provided in Merloni et al. in prep., where the SPIDERS sample targeting and data are described.

\subsubsection{The index diagram as a tool to disentangle the influence of the host galaxy in quasar spectra}\label{vac:host_galaxy_spectra}

It is well known that the decoupling between active galactic nuclei (AGN) and their host galaxies (HG) remains a challenging task due to the degeneracy between the stellar continuum and the AGN power-law emission. Therefore, it is crucial to develop methods that allow the identification of AGN spectra with significant stellar contributions, while simultaneously quantifying the level of HG contamination affecting the AGN continuum. The availability of spectra free of HG emission permits a more precise determination of the intrinsic properties of active galactic nuclei.

The main objective of the index diagram introduced in \citet{Negrete2025MNRAS.543.4272N} is to provide a novel, model-independent diagnostic tool based only on redshift-corrected spectra, designed to efficiently identify sources with significant stellar contributions from the host galaxy.

\citet{Negrete2025MNRAS.543.4272N} presents a detailed description of the data processing, the construction of the index diagram using the Ca II $H$-band index and the H$\beta$ Lick index, and the subsequent analysis in the context of the so-called Quasar Main Sequence (QMS), which relates the FWHM of the broad H$\beta$ component (H$\beta_{BC}$) with the relative intensities of FeII and emission H$\beta_{BC}$ (R$_{FeII}$ = I(FeII)/I(H$\beta_{BC}$)). The analysis is further complemented by trends observed at radio, X-ray, and infrared wavelengths.

In brief, after identifying spectra with broad emission lines through a line-ratio method, a sample of $\sim$ 3000 high S/N sources ($>$ 20 in the continuum) with redshift $z <$ 0.9 was selected from the northern hemisphere. The index diagram was then constructed and used to classify AGN spectra into three categories according to the relative dominance of AGN and HG emission: HG-dominated (HGD), intermediate (INT), and AGN-dominated (AGND) sources. A colour–$z$ diagram was further employed to refine this classification. Following stellar subtraction, “pure” AGN spectra were obtained. Spectral modelling of the AGN components provided measurements for the construction of the QMS. Additionally, black hole masses, continuum luminosities, and Eddington ratios (defined as the ratio between bolometric and Eddington luminosities) were derived.

The QMS analysis shows that the majority of HGD sources exhibit negligible FeII emission (R$_{\rm FeII} \sim$ 0), with a few outliers (R$_{\rm FeII} >$ 1) likely affected by residuals from the HG subtraction and a weak H$\beta_{BC}$ contribution. INT and AGND sources show similar distributions in the QMS for FWHM(H$\beta_{BC}$) $<$ 4000 km s$^{-1}$, while a tail of AGND sources becomes evident at larger FWHM(H$\beta_{BC}$). Cross-matching with radio, infrared, and X-ray catalogues reveals that the strongest radio emitters are predominantly associated with HGD and INT sources, whereas strong X-ray emitters are mainly found among INT and AGND objects, which also occupy the AGN region in infrared colour diagrams.

The value-added catalogue associated with \citet{Negrete2025MNRAS.543.4272N} consists of two datasets. The first dataset includes spectral measurements for the sample described above, such as line fluxes used to identify broad emission line spectra, index measurements, QMS parameters, derived quasar properties (including black hole mass and Eddington ratio), and multiwavelength information. The second dataset comprises $\sim$ 20,000 objects selected from the BHM sample, classified as QSOs, with $z < 0.9$ and SNR(median) $>$ 15. This dataset provides measurements of the Ca II $H$ and H$\beta_{BC}$ indices for an extended sample, including sources from the southern hemisphere. This VAC enables systematic studies of AGN–HG decomposition and provides a robust framework for future investigations of quasar properties and their role in galaxy evolution, which will be explored in detail in forthcoming papers.

\subsubsection{Updated redshifts and classifications of Fermi-detected sources within SDSS-V}\label{vac:fermi_redshifts}

The SDSS automated spectroscopic pipeline classifies sources using stellar, galaxy, and quasar templates without explicitly accounting for any non-thermal jet continuum component. This often leads to incorrect source classification and incorrect redshift of jet-dominated sources, particularly in the cases of blazars, a special subclass of radio-loud AGN where one of the collimated relativistic jets arising from the proximity of the black hole is oriented close to our line of sight, leading to a Doppler-boosted non-thermal emission across the entire electromagnetic spectrum. In essence, a large number of strongly jet-dominated sources are misidentified as white dwarfs (that produce similarly blue and relatively featureless optical spectra) and assigned to the STAR class. 

Blazars are traditionally classified by optical emission line strength into two main groups: Flat Spectrum Radio Quasars (FSRQs); which exhibit strong broad-line region emission (rest-frame ${\rm |EW|} > 5$ \AA) and  BL Lacertae objects (BL Lacs), which show weak or absent emission features (${\rm |EW| < 5}$ \AA), in both cases associated with a strong, very blue, power-law continuum produced by the jet \citep{stickel1991complete}. In Nlowie et al. (in preparation), we cross-match the SDSS-V DR20 (spAll-lite-v6\_ 2\_1) spectroscopic catalog with the optical counterparts to $\gamma$-ray sources from the \textit{Fermi} 4FGL-DR4 catalog \citep{ballet2023fermi}, yielding 746 unique counterparts with SDSS-V spectroscopy. Examination of their mid-infrared WISE colors confirms that the vast majority occupy the canonical blazar locus \citep{d2012infrared, massaro2011identification}, demonstrating their extragalactic origin despite frequent SDSS stellar or galaxy classifications.    

We implement a multi-component spectral fitting approach combining a flexible dual power-law (jet), galaxy templates from SWIRE \citep{polletta2007spectral}, and synthetic QSO templates \citep{temple2021modelling} to model jet, host 
galaxy, and broad-line region contributions simultaneously. Six model families (Galaxy, QSO, Powerlaw, Powerlaw+Galaxy, Powerlaw+QSO, and Powerlaw+emission-lines) are evaluated over a logarithmic redshift grid ($z = 0.01$--$5.0$) using Bayesian marginalisation and AICc model selection. The jet continuum is described by a flexible dual power-law of the form $F(\lambda) \propto (\lambda/\lambda_0)^{-\alpha}$, parameterised by a base slope $\alpha$ and a curvature parameter $\delta$, which controls the transition between two spectral regimes: $\delta > 0$ produces a spectrum that flattens towards shorter wavelengths, while $\delta < 0$ produces a steepening, allowing the model to capture both the canonical flat-spectrum and steep-spectrum radio source behaviour in the optical. 

Of the 746 \textit{Fermi}-matched sources, 732 achieve well-constrained fits ($\chi^2_{\rm r,\,lmfit} \leq 3\times \chi^2_{\rm r,\,SDSS}$), of which 707 are confirmed \textit{Fermi} blazar candidates. We classify 425 ($60.1\%$) as BL Lac candidates (420 Powerlaw+Galaxy, 5 Powerlaw) and 266 ($37.6\%$) as FSRQ candidates (106 Powerlaw+QSO, 160 Powerlaw+emission-lines), with 16 ($2.3\%$) best fitted by pure Galaxy or QSO models indicating negligible jet contribution at optical wavelengths. Of the 707 classified sources, 597 ($84.4\%$) have median spectral ${\rm S/N} \geq 3$ as reported by the SDSS pipeline; the remaining 110 sources ($15.6\%$) have ${\rm S/N} < 3$ and are retained in the catalogue but carry lower confidence in their redshift and jet fraction estimates. We validate a sample of sources ($111/707$) with pre-existing redshifts and also make the quality threshold cut ($\chi^2_{\rm r,\,lmfit} \leq 3\times \chi^2_{\rm r,\,SDSS}$; S/N $\geq$ 3) in the Third Catalog of Hard \textit{Fermi}-LAT Sources \citep[3FHL;][]{ajello20173fhl}, which achieves a 4.4\%  reduction in the catastrophic outlier fraction ($\eta = 0.387$ versus $\eta = 0.432$ for the SDSS pipeline). 

In this value-added catalogue (VAC) we provide all 707 
\textit{Fermi}-classified blazar sources, including: (1) best-fit spectral model and blazar subclass; (2) updated spectroscopic redshift with uncertainty; (3) reduced $\chi^2$ for the multi-component model fit; (4) optical jet flux fraction; (5) power-law slope parameters $\alpha$ and $\delta$ describing the jet continuum shape. This work represents the largest systematic reclassification of \textit{Fermi} blazars within the SDSS-V DR20 spectroscopic sample.

\subsubsection{Visual Inspection of DR20 BOSS spectra}\label{vac:inspect_boss}
In general, the \texttt{idlspec1d} pipeline used to determine redshifts from BOSS 1D
spectra is robust and highly reliable. However, in a small but significant fraction of cases, 
inspection reveals that the fitted redshifts are actually incorrect. 
Most failures can be attributed to one or more of the following reasons: i) low SNR data, ii) instrumental or reduction artifacts that imprint spurious features in the reduced data (sky subtraction, flux calibration etc), and iii) astrophysical objects that exhibit spectral shapes/features (or a lack of features) that cannot be effectively reconstructed by the template library (and polynomial terms) used by \texttt{idlspec1d}.  
In some, but not all, of these problematic cases, one can identify spectra with less reliable pipeline redshifts via the \texttt{ZWARNING} flags reported by \texttt{idlspec1d}. 
To increase the reliability of redshifts in
general, and in the BHM samples in particular, we have conducted a dedicated program of visual
inspection of DR20 BOSS spectra. The inspections were carried out by a team of more than 20 volunteers from the BHM team, 
via a dedicated web app (adapted from a tool previously used for inspecting SDSS spectra in the eFEDS field, \citealt{Salvato2022}).
We limited our inspections to a subset of the DR20 `allepoch' coadded BOSS
spectra. We initially inspected a diverse selection of the available spectra, but then focused efforts on a subset of spectra lying in parts of parameter space where we expected a relatively high pipeline redshift failure rate but where redshifts were frequently recoverable via visual inspection.  When possible, we supplemented our own inspections with visually inspected redshifts taken from the SDSS DR16Q QSO catalog \citep{Lyke2020}, but only in cases where the redshift provided in \citet{Lyke2020} differed by less than 0.01 from the `allepoch' pipeline redshift. 
In total, we have collated 50\,758 visual inspections of 45\,780 DR20 `allepoch' BOSS spectra (9\% of the parent sample).

This VAC provides a supplement to the
DR20 allepoch spAll catalog. For each entry in the VAC, we flag if the source has been
visually inspected, give the redshift from the visual inspection, the
confidence on that redshift, and the visual classification. In addition, using
the visually inspected spectra as a training sample, and taking only features 
from the \texttt{spAll-v6\_2\_1-allepoch.fits} catalog, we built a
machine learning (ML) classifier to predict the reliability
of the remaining un-inspected spectra.
We judge a pipeline redshift ($z_\mathrm{pipe}$) to be `reliable' when it is within 1\% (specifically $|z_\mathrm{pipe} - z_\mathrm{VI}|/(1+z_\mathrm{VI}) < 0.01$) of a secure visually inspected redshift ($z_\mathrm{VI}$).
Those ML Classifier quality flags are included in this VAC.
By combining visual inspections and the ML Classifier, we estimate that 448449/506550 (89\%) of the redshifts derived from the `allepoch' spectra are reliable. Within the subset flagged as reliable, we estimate a false positive rate of around 1.3\%.
This VAC will be described in more detail by Merloni et al. (in prep.).

\subsection{MWM VACs}

\subsubsection{Catalog of DA White Dwarf Binary Candidates from SDSS-V DR19}\label{vac:wd_binary}
The selection of unresolved short-period double white dwarf binaries (DWDs) using radial velocity (RV) measurement of multi-epoch SDSS spectroscopic data has been successfully carried out over the past iterations of SDSS and we extend this to SDSS-V in this VAC \citep{badenes_first_2009,yan_search_2024}. The details of the selection are presented in \cite{adamane_pallathadka_double_2026}.

We first select WD exposures classified as hydrogen atmosphere (DA) WDs by \texttt{SnowWhite} \citep{sdssVdr19}. To ensure accurate RV measurement, we discard exposures with SNR $<$ 3, and those with \texttt{SnowWhite} classification probability $>10\%$ to be a featureless DC WD, an extremely hot DO WD, a WD+main-sequence binary, or likely to be a cataclysmic variable (CV). We measure the RV of each SDSS exposure using Compact Object Radial Velocity package \citep[\texttt{CORV},][]{chandra_corv_2023,arseneau_measuring_2024}. We simultaneously fit the first four Balmer lines to 3D DA WD templates by \cite{tremblay_spectroscopic_2013} to measure the RVs and associated RV errors. We discard exposures with internally inconsistent RVs at 3$\sigma$ level by comparing the RV measured with H$_{\alpha}$ alone and the RV measured with H$_{\beta}$, H$_{\gamma}$, and H$_{\delta}$. We then perform pair-subtraction test to calibrate the RV errors and bring the measured RV error accuracy to within 10\% agreement with the expected RV errors based on the variation of RVs across different exposures of each WD (Adamane Pallathadka et al. 2026, accepted). We remove exposures with RV errors greater than 90 km s$^{-1}$, beyond which we cannot reliably calibrate RV errors.

Using the measured RVs and the well-calibrated RV errors, we look for statistically significant RV variation across different exposures. Following \cite{maxted_radial_2002} and \cite{breedt_using_2017}, we compute $\chi_m^2$, the chi-squared statistic for the RVs about their weighted mean for each WD. Under the null hypothesis that any RV scatter is due to spectral noise and no real motion, $\chi_m^2$ should follow a $\chi^2$ distribution with n-1 degrees of freedom, where n is the number of sub-exposures. From the measured $\chi_m^2$, we estimate the false-alarm probability P($\chi^2 > \chi_m^2$), and then define the RV variability parameter $\eta = -\log{[P(\chi^2 > \chi_m^2)]}$, which is always positive and larger $\eta$ values indicate a higher likelihood that the WD is in a binary.

We use $\eta$ to select binary candidates. We simulate the SDSS-V observation pattern, with realistically drawn cadence and RV errors, and calculate the resulting $\eta$ distribution due to statistical variation. We find that $\eta > 3$ is a good cutoff to select the binaries while minimizing false-positives. We report 60 high-confidence DWD binary candidates and expect about five false-positives.

In this VAC, we report the results of this analysis for all the systems selected, after selection cuts. For each object, we report the SDSS-ID, Gaia DR3 ID, and $\eta$. We also include the measured RV, RV error, and observation time. Full details of the selection, binary candidates, and the subsequent constraints on the binary population are presented in \cite{adamane_pallathadka_double_2026}.

\subsubsection{Catalog of eROSITA observed CVs in SDSS DR20}\label{vac:erosita_cvs}

SDSS-V optical spectroscopic follow-up observations of eROSITA identified X-ray sources provide a unique opportunity to uncover the nature of these objects. The vast majority of these sources are expected to be extragalactic in nature, such as AGN and QSOs, however a significant number of these objects will be galactic, most of which are expected to be coronal emitting stars. However, amongst the galactic sources we will find accreting compact objects, specifically cataclysmic variables (CVs). CVs are semi-detached binaries, consisting of a white dwarf (WD) primary, and a low-mass main sequence companion, which overflows its Roche-lobe, resulting in material from the companion accreting onto the WD, and in the process producing X-rays. As accretion occurs in all CVs, it therefore implies that all CVs are X-ray emitters. Here we present the methodology followed in creating the VAC `Catalog of eROSITA observed CVs in SDSS DR20', while the results will be published in Brink et al. (submitted).

Several strategies were followed to generate comprehensive target lists for spectroscopic follow-up with BOSS. The first strategy was to choose all eROSITA pointlike sources that were detected in eRASS1 (i.e. the first eROSITA survey) with a detection likelihood $>$ 8, in either of the 0.2-0.6 keV, 0.6-2.3 keV, and 2.3-5.0 keV bands, thus ensuring the inclusion of objects that are purely soft or purely hard X-ray emitters. We then identified as the optical counterpart to the X-ray source the closest Gaia object to the eROSITA X-ray position (within a radius of 30") that pass certain optical and X-ray criteria that is indicative of CV characteristics, including X-ray to optical flux ratio log(F$_\text{X}$/F$_\text{opt}$) $>$ -2.7 (as defined in \citealp{schwope24}). Lastly we imposed a distance cut to d $<$ 3000pc. A total of 78,252 targets were submitted to SDSS-V and formed the mwm\_erosita\_compact\_gen carton. 

The second approach was very similar to the first, however, instead of choosing the closest Gaia source fulfilling certain optical and X-ray criteria, now the most optically variable Gaia source within 30" from the X-ray position was identified as the likely optical counterpart. We used the same variability and non-variability relation as \citep{varproxy}, and identified 17,058 targets, which were submitted as the \texttt{mwm\_erosita\_compact\_var} carton. 

For the last approach, we used a training sample of known CVs to train a Bayesian machine learning algorithm (based on NWAY \citealp{NWAY}) to identify the likely Gaia optical counterpart to the eRASS:3 source (i.e., the stack of the first 3 eROSITA surveys). This was submitted as the \texttt{mwm\_erosita\_compact\_boss} carton, and consists of 11,113 CV candidates. We applied no distance cut to this carton. All the spectra obtained in these three cartons were visually inspected by us, and those being CVs were identified. We also attempted to subclassify the CVs with the aid of additional data available in the public domain, such as ATLAS and CRTS/ZTF light curves, as well as color-magnitude and color-color plots. A VAC of these CVs is released with SDSS-V DR20, and contains, for each object, the SDSS-ID, eRO-DETUID, Gaia DR3 ID, eRASS:3 coordinates, Gaia DR3 coordinates, flux measurements in eRASS:3, Gaia mean G magnitude, Gaia BP-RP magnitude, r$_{\text{geo}}$ distance, absolute G magnitude, X-ray luminosity, log(F$_{\text{x}}$/F$_{\text{opt}}$), X-ray hardness ratios, and, if known, the CV subtype and orbital period.

\subsubsection{Gyrochronology Ages for MWM stars}\label{vac:gyrochronology_mwm}
Gyrochronology \citep{Barnes2003}, which estimates the ages of dwarf stars from their rotation periods and temperatures, is currently one of the most reliable methods for determining ages of main-sequence stars. 
However, an incomplete theoretical understanding of stellar dynamos prevents gyrochronology from being derived purely from first principles. 
As a result, gyrochronology relations are primarily calibrated empirically.

To infer gyrochronology ages, we applied two empirically calibrated relations. 
The first, \texttt{GPgyro}\footnote{Available at \url{http://github.com/lyx12311/GPgyro}.} \citep{Lu2024}, is calibrated using open clusters \citep{Curtis2020} and gyro-kinematic ages \citep{Lu2021}. 
It is currently the only gyrochronology relation calibrated for stars older than 4 Gyr and applicable to both partially and fully convective stars. However, because the age–velocity-dispersion relation is unreliable for young stars, \texttt{GPgyro} is applicable only to stars older than 1.5 Gyr.
The second relation, \texttt{gyro-interp}\footnote{Available at \url{https://github.com/lgbouma/gyro-interp}.} \citep{Bouma2023}, is calibrated using open clusters. 
This method accounts for the bimodality of fast- and slow-rotating stars at early times where stars initially rotate rapidly due to angular momentum conservation and later converge onto the slower rotating sequence. 
Because it is calibrated solely on open clusters, \texttt{gyro-interp} is applicable only to stars younger than 4 Gyr, corresponding to the age of the oldest open cluster with rotation period measurements (M67).

\begin{figure}
    \centering
    \includegraphics[width=\columnwidth]{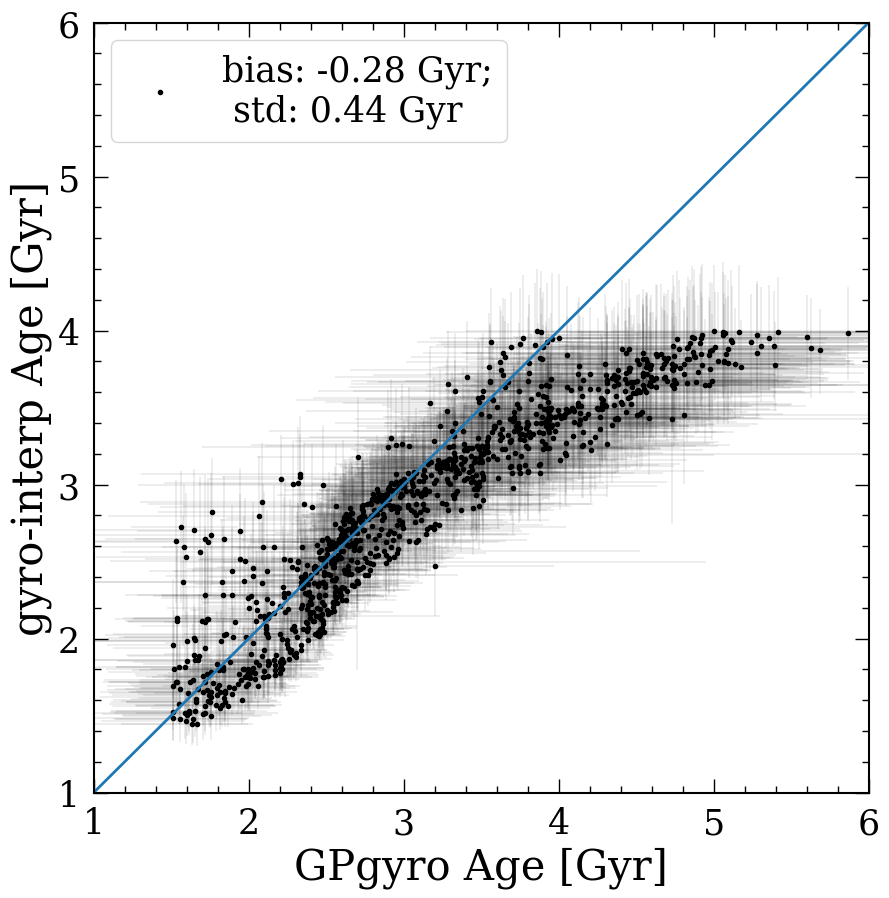}
    \caption{Comparison between gyrochronology ages obtained from \texttt{gyro-interp} and \texttt{GPgyro} in their applicable regime.
    The ages agree with each other within uncertainty.}
    \label{fig:vac_gyro}
\end{figure}

To construct the sample, we cross-matched the SDSS-V MWM catalog with rotation period catalogs \citep{Irwin2011, Berta2012, McQuillan2014, Santos2021, Holcomb2022, Lu2022, Colman2024} from the Zwicky Transient Facility Website \citep[ZTF;][]{Bellm2019, ztftime, ztfdata}, the Transiting Exoplanet Survey Satellite \citep[TESS;][]{TESS}, Kepler \citep{kepler}, and MEarth \citep{MEarth}.
Since gyrochronology can only be applied to dwarf stars, we only selected stars with absolute Gaia $g$ magnitude $>$ 4.2.
This left us with $\sim$11,000 stars.
We then applied both gyrochronology relations using the SDSS-V effective temperatures and the rotation periods described above.
\autoref{fig:vac_gyro} shows the comparison for the ages determined from both relations for the age range they are applicable for (\texttt{GPgyro} $>$ 1.5 Gyr; \texttt{gyro-interp} $<$ 4 Gyr).
The ages agree within uncertainty, with a bias of $-$0.25 Gyr and a standard deviation of 0.44 Gyr.

\subsubsection{MINESweeper Stellar Parameters for BOSS Halo Stars}\label{vac:minesweeper_params}

The Milky Way Mapper halo survey is expanding the Galactic frontier towards the most distant and metal-poor stars. 
For these stars, it is vital to incorporate all available information about a star when inferring their fundamental parameters---photometry, spectra, and astrometry. 
Furthermore, a homogeneous sample of stars with spectro-photometric distances is required to map the all-sky dynamics of the halo on the largest scales. 
The \texttt{MINESweeper} VAC of stellar parameters for BOSS halo stars is designed to address this. 

\cite{Chandra2026} describe the MWM \texttt{MINESweeper} VAC catalog in detail, including validation and preliminary results. 
Briefly, this catalog uses the \texttt{MINESweeper} code to infer stellar parameters for a subset of stars targeted by the halo survey. 
The \texttt{MINESweeper} code utilizes all available information about a star---the SDSS-V spectrum, broadband photometry from public surveys, and the \textit{Gaia} parallax---to find the best-matching stellar parameters \citep{Cargile2020}. Solutions are constrained to lie on isochrones, limiting the fit to physically-plausible regions of the parameter space. Priors can be placed on various explicit and implicit variables in the stellar model. \texttt{MINESweeper} delivers posterior distributions of the radial velocity $v_\mathrm{r}$, effective temperature $T_\mathrm{eff}$, surface gravity $\log{g}$, metallicity [Fe/H], [$\alpha$/Fe] abundance, and heliocentric distance.

\subsubsection{NLTE Abundances and Corrections for APOGEE RGB stars}\label{vac:nlte_apogee}

The spectral grids used by ASPCAP to fit DR19 APOGEE spectra \citep{Meszaros25_astra_aspcap} assume local thermodynamic equilibrium (LTE), that the level populations of the atoms in the stellar atmosphere can be expressed purely as a function of the local temperature. However, for luminous, red giant stars, interactions with non-local photons can violate this assumption, requiring calculation of non-LTE (NLTE) departures from level populations, which affect the strengths of atomic absorption features.

In optical spectra, NLTE chemical abundance trends have been shown to substantially differ from those derived with LTE \citep[e.g.,][]{bergemann_non-local_2017}, motivating the effort to understand how NLTE would change abundance trends in APOGEE H-band spectra. In DR17, \citet{2020A&A...637A..80O} calculated new spectral grids using TLUSTY and SYNSPEC \citep{2017arXiv170601859H,2021arXiv210402829H} and derived NLTE abundances for Mg, K, Na, and Ca, finding that the abundances for these elements did not differ by more than 0.1 dex between NLTE and LTE \citep{Abdurrouf_2022}. All abundances in DR19 use the LTE approximation \citep{Meszaros25_astra_aspcap}. 

We provide a VAC for DR20 with 1D-NLTE abundances and corrections for Mg, Mn, Al, Si, Ca, Ti, Ni, and Na for red giant spectra from DR19. We use the TSFitPy wrapper \citep{storm_observational_2023} of Turbospectrum NLTE \citep{gerber_non-lte_2023} to synthesize a grid of H-band spectra. The NLTE departure coefficients are precalculated from several analyses: H \citep{mashonkina_non-lte_2008}; Mg \citep{bergemann_non-local_2017}; Ca \citep{mashonkina_influence_2017,semenova_gaia-eso_2020}; Mn \citep{bergemann_observational_2019}; Na \citep{ezzeddine_empirical_2018}; Al \citep{ezzeddine_empirical_2018}; Ni \citep{bergemann_solar_2021, voronov_inelastic_2022}; Ti \citep{bergemann_ionization_2011}; Si \citep{bergemann_red_2013, magg_observational_2022}. We use the atomic and molecular linelists developed for APOGEE \citep{2021AJ....161..254S}. In ASPCAP, the atmospheric grids vary in \teff, \logg, [M/H], [C/M], [N/M], and [$\alpha$/M]. We use the standard MARCS stellar atmospheres \citep{gustafsson_grid_2008}, which only vary in \teff, \logg, and [M/H] ([C/M] and [N/M] are varied as a function of [M/H]).
Our abundances are on the \citet{2007SSRv..130..105G} solar abundance scale.

Our parameter range spans \feh\,from $-3.0$ to $+0.5$, \teff\,from $\sim3500$ K to $\sim5900$ K, and \logg\,from 3.5 to 1.0. We vary the NLTE elements above as well as C, N and O. Our grid is limited in \logg\, due to a lack of NLTE departure coefficients below $\logg < 1$. To interpolate the grid, following The Payne \citep{ThePayne2019ApJ...879...69T} we use a fully connected neural network emulator, which we call the Payne4GAIN (\code{p4g}, \textbf{G}iants in \textbf{A}POGEE \textbf{I}ncluding \textbf{N}LTE). We train an NLTE and LTE Payne4GAIN on $\sim 20,000$ spectra randomly sampled from within the parameter ranges, resulting in an emulator for Turbospectrum NLTE with a median pixel accuracy of $1\%$, i.e., only $\sim 1\%$ of pixels deviate by more than $1\%$ in normalized flux. We then use chi-squared minimization to fit DR19 APOGEE normalized spectra whose raw ASPCAP values fall within our grid range. We fit each spectrum with both the NLTE and LTE Payne4GAIN to derive NLTE-LTE corrections for each star. As the fits are noisy but the corrections are expected to be smooth, we fit the per-star corrections with polynomial functions of \teff\,, \logg\,, \feh\,, [X/H], and we apply these corrections to the ASPCAP raw abundances. Compared to ASPCAP, this fitting is performed simultaneously, instead of iteratively, and only takes seconds to converge per spectrum.

We present a VAC of $\sim 360,000$ stars, covering DR19 stars whose ASPCAP stellar parameters fall within our grid range. For each star, the VAC provides:
\begin{itemize}
    \item Payne4GAIN LTE stellar parameters, chemical abundances, and fitting flags
    \item Payne4GAIN NLTE stellar parameters, chemical abundances, and fitting flags
    \item Payne4GAIN polynomial NLTE-LTE corrected raw ASPCAP abundances
\end{itemize}
As well as the raw ASPCAP stellar parameters, chemical abundances, fitting flags and data flags for comparison. The stellar parameters and Fe, C, N, and O abundances do not differ significantly between the LTE and NLTE fits, and broadly agree with ASPCAP, but we include them for completeness. We do not find profound differences for Mg, Ca, Na, or Ni trends. We do find significant trend differences for Al, Mn, Ti, and Si.  
More details will be provided in \citet{thibodeaux2026}.

\subsubsection{Open Cluster Chemical Abundance and Mapping (OCCAM) catalog}\label{vac:occam}

Open clusters are important objects for tracing Galactic properties as well as calibrating survey parameters. 
The Open Cluster Chemical Abundances and Mapping (OCCAM) survey has been creating uniform infrared-based spectroscopic catalogs of open cluster member stars within SDSS surveys from SDSS-III/APOGEE-1 DR10 through to the present SDSS-V/MWM DR20 \citep[e.g.,][]{frinchaboy_13,cunha_16_grads,donor_18,donor_20,myers_22,otto_26}. 

As DR20 does not present any new APOGEE data, the previous APOGEE-OCCAM DR19 VAC\footnote{\url{https://www.sdss.org/dr19/data_access/value-added-catalogs/?vac_id=10017}} \citep{otto_26} is still the preferred high quality, multi-element catalog of cluster member stars and bulk parameters, due to it being based solely on stellar parameters and 15+ abundances measured from the high quality, higher resolution (R$\sim$22,500) APOGEE data for those clusters/stars that have APOGEE data. 

However, as a part of SDSS-V/MWM DR20, hundreds of thousands of stars that have been observed with the optical low-resolution (R$\sim$2,000) BOSS spectrograph are being released.  Many of these can be found in more open clusters than the previous APOGEE-based sample and still can provide new useful data and results, despite being derived from lower resolution data. 

This 1st BOSS-OCCAM VAC (the 5th VAC from the OCCAM survey team) will be the first to exploit the large MWM dataset of BOSS spectra of stars in open clusters.  The BOSS-OCCAM VAC presents the determination of RV and [Fe/H] membership probabilities for thousands of {\it Gaia} \citep[][]{Gaia_2016} determined 5-D astrometry cluster member stars. This new catalog serves as a supplement to the previous APOGEE-based DR19 sample, as it includes new data {\em for over 100 clusters}, including 95 clusters without APOGEE-based parameters, present in the BOSS DR20 dataset, with the majority of new clusters observed by the Du Pont telescope in the southern hemisphere, that has only been utilized by SDSS surveys starting with the last three years of SDSS-IV/APOGEE-2. 

For this BOSS-OCCAM VAC analysis, we start with the catalog of member stars compiled by \citet{huntIII_24} which used {\it Gaia} Data Release 3 \citep[][]{gaiadr3} observations and 5-D astrometry to determine a joint membership probability. 
We then compute RV and [Fe/H] membership probabilities for each star using the stellar parameters from the BOSS spectra, determined by the CLAM (see Section~\ref{clam}; \citealp{medan2026clam});  for [Fe/H] and PYXCSAO \citep[][]{marina_kounkel_2022_6998993} for RV.  These individual stellar member data are the second of the BOSS-OCCAM VAC files.  

\begin{figure}
    \centering
    \includegraphics[width=0.95\columnwidth]{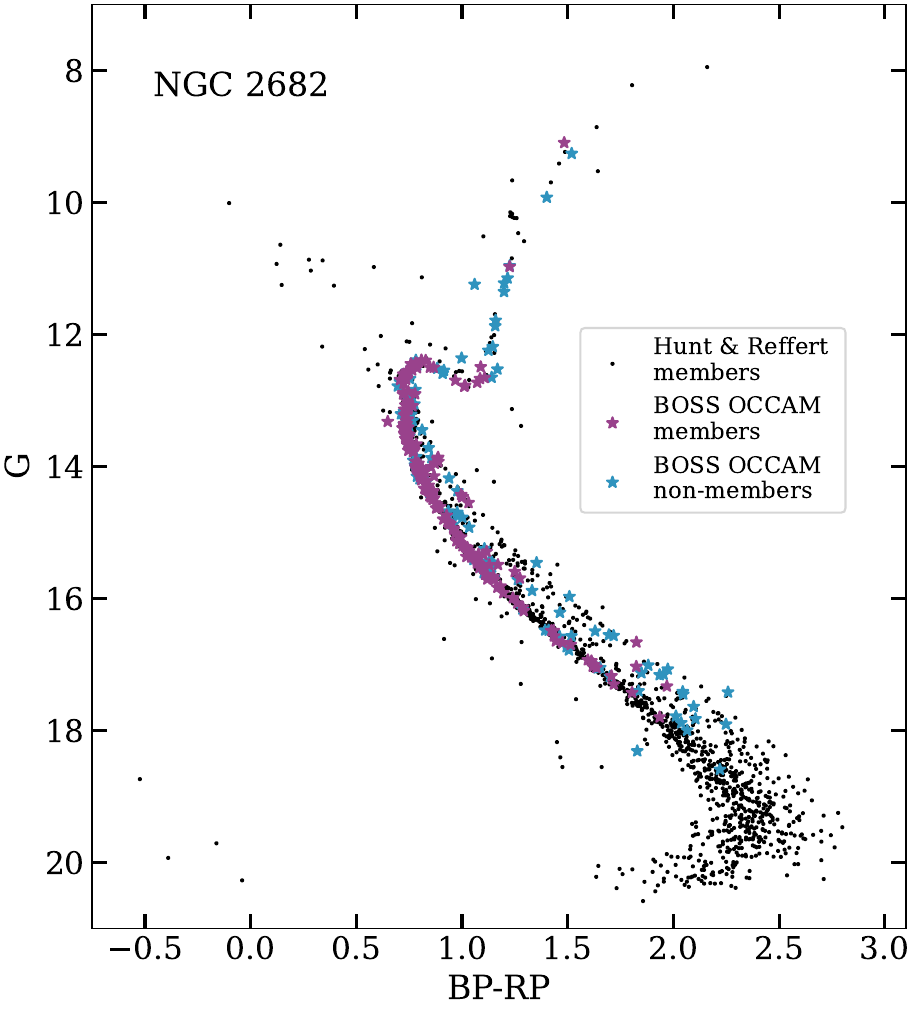}
    \caption{Example cluster {NGC 2682} color-magnitude diagram from the BOSS-OCCAM VAC showing spectroscopically identified members stars (Prob$_{RV} > 0.05$ and Prob$_{[Fe/H]} > 0.05$) in purple, non-members in light blue, and other non-BOSS observed  proper motion members \citep{huntIII_24} in black.}
    \label{fig:boss_occam}
\end{figure}

The BOSS-OCCAM bulk cluster parameters are determined using stars that are within $2\sigma$ of the cluster mean, operationally this translates to stars with probability $>5\%$ for each of the computed membership probabilities. 
The method for determining the membership probabilities closely follow the process outlined in \citet{otto_26}, however, we do update the maximum and minimum widths for the Gaussian fitting to accommodate the lower resolution (R$\sim$2,000) data.  This bulk analysis data are the first of the BOSS-OCCAM VAC files.  An example color-magnitude diagram (CMD) showing member stars from the VAC for the cluster {NGC 2682} is shown in Figure \ref{fig:boss_occam}.
A more thorough description of the methodology and differences to previous OCCAM iterations will be discussed in Otto et al. (2026, {\em in prep}). 

The 1st BOSS-OCCAM  (5th OCCAM) VAC comprises two datasets including, (1) bulk cluster motions, orbital parameters, [Fe/H] abundance and [$\alpha$/M] abundance for over 100 clusters, and (2) the DR20 data used to determine membership probabilities and bulk cluster parameters, including positional data, survey IDs, kinematics as well as, metallicity and $\alpha$-abundance (derived using the BOSS-CLAM; Section~\ref{clam}) for individual stars.
Otto et al. (2026, {\em in prep}) will present results and discuss caveats from this SDSS-V VAC. 

\subsubsection{Pseudo-Continua for M Dwarf BOSS Spectra}\label{vac:mdwarf_continua}

M dwarfs make up a significant portion of the BOSS optical spectra in MWM. They have complex atmospheres, and their spectra contain
many wide and densely overlapping molecular features, which cause them to be almost entirely in absorption in the BOSS wavelength range \citep{allard1990, allard1995}. Additionally, it has been shown in \citet{medan_mdwarf_cont} that there are significant flux calibration issues in some of the BOSS spectra. This means that M dwarfs of similar stellar parameters can have significantly different SED shapes. These calibration issues have to be accounted for in order to \textit{standardize} the BOSS spectra of M dwarfs. This is different than simply defining the thermal continuum of the sources to normalize the spectra. Instead the method seeks to define a ``pseudo-continuum" that, when divided out from the BOSS spectra, will produce M dwarf spectra that are consistent for stars of the same stellar parameters.

The details of the method can be found in \citet{medan_mdwarf_cont}, but here we provide a brief summary of the work for the reader. To fit this pseudo-continuum, we use the spectrum's alpha shape \citep{alpha_shape_def} to find the points which lie between 
the absorption features and apply local polynomial regression to find 
this pseudo-continuum. To tune the hyperparameters of this method, we create BOSS-like spectra from BT-NextGen models \citep{Allard2011} to replicate instrumental, signal-to-noise, and reddening effects. We find that in both this generated set and a validation set of the 
SDSS-V data, our method performs better than alternative standardizations
 by producing spectra that are both more uniform for M dwarfs with 
similar stellar parameters and more easily distinguished compared to M 
dwarfs of differing parameters. This is especially crucial for pipelines that will be developed to estimate stellar parameters of M dwarfs from the BOSS spectra, as this improvement will result in more accurate and precise stellar parameters.

This VAC then uses this method to calculate the pseudo-continuum for all M dwarf \texttt{mwmVisit} spectra in DR20. We define an M dwarf as a target with $M_G > 8.16$, $G-RP > 0.56$, $M_G < 10 (G-RP) + 5$, and either in the mappers \texttt{mwm} or  \texttt{open} (as defined by the \texttt{SDSS5\_TARGET\_FLAGS}). The VAC file is in an HDF5 format and is broadly in two groups. The first is \texttt{meta}, which is a copy of the subset of \texttt{mwmAllVisit} that meets the above criteria. The second group is \texttt{spectra}, which is the corresponding fluxes, inverse variances and pseudo-continua calculated with the method described above, that are row matched to the data in \texttt{meta}. This means this single file can be used to examine the targeting data, spectra meta data and spectra of all M dwarfs in SDSS-V DR20.

\subsubsection{Sodium ISM Absorption in MWM BOSS}\label{vac:sodium_ism_mwm}

Interstellar \ion{Na}{1} D absorption features imprinted on stellar spectra provide a powerful means of tracing diffuse, cold neutral gas. We present a VAC of these measurements from MWM BOSS spectra, designed to enable 3D tomographic studies of the cold neutral ISM.

The catalog is constructed from the BOSS stellar sample current to MJD 61016, comprising the vast majority of DR20 stellar optical sources. To isolate the stellar atmospheric absorption, we model each BOSS spectrum using a grid of MaStar empirical templates for which the Na I D region has been cleaned of contaminating interstellar absorption \citep{Rubin2025}. For each BOSS spectrum, we determine the best-fitting model using a least-squares procedure that simultaneously solves for a scalar normalization and foreground extinction assuming an \citet{Fitzpatrick_1999} reddening law. We adopt the best-fitting model according to the minimum reduced $\chi^2$. After dividing the observed spectrum by the best-fitting model, we measure the interstellar Na I equivalent width from the residual spectrum. The continuum fitting also yields line-of-sight extinction estimates, $A_V$. We obtain a final sample of 1,373,392 measurements, with 192,396 possessing a $\chi{^2}_{\nu} < 1$.

The resulting all-sky distribution of \ion{Na}{1} absorption recovers the expected large-scale structure of the Milky Way, with stronger absorption towards the Galactic midplane and weaker absorption at higher latitudes. The \ion{Na}{1} map broadly follows the fitted extinction distribution, consistent with \ion{Na}{1} tracing cold neutral gas associated with dust. When combined with external distance estimates, such as the Gaia-based photogeometric distances of \citet{BailerJones_2021}, the catalog can be used to trace the 3D distribution of cold gas along the line of sight.

The VAC includes 1,373,392 rows and provides \ion{Na}{1} D equivalent widths, fitted extinctions, associated uncertainties, and $\chi{^2}_{\nu}$ values for the stellar-template fits, ICRS coordinates, and \texttt{sdssid} identifier. A full description of the methodology, validation tests, and initial science applications is presented in \citet{McQuaid2026} 

\subsubsection{Spectral subtypes, morphology, and H{$\alpha$} emission for late-K and M dwarfs observed with BOSS}\label{vac:classify_kmdwarfs}

Low-mass stars, particularly K/M-dwarfs, make up a significant fraction ($\sim$75\%; \citealp{Henry2006}) of the stellar population in the Milky Way. As such, M-dwarfs contribute a large portion of the BOSS observations as part of the MWM, with $\approx$400,000 K dwarfs and $\approx$350,000 M dwarfs in DR20. Low-mass stars were notably targeted in large numbers as part of the Young Stellar Objects (YSO), Solar Neighborhood Census (SNC), and Local Halo (LH) cartons. Stellar parameters for M dwarfs are notoriously difficult to obtain from optical spectra through standard atmospheric model fits, and the parameters estimated from SDSS-V pipelines are generally unreliable. Stellar classification using empirical spectral templates therefore remains extremely useful. 

This VAC compares BOSS spectra of low-mass stars to a set of 536 empirical spectral templates assembled from earlier SDSS optical spectra \citep{Galligan2026}, which span a broad range of effective temperatures and metallicities. BOSS spectra are matched to their best-fit templates to assign: (1) a spectral subtype, which ranges from K5.0-M8.5 with half-subtype resolution, (2) a “morphology class”, which ranges from 0.5 to 12.5 in half-integer steps, (3) H$\alpha$ equivalent widths, and (4) calculated UVW velocities relative to the Sun. 

The templates are adapted from \citet{Zhong2015}, where low-resolution spectra of K/M dwarfs were classified using the set of spectral indices from \citet{Lepine2003} to determine spectral sub-type. The “morphology class” (MC) is based on TiO/CaH molecular band ratio, which is strongly correlated with metallicity in M dwarfs. The MC parameter presented here is an updated and interpolated version of the four metallicity classes (d/sd/esd/usd) defined in \citet{Lepine2007}, which describe the relative strengths of the TiO and CaH absorption features that dominate M-dwarf spectra in the optical regime. Though not directly related to metallicity, the MC parameter does correlate strongly with [M/H]. Stars with MC=0.5 are consistent with super-metal-rich abundances, while stars with MC=12.5 appear to be the most metal-poor, with solar metallicity (thin disk) objects near MC=2.0.

We find the best fit spectral template to each BOSS spectrum by fitting and dividing second order polynomials to the templates and spectra to obtain rectified templates/spectra. Each template is then cross-correlated with the BOSS spectrum, and the $\chi^2$ value is calculated. The spectral subtype and MC are recorded as the template with the minimum $\chi^2$ value. A more detailed description of the classification procedure and validations of the ST and MC parameter are presented in \citet{Galligan2026}. 

H$\alpha$ equivalent widths notably serve to identify very active and potentially young M dwarfs, for which MC index may not strictly correlate with metallicity due to the possible dependence of TiO/CaH on gravity. The UVW components of motion are most useful at validating the MC parameter by associating the stars with their most likely parent population (disk, old disk, halo).

\subsubsection{Stellar Orbits for SDSS-V DR20}
\label{vac:galactic_orbital_elements}

We present the GravPot16 value-added catalog (VAC) for DR20, which provides a homogeneous entry for every star observed by the SDSS-V survey in both the northern and southern hemispheres that has a counterpart in the Gaia DR3 catalog \citep{gaiadr3_23}. The catalog includes a comprehensive set of stellar parameters and, in particular, precise Galactic orbital elements derived for each source. The VAC will be fully described in Fern\'andez-Trincado et al. (2026, in preparation). 

Stellar orbits were integrated in a non-axisymmetric Galactic potential that includes a realistic (as far as possible) Galactic bar, using the \texttt{GravPot16}\footnote{\url{https://gravpot.utinam.cnrs.fr}} galaxy modeling code. This code follows the mass-density profiles of the Besan\c{}con\footnote{\url{https://model.obs-besancon.fr}} Galaxy Model \citep{robin2003, robin2012, robin2014} and incorporates the most up-to-date structural and dynamical constraints available for the Milky Way. For each star, we employed a Monte Carlo approach consisting of 500,000 orbital realizations to properly propagate the observational uncertainties. Initial conditions were constructed by sampling the radial velocities from SDSS-V DR20, the absolute proper motions from \textit{Gaia} DR3 \citep{gaiadr3_23}, and the heliocentric distances from SDSS-V DR20. This procedure allows us to quantify the impact of measurement errors on the derived orbital properties and to assess the influence of the Galactic bar on the stellar orbits.

From these realizations, we computed key orbital parameters for each star, including the orbital eccentricity, apogalactocentric and perigalactocentric distances, maximum vertical excursions from the Galactic plane ($Z_{\rm max}$), the angular momentum vector, the specific orbital energy ($E$), and the Jacobi constant ($E_{\rm J}$) in the reference frame where the bar is at rest. This unprecedented homogeneous dataset allows us to give a short overview of these quantities in this paper. 

For reference, the solar position $R_{\odot}$ = 8.178 kpc \citep{gravity2019}, and $Z_{\odot}$ = 25 pc \citep{juric2008}, and solar motion of [$U_{\odot}, V_{\rm LSR} + V_{\odot}, W_{\odot}] = [11.10, 248.5, 7.25]$ km/s, in line with \citet{brunthaler2011} and \citep{reid2020}. We assumed a bar angle of 20 degrees, and a bar mass of 11 billions Solar masses, in line with \citet{fernandez2017} and \citet{fernandez2020}. For each star, the VAC provides the median values and associated statistics derived from half a million representative orbital ensembles integrated over a 3 Gyr timespan. Orbits were computed with a “boxy/peanut” bar structure with a pattern speed of 41 $\pm$ 10 km s$^{-1}$ kpc$^{-1}$ \citep{sanders2019}.

\subsubsection{Stellar Parameters from BOSS Spectra with The CLAM\label{clam}}\label{vac:clam_parameters}

This is a reimplementation of the CLAM, which we dub ``BOSS-CLAM". BOSS-CLAM is a data driven method that seeks to forward model stellar parameters (the labels) to continuum normalized BOSS spectra. To facilitate this, we need data with good labels to train our model. To cover a wide range of the HR diagram, we gather data from four sources: (1) ASPCAP calibrated stellar parameters from SDSS-V DR19 \citep{sdssVdr19, aspcapdr19} transferred to the DR20 BOSS spectra, (2) the \texttt{BOSS-MINESweeper} value-added catalog \citep{Cargile2020, minesweeper}, (3) wide binaries from \citet{elbadry2021} where ASPCAP labels of primaries are transferred to M dwarf secondaries with a DR20 BOSS spectrum, and (4) a hot star validation sample from validation sample from \citet{Tkachenko2026}. All sources have various quality cuts, which are discussed in more detail in \citet{medan2026clam}.

The BOSS-CLAM method then consists of training, followed by an inference stage. For the training stage, let $\boldsymbol{\ell}_n = (\ell_{n,1}, \ldots, \ell_{n,L})^\top$ denote the $L$ stellar labels (in this case $L=4$; $T_\mathrm{eff}$, $\log g$, $[\mathrm{Fe/H}]$, $[\alpha/\mathrm{M}]$) for star $n$, standardized as $\tilde{\ell}_{n,l} = (\ell_{n,l} - \mu_l) / \sigma_l$, where $\mu_l$ and $\sigma_l$ are the mean and standard deviation of label $l$ in the training set. Each star's standardized labels are mapped to a quadratic feature vector $\boldsymbol{d}_n \in \mathbb{R}^P$ containing a bias, $L$ linear terms, $L$ quadratic
terms, and $L(L-1)/2$ cross-terms ($P = 15$ for $L = 4$). Stacking over $N$ stars yields the design matrix $\mathbf{D} \in \mathbb{R}^{N \times P}$.

A learned coefficient matrix $\boldsymbol{\Theta} \in \mathbb{R}^{P \times K}$ then maps
$\mathbf{D}$ to non-negative NMF weights \citep{NMF} via.~$\mathbf{W} = \mathrm{softplus}(\mathbf{D}\,\boldsymbol{\Theta})$. This then combines $K$ non-negative basis spectra $\mathbf{H} \in \mathbb{R}^{K \times M}_{\geq 0}$ (parameterized in log-space to enforce positivity) to produce the predicted normalized flux: $\hat{\mathbf{F}} = \mathbf{1} - \mathbf{W}\mathbf{H}$.

During the training stage, the stellar labels ($\boldsymbol{\ell}_n$), polynomial mapping terms ($\boldsymbol{\Theta}$), NMF basis spectra ($\mathbf{H}$) and a per-pixel scatter ($s_\lambda$; see below) are to be optimized, resulting in a total of 
$N \times L + P \times K + K \times M$ parameters. These are optimized using adam \citep{kingma2014adam}, minimizing the loss function,
\begin{eqnarray}
    \mathcal{L} &=& \frac{1}{NM}\sum_{n,\lambda}
    \left[
        \frac{(f_{n\lambda} - \hat{f}_{n\lambda})^2}{\sigma_{n\lambda}^2 + s_\lambda^2}
        + \ln(\sigma_{n\lambda}^2 + s_\lambda^2)
    \right] \nonumber \\
    && + \frac{\lambda_\ell}{N_c}\sum_{n,l} w_{nl}(\tilde{\ell}_{nl} - \tilde{\ell}_{nl}^{(0)})^2,
\end{eqnarray}
where $\sigma_{n\lambda}^2$ is the per-pixel flux variance, $s_\lambda$ is a learned
per-wavelength scatter term, $\tilde{\ell}_{nl}^{(0)}$ are the initial standardized labels,
$w_{nl}$ are per-label weights, $N_c = \sum_{n,l}\mathbf{1}[w_{nl}>0]$ is the number of
contributing label terms, and $\lambda_\ell$ controls the strength of label regularization ($\lambda_\ell = 1$ in this implementation).

During the inference stage, only the stellar labels are left as free parameters (the unknown quantity), and the loss becomes:
\begin{eqnarray}
    \mathcal{L} &=& \frac{1}{2}\sum_{n,\lambda}
    \left[
        \frac{(f_{n\lambda} - \hat{f}_{n\lambda})^2}{\sigma_{n\lambda}^2 + s_\lambda^2}
    \right]
\end{eqnarray}
First, we train a Multi-layer Perceptron (MLP) regressor \citep{MLP, scikit-learn} with the training set to estimate labels from the basis weights of the spectra.  Then, adam is used with a starting value based on the prediction from the MLP regressor. Finally, we run an additional 100 iterations with L-BFGS-B \citep{byrd1995limited, zhu1997algorithm} to settle on the optimal labels for the spectrum.

\begin{figure}
    \centering
    \includegraphics[width=\columnwidth]{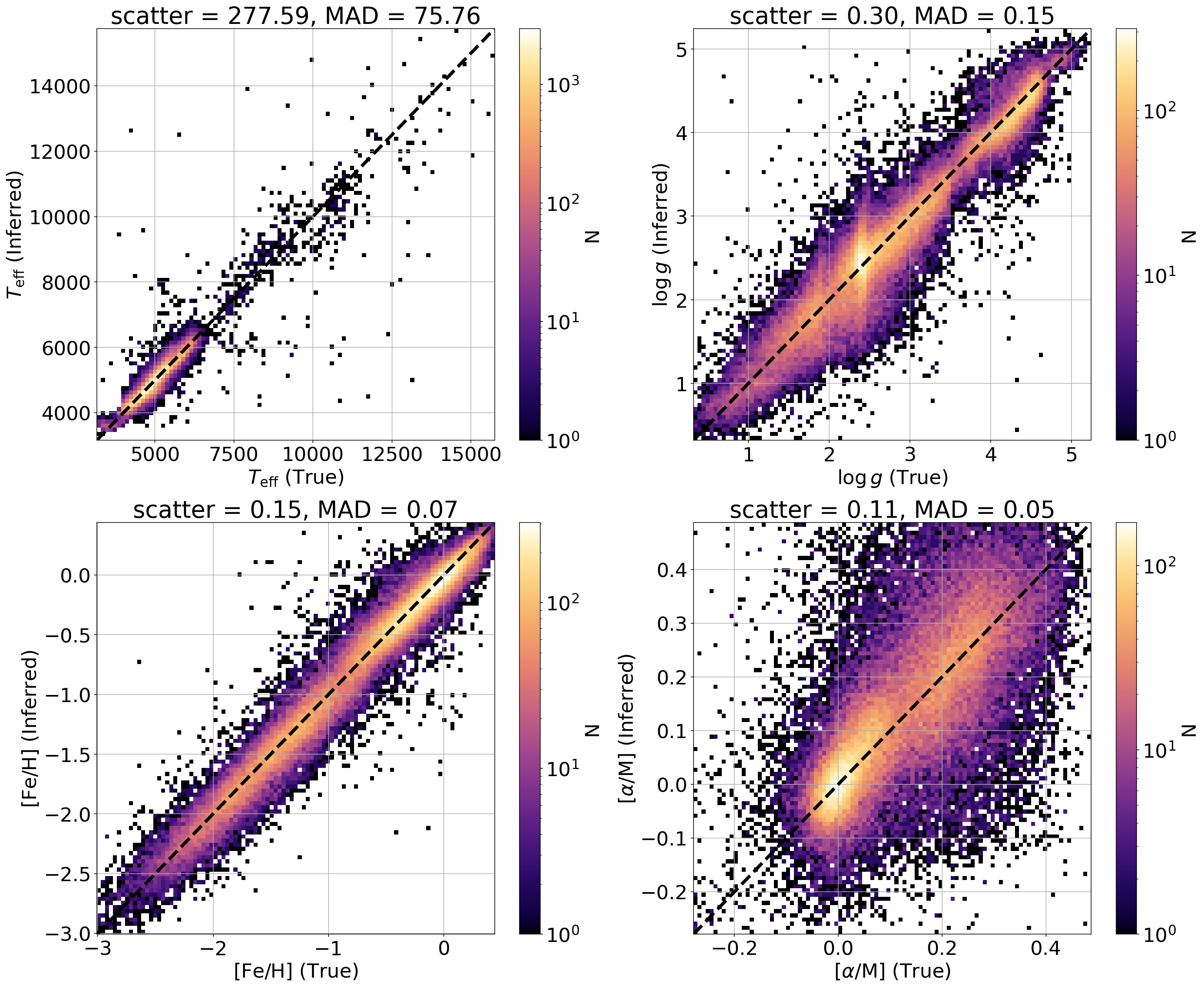}
    \caption{Inference results for the testing set from the BOSS-CLAM pipeline. On all plots, the x-axis shows the true parameters and the y-axis is the BOSS-CLAM parameters. The black dashed line is the one-to-one relation. The title for each panel gives the scatter and median absolute deviation between the true and inferred parameters. In all cases, a good agreement is observed between the true and inferred parameters.}
    \label{fig:clam_results}
\end{figure}

We run this inference step on the data from our four groups. Figure \ref{fig:clam_results} shows the comparison between the inferred and true parameters for the data. In all cases, a good agreement is observed between the parameters across the full range. Additionally, above each panel we list the scatter and median absolute deviation between the true and inferred parameters. This gives the overall precision that we can achieve with this method.

This VAC then uses the above method to infer the stellar parameters for all BOSS DR20 \texttt{mwmStar} files with \texttt{snr > 10}. These results are stored in two HDF5 files; a nominal one and a lite one. This HDF5 file is split in two groups. The first is \texttt{meta}, which in the nominal file includes all columns from \texttt{mwmAllStar} file (so all targeting and spectrum related information). The lite file only includes a subset of those columns. The second group is \texttt{clam}, which includes the BOSS-CLAM inferred results: $T_\mathrm{eff}$, $\log g$, $[\mathrm{Fe/H}]$, $[\alpha/\mathrm{M}]$. Additionally, we include a set of flags and warnings, to identify questionable parameters. Details on these flags can be found in Medan et al. (in prep) and the corresponding datamodel for the VAC.

\subsubsection{The large-scale kinematics of young stars in the Milky Way disc}\label{vac:yso_kinematics}
The construction and validation of the VAC “The large-scale kinematics of young stars in the Milky Way disc” are presented in \cite{2025A&A...703A.303Z}. Here we summarize the main steps in the construction of the VAC and refer the reader to \cite{2025A&A...703A.303Z} for an example of its scientific application.

To construct the catalog, we began by selecting all stars belonging to the MWM \texttt{mwm\_ob} carton. We then filtered this parent sample based on four parameters: the signal-to-noise ratio of the observed single-epoch spectra, the parallax error, the renormalized unit weight error \citep[RUWE,][]{2021A&A...649A...2L}, and the effective temperature derived from the BOSS spectra.

We required the median signal-to-noise ratio (\texttt{sn\_median\_all}) of the single-epoch spectra in the BOSS pipeline summary file (reduction version \texttt{v6\_2\_1}, Morrison et al., in prep.\footnote{In \cite{2025A&A...703A.303Z} we used reduction version \texttt{v6\_2\_0}.}) to satisfy S/N $>$ 20 in order to ensure reliable spectral parameter estimates. We further selected sources with relative parallax error $\sigma_{\varpi}/\varpi < 0.2$ and RUWE $<$ 1.4. The former criterion selects sources with well-determined distances, while the latter removes sources with potentially spurious parallaxes or proper motions \citep[and possibly unresolved binaries;][]{2020MNRAS.496.1922B}. Finally, we retained stars with $T_{\mathrm{eff}} > 10,000$ K according to the BOSSNet estimates \citep{BOSSNET2024AJ....167..173S}.
The VAC contains estimates of the line-of-sight velocities ($v_{l.o.s.}$) for all targets. For stars with correlation strength \texttt{XCSAO\_RVX} $>$ 6 (95\% of the sample), we adopted the $v_{\mathrm{l.o.s.}}$ values derived with the XCSAO pipeline \citep{marina_kounkel_2022_6998993}. When the correlation strength fell below this threshold, we used the BOSSNet $v_{l.o.s.}$ estimate instead.
For targets with multi-epoch observations, we determined $v_{l.o.s.}$ as the average of the available measurements. For most sources the number of epochs is insufficient to derive orbital solutions and therefore to determine the systemic velocity of binary systems that were not excluded by the previous selection criteria. For single stars, averaging multiple observations improves the precision of $v_{l.o.s.}$; for binaries or higher-order multiples, the resulting value more closely approximates the centre-of-mass motion.
For stars with a single observation, we adopted the uncertainties provided by XCSAO or BOSSNet. For stars with multiple observations, we computed the uncertainty as $\sigma_{v_{\mathrm{l.o.s.}}} = \mathrm{std}(v_{l.o.s.})/\sqrt{N_{\mathrm{obs}}}$. 
In the VAC, the line-of-sight velocity is reported under the column name \texttt{radial\_velocity} for consistency with other data products.
Similarly, the VAC provides average estimates of the BOSSNet $T_{\mathrm{eff}}$ (\texttt{teff}), $\log g$ (\texttt{logg}), and [Fe/H] (\texttt{feh}) for stars with multi-epoch observations, and their errors (\texttt{e\_teff}, \texttt{e\_logg}, \texttt{e\_feh}).

The VAC also includes \textit{Gaia} DR3 identifiers (\texttt{gaia\_dr3\_source\_id}), coordinates (\texttt{ra}, \texttt{dec}), and proper motions (\texttt{pmra}, \texttt{pmdec}), for all targets, together with the Bailer-Jones photo-geometric distance estimates \citep[][column name: \texttt{distance}]{BailerJones_2021}. These parameters were used to derive Galactocentric Cartesian coordinates for all targets (columns: \texttt{xg}, \texttt{yg}, \texttt{zg}), the Galactocentric distance (\texttt{R}), and the Galactocentric radial velocity (\texttt{vR}) along with its uncertainty (\texttt{e\_vR}), as described in Section 3.1 of \cite{2025A&A...703A.303Z}.
Finally, the VAC includes the stellar age estimates (\texttt{median\_log\_age}) derived in Section 3.2 of \cite{2025A&A...703A.303Z}, together with the 16th and 84th percentiles (\texttt{lo\_median\_age} and \texttt{hi\_median\_age}, respectively).

\subsection{LVM VAP}

\subsubsection{HiPS RGB Map}\label{vac:LVM_map}


As part of the DR20 value-added products, we provide a color RGB HiPS map constructed from LVM measurements of $H\alpha$ (green channel), [O{\sc iii}]$\lambda5007$ (blue channel), and [S{\sc ii}]$\lambda\lambda6717,6731$ (red channel).
The map is based on all LVM observations obtained before MJD 61086 (February 15, 2026).
It is intended to provide a representative preview of the scientific data products that LVM is expected to deliver throughout the course of the survey, illustrating the anticipated data quality, spatial scales, and sky coverage that will be achieved by future observations.
The map is optimized for visualization rather than quantitative analysis: emission-line flux cuts are scaled using an asinh stretching function to enhance structures over a wide dynamic range, from faint diffuse emission to very bright regions.

The RGB HiPS map is available directly in the \lvmvis\ interface\footnote{\url{https://dr20.sdss.org/lvmvis/}}, where it can be displayed together with the wide-field LVM pointings and individual fiber locations.
It is also directly accessible through the SDSS SAS at \url{https://dr20.sdss.org/sas/dr20/vap/lvm/hips/rgb_Halpha_OIII_SII/v1/}.

Future releases will include HiPS maps in FITS format based on LVM DAP products, enabling quantitative measurements of physical parameters.
A detailed description of the construction of HiPS maps in individual emission lines, including the aggregation scheme and the averaging of measurements in regions covered by overlapping fibers, will be presented in Katkov et al. (in prep.).

\section{Summary and Future}\label{sec:summary}

This paper presents the twentieth data release (DR20) of the fifth generation of the Sloan Digital Sky Survey---the third data release of SDSS-V and the first release of SDSS-V spectra from LCO. Most notably, with DR20 we release all optical spectra from the northern and southern hemispheres observed through 2 February 2025. This manuscript presents the scope of DR20 in Section~\ref{sec:scope} and discusses data access through archive servers and data visualization platforms in Section~\ref{sec:data_access}---introducing the Lite version of LVMvis. In Section~\ref{sec:pipelines} we discuss updates to the data reduction and data analysis pipelines used by all mappers. With DR20, we release new data for all three mappers, including:
\begin{itemize}
    \item Over 3M new low-resolution, optical BOSS spectra, comprising over 1M spectra of 0.5M galaxies from the BHM (Section~\ref{sec:bhm}) and over 2M spectra of 1.5M stars from the MWM (Section~\ref{sec:mwm}). The extra-galactic spectra span all BHM core programs and many ancillary programs, with major progress in SPIDERS fields, following up eROSITA X-ray sources. The stellar, optical spectra released with DR20 span core MWM programs including YSO, Halo, SNC, WD, and Galactic eROSITA sources.
    \item  169 tiles of six targets from LVM, spanning three Galactic HII regions, a planetary nebula, and two nearby galaxies (Section~\ref{sec:lvm}). With DR20, we release LVM DAP and DRP data products for the first time, providing calibrated spectroscopic data and measurements of the physical conditions in the ISM.
    \item One Target of Opportunity spectrum of a nearby AGN. In Section~\ref{sec:ToO}, we describe this new mode of SDSS operations and implementation within the survey.
    \item Updates to the targeting catalog and database, including new parent catalogs, implementation of an updated cross-matching algorithm, and updated targeting cartons released to the MOS target product (Section~\ref{sec:targeting}). 
    \item 18 Value Added Catalogs based on BHM and MWM data (Section~\ref{sec:vacs}), including 14 new VACs for DR20. These additional catalogs offer supplemental data analysis products that enhance the core SDSS data sets.
    \item A Value Added Product releasing a RGB HiPS map from LVM data, constructed from $H\alpha$, [O{\sc iii}]$\lambda5007$, and [S{\sc ii}]$\lambda\lambda6717,6731$.
\end{itemize}

With SDSS-V observations ongoing through the end of 2026 at APO and Summer of 2027 at LCO, much more is in store for the final two data releases of SDSS-V, planned for 2027 (DR21) and 2028 (DR22). DR21 will contain all-sky high-resolution IR spectra and data products, including the first SDSS-V APOGEE spectra from the southern hemisphere, a large sample of LVM tiles, and updates to the targeting catalog. Finally, DR22 will complete SDSS-V and release all observations taken as a part of the BHM, MWM, and LVM surveys.

\section*{Acknowledgments}

Funding for the Sloan Digital Sky Survey V has been provided by the Alfred P. Sloan Foundation, the Heising-Simons Foundation, the National Science Foundation, and the Participating Institutions. SDSS acknowledges support and resources from the Center for High-Performance Computing at the University of Utah. SDSS telescopes are located at Apache Point Observatory, funded by the Astrophysical Research Consortium and operated by New Mexico State University, and at Las Campanas Observatory, operated by the Carnegie Institution for Science. The SDSS web site is \url{www.sdss.org}.

SDSS is managed by the Astrophysical Research Consortium for the Participating Institutions of the SDSS Collaboration, including the Carnegie Institution for Science, Chilean National Time Allocation Committee (CNTAC) ratified researchers, Caltech, the Gotham Participation Group, Harvard University, Heidelberg University, The Flatiron Institute, The Johns Hopkins University, L'Ecole polytechnique f\'{e}d\'{e}rale de Lausanne (EPFL), Leibniz-Institut f\"{u}r Astrophysik Potsdam (AIP), Max-Planck-Institut f\"{u}r Astronomie (MPIA Heidelberg), Max-Planck-Institut f\"{u}r Extraterrestrische Physik (MPE), Nanjing University, National Astronomical Observatories of China (NAOC), New Mexico State University, The Ohio State University, Pennsylvania State University, Smithsonian Astrophysical Observatory, Space Telescope Science Institute (STScI), the Stellar Astrophysics Participation Group, Universidad Nacional Aut\'{o}noma de M\'{e}xico, University of Arizona, University of Colorado Boulder, University of Illinois at Urbana-Champaign, University of Toronto, University of Utah, University of Virginia, Yale University, and Yunnan University.

The research behind these results has (partially) received funding from the BELgian federal Science Policy Office (BELSPO) through PRODEX grant PLATO: ZKE8588 and from the Flemish Government under the long-term structural Methusalem funding program by means of the project SOUL: Stellar evolution in full glory, grant METH/24/012 at KU Leuven.

This work is based on data from eROSITA, the primary instrument aboard SRG, a joint Russian-German
science mission supported by the Russian Space Agency
(Roskosmos), in the interests of the Russian Academy
of Sciences represented by its Space Research Institute
(IKI), and the Deutsches Zentrum f\"{u}r Luftund Raumfahrt (DLR). The SRG spacecraft was built by Lavochkin Association (NPOL) and its subcontractors, and
is operated by NPOL with support from the Max Planck
Institute for Extraterrestrial Physics (MPE). The development and construction of the eROSITA X-ray instrument was led by MPE, with contributions from the
Dr. Karl Remeis Observatory Bamberg \& ECAP (FAU
Erlangen-N\"{u}rnberg), the University of Hamburg Observatory, the Leibniz Institute for Astrophysics Potsdam (AIP), and the Institute for Astronomy and Astrophysics of the University of T\"{u}bingen, with the support of DLR and the Max Planck Society. The Argelander Institute for Astronomy of the University of Bonn
and the Ludwig Maximilians Universit\"{a}t Munich also
participated in the science preparation for eROSITA.
The eROSITA data shown here were processed using
the eSASS software system developed by the German
eROSITA consortium.

\section*{Author Contributions}
All authors are listed alphabetically according to the SDSS-V publication policy.

\appendix
\label{sec:appendix}

\section{LVM DRP File Structure}\label{appendix:LVM_DRP}

\begin{table*}
\centering
\caption{Summary of DRP \texttt{lvmSFrame} FITS extensions.}
\begin{tabular}{c l l}
\hline
Ext. & Name & Description \\
\hline
0 & PRIMARY & Header with observational and reduction metadata \\
1 & FLUX & Flux-calibrated spectra [$10^{-17}$ erg s$^{-1}$ cm$^{-2}$ \AA$^{-1}$ fiber$^{-1}$] \\
2 & IVAR & Inverse variance of the flux \\
3 & MASK & Pixel mask (data quality flags) \\
4 & WAVE & Wavelength array (rectified grid) \\
5 & LSF & Line spread function (spectral resolution) \\
6 & SKY & Sky model subtracted from each fiber \\
7 & SKY\_IVAR & Inverse variance of the sky model \\
8 & FLUXCAL\_STD & Calibration vectors from standard stars \\
9 & FLUXCAL\_SCI & Applied flux calibration \\
10 & FLUXCAL\_MOD & Model calibration curves \\
11 & SLITMAP & Fiber mapping and instrumental metadata \\
\hline
\end{tabular}
\label{tab:drp_extensions}
\end{table*}

The DRP provides fully reduced, wavelength-calibrated, and sky-subtracted spectra in the form of a multi-extension FITS file denoted as lvmSFrame-file (see Section \ref{sec:lvmdrp}). In addition to these 169 individual files, one for each delivered exposure (following the format \texttt{lvmSFrame-\textit{expnum}.fits}), we distribute a summary FITS file (\texttt{drpall-\textit{drpver}.fits}, with current drpver=1.2.0) that collects metadata for all
reduced exposures. Both the science frames and the summary file are stored in the following directory: \url{https://dr20.sdss.org/sas/dr20/spectro/lvm/redux/1.2.0/}

Table~\ref{tab:drp_extensions} summarizes the content of the twelve extensions included in each lvmSFrame-file. The \texttt{PRIMARY} extension (Ext.~0) contains no science data but carries the full observational and reduction metadata in its header, including provenance information inherited from the raw frames and processing flags set by the pipeline. The core science data are stored in the \texttt{FLUX} extension (Ext.~1), which contains the sky-subtracted, flux-calibrated spectra for all science fibers in units of $10^{-17}$\,erg\,s$^{-1}$\,cm$^{-2}$\,\AA$^{-1}$\,fiber$^{-1}$together with the associated \texttt{IVAR} extension (Ext.~2) providing the inverse variance of those fluxes for use in weighted analyses and error propagation, organized as a row-stacked spectra (RSS) with 1801 spectra (one for each fiber) and $\sim$12,400 wavelength elements the wavelength range 3600--9800~\AA. Pixels affected by detector defects, cosmic rays, or other reduction artifacts are identified through per-pixel data-quality flags stored in the \texttt{MASK} extension (Ext.~3). 

The \texttt{WAVE} extension (Ext.~4) provides the rectified, common wavelength grid onto which all fiber spectra have been resampled, covering 3600--9800\,\AA\ at a sampling of $\approx0.5$\,\AA\ per pixel, while the \texttt{LSF} extension (Ext.~5) characterises the spectrally resolved line spread function as a function of fiber and wavelength, enabling accurate convolution or forward-modelling of spectral features. The sky background subtracted from each fiber is preserved in the \texttt{SKY} extension (Ext.~6), with its corresponding inverse variance in \texttt{SKY\_IVAR} (Ext.~7), allowing users to reconstruct the un-subtracted spectra or to assess the quality of the sky removal independently. Flux calibration information is provided across three dedicated extensions: \texttt{FLUXCAL\_STD} (Ext.~8) contains the sensitivity vectors derived from the spectrophotometric standard stars observed simultaneously via telescope~T4, \texttt{FLUXCAL\_SCI} (Ext.~9) records the calibration solution actually applied to the science fibers, and \texttt{FLUXCAL\_MOD} (Ext.~10) stores the model calibration curves used in the fitting procedure, together providing full traceability of the photometric calibration chain. 

Finally, the \texttt{SLITMAP} extension (Ext.~11) contains a fiber-by-fiber mapping table that records the on-sky position (right ascension and declination) of each of the 1801 science fibers, together with instrumental metadata such as fiber identifiers, spectrograph channel assignments, and quality indicators, enabling direct association of each spectrum with its corresponding spatial location on the sky.

\section{LVM DAP File Structure}\label{appendix:LVM_DAP}

\begin{table*}[ht]
\centering
\caption{Description of the DAP file containing the data products of the analysis}
\label{tab:dapfile}
\begin{tabular}{clll}
\hline
HDU & EXTENSION & \# Rows & \# Columns \\
\hline
0  & PRIMARY             &              &                  \\
1  & PT                  & \#spec       & 6                \\
2  & RSP                 & \#spec       & 1 + \#RSP        \\
3  & COEFFS              & \#spec $\times$ \#RSP & 13     \\
4  & PM\_ELINES          & \#spec $\times$ \#PM\_EL & 10  \\
5  & NP\_ELINES\_B       & \#spec       & 1 + \#NP\_EL\_B $\times$ 8 \\
6  & NP\_ELINES\_R       & \#spec       & 1 + \#NP\_EL\_R $\times$ 8 \\
7  & NP\_ELINES\_I       & \#spec       & 1 + \#NP\_EL\_I $\times$ 8 \\
8  & PM\_KEL             & \#spec $\times$ \#PM\_EL & 10  \\
9  & ELINES\_SIGMA\_CHI  & \#spec       & 2(2\#PM\_EL + \#NP\_EL) \\
10 & ELINES\_CHI2\_AVG   & 2\#PM\_EL + \#NP\_EL & 4     \\
11 & INFO                & \#param      & 2                \\
\hline
\end{tabular}

\vspace{0.2cm}
\noindent{\footnotesize
Structure of extensions included in the delivered DAP file, where: (i) \#spec is the number of science spectra (or fibers) included in the analyzed {\tt Tile} or RSS (row-stacked spectra) frame; (ii) \#RSP is the number of templates/spectra included in the stellar library; (iii) \#PM\_EL is the number of individual models (emission lines) included in the parametric analysis of the ionized gas emission lines; (iv) \#NP\_EL\_BAND is the number of emission lines included in the non-parametric analysis for each band (B, R, and I), corresponding to each arm of the spectrograph; and (v) \#NP\_EL is the total number of emission lines analyzed using the non-parametric procedure.
}
\end{table*}

The LVM-DAP delivers the results of the analysis as a multi-extension FITS file too, one for each analyzed exposure (i.e., each \texttt{lvmSFrame}), comprising the measured quantities and their corresponding errors for each fiber, called the DAP-file. Each extension in this file consists of a FITS table, mapping each individual science fiber, comprising a specific group of measurements as described in detail in \citet{lvmdap} and updated in \citet{HelixDR19}.

Table~\ref{tab:dapfile} describes the structure of delivered DAP-files, which content is summarized in here, extension by extension: (i) the {\tt PRIMARY} extension contains only header information describing the data provenance and processing, being inherited from the corresponding \texttt{lvmSFrame}; (ii) the {\tt PT} table provides with the position table of the science fibers, i.e., their location in the sky; (iii) {\tt RSP} stores the average stellar properties derived from the continuum modelling based on the adopted RSP template library (e.g., Teff, log(g)...), their corresponding kinematics parameters ($v_\star$, $\sigma_\star$), and dust attenuation (A$_{V,\star}$); (iv) {\tt COEFFS}, on the other hand, lists the corresponding light-weights at 5500\AA of the RSP decomposition; (v) the ionized gas emission is characterized through both parametric and non-parametric approaches, with {\tt PM\_ELINES} and {\tt PM\_KEL} extensions contain the results of the Gaussian-based modeling of emission lines, whereas (vi) the {\tt NP\_ELINES\_B}, {\tt NP\_ELINES\_R}, and {\tt NP\_ELINES\_I} tables provide non-parametric measurements of the emission line properties across the blue, red, and infrared spectral ranges, respectively; (vii) the quality and consistency of the emission-line analysis are quantified in the {\tt ELINES\_SIGMA\_CHI} and {\tt ELINES\_CHI2\_AVG} extensions, which include instrumental corrected velocity dispersions and goodness-of-fit indicators derived for both methodologies; (viii) finally, the {\tt INFO} extension compiles the parameters included in the configuration file adopted for the analysis, enabling full traceability of the DAP processing.

In DR20 we deliver a total of 172 DAP-files, 169 of them corresponding to the standard analysis performed on each lvmSFrame-file described in App. \ref{appendix:LVM_DRP}, and three more corresponding to the analysis of the Local Volume galaxies described in Sec. \ref{fig:lvm_DR20_LV}.
We adopted a convention for naming the DAP-files that codifies (i) the number of templates included in the RSP decomposition (\#RSP), (ii) the minimum signal-to-noise adopted for a multi-RSP fitting (SN), and (iii) the exposure unique identification number of the observed (and reduced) frame (EXPNUM), following {\tt dap-rsp\{\#RSP\}-sn\{SN\}-\{EXPNUM\}.dap.fits} format. For instance,
{\tt dap-rsp108-sn20-000301999.dap.fits} corresponds to the result of an analysis using the RSP library comprising 108 templates, with a minimum signal-to-noise of 20, for the LVM exposure 301999, i.e., the spectra comprised in the \texttt{lvmSFrame-301999.fits} DRP-file. 
We note that of the 1801 science spectra comprised in the DRP-file, only 1754 are analyzed and delivered in the corresponding DAP-file. Those correspond to the good quality (not broken) science fibers, i.e., those that verify that \texttt{targettype = 'science'} and \texttt{fibstatus = 0} in the \texttt{SLITMAP} extension of the \texttt{lvmSFrame}.

In addition to the DAP-file described above, we deliver for each exposure a complementary FITS file containing the spectral models derived during the analysis, hereafter referred to as the DAP-model-file. These are named following the same convention as the DAP-files, replacing the \texttt{.dap.fits} suffix with \texttt{.model.fits}, e.g., \texttt{dap-rsp108-sn20-\{EXPNUM\}.model.fits}. Each model-file consists of a single \texttt{PRIMARY} extension containing a three-dimensional data cube of dimensions $N_\lambda \times N_{\rm spec} \times 9$, where $N_\lambda = 12401$ is the number of wavelength pixels spanning the full LVM spectral range ($\sim$3600--9800\,\AA), and $N_{\rm spec} = 1754$ is the number of good-quality science fibers delivered in the corresponding DAP-file (i.e., those satisfying \texttt{targettype = 'science'} and \texttt{fibstatus = 0} in the \texttt{SLITMAP} extension of the \texttt{lvmSFrame}). The nine slices of this cube store an RSS array, in which, for each fiber, the following spectral quantities derived by the LVM-DAP analysis \citep[Sec. \ref{sec:lvmdap}][]{lvmdap}: (1) the original observed spectrum (\texttt{org\_spec}); (2) the best-fitting stellar continuum model from the full RSP decomposition (\texttt{model\_spec}); (3) the joint model combining the stellar continuum and the 1st parametric emission-line fit (\texttt{mod\_joint\_spec}); (4) the gas-pure spectrum obtained by subtracting the stellar model from the observed spectrum (\texttt{gas\_spec}); (5) the residual spectrum after subtraction of the joint stellar and emission-line model (\texttt{res\_joint\_spec}); (6) the gas-free spectrum, i.e., the observed spectrum with the 1st parametric emission-line model subtracted (\texttt{no\_gas\_spec}); (7) the non-parametric model of the emission lines (\texttt{gas\_model\_NP}); (8) the emission-line model from the 2nd parametric fit (\texttt{gas\_model\_PEK}); and (9) the joint model combining the stellar continuum and the 2nd parametric emission-line fit (\texttt{mod\_joint\_spec\_PEK}). For the three exposures on extragalactic targets that were analyzed twice, as described in Sec. \ref{sec:lvm_data_product}, the corresponding model was created following a similar nomenclature as the DAP-files, i.e., adding an \texttt{LV\_} prefix to the model-file. The model-file is stored alongside the individual DAP-files in the corresponding subdirectories at \url{https://dr20.sdss.org/sas/dr20/spectro/lvm/analysis/1.2.0/1.2.0.251218/}.

Like in the case of the LVM DRP products we deliver a summary FITS file that collects the information of each science exposure together with a subset of average properties across the entire FoV covered by the LVM science IFU extracted from the individual DAP-files, using the convention \texttt{dapall-\textit{drpver}-\textit{dapver}.fits} (with current dapver=1.2.0.251218). The individual DAP-files and this summary file are stored distributed in the following link: \url{https://dr20.sdss.org/sas/dr20/spectro/lvm/analysis/1.2.0/1.2.0.251218/}

\section{New MWM target cartons}
\label{sec:appendix_mwm}
\begin{longrotatetable}
\begin{deluxetable*}{llll}
\tabletypesize{\scriptsize}
\label{tab:mwm_cartons}
\tablecaption{Milky Way Mapper Cartons}
\tablehead{\colhead{\multirow{2}{*}{Carton Name}} & \colhead{\multirow{2}{*}{Selection Summary\tablenotemark{1}}} & 
\colhead{\multirow{2}{*}{Instrument}} & \colhead{Available} \\ [-0.2cm]
\colhead{} & \colhead{} & \colhead{} & \colhead{Targets\tablenotemark{2}}}

\startdata
         manual\_mwm\_boss\_only\_cluster\_jan\_2025\_high\_priority & \parbox[t]{11cm}{Targets selected from globular and open clusters (according to Hunt \& Reffert 2024) high priority} & BOSS & 5444 \\ 
        manual\_mwm\_boss\_only\_cluster\_jan\_2025\_low\_priority & \parbox[t]{11cm}{Targets selected from globular and open clusters (according to Hunt \& Reffert 2024) low priority} & BOSS & 12537 \\ 
        manual\_mwm\_crosscalib\_apogee & \parbox[t]{11cm}{Validation -- Cross-calibration sample of stars from WEAVE, 4MOST, DESI, etc with APOGEE} & APOGEE & 5552 \\ 
        manual\_mwm\_crosscalib\_yso\_apogee & \parbox[t]{11cm}{MWM - science cross-validation targets SDSS YSO program and 4MOST Young Stars Program, to be observed with APOGEE} & APOGEE & 3198 \\ 
        manual\_mwm\_crosscalib\_yso\_boss & \parbox[t]{11cm}{MWM - science cross-validation targets SDSS YSO program and 4MOST Young Stars Program, to be observed with BOSS} & BOSS & 3198 \\ 
        manual\_mwm\_dust\_gaia\_madgics\_dib\_dr3 & \parbox[t]{11cm}{All DIBs found in Gaia DR3 RVS spectra found by MADGICS reduction of public data} & APOGEE & 7789 \\ 
        manual\_mwm\_dust\_ullyses & \parbox[t]{11cm}{All ULLYSES targets with Gaia DR3 ids} & APOGEE & 364 \\ 
        manual\_mwm\_halo\_distant\_bhb\_boss & \parbox[t]{11cm}{Galactic Halo Targets (Distant BHB from GALEX+Gaia), dark time cadence two epochs} & BOSS & 49900 \\ 
        manual\_mwm\_halo\_distant\_bhb\_boss\_single & \parbox[t]{11cm}{Galactic Halo Targets (Distant BHB from GALEX+Gaia), bright time cadence one epoch} & BOSS & 49900 \\ 
        manual\_mwm\_halo\_distant\_bhb\_boss\_triple & \parbox[t]{11cm}{Galactic Halo Targets (Distant BHB from GALEX+Gaia), dark time cadence three epochs} & BOSS & 49900 \\ 
        manual\_mwm\_halo\_distant\_kgiant\_far\_boss & \parbox[t]{11cm}{Galactic Halo Targets (Distant K Giants from Gaia+WISE, >8 kpc), dark time cadence two epochs} & BOSS & 46912 \\ 
        manual\_mwm\_halo\_distant\_kgiant\_far\_boss\_single & \parbox[t]{11cm}{Galactic Halo Targets (Distant K Giants from Gaia+WISE, >8 kpc), bright time cadence one epoch} & BOSS & 46912 \\ 
        manual\_mwm\_halo\_distant\_kgiant\_far\_boss\_triple & \parbox[t]{11cm}{Galactic Halo Targets (Distant K Giants from Gaia+WISE, >8 kpc), dark time cadence three epochs} & BOSS & 46912 \\ 
        manual\_mwm\_halo\_distant\_kgiant\_near\_boss & \parbox[t]{11cm}{Galactic Halo Targets (Distant K Giants from Gaia+WISE, <8 kpc), dark time cadence two epochs} & BOSS & 305846 \\ 
        manual\_mwm\_halo\_distant\_kgiant\_near\_boss\_single & \parbox[t]{11cm}{Galactic Halo Targets (Distant K Giants from Gaia+WISE, <8 kpc), bright time cadence one epoch} & BOSS & 305846 \\ 
        manual\_mwm\_halo\_distant\_kgiant\_near\_boss\_triple & \parbox[t]{11cm}{Galactic Halo Targets (Distant K Giants from Gaia+WISE, <8 kpc), dark time cadence three epochs} & BOSS & 305846 \\ 
        manual\_mwm\_halo\_mp\_wise\_apogee & \parbox[t]{11cm}{Galactic Halo Targets (Metal-Poor Giants from Gaia+WISE) observed with APOGEE, dark time cadence two epoch} & APOGEE & 8516 \\ 
        manual\_mwm\_halo\_mp\_wise\_apogee\_single & \parbox[t]{11cm}{Galactic Halo Targets (Metal-Poor Giants from Gaia+WISE) observed with APOGEE, bright time cadence one epoch} & APOGEE & 8516 \\ 
        manual\_mwm\_halo\_mp\_wise\_apogee\_triple & \parbox[t]{11cm}{Galactic Halo Targets (Metal-Poor Giants from Gaia+WISE) observed with APOGEE, dark time cadence three epoch} & APOGEE & 8516 \\ 
        manual\_mwm\_halo\_mp\_wise\_boss & \parbox[t]{11cm}{Galactic Halo Targets (Metal-Poor Giants from Gaia+WISE) observed with BOSS, dark time cadence two epoch} & BOSS & 25277 \\ 
        manual\_mwm\_halo\_mp\_wise\_boss\_single & \parbox[t]{11cm}{Galactic Halo Targets (Metal-Poor Giants from Gaia+WISE) observed with BOSS, bright time cadence one epoch} & BOSS & 25277 \\ 
        manual\_mwm\_halo\_mp\_wise\_boss\_triple & \parbox[t]{11cm}{Galactic Halo Targets (Metal-Poor Giants from Gaia+WISE) observed with BOSS, dark time cadence three epoch} & BOSS & 25277 \\ 
        manual\_mwm\_halo\_vmp\_wise\_apogee & \parbox[t]{11cm}{Galactic Halo Targets (Very metal-Poor Giants from Gaia+WISE) observed with APOGEE, dark time cadence two epoch} & APOGEE & 5133 \\ 
        manual\_mwm\_halo\_vmp\_wise\_apogee\_single & \parbox[t]{11cm}{Galactic Halo Targets (Very metal-Poor Giants from Gaia+WISE) observed with APOGEE bright time candence one epoch} & APOGEE & 5133 \\ 
        manual\_mwm\_halo\_vmp\_wise\_apogee\_triple & \parbox[t]{11cm}{Galactic Halo Targets (Very metal-Poor Giants from Gaia+WISE) observed with APOGEE, dark time cadence three epoch} & APOGEE & 5133 \\ 
        manual\_mwm\_halo\_vmp\_wise\_boss & \parbox[t]{11cm}{Galactic Halo Targets (Very metal-Poor Giants from Gaia+WISE) observed with BOSS, dark time cadence two epoch} & BOSS & 11748 \\ 
        manual\_mwm\_halo\_vmp\_wise\_boss\_single & \parbox[t]{11cm}{Galactic Halo Targets (Very metal-Poor Giants from Gaia+WISE) observed with BOSS, bright time candece one epoch} & BOSS & 11748 \\ 
        manual\_mwm\_halo\_vmp\_wise\_boss\_triple & \parbox[t]{11cm}{Galactic Halo Targets (Very metal-Poor Giants from Gaia+WISE) observed with BOSS, dark time cadence three epoch} & BOSS & 11748 \\ 
        manual\_mwm\_magcloud\_massive\_apogee & \parbox[t]{11cm}{Magellanic Clouds massive stars (APOGEE)} & APOGEE & 1000 \\ 
        manual\_mwm\_magcloud\_massive\_boss & \parbox[t]{11cm}{Magellanic Clouds massive stars (BOSS)} & BOSS & 1000 \\ 
        manual\_mwm\_magcloud\_symbiotic\_apogee & \parbox[t]{11cm}{Magellanic Clouds symbiotic stars (APOGEE)} & APOGEE & 24 \\ 
        manual\_mwm\_nsbh\_apogee & \parbox[t]{11cm}{compact NS/BH binaries APOGEE} & APOGEE & 80 \\ 
        manual\_mwm\_nsbh\_boss & \parbox[t]{11cm}{compact NS/BH binaries BOSS} & BOSS & 272 \\ 
        manual\_mwm\_planet\_ca\_legacy\_apogee & \parbox[t]{11cm}{CA Legacy sample of Doppler-surveyed stars} & APOGEE & 719 \\ 
        manual\_mwm\_planet\_gaia\_astrometry\_apogee & \parbox[t]{11cm}{subsample of Gaia EDR3 100-pc sample that Gaia DR5 will have sufficient astrometric precision to detect a planet-mass object} & APOGEE & 35383 \\ 
        manual\_mwm\_planet\_gpi\_apogee & \parbox[t]{11cm}{GPIES sample} & APOGEE & 624 \\ 
        manual\_mwm\_planet\_harps\_apogee & \parbox[t]{11cm}{HARPS sample of Doppler-surveyed stars} & APOGEE & 684 \\ 
        manual\_mwm\_planet\_known\_apogee & \parbox[t]{11cm}{known exoplanets as of 3 January 2023} & APOGEE & 3807 \\ 
        manual\_mwm\_planet\_sophie\_apogee & \parbox[t]{11cm}{SOPHIE sample of Doppler-surveyed stars} & APOGEE & 425 \\ 
        manual\_mwm\_planet\_sphere\_apogee & \parbox[t]{11cm}{stars observed by SPHERE} & APOGEE & 781 \\ 
        manual\_mwm\_planet\_tess\_eb\_apogee & \parbox[t]{11cm}{TESS eclipsing binary catalog} & APOGEE & 4559 \\ 
        manual\_mwm\_planet\_tess\_pc\_apogee & \parbox[t]{11cm}{TESS TOIs as of 3 January 2023} & APOGEE & 9784 \\ 
        manual\_mwm\_planet\_transiting\_bd\_apogee & \parbox[t]{11cm}{known brown dwarfs transiting solar-type stars} & APOGEE & 56 \\ 
        manual\_mwm\_tess\_ob\_apogee & \parbox[t]{11cm}{OB(AF)-type pulsating stars in eclipsing binaries (TESS Continuous Viewing Zones)} & APOGEE & 329 \\ 
        manual\_mwm\_validation\_cool\_apogee & \parbox[t]{11cm}{Validation -- FGKM stars with known fundamental parameters, apogee} & APOGEE & 19733 \\ 
        manual\_mwm\_validation\_cool\_boss & \parbox[t]{11cm}{Validation -- FGKM stars with known fundamental parameters, boss} & BOSS & 130528 \\ 
        manual\_mwm\_validation\_hot\_apogee & \parbox[t]{11cm}{Science validation targets for observations with APOGEE} & APOGEE & 1091 \\ 
        manual\_mwm\_validation\_hot\_boss & \parbox[t]{11cm}{Science validation targets for observations with BOSS} & BOSS & 3535 \\ 
        manual\_mwm\_validation\_rv\_apogee & \parbox[t]{11cm}{Validation -- Stars observed by extreme-precision RV instruments} & APOGEE & 204 \\ 
        mwm\_astar\_core\_boss & \parbox[t]{11cm}{A stars in the Milky Way and Magellanic clouds with BOSS} & BOSS & 373041 \\ 
        mwm\_astar\_core\_boss\_single & \parbox[t]{11cm}{A stars in the Milky Way and Magellanic clouds with BOSS, single exposure} & BOSS & 373041 \\ 
        mwm\_bin\_gaia\_astb\_apogee & \parbox[t]{11cm}{Astrometry binaries from Gaia APOGEE} & APOGEE & 125476 \\ 
        mwm\_bin\_gaia\_astb\_boss & \parbox[t]{11cm}{Astrometry binaries from Gaia BOSS} & BOSS & 174940 \\ 
        mwm\_bin\_gaia\_sb\_apogee & \parbox[t]{11cm}{SB1 binaries from Gaia APOGEE} & APOGEE & 51810 \\ 
        mwm\_bin\_gaia\_sb\_boss & \parbox[t]{11cm}{SB1 binaries from Gaia BOSS} & BOSS & 111852 \\ 
        mwm\_bin\_rv\_long\_apogee & \parbox[t]{11cm}{RV monitoring of stars with previous APOGEE observations} & APOGEE & 23974 \\ 
        mwm\_bin\_rv\_short\_mdwarf\_apogee\_08epoch & \parbox[t]{11cm}{RV monitoring of m-dwarfs stars with no previous APOGEE observations, 8 epoch cadence} & APOGEE & 156565 \\ 
        mwm\_bin\_rv\_short\_mdwarf\_apogee\_12epoch & \parbox[t]{11cm}{RV monitoring of m-dwarfs stars with no previous APOGEE observations, 12 epoch cadence} & APOGEE & 156565 \\ 
        mwm\_bin\_rv\_short\_mdwarf\_apogee\_18epoch & \parbox[t]{11cm}{RV monitoring of m-dwarfs stars with no previous APOGEE observations, 18 epoch cadence} & APOGEE & 156565 \\ 
        mwm\_bin\_rv\_short\_rgb\_apogee & \parbox[t]{11cm}{RV monitoring of red giants with no previous APOGEE observations of red giants, 18 epoch cadence} & APOGEE & 2927092 \\ 
        mwm\_bin\_rv\_short\_subgiant\_apogee & \parbox[t]{11cm}{RV monitoring of subgiants with no previous APOGEE observations of subgiant stars, 18 epoch cadence} & APOGEE & 590033 \\ 
        mwm\_bin\_rv\_short\_subgiant\_apogee\_12epoch & \parbox[t]{11cm}{RV monitoring of subgiants with no previous APOGEE observations of subgiant stars, 12 epoch cadence} & APOGEE & 590033 \\ 
        mwm\_bin\_vis\_apogee & \parbox[t]{11cm}{Gaia wide binaries for APOGEE} & APOGEE & 287891 \\ 
        mwm\_bin\_vis\_boss & \parbox[t]{11cm}{Gaia wide binaries for BOSS} & BOSS & 2602244 \\ 
        mwm\_cb\_galex\_mag\_boss & \parbox[t]{11cm}{Compact binaries from Gaia+GALEX FUV, magnitude limited} & BOSS & 234815 \\ 
        mwm\_cb\_galex\_vol\_boss & \parbox[t]{11cm}{Compact binaries from Gaia+GALEX NUV, volume limited} & BOSS & 402894 \\ 
        mwm\_cb\_swiftuvot\_boss & \parbox[t]{11cm}{Compact binaries from Gaia+Swift UVOT} & BOSS & 5600 \\ 
        mwm\_cb\_xmmom\_boss & \parbox[t]{11cm}{Compact binaries from Gaia+XMM-Newton OM} & BOSS & 3112 \\ 
        mwm\_dust\_core\_dist\_apogee & \parbox[t]{11cm}{Dust program -- Sample of bright, nearby, mid-plane giants complimentary with galactic\_core\_dist} & APOGEE & 26262 \\ 
        mwm\_erosita\_compact\_boss & \parbox[t]{11cm}{eROSITA/Gaia fainter compact binary candidates} & BOSS & 11113 \\ 
        mwm\_erosita\_compact\_boss\_d3 & \parbox[t]{11cm}{eROSITA/Gaia fainter compact binary candidates, 3 epoch cadence} & BOSS & 11113 \\ 
        mwm\_erosita\_compact\_boss\_shallow & \parbox[t]{11cm}{eROSITA/Gaia compact binary candidates} & BOSS & 22226 \\ 
        mwm\_erosita\_stars\_boss & \parbox[t]{11cm}{eROSITA/Gaia coronal emitters, bright cadence} & BOSS & 222406 \\ 
        mwm\_galactic\_core\_dist\_apogee & \parbox[t]{11cm}{Galactic Genesis -- Luminous Giants across the sky based on parallax, magnitude, and G-H color cuts (depreciated)} & APOGEE & 6002106 \\ 
        mwm\_galactic\_core\_dist\_apogee\_extra & \parbox[t]{11cm}{Galactic Genesis -- Luminous Giants across the sky based on parallax, magnitude, and G-H color cuts} & APOGEE & 2000702 \\ 
        mwm\_galactic\_core\_dist\_apogee\_sparse & \parbox[t]{11cm}{Galactic Genesis -- Luminous Giants across the sky based on parallax, magnitude, and G-H color cuts} & APOGEE & 4001404 \\ 
        mwm\_halo\_distant\_rrl\_boss & \parbox[t]{11cm}{Galactic Halo Targets (Distant RRL from Gaia), dark time cadence two epochs} & BOSS & 129733 \\ 
        mwm\_halo\_distant\_rrl\_boss\_single & \parbox[t]{11cm}{Galactic Halo Targets (Distant RRL from Gaia), bright time cadence one epoch} & BOSS & 129733 \\ 
        mwm\_halo\_distant\_rrl\_boss\_triple & \parbox[t]{11cm}{Galactic Halo Targets (Distant RRL from Gaia), dark time cadence three epochs} & BOSS & 129733 \\ 
        mwm\_halo\_local\_high\_apogee & \parbox[t]{11cm}{Galactic Halo Targets (Local Halo Stars from Gaia with high tangential velocity), dark time cadence two epochs, apogee} & APOGEE & 30082 \\ 
        mwm\_halo\_local\_high\_apogee\_single & \parbox[t]{11cm}{Galactic Halo Targets (Local Halo Stars from Gaia with high tangential velocity), bright time cadence one epoch, apogee} & APOGEE & 30082 \\ 
        mwm\_halo\_local\_high\_apogee\_triple & \parbox[t]{11cm}{Galactic Halo Targets (Local Halo Stars from Gaia with high tangential velocity), dark time cadence three epochs, apogee} & APOGEE & 30082 \\ 
        mwm\_halo\_local\_high\_boss & \parbox[t]{11cm}{Galactic Halo Targets (Local Halo Stars from Gaia with high tangential velocity), dark time cadence two epochs, boss} & BOSS & 954847 \\ 
        mwm\_halo\_local\_high\_boss\_single & \parbox[t]{11cm}{Galactic Halo Targets (Local Halo Stars from Gaia with high tangential velocity), brigth time cadence one epoch, boss} & BOSS & 954847 \\ 
        mwm\_halo\_local\_high\_boss\_triple & \parbox[t]{11cm}{Galactic Halo Targets (Local Halo Stars from Gaia with high tangential velocity), dark time cadence three epochs, boss} & BOSS & 954847 \\ 
        mwm\_halo\_local\_low\_apogee & \parbox[t]{11cm}{Galactic Halo Targets (Local Halo Stars from Gaia with low tangential velocity), dark time cadence two epochs, apogee} & APOGEE & 19 \\ 
        mwm\_halo\_local\_low\_apogee\_single & \parbox[t]{11cm}{Galactic Halo Targets (Local Halo Stars from Gaia with low tangential velocity), bright time cadence one epoch, apogee} & APOGEE & 19 \\ 
        mwm\_halo\_local\_low\_apogee\_triple & \parbox[t]{11cm}{Galactic Halo Targets (Local Halo Stars from Gaia with low tangential velocity), dark time cadence three epochs, apogee} & APOGEE & 19 \\ 
        mwm\_halo\_local\_low\_boss & \parbox[t]{11cm}{Galactic Halo Targets (Local Halo Stars from Gaia with low tangential velocity), dark time cadence two epochs, boss} & BOSS & 710338 \\ 
        mwm\_halo\_local\_low\_boss\_single & \parbox[t]{11cm}{Galactic Halo Targets (Local Halo Stars from Gaia with low tangential velocity), bright time cadence one epoch, boss} & BOSS & 710338 \\ 
        mwm\_halo\_local\_low\_boss\_triple & \parbox[t]{11cm}{Galactic Halo Targets (Local Halo Stars from Gaia with low tangential velocity), dark time cadence three epochs, boss} & BOSS & 710338 \\ 
        mwm\_halo\_mp\_xp\_apogee & \parbox[t]{11cm}{Galactic Halo Targets (Metal-Poor Giants from Gaia XP Spectra), dark time cadence two epochs, apogee} & APOGEE & 15425 \\ 
        mwm\_halo\_mp\_xp\_apogee\_single & \parbox[t]{11cm}{Galactic Halo Targets (Metal-Poor Giants from Gaia XP Spectra), bright time cadence one epoch, apogee} & APOGEE & 15425 \\ 
        mwm\_halo\_mp\_xp\_apogee\_triple & \parbox[t]{11cm}{Galactic Halo Targets (Metal-Poor Giants from Gaia XP Spectra), dark time cadence three epochs, apogee} & APOGEE & 15425 \\ 
        mwm\_halo\_mp\_xp\_boss & \parbox[t]{11cm}{Galactic Halo Targets (Metal-Poor Giants from Gaia XP Spectra), dark time cadence two epochs, boss} & BOSS & 128476 \\ 
        mwm\_halo\_mp\_xp\_boss\_single & \parbox[t]{11cm}{Galactic Halo Targets (Metal-Poor Giants from Gaia XP Spectra), bright time cadence one epoch, boss} & BOSS & 128476 \\ 
        mwm\_halo\_mp\_xp\_boss\_triple & \parbox[t]{11cm}{Galactic Halo Targets (Metal-Poor Giants from Gaia XP Spectra), dark time cadence three epochs, boss} & BOSS & 128476 \\ 
        mwm\_halo\_nmp\_xp\_apogee & \parbox[t]{11cm}{Galactic Halo Targets (Non Metal-Poor Giants from Gaia XP Spectra), dark time cadence two epochs, apogee} & APOGEE & 39255 \\ 
        mwm\_halo\_nmp\_xp\_apogee\_single & \parbox[t]{11cm}{Galactic Halo Targets (Non Metal-Poor Giants from Gaia XP Spectra), bright time cadence one epoch apogee} & APOGEE & 39255 \\ 
        mwm\_halo\_nmp\_xp\_apogee\_triple & \parbox[t]{11cm}{Galactic Halo Targets (Non Metal-Poor Giants from Gaia XP Spectra), dark time cadence three epochs, apogee} & APOGEE & 39255 \\ 
        mwm\_halo\_nmp\_xp\_boss & \parbox[t]{11cm}{Galactic Halo Targets (Non Metal-Poor Giants from Gaia XP Spectra), dark time cadence two epochs, boss} & BOSS & 450725 \\ 
        mwm\_halo\_nmp\_xp\_boss\_single & \parbox[t]{11cm}{Galactic Halo Targets (Non Metal-Poor Giants from Gaia XP Spectra), bright time cadence one epoch, boss} & BOSS & 450725 \\ 
        mwm\_halo\_nmp\_xp\_boss\_triple & \parbox[t]{11cm}{Galactic Halo Targets (Non Metal-Poor Giants from Gaia XP Spectra), dark time cadence three epochs, boss} & BOSS & 450725 \\ 
        mwm\_halo\_vmp\_xp\_apogee & \parbox[t]{11cm}{Galactic Halo Targets (Very Metal-Poor Giants from Gaia XP Spectra), dark time cadence two epochs, apogee} & APOGEE & 5287 \\ 
        mwm\_halo\_vmp\_xp\_apogee\_single & \parbox[t]{11cm}{Galactic Halo Targets (Very Metal-Poor Giants from Gaia XP Spectra), bright time cadence one epoch apogee} & APOGEE & 5287 \\ 
        mwm\_halo\_vmp\_xp\_apogee\_triple & \parbox[t]{11cm}{Galactic Halo Targets (Very Metal-Poor Giants from Gaia XP Spectra), dark time cadence three epochs, apogee} & APOGEE & 5287 \\ 
        mwm\_halo\_vmp\_xp\_boss & \parbox[t]{11cm}{Galactic Halo Targets (Very Metal-Poor Giants from Gaia XP Spectra), dark time cadence two epochs, boss} & BOSS & 32122 \\ 
        mwm\_halo\_vmp\_xp\_boss\_single & \parbox[t]{11cm}{Galactic Halo Targets (Very Metal-Poor Giants from Gaia XP Spectra), bright time cadence one epoch, boss} & BOSS & 32122 \\ 
        mwm\_halo\_vmp\_xp\_boss\_triple & \parbox[t]{11cm}{Galactic Halo Targets (Very Metal-Poor Giants from Gaia XP Spectra), dark time cadence three epochs, boss} & BOSS & 32122 \\ 
        mwm\_legacy\_ir2opt\_boss & \parbox[t]{11cm}{Follow-up of APOGEE-1/2 targets with BOSS -- Filler Carton} & BOSS & 626085 \\ 
        mwm\_magcloud\_agb\_apogee & \parbox[t]{11cm}{Magellanic Clouds AGB stars (APOGEE)} & APOGEE & 28156 \\ 
        mwm\_magcloud\_rgb\_boss & \parbox[t]{11cm}{Magellanic Clouds RGB stars (BOSS)} & BOSS & 652149 \\ 
        mwm\_monitor\_m15\_apogee\_long & \parbox[t]{11cm}{Validation -- Long Term Instrument Monitoring targets M15} & APOGEE & 95 \\ 
        mwm\_monitor\_m15\_apogee\_short & \parbox[t]{11cm}{Validation -- Long Term Instrument Monitoring targets M15} & APOGEE & 184 \\ 
        mwm\_monitor\_m67\_apogee\_long & \parbox[t]{11cm}{Validation -- Long Term Instrument Monitoring targets M67} & APOGEE & 142 \\ 
        mwm\_monitor\_m67\_apogee\_short & \parbox[t]{11cm}{Validation -- Long Term Instrument Monitoring targets M67} & APOGEE & 210 \\ 
        mwm\_monitor\_n188\_apogee\_long & \parbox[t]{11cm}{Validation -- Long Term Instrument Monitoring targets NGC 188} & APOGEE & 264 \\ 
        mwm\_monitor\_n188\_apogee\_short & \parbox[t]{11cm}{Validation -- Long Term Instrument Monitoring targets NGC 188} & APOGEE & 123 \\ 
        mwm\_ob\_cepheids\_boss & \parbox[t]{11cm}{Cepheids from Inno+21 with BOSS} & BOSS & 2357 \\ 
        mwm\_ob\_core\_boss & \parbox[t]{11cm}{OB stars in the Milky Way and Magellanic clouds with BOSS} & BOSS & 139918 \\ 
        mwm\_ob\_core\_boss\_single & \parbox[t]{11cm}{OB stars in the Milky Way and Magellanic clouds with BOSS, single epoch cadence} & BOSS & 139918 \\ 
        mwm\_snc\_100pc\_boss\_single & \parbox[t]{11cm}{Volume limited census of the solar neighbourhood G>=16, single epoch cadence} & BOSS & 264085 \\ 
        mwm\_snc\_100pc\_bright\_boss\_single & \parbox[t]{11cm}{Volume limited census of the solar neighbourhood G>=16, single epoch cadence} & BOSS & 264085 \\ 
        mwm\_snc\_100pc\_low\_lum\_boss\_single & \parbox[t]{11cm}{Volume limited census of the solar neighbourhood G>=16, single epoch cadence} & BOSS & 6492 \\ 
        mwm\_snc\_ext\_filler\_apogee & \parbox[t]{11cm}{Extended sample of the solar neighborhood for FGK stellar types, G < 16, filler} & APOGEE & 1520090 \\ 
        mwm\_snc\_ext\_filler\_boss & \parbox[t]{11cm}{Extended sample of the solar neighborhood for FGK stellar types, G >=16, filler} & BOSS & 5689667 \\ 
        mwm\_snc\_ext\_main\_apogee & \parbox[t]{11cm}{Extended sample of the solar neighborhood for FGK stellar types, G < 16} & APOGEE & 100903 \\ 
        mwm\_snc\_ext\_main\_boss & \parbox[t]{11cm}{Extended sample of the solar neighborhood for FGK stellar types, G >=16} & BOSS & 299190 \\ 
        mwm\_tess\_2min\_apogee & \parbox[t]{11cm}{Planet -- TESS TOI and 2 minute Cadence Follow-up} & APOGEE & 335231 \\ 
        mwm\_tess\_rgb\_apogee & \parbox[t]{11cm}{TESS Observed Red Giant Astroseismology Follow-up} & APOGEE & 6921315 \\ 
        mwm\_wd\_gaia\_boss & \parbox[t]{11cm}{WDs from Gaia DR3 alone, HWR algorithm} & BOSS & 411286 \\ 
        mwm\_wd\_gaia\_boss\_1x2 & \parbox[t]{11cm}{WDs from Gaia DR3 alone, HWR algorithm} & BOSS & 205643 \\ 
        mwm\_wd\_gaia\_boss\_1x3 & \parbox[t]{11cm}{WDs from Gaia DR3 alone, HWR algorithm} & BOSS & 205643 \\ 
        mwm\_wd\_gaia\_boss\_2x1 & \parbox[t]{11cm}{WDs from Gaia DR3 alone, HWR algorithm} & BOSS & 205643 \\ 
        mwm\_wd\_pwd\_boss & \parbox[t]{11cm}{WDs from Gaia DR3 + prob Gentile Fusilo paper} & BOSS & 450656 \\ 
        mwm\_wd\_pwd\_boss\_1x2 & \parbox[t]{11cm}{WDs from Gaia DR3 + prob Gentile Fusilo paper} & BOSS & 225328 \\ 
        mwm\_wd\_pwd\_boss\_1x3 & \parbox[t]{11cm}{WDs from Gaia DR3 + prob Gentile Fusilo paper} & BOSS & 225328 \\ 
        mwm\_wd\_pwd\_boss\_2x1 & \parbox[t]{11cm}{WDs from Gaia DR3 + prob Gentile Fusilo paper} & BOSS & 225328 \\ 
        mwm\_yso\_cluster\_apogee\_single & \parbox[t]{11cm}{Candidates identified through phase space clustering, targeted with APOGEE, 1x1 cadence} & APOGEE & 45461 \\ 
        mwm\_yso\_cluster\_boss\_single & \parbox[t]{11cm}{Candidates identified through phase space clustering, targeted with BOSS, 1x1 cadence} & BOSS & 59065 \\ 
        mwm\_yso\_cmz\_apogee\_single & \parbox[t]{11cm}{Candidates located in Central Molecular Zone+inner Galactic disk, targeted with APOGEE, 1x1 cadence} & APOGEE & 12818 \\ 
        mwm\_yso\_disk\_apogee\_single & \parbox[t]{11cm}{Optically bright candidates identified through infrared excess, targeted with APOGEE, 1x1 cadence} & APOGEE & 31440 \\ 
        mwm\_yso\_disk\_boss\_single & \parbox[t]{11cm}{Optically bright candidates identified through infrared excess, targeted with BOSS, 1x1 cadence} & BOSS & 39414 \\ 
        mwm\_yso\_embedded\_apogee\_single & \parbox[t]{11cm}{Optically faint candidates identified through infrared excess, targeted with APOGEE, 1x1 cadence} & APOGEE & 5454 \\ 
        mwm\_yso\_nebula\_apogee\_single & \parbox[t]{11cm}{Candidates found towards areas of high nebulosity, targeted with APOGEE, 1x1 cadence} & APOGEE & 1113 \\ 
        mwm\_yso\_ob\_apogee & \parbox[t]{11cm}{Candidate OB stars, targeted with APOGEE (depreciated)} & APOGEE & 8670 \\ 
        mwm\_yso\_ob\_boss & \parbox[t]{11cm}{Candidate OB stars, targeted with BOSS (depreciated)} & BOSS & 8670 \\ 
        mwm\_yso\_pms\_apogee\_sagitta\_edr3 & \parbox[t]{11cm}{Low mass candidates located above main sequence on HR diagram, targeted with APOGEE, cataloged by McBride+21 from Gaia EDR3} & APOGEE & 63532 \\ 
        mwm\_yso\_pms\_apogee\_sagitta\_edr3\_single & \parbox[t]{11cm}{Low mass candidates located above main sequence on HR diagram, targeted with APOGEE, cataloged by McBride+21 from Gaia EDR3, 1x1 cadence} & APOGEE & 63532 \\ 
        mwm\_yso\_pms\_apogee\_zari18pms & \parbox[t]{11cm}{Low mass candidates located above main sequence on HR diagram, targeted with APOGEE, cataloged by Zari+21 from Gaia DR2} & APOGEE & 32534 \\ 
        mwm\_yso\_pms\_apogee\_zari18pms\_single & \parbox[t]{11cm}{Low mass candidates located above main sequence on HR diagram, targeted with APOGEE, cataloged by Zari+21 from Gaia DR2, 1x1 cadence} & APOGEE & 32534 \\ 
        mwm\_yso\_pms\_boss\_sagitta\_edr3 & \parbox[t]{11cm}{Low mass candidates located above main sequence on HR diagram, targeted with BOSS, cataloged by McBride+21 from Gaia EDR3} & BOSS & 56938 \\ 
        mwm\_yso\_pms\_boss\_sagitta\_edr3\_single & \parbox[t]{11cm}{Low mass candidates located above main sequence on HR diagram, targeted with BOSS, cataloged by McBride+21 from Gaia EDR3, 1x1 cadence} & BOSS & 56938 \\ 
        mwm\_yso\_pms\_boss\_zari18pms & \parbox[t]{11cm}{Low mass candidates located above main sequence on HR diagram, targeted with APOGEE, cataloged by Zari+21 from Gaia DR2} & BOSS & 32442 \\ 
        mwm\_yso\_pms\_boss\_zari18pms\_single & \parbox[t]{11cm}{Low mass candidates located above main sequence on HR diagram, targeted with APOGEE, cataloged by Zari+21 from Gaia DR2, 1x1 cadence} & BOSS & 32442 \\ 
        mwm\_yso\_variable\_apogee\_single & \parbox[t]{11cm}{Low mass candidates identified through Gaia variability, targeted with APOGEE, 1x1 cadence} & APOGEE & 42990 \\ 
        mwm\_yso\_variable\_boss\_single & \parbox[t]{11cm}{Low mass candidates identified through Gaia variability, targeted with BOSS, 1x1 cadence} & BOSS & 41124 \\ 
\enddata

\tablecomments{\tablenotetext{1}{See online documentation for complete selection details: \url{https://www.sdss.org/dr20/mwm/programs/cartons/}} 
\tablenotetext{2}{\textbf{Available targets} is the number of targets that satisfy the carton selection function in the targeting database. The number of targets ultimately observed for each carton will be smaller than this value.}}
\end{deluxetable*}
\end{longrotatetable}
In Table \ref{tab:mwm_cartons}, we list all of the new MWM cartons associated with DR20 with a short description of their selection functions.  
While the number of ``new'' cartons is comparatively large, relatively few cartons had substantial changes in the fundamental selection criteria. Some cartons changed in name only (e.g., ``mwm\_legacy\_ir2opt'' to ``mwm\_legacy\_ir2opt\_boss''), while other cartons are duplicates in selection function but have varying combinations of priorities and cadences to assist with planning. E.g., a carton with the desired, but more difficult to complete cadence is given a high priority, while the same targets are duplicated in another carton with a simpler cadence at lower priority. Because the \texttt{robostrategy} software \citep{Blanton2025} already handles targets belonging to multiple cartons, there is no risk of unnecessary duplicate observing with this strategy.  An example of this is the addition of the ``mwm\_snc\_100pc\_boss\_single'' carton, which is a duplicate of the ``mwm\_snc\_100pc\_boss'' carton but requests a single epoch of observation instead of two epochs.

Other cartons had more substantial changes to adjust the target selection criteria (e.g.,  \ref{ex1_mwm_gg} and \ref{ex2_mwm_rv_short}) or to add entirely new but complementary selections to existing programs (e.g., \ref{ex3_mwm_wd}).
For each of these three examples, we provide the following information:
The {\bf Description of selection criteria} provides a short summary of the carton selection and relationship to previous carton names.  It also includes a list of the limits in color, magnitude, parallax, and other quantities applied to the carton's target candidates, with changes compared to previous iterations bolded.
{\bf Data Sources} gives the catalogs from which these quantities are drawn. 
In {\bf Target priority options}, we indicate which priority is given to targets in this carton for observing; smaller priorities are more likely to be assigned fibers. The {\bf Cadence options} describe the exposure time requirement(s) assigned to sources in the carton. 
{\bf Related Cartons} describes the relationship to other cartons appearing in DR20. 
The {\bf DR18 Counterpart} refers the reader to the corresponding science carton that was released in \citet{almeida2023}. 

\begin{center}\rule{0.5\linewidth}{0.5pt}\end{center}

\subsubsection{mwm\_galactic\_core\_dist\_apogee}\label{ex1_mwm_gg}
\begin{description}
    \item[Description of selection criteria] A simple color-magnitude cut effectively targets luminous cool giant stars (${\rm median}(\log g) \sim 1.0-1.5$) with little ($<$6\%) contamination from dwarf stars. Relative to its DR18 counterpart, this carton imposes a redder color cut and introduces an absolute magnitude cut.  The carton was later split into two cartons. 
    \begin{itemize}
        \item $H < 11$
        \item$\mathbf{ G - H > 5.0}$ \textbf{or Gaia non-detection}
         \item gal\_contam == 0
         \item cc\_flg == 0
        \item 0 $<$ rd\_flag $<= 3$
         \item ph\_qual flag is A or B
         \item $\mathbf{\varpi < 10^{0.2(-2.5-H)}}$  
    \end{itemize}
\item[Data Sources] Gaia DR3 (G, $\varpi$), 2MASS PSC (H, gal\_contam, cc\_flg, rd\_flag, ph\_qual)
\item[Target priority options] 2710
\item[Cadence options] bright\_1x1
\item[Related Cartons]  This core carton replaces the original mwm\_galactic\_core carton. It was later split into two separate cartons by random downsampling.  Two-thirds of the parent sample was used to create a more sparsely sampled core carton, resulting in the ``mwm\_galactic\_core\_dist\_apogee\_sparse'' carton. The remaining one-third of the parent sample became a lower priority carton to be observed only if resources allowed (```mwm\_galactic\_core\_dist\_apogee\_extra'').
\item[DR18 Counterpart] \textit{mwm\_galactic\_core} -- Section A.19 of \citet{almeida2023}.
\end{description}

\begin{center}\rule{0.5\linewidth}{0.5pt}\end{center}

\subsubsection{mwm\_bin\_rv\_short\_rgb\_apogee}\label{ex2_mwm_rv_short}
\begin{description}
    \item[Description of selection criteria] This carton selects stars never previously observed with the APOGEE instrument and applies the same color and quality cuts as the main APOGEE surveys. Previously,  temperature and surface gravity from the TIC catalog were used to prioritize targets to give precedence to those that might be subgiant or M dwarf stars (since they are rarer).  In v1.0, each stellar type is now selected for explicitly.  Furthermore, early SDSS-V observations were dominated by upper red giant branch stars; thus,  a lower limit to surface gravity was also imposed. The carton name was also changed to add explicit mention of the instrument used (the ``apogee'' suffix) and to reflect a change to the scope of the parent ``Binary'' science program, which newly includes visual binary cartons (``mwm\_bin\_vis\_*'') in addition to the cartons that will identify binarity with many RV epochs (``mwm\_bin\_rv\_*'').
    \begin{itemize}
        \item $H< 10.8$
        \item $J - K_s - (1.5 \cdot 0.918 (H - W2 - 0.05)) >= 0.05$
           \item $J\_msigcom,H\_msigcom,K_s\_msigcom   \leq 0.1$
        \item $W2\_sigmpro \leq 0.1$
        \item 2MASS ph\_qual any of the following: AAA, AAB, ABA, BAA, ABB, BAB, BBA, BBB
         \item 2MASS gal\_contam = 0
         \item  2MASS cc\_flg = 0
         \item 2MASS rd\_flg any of the following: 111, 112, 121, 211, 122, 212, 221, 222
         \item 2MASS prox $\geq$ 6
          \item Does not match a source in the 2MASS extended catalog
          \item Gaia \textbf{DR3} parallax ($\varpi$) exists
         \item $\mathbf{\varpi/\sigma_\varpi < 0.2 }$
         \item $\mathbf{T_{\rm eff} < 5000}$
         \item $\mathbf{1.2 < log~g \le 3.0}$
    \end{itemize}
\item[Data Sources] 2MASS PSC (J, H, $K_s$, J\_msigcom, H\_msigcom, K$_s$\_msigcom, ph\_qual, gal\_contam, cc\_flg, rd\_flg, prox), Gaia DR 3 ( $\varpi$, $\sigma_\varpi$, $T_{\rm eff}$, $\log g$), AllWise (W2)
\item[Target priority options] 2535
\item[Cadence options]  bright\_18x1
\item[Related Cartons]  The rarer subgiant and M dwarf stars identified with this selection function now belong to their own cartons, and duplicate cartons at lower priority but with easier-to-achieve 12 epoch or 8 epoch cadences were also created (mwm\_bin\_rv\_short\_subgiant\_apogee, mwm\_bin\_rv\_short\_subgiant\_apogee\_12epoch, mwm\_bin\_rv\_short\_mdwarf\_apogee\_18epoch,  mwm\_bin\_rv\_short\_mdwarf\_apogee\_12epoch,  and mwm\_bin\_rv\_short\_mdwarf\_apogee\_08epoch).
\item[DR18 Counterpart] \textit{mwm\_rv\_short\_fps} -- Section A.24 of \citet{almeida2023}.
\end{description}

\begin{center}\rule{0.5\linewidth}{0.5pt}\end{center}

\subsubsection{mwm\_wd\_gaia\_boss}\label{ex3_mwm_wd}
\begin{description}
    \item[Description of selection criteria] This carton is entirely new to DR20 and selects all white dwarf candidates using Gaia color magnitude cuts.
    \begin{itemize}
        \item $G < 20.0$
       \item  $\varpi \cdot 10^{0.2(22-G)} > 3$
        \item $G + 5 \log_{10}(\varpi/100) -0.5-5.5\cdot(G_{BP}-G_{RP}+0.5)> 0$
        \item $(G_{BP}-G_{RP} < 0.4)$ OR ($G_{BP}-G_{RP} > 0.4$ AND $G + 5 \log_{10}(\varpi/100) > 11 + 2.2 \cdot (G_{BP}-G_{RP} - 0.5)$)
       \item $ (G + 5 \log_{10}(\varpi/100) - 6.7 -2.2 (G_{BP}-G_{RP})) > 0 $ OR $(G_{BP}-G_{RP} < -0.6)$
       \item \texttt{fidelity\_v2} $> 0.9$
    \end{itemize}
\item[Data Sources]  Gaia DR3 ($\varpi$, $G$, $G_{BP}$, $G_{RP}$), \cite{2022MNRAS.510.2597R} (\texttt{fidelity\_v2})
\item[Target priority options] 1401
\item[Cadence options] dark\_2x1
\item[Related Cartons]  This carton supplements the selection of white dwarfs from \cite{Gentile_2021} (the mwm\_wd\_pwd\_boss carton) but at lower priority.
\item[DR18 Counterpart] None. New for DR20.
\end{description}

\section{New BHM target cartons}
\label{sec:appendix_bhm}
We describe here all new BHM (and extragalactic open fiber) target cartons (summarized in table~\ref{tab:bhm_cartons}) that we are releasing as part of DR20 and that have not been documented in previous SDSS-V data releases (namely DR18 and DR19).
These new cartons are associated with the \vonex\ catalog cross-match which underpins target selection for the main FPS survey.\\

\begin{table}[ht]
\centering
\caption{New BHM (and extragalactic openfiber) cartons in DR20.
The plan column specifies the version of \texttt{target\_selection} code used to generate each
carton\footnote{\url{https://github.com/sdss/target_selection}}
$N_\mathrm{targets}$ gives the number of targets in each carton.}
\label{tab:bhm_cartons}
{\scriptsize
\begin{tabular}{lcr}
Carton name & plan & $N_\mathrm{targets}$ \\
\hline
\hyperref[openfibertargets_nov2020_11_plan1.0.7]{\texttt{openfibertargets\_nov2020\_11}} & 1.0.7 & 39684 \\
\hyperref[openfibertargets_nov2020_18_plan1.0.7]{\texttt{openfibertargets\_nov2020\_18}} & 1.0.7 & 10090 \\
\hyperref[openfibertargets_nov2020_26_plan1.0.7]{\texttt{openfibertargets\_nov2020\_26}} & 1.0.7 & 2105 \\
\hyperref[openfibertargets_nov2020_27_plan1.0.7]{\texttt{openfibertargets\_nov2020\_27}} & 1.0.7 & 814965 \\
\hyperref[openfibertargets_nov2020_30_plan1.0.7]{\texttt{openfibertargets\_nov2020\_30}} & 1.0.7 & 237 \\
\hyperref[openfibertargets_nov2020_33_plan1.0.10]{\texttt{openfibertargets\_nov2020\_33}} & 1.0.10 & 22246 \\
\hyperref[bhm_rm_core_plan1.0.12]{\texttt{bhm\_rm\_core}} & 1.0.12 & 3741 \\
\hyperref[bhm_rm_known_spec_plan1.0.12]{\texttt{bhm\_rm\_known\_spec}} & 1.0.12 & 3206 \\
\hyperref[bhm_rm_var_plan1.0.12]{\texttt{bhm\_rm\_var}} & 1.0.12 & 871 \\
\hyperref[bhm_rm_ancillary_plan1.0.12]{\texttt{bhm\_rm\_ancillary}} & 1.0.12 & 1114 \\
\hyperref[bhm_rm_xrayqso_plan1.0.12]{\texttt{bhm\_rm\_xrayqso}} & 1.0.12 & 43 \\
\hyperref[bhm_csc_boss_plan1.0.37]{\texttt{bhm\_csc\_boss}} & 1.0.37 & 135154 \\
\hyperref[bhm_csc_apogee_plan1.0.37]{\texttt{bhm\_csc\_apogee}} & 1.0.37 & 54593 \\
\hyperref[bhm_gua_dark_plan1.0.37]{\texttt{bhm\_gua\_dark}} & 1.0.37 & 1936784 \\
\hyperref[bhm_gua_bright_plan1.0.37]{\texttt{bhm\_gua\_bright}} & 1.0.37 & 185644 \\
\hyperref[bhm_spiders_clusters_lsdr10_plan1.0.37]{\texttt{bhm\_spiders\_clusters\_lsdr10}} & 1.0.37 & 440276 \\
\hyperref[bhm_spiders_agn_lsdr10_plan1.0.37]{\texttt{bhm\_spiders\_agn\_lsdr10}} & 1.0.37 & 975978 \\
\hyperref[bhm_spiders_agn_hard_plan1.0.37]{\texttt{bhm\_spiders\_agn\_hard}} & 1.0.37 & 2603 \\
\hyperref[bhm_spiders_agn_gaiadr3_plan1.0.37]{\texttt{bhm\_spiders\_agn\_gaiadr3}} & 1.0.37 & 717805 \\
\hyperref[bhm_spiders_agn_tda_plan1.0.37]{\texttt{bhm\_spiders\_agn\_tda}} & 1.0.37 & 9627 \\
\hyperref[bhm_spiders_agn_sep_plan1.0.37]{\texttt{bhm\_spiders\_agn\_sep}} & 1.0.37 & 1681 \\
\hyperref[bhm_aqmes_med_plan1.0.37]{\texttt{bhm\_aqmes\_med}} & 1.0.37 & 2663 \\
\hyperref[bhm_aqmes_med_faint_plan1.0.37]{\texttt{bhm\_aqmes\_med\_faint}} & 1.0.37 & 16853 \\
\hyperref[bhm_aqmes_wide2_plan1.0.37]{\texttt{bhm\_aqmes\_wide2}} & 1.0.37 & 24142 \\
\hyperref[bhm_aqmes_wide2_faint_plan1.0.37]{\texttt{bhm\_aqmes\_wide2\_faint}} & 1.0.37 & 99586 \\
\hyperref[bhm_aqmes_bonus_core_plan1.0.37]{\texttt{bhm\_aqmes\_bonus\_core}} & 1.0.37 & 83163 \\
\hyperref[bhm_aqmes_bonus_bright_plan1.0.37]{\texttt{bhm\_aqmes\_bonus\_bright}} & 1.0.37 & 10848 \\
\hyperref[bhm_aqmes_bonus_faint_plan1.0.37]{\texttt{bhm\_aqmes\_bonus\_faint}} & 1.0.37 & 424163 \\
\hyperref[bhm_colr_galaxies_lsdr10_plan1.0.38]{\texttt{bhm\_colr\_galaxies\_lsdr10}} & 1.0.38 & 6246703 \\
\hyperref[openfibertargets_bhm_racsradio_boss_plan1.2.0]{\texttt{openfibertargets\_bhm\_racsradio\_boss}} & 1.2.0 & 149989 \\
\hyperref[openfibertargets_bhm_varagn_boss_plan1.2.0]{\texttt{openfibertargets\_bhm\_varagn\_boss}} & 1.2.0 & 6966 \\
\hyperref[openfibertargets_bhm_morexray_boss_plan1.2.0]{\texttt{openfibertargets\_bhm\_morexray\_boss}} & 1.2.0 & 103105 \\
\hyperref[openfibertargets_bhm_hightderates_psb_boss_plan1.2.0]{\texttt{openfibertargets\_bhm\_hightderates\_psb\_boss}} & 1.2.0 & 8934 \\
\hyperref[openfibertargets_bhm_hightderates_qbs_boss_plan1.2.0]{\texttt{openfibertargets\_bhm\_hightderates\_qbs\_boss}} & 1.2.0 & 29659 \\
\hyperref[openfibertargets_bhm_quaia_boss_plan1.2.0]{\texttt{openfibertargets\_bhm\_quaia\_boss}} & 1.2.0 & 999906 \\
\hyperref[openfibertargets_bhm_lofarradio_boss_plan1.2.1]{\texttt{openfibertargets\_bhm\_lofarradio\_boss}} & 1.2.1 & 210483 \\
\hyperref[bhm_csc_boss_d3_plan1.2.4]{\texttt{bhm\_csc\_boss\_d3}} & 1.2.4 & 97267 \\
\hyperref[bhm_spiders_clusters_lsdr10_d3_plan1.2.4]{\texttt{bhm\_spiders\_clusters\_lsdr10\_d3}} & 1.2.4 & 427445 \\
\hyperref[bhm_spiders_agn_gaiadr3_d3_plan1.2.4]{\texttt{bhm\_spiders\_agn\_gaiadr3\_d3}} & 1.2.4 & 398493 \\
\hyperref[bhm_spiders_agn_sep_d3_plan1.2.4]{\texttt{bhm\_spiders\_agn\_sep\_d3}} & 1.2.4 & 863 \\
\hyperref[bhm_aqmes_wide1_plan1.2.4]{\texttt{bhm\_aqmes\_wide1}} & 1.2.4 & 24142 \\
\hyperref[bhm_aqmes_wide1_faint_plan1.2.4]{\texttt{bhm\_aqmes\_wide1\_faint}} & 1.2.4 & 99586 \\
\hyperref[bhm_gua_dark_d3_plan1.2.4]{\texttt{bhm\_gua\_dark\_d3}} & 1.2.4 & 1936784 \\
\hyperref[bhm_spiders_agn_hard_d3_plan1.2.4]{\texttt{bhm\_spiders\_agn\_hard\_d3}} & 1.2.4 & 798 \\
\hyperref[bhm_spiders_agn_tda_d3_plan1.2.4]{\texttt{bhm\_spiders\_agn\_tda\_d3}} & 1.2.4 & 8141 \\
\hyperref[bhm_spiders_agn_lsdr10_d3_plan1.2.4]{\texttt{bhm\_spiders\_agn\_lsdr10\_d3}} & 1.2.4 & 851160 \\
\hyperref[bhm_colr_galaxies_lsdr10_d3_plan1.2.16]{\texttt{bhm\_colr\_galaxies\_lsdr10\_d3}} & 1.2.16 & 6091854 \\
\hyperref[bhm_rm_core_plan1.0.48]{\texttt{bhm\_rm\_core}} & 1.0.48 & 3757 \\
\hyperref[bhm_rm_known_spec_plan1.0.48]{\texttt{bhm\_rm\_known\_spec}} & 1.0.48 & 3212 \\
\hyperref[bhm_rm_var_plan1.0.48]{\texttt{bhm\_rm\_var}} & 1.0.48 & 871 \\
\hyperref[bhm_rm_ancillary_plan1.0.48]{\texttt{bhm\_rm\_ancillary}} & 1.0.48 & 1114 \\
\hline
\end{tabular}
}
\end{table}

\hypertarget{openfibertargets_nov2020_11_plan1.0.7}{%
\subsection{openfibertargets\_nov2020\_11}\label{openfibertargets_nov2020_11_plan1.0.7}}

\noindent\textbf{target\_selection plan:} 1.0.7

\noindent\textbf{target\_selection tag:}
\href{https://github.com/sdss/target_selection/tree/1.0.7/}{1.0.7}

\noindent\textbf{crossmatch plan:} 1.0.0

\noindent\textbf{Summary:} Quasars observed during SDSS-I and -II which had only
a single epoch to changes in emission lines and absorption line troughs
over timescales of years to decades

\noindent\textbf{Simplified description of selection criteria:} manual

\noindent\textbf{Target priority options:} 6080

\noindent\textbf{Cadence options:} \texttt{bright\_1x1,\ dark\_1x1}

\noindent\textbf{Implementation:} n/a

\noindent\textbf{Number of targets:} 39684

\begin{center}\rule{0.5\linewidth}{0.5pt}\end{center}

\hypertarget{openfibertargets_nov2020_18_plan1.0.7}{%
\subsection{openfibertargets\_nov2020\_18}\label{openfibertargets_nov2020_18_plan1.0.7}}

\noindent\textbf{target\_selection plan:} 1.0.7

\noindent\textbf{target\_selection tag:}
\href{https://github.com/sdss/target_selection/tree/1.0.7/}{1.0.7}

\noindent\textbf{crossmatch plan:} 1.0.0

\noindent\textbf{Summary:} Hard X-ray emitting sources (primarily AGNs) from
XMM-Newton and Swift serendipitous surveys

\noindent\textbf{Simplified description of selection criteria:} manual

\noindent\textbf{Target priority options:} 6080

\noindent\textbf{Cadence options:} \texttt{bright\_1x1,\ dark\_1x1}

\noindent\textbf{Implementation:} n/a

\noindent\textbf{Number of targets:} 10090

\begin{center}\rule{0.5\linewidth}{0.5pt}\end{center}

\hypertarget{openfibertargets_nov2020_26_plan1.0.7}{%
\subsection{openfibertargets\_nov2020\_26}\label{openfibertargets_nov2020_26_plan1.0.7}}

\noindent\textbf{target\_selection plan:} 1.0.7

\noindent\textbf{target\_selection tag:}
\href{https://github.com/sdss/target_selection/tree/1.0.7/}{1.0.7}

\noindent\textbf{crossmatch plan:} 1.0.0

\noindent\textbf{Summary:} Quasars monitored in SDSS-I and II as a part of
ultraviolet Broad Absorption Line (BAL) Quasar Variability Survey
(\citealt{Gibson2009}) to receive an additional epoch of observations to monitor
wind evolution

\noindent\textbf{Simplified description of selection criteria:} manual

\noindent\textbf{Target priority options:} 6080

\noindent\textbf{Cadence options:} \texttt{dark\_1x1}

\noindent\textbf{Implementation:} n/a

\noindent\textbf{Number of targets:} 2105

\begin{center}\rule{0.5\linewidth}{0.5pt}\end{center}

\hypertarget{openfibertargets_nov2020_27_plan1.0.7}{%
\subsection{openfibertargets\_nov2020\_27}\label{openfibertargets_nov2020_27_plan1.0.7}}

\noindent\textbf{target\_selection plan:} 1.0.7

\noindent\textbf{target\_selection tag:}
\href{https://github.com/sdss/target_selection/tree/1.0.7/}{1.0.7}

\noindent\textbf{crossmatch plan:} 1.0.0

\noindent\textbf{Summary:} Bright quasar candidates with quasar-like Gaia and
unWISE photometry, supplemented with multi-epoch variability observed by
DES, excluding regions of high confusion

\noindent\textbf{Simplified description of selection criteria:} manual

\noindent\textbf{Target priority options:} 6085

\noindent\textbf{Cadence options:} \texttt{bright\_1x1,\ dark\_1x1}

\noindent\textbf{Implementation:} n/a

\noindent\textbf{Number of targets:} 814965

\begin{center}\rule{0.5\linewidth}{0.5pt}\end{center}

\hypertarget{openfibertargets_nov2020_30_plan1.0.7}{%
\subsection{openfibertargets\_nov2020\_30}\label{openfibertargets_nov2020_30_plan1.0.7}}

\noindent\textbf{target\_selection plan:} 1.0.7

\noindent\textbf{target\_selection tag:}
\href{https://github.com/sdss/target_selection/tree/1.0.7/}{1.0.7}

\noindent\textbf{crossmatch plan:} 1.0.0

\noindent\textbf{Summary:} Bright sources in the JWST North Ecliptic Pole
Time-Domain Field

\noindent\textbf{Simplified description of selection criteria:} manual

\noindent\textbf{Target priority options:} 6080

\noindent\textbf{Cadence options:} \texttt{dark\_1x1}

\noindent\textbf{Implementation:} n/a

\noindent\textbf{Number of targets:} 237

\begin{center}\rule{0.5\linewidth}{0.5pt}\end{center}

\hypertarget{openfibertargets_nov2020_33_plan1.0.10}{%
\subsection{openfibertargets\_nov2020\_33}\label{openfibertargets_nov2020_33_plan1.0.10}}

\noindent\textbf{target\_selection plan:} 1.0.10

\noindent\textbf{target\_selection tag:}
\href{https://github.com/sdss/target_selection/tree/1.0.10/}{1.0.10}

\noindent\textbf{crossmatch plan:} 1.0.0

\noindent\textbf{Summary:} Candidate AGN, QSOs, and Blazars identified based on
their ZTF variability flagged by ALeRCE light curve classifier
(\citealt{SanchezSaez2020})

\noindent\textbf{Simplified description of selection criteria:} manual

\noindent\textbf{Target priority options:} 6080

\noindent\textbf{Cadence options:} \texttt{bright\_1x1,\ dark\_1x1}

\noindent\textbf{Implementation:} n/a

\noindent\textbf{Number of targets:} 22246

\begin{center}\rule{0.5\linewidth}{0.5pt}\end{center}

\hypertarget{bhm_rm_core_plan1.0.12}{%
\subsection{bhm\_rm\_core}\label{bhm_rm_core_plan1.0.12}}

\noindent\textbf{target\_selection plan:} 1.0.12

\noindent\textbf{target\_selection tag:}
\href{https://github.com/sdss/target_selection/tree/1.0.12/}{1.0.12}

\noindent\textbf{crossmatch plan:} 1.0.0

\noindent\textbf{Summary:} A sample of candidate QSOs selected via the methods
presented by
\citet{Yang2022}. These targets are located within five (+1 backup) well
known survey fields (SDSS-RM, COSMOS, XMM-LSS, ECDFS, CVZ-S/SEP, and
ELIAS-S1).

\noindent\textbf{Simplified description of selection criteria:} Starting from a
parent catalog of optically selected objects in the RM fields (as
presented by
\citealt{Yang2022}), select candidate QSOs that satisfy all of the
following: i) are identified via the Skew-T algorithm
(\texttt{skewt\_qso\ ==\ 1}); ii) have
17\textless{}\texttt{psfmag\_i}\textless21.5~AB
(16\textless{}\emph{G}\textless21.7~AB in the CVZ-S/SEP field); iii) do
not have significant detections (\textgreater3$\sigma$) of parallax and/or
proper motion in Gaia DR2; iv) are not vetoed due to results of visual
inspections of recent spectroscopy; vi) have detections in all of the
gri bands (a Gaia detection is sufficient in the CVZ-S/SEP field);and
vii) do not lie in the SDSS-RM field.

\noindent\textbf{Target priority options:} 1000-1100

\noindent\textbf{Cadence options:} \texttt{dark\_100x8,\ dark\_174x8}

\noindent\textbf{Implementation:}
\href{https://github.com/sdss/target_selection/blob/1.0.12/python/target_selection/cartons/bhm_rm.py}{bhm\_rm.py}

\noindent\textbf{Number of targets:} 3741

\begin{center}\rule{0.5\linewidth}{0.5pt}\end{center}

\hypertarget{bhm_rm_known_spec_plan1.0.12}{%
\subsection{bhm\_rm\_known\_spec}\label{bhm_rm_known_spec_plan1.0.12}}

\noindent\textbf{target\_selection plan:} 1.0.12

\noindent\textbf{target\_selection tag:}
\href{https://github.com/sdss/target_selection/tree/1.0.12/}{1.0.12}

\noindent\textbf{crossmatch plan:} 1.0.0

\noindent\textbf{Summary:} A sample of known QSOs identified through optical
spectroscopy from various projects, as collated by
\citet{Yang2022}. These targets are located within five (+1 backup) well
known survey fields (SDSS-RM, COSMOS, XMM-LSS, ECDFS, CVZ-S/SEP, and
ELIAS-S1).

\noindent\textbf{Simplified description of selection criteria:} Starting from a
parent catalog of optically selected objects in the RM fields (as
presented by
\citealt{Yang2022}), select targets which satisfy all of the following: i)
are flagged as having a spectroscopic identification (in the parent
catalog); ii) have 15\textless{}\texttt{psfmag\_i}\textless21.7~AB
(SDSS-RM, CDFS, ELIAS-S1 field), 16\textless{}\emph{G}\textless21.7~Vega
(CVZ-S/SEP field), 15\textless{}\texttt{psfmag\_i}\textless21.5~AB
(COSMOS and XMM-LSS fields); iii) have a spectroscopic redshift in the
range 0.005\textless{}\emph{z}\textless7; iv) are not vetoed due to
results of visual inspections of recent spectroscopy.

\noindent\textbf{Target priority options:} 1000-1100

\noindent\textbf{Cadence options:} \texttt{dark\_100x8,\ dark\_174x8}

\noindent\textbf{Implementation:}
\href{https://github.com/sdss/target_selection/blob/1.0.12/python/target_selection/cartons/bhm_rm.py}{bhm\_rm.py}

\noindent\textbf{Number of targets:} 3206

\begin{center}\rule{0.5\linewidth}{0.5pt}\end{center}

\hypertarget{bhm_rm_var_plan1.0.12}{%
\subsection{bhm\_rm\_var}\label{bhm_rm_var_plan1.0.12}}

\noindent\textbf{target\_selection plan:} 1.0.12

\noindent\textbf{target\_selection tag:}
\href{https://github.com/sdss/target_selection/tree/1.0.12/}{1.0.12}

\noindent\textbf{crossmatch plan:} 1.0.0

\noindent\textbf{Summary:} A sample of candidate QSOs selected via their optical
variability properties, as presented by
\citet{Yang2022}. These targets are located within five (+1 backup) well
known survey fields (SDSS-RM, COSMOS, XMM-LSS, ECDFS, CVZ-S/SEP, and
ELIAS-S1).

\noindent\textbf{Simplified description of selection criteria:} Starting from a
parent catalog of optically selected objects in the RM fields (as
presented by
\citealt{Yang2022}), select candidate QSOs that satisfy all of the
following: i) have significant variability in the DES or PanSTARRS1
multi-epoch photometry (\texttt{var\_sn{[}g{]}}\textgreater3 and
\texttt{var\_rms{[}g{]}}\textgreater0.05); ii) have
17\textless{}\texttt{psfmag\_i}\textless20.5~AB
(16\textless{}\emph{G}\textless21.7~AB in the CVZ-S/SEP field); iii) do
not have significant detections (\textgreater3$\sigma$) of parallax and/or
proper motion in Gaia DR2; iv) are not vetoed due to results of visual
inspections of recent spectroscopy; and vi) do not lie in the SDSS-RM
field.

\noindent\textbf{Target priority options:} 1000-1100

\noindent\textbf{Cadence options:} \texttt{dark\_174x8}

\noindent\textbf{Implementation:}
\href{https://github.com/sdss/target_selection/blob/1.0.12/python/target_selection/cartons/bhm_rm.py}{bhm\_rm.py}

\noindent\textbf{Number of targets:} 871

\begin{center}\rule{0.5\linewidth}{0.5pt}\end{center}

\hypertarget{bhm_rm_ancillary_plan1.0.12}{%
\subsection{bhm\_rm\_ancillary}\label{bhm_rm_ancillary_plan1.0.12}}

\noindent\textbf{target\_selection plan:} 1.0.12

\noindent\textbf{target\_selection tag:}
\href{https://github.com/sdss/target_selection/tree/1.0.12/}{1.0.12}

\noindent\textbf{crossmatch plan:} 1.0.0

\noindent\textbf{Summary:} A supporting sample of candidate QSOs which have been
selected by the Gaia-unWISE AGN catalog
(\citealt{Shu2019}) and/or the SDSS XDQSO catalog
(\citealt{Bovy2011}). These targets are located within five (+1 backup) well
known survey fields (SDSS-RM, COSMOS, XMM-LSS, ECDFS, CVZ-S/SEP, and
ELIAS-S1).

\noindent\textbf{Simplified description of selection criteria:} Starting from a
parent catalog of optically selected objects in the RM fields (as
presented by
\citealt{Yang2022}), select candidate QSOs that satisfy all of the
following: i) are identified via external ancillary methods
(\texttt{photo\_bitmask\ \&\ 3\ !=\ 0}); ii) have
15\textless{}\texttt{psfmag\_i}\textless21.5~AB
(16\textless{}\emph{G}\textless21.7~AB in the CVZ-S/SEP field); iii) do
not have significant detections (\textgreater3$\sigma$) of parallax and/or
proper motion in Gaia DR2; iv) are not vetoed due to results of visual
inspections of recent spectroscopy; and v) do not lie in the SDSS-RM
field.

\noindent\textbf{Target priority options:} 1000-1100

\noindent\textbf{Cadence options:} \texttt{dark\_100x8,\ dark\_174x8}

\noindent\textbf{Implementation:}
\href{https://github.com/sdss/target_selection/blob/1.0.12/python/target_selection/cartons/bhm_rm.py}{bhm\_rm.py}

\noindent\textbf{Number of targets:} 1114

\begin{center}\rule{0.5\linewidth}{0.5pt}\end{center}

\hypertarget{bhm_rm_xrayqso_plan1.0.12}{%
\subsection{bhm\_rm\_xrayqso}\label{bhm_rm_xrayqso_plan1.0.12}}

\noindent\textbf{target\_selection plan:} 1.0.12

\noindent\textbf{target\_selection tag:}
\href{https://github.com/sdss/target_selection/tree/1.0.12/}{1.0.12}

\noindent\textbf{crossmatch plan:} 1.0.0

\noindent\textbf{Summary:} A small supporting sample of candidate QSOs in two of
the SDSS-V Reverberation Mapping fields (CDFS, ELIAS-S1), identified via
X-ray detection and SED modelling by
\citealt{Ni2021}.

\noindent\textbf{Simplified description of selection criteria:} Starting from the
sample of candidate QSOs identified by
\citealt{Ni2021}, select all that satisfy the following criteria: i) are
\textless1.5~deg from the centre of at least one RM field, ii) have
optical magnitudes in the range
16.5\textless{}\texttt{mag\_i}\textless21.5, iii) are not vetoed due to
results of visual inspections of recent spectroscopy.

\noindent\textbf{Target priority options:} 1000-1100

\noindent\textbf{Cadence options:} \texttt{dark\_174x8}

\noindent\textbf{Implementation:}
\href{https://github.com/sdss/target_selection/blob/1.0.12/python/target_selection/cartons/bhm_rm.py}{bhm\_rm.py}

\noindent\textbf{Number of targets:} 43

\begin{center}\rule{0.5\linewidth}{0.5pt}\end{center}

\hypertarget{bhm_csc_boss_plan1.0.37}{%
\subsection{bhm\_csc\_boss}\label{bhm_csc_boss_plan1.0.37}}

\noindent\textbf{target\_selection plan:} 1.0.37

\noindent\textbf{target\_selection tag:}
\href{https://github.com/sdss/target_selection/tree/1.0.37/}{1.0.37}

\noindent\textbf{crossmatch plan:} 1.0.0

\noindent\textbf{Summary:} X-ray sources from the CSC2.1 source catalog with
counterparts in LegacySurvey DR10, Panstarrs1-DR1 or Gaia DR3.

\noindent\textbf{Simplified description of selection criteria:} Starting from the
parent catalog of CSC sources with optical/IR counterparts
(\texttt{bhm\_csc\_v3}). Select entries satisfying the following
criteria: i) optical counterpart is from the LegacySurvey DR10,
PanSTARRS1 or Gaia DR3 catalogs, ii) optical flux/magnitude is in the
accepted range for SDSS-V: all \texttt{fibermag\_{[}g,r,i,z{]}}
\textgreater14.5 and at least one \texttt{fibermag\_{[}g,r,i,z{]}}
\textless21.0~AB (objects with LegacySurvey DR10 counterparts); all
\texttt{psfmag\_{[}g,r,i,z{]}} \textgreater{} 14.0 and at least one
\texttt{psfmag\_{[}g,r,i,z{]}} \textless{} 22.0~AB (objects with
PanSTARRS1 counterparts); 14.0 \textless{} \emph{G}\textless20.0~Vega
(Gaia DR3 counterparts). De-prioritize targets which already have good
quality SDSS spectroscopy. Allocate cadence (exposure time) requests
based on optical brightness (LegacySurvey \texttt{fibermag\_r}, PS1
\texttt{psfmag\_i} or Gaia \emph{G}).

\noindent\textbf{Target priority options:} 1920-1939, 2920-2939

\noindent\textbf{Cadence options:}
\texttt{bright\_1x1,\ dark\_flexible\_2x1,\ dark\_flexible\_2x2}

\noindent\textbf{Implementation:}
\href{https://github.com/sdss/target_selection/blob/1.0.37/python/target_selection/cartons/bhm_csc.py}{bhm\_csc.py}

\noindent\textbf{Number of targets:} 135154

\begin{center}\rule{0.5\linewidth}{0.5pt}\end{center}

\hypertarget{bhm_csc_apogee_plan1.0.37}{%
\subsection{bhm\_csc\_apogee}\label{bhm_csc_apogee_plan1.0.37}}

\noindent\textbf{target\_selection plan:} 1.0.37

\noindent\textbf{target\_selection tag:}
\href{https://github.com/sdss/target_selection/tree/1.0.37/}{1.0.37}

\noindent\textbf{crossmatch plan:} 1.0.0

\noindent\textbf{Summary:} X-ray sources from the CSC2.1 source catalog with NIR
counterparts in 2MASS PSC

\noindent\textbf{Simplified description of selection criteria:} Starting from the
parent catalog of CSC sources with optical/IR counterparts
(\texttt{bhm\_csc\_v2}). Select entries satisfying the following
criteria: i) NIR counterpart is from the 2MASS catalog, ii) 2MASS
\emph{H}-band magnitude measurement is not null and is within a
reasonable range for SDSS-V: 7.0\textless{}\emph{H}\textless14.0.
Allocate cadence (exposure time) requests based on \emph{H}-band
magnitude.

\noindent\textbf{Target priority options:} 2930-2939

\noindent\textbf{Cadence options:} \texttt{bright\_1x1,\ bright\_flexible\_3x1}

\noindent\textbf{Implementation:}
\href{https://github.com/sdss/target_selection/blob/1.0.37/python/target_selection/cartons/bhm_csc.py}{bhm\_csc.py}

\noindent\textbf{Number of targets:} 54593

\begin{center}\rule{0.5\linewidth}{0.5pt}\end{center}

\hypertarget{bhm_gua_dark_plan1.0.37}{%
\subsection{bhm\_gua\_dark}\label{bhm_gua_dark_plan1.0.37}}

\noindent\textbf{target\_selection plan:} 1.0.37

\noindent\textbf{target\_selection tag:}
\href{https://github.com/sdss/target_selection/tree/1.0.37/}{1.0.37}

\noindent\textbf{crossmatch plan:} 1.0.0

\noindent\textbf{Summary:} A sample of optically faint candidate AGN lacking
spectroscopic confirmations, derived from the parent sample presented by
\citet{Shu2019}, who applied a machine-learning approach to select QSO
candidates from a combination of the Gaia DR2 and unWISE catalogs.

\noindent\textbf{Simplified description of selection criteria:} Starting with the
\citet{Shu2019} catalog, select targets which satisfy the following
criteria: i) have a Random Forest probability of being a QSO
of\textgreater0.8, ii) are in the magnitude range suitable for BOSS
spectroscopy in dark time (\emph{G}\textgreater16.5 and
\emph{RP}\textgreater16.5, as well as \emph{G}\textless21.2 or
\emph{RP}\textless21.0,~Vega mag), iii) do not yet have good SDSS
optical spectroscopic measurements. Note that the selection criteria are
based on apparent magnitudes, rather than the dereddened magnitudes that
were used in an earlier iteration of this carton.

\noindent\textbf{Target priority options:} 3400

\noindent\textbf{Cadence options:} \texttt{dark\_flexible\_2x2}

\noindent\textbf{Implementation:}
\href{https://github.com/sdss/target_selection/blob/1.0.37/python/target_selection/cartons/bhm_gua.py}{bhm\_gua.py}

\noindent\textbf{Number of targets:} 1936784

\begin{center}\rule{0.5\linewidth}{0.5pt}\end{center}

\hypertarget{bhm_gua_bright_plan1.0.37}{%
\subsection{bhm\_gua\_bright}\label{bhm_gua_bright_plan1.0.37}}

\noindent\textbf{target\_selection plan:} 1.0.37

\noindent\textbf{target\_selection tag:}
\href{https://github.com/sdss/target_selection/tree/1.0.37/}{1.0.37}

\noindent\textbf{crossmatch plan:} 1.0.0

\noindent\textbf{Summary:} A sample of optically bright candidate AGN lacking
spectroscopic confirmations, derived from the parent sample presented by
\citet{Shu2019}, who applied a machine-learning approach to select QSO
candidates from a combination of the Gaia DR2 and unWISE catalogs.

\noindent\textbf{Simplified description of selection criteria:} Starting with the
\citet{Shu2019} catalog, select targets which satisfy the following
criteria: i) have a Random Forest probability of being a QSO
of\textgreater0.8, ii) are in the magnitude range suitable for BOSS
spectroscopy in bright time (\emph{G}\textgreater13.0 and
\emph{RP}\textgreater13.5, as well as \emph{G}\textless18.5 or
\emph{RP}\textless18.5,~Vega mag), iii) do not yet have good SDSS
optical spectroscopic measurements. Note that the selection criteria are
based on apparent magnitudes, rather than the dereddened magnitudes that
were used in an earlier iteration of this carton.

\noindent\textbf{Target priority options:} 4040

\noindent\textbf{Cadence options:} \texttt{bright\_flexible\_2x1}

\noindent\textbf{Implementation:}
\href{https://github.com/sdss/target_selection/blob/1.0.37/python/target_selection/cartons/bhm_gua.py}{bhm\_gua.py}

\noindent\textbf{Number of targets:} 185644

\begin{center}\rule{0.5\linewidth}{0.5pt}\end{center}

\hypertarget{bhm_spiders_clusters_lsdr10_plan1.0.37}{%
\subsection{bhm\_spiders\_clusters\_lsdr10}\label{bhm_spiders_clusters_lsdr10_plan1.0.37}}

\noindent\textbf{target\_selection plan:} 1.0.37

\noindent\textbf{target\_selection tag:}
\href{https://github.com/sdss/target_selection/tree/1.0.37/}{1.0.37}

\noindent\textbf{crossmatch plan:} 1.0.0

\noindent\textbf{Summary:} This is the main carton for SPIDERS galaxy cluster
wide area follow up. The carton provides a catalog of galaxies which are
candidate members of clusters selected from early reductions of the
first 24-months of eROSITA all sky survey data (eRASS:4). The X-ray
clusters have been associated by the eROSITA-DE team to
\href{https://www.legacysurvey.org/dr10/}{legacysurvey.org/dr10}
optical/IR counterparts using the eROMAPPER red-sequence finder
algorithm
(\citealt{Rykoff2014};
\citealt{IderChitham2020}). All targets are located in the sky hemisphere
where MPE controls the data rights (approx.
180\textless{}\emph{l}\textless360~deg). Targets in this carton are
located at high Galactic latitudes
\textbar{}\emph{b}\textbar\textgreater20~deg.

\noindent\textbf{Simplified description of selection criteria:} Starting from a
parent catalog of eRASS:4 $\rightarrow$ legacysurvey.org/dr10 eROMAPPER cluster
associations, select targets which meet all of the following criteria:
i) have 13.5\textless{}\texttt{fibertotmag\_i}\textless21.0 or
13.5\textless{}\texttt{fibertotmag\_z}\textless20.5~AB, ii) if detected
by Gaia DR3 then have \emph{G}\textgreater13.5 and
\emph{RP}\textgreater13.5~Vega, and iii) do not have existing good
quality SDSS spectroscopy. We assign a range of priorities to targets in
this carton, with Brightest Cluster Galaxies (BCGs) top ranked, followed
by candidate member galaxies according their probability of membership
and whether they are within the 'core'
16\textless{}\texttt{fibertotmag\_i}\textless20.5 or
16\textless{}\texttt{fibertotmag\_z}\textless20.0~AB). We assign
cadences (exposure time requests) based on optical brightness.

\noindent\textbf{Target priority options:} 1501, 1631-1634, 3501, 3631-3635

\noindent\textbf{Cadence options:}
\texttt{bright\_flexible\_2x1,\ dark\_flexible\_2x1,\ dark\_flexible\_2x2}

\noindent\textbf{Implementation:}
\href{https://github.com/sdss/target_selection/blob/1.0.37/python/target_selection/cartons/bhm_spiders_clusters.py}{bhm\_spiders\_clusters.py}

\noindent\textbf{Number of targets:} 440276

\begin{center}\rule{0.5\linewidth}{0.5pt}\end{center}

\hypertarget{bhm_spiders_agn_lsdr10_plan1.0.37}{%
\subsection{bhm\_spiders\_agn\_lsdr10}\label{bhm_spiders_agn_lsdr10_plan1.0.37}}

\noindent\textbf{target\_selection plan:} 1.0.37

\noindent\textbf{target\_selection tag:}
\href{https://github.com/sdss/target_selection/tree/1.0.37/}{1.0.37}

\noindent\textbf{crossmatch plan:} 1.0.0

\noindent\textbf{Summary:} This is the highest priority carton for SPIDERS AGN
wide area follow up. The carton provides optical counterparts to
point-like (unresolved) X-ray sources detected in early reductions of
the first 18~months of eROSITA all sky survey data (eRASS:3). The sample
is expected to contain a mixture of QSOs, AGN, stars and compact
objects. The X-ray sources have been cross-matched by the eROSITA-DE
team to \href{https://www.legacysurvey.org/dr10/}{legacysurvey.org/DR10}
optical/IR counterparts (supplemented with the DR9 catalog at
Dec\textgreater+32.375~deg). All targets are located in the sky
hemisphere where MPE controls the data rights (approx.
180\textless{}\emph{l}\textless360~deg). Due to the LegacySurvey
footprint, nearly all targets in this carton are located at high
Galactic latitudes \textbar{}\emph{b}\textbar\textgreater15~deg.

\noindent\textbf{Simplified description of selection criteria:} Starting from a
parent catalog of eRASS:3 point source $\rightarrow$ legacysurvey.org/DR10(+DR9)
associations (method: NWAY assisted by optical/IR priors computed via a
pre-trained Random Forest, building on
\citealt{Salvato2022}), select targets which meet all of the following criteria:
i) have eROSITA detection likelihood\textgreater6.0, ii) have an X-ray $\rightarrow$
optical/IR cross-match probability (NWAY) of
\texttt{p\_any}\textgreater0.1, iii) have
13.5\textless{}\texttt{fibermag\_r}\textless22.5 or
13.5\textless{}\texttt{fibermag\_i}\textless22.3 or
13.5\textless{}\texttt{fibermag\_z}\textless21.5, iv) are not saturated
in LegacySurvey imaging, v) if detected by Gaia DR2/DR3 then have
\emph{G}\textgreater13.5 and \emph{RP}\textgreater13.5~Vega. We
\emph{deprioritise} targets if any of the following criteria are met: a)
the target already has existing good quality SDSS spectroscopy, b) the
X-ray detection likelihood is \textless8.0, c) the target is a secondary
X-ray$\rightarrow$optical/IR association, d) the target falls outside the core
magnitude range (i.e. doesn't satisfy one of
16.5\textless{}\texttt{fibermag\_r}\textless22.0 or
16.5\textless{}\texttt{fibermag\_i}\textless21.8 or
16.5\textless{}\texttt{fibermag\_z}\textless21.0), e) the target lies at
\textbar{}\emph{b}\textbar\textless15~deg or at Dec\textless-75~deg, f)
the X-ray flux is
\textless2x10\textsuperscript{-14}~erg~s\textsuperscript{-1}~cm\textsuperscript{-2}.
We slightly boost the priorities of targets which have hard X-ray
detections. Cadence choices (exposure time requests per target) are
based on optical brightness.

\noindent\textbf{Target priority options:} 1520-1527, 1720-1727, 3520-3527,
3720-3727

\noindent\textbf{Cadence options:}
\texttt{bright\_flexible\_2x1,\ dark\_flexible\_2x1,\ dark\_flexible\_2x2}

\noindent\textbf{Implementation:}
\href{https://github.com/sdss/target_selection/blob/1.0.37/python/target_selection/cartons/bhm_spiders_agn.py}{bhm\_spiders\_agn.py}

\noindent\textbf{Number of targets:} 975978

\begin{center}\rule{0.5\linewidth}{0.5pt}\end{center}

\hypertarget{bhm_spiders_agn_hard_plan1.0.37}{%
\subsection{bhm\_spiders\_agn\_hard}\label{bhm_spiders_agn_hard_plan1.0.37}}

\noindent\textbf{target\_selection plan:} 1.0.37

\noindent\textbf{target\_selection tag:}
\href{https://github.com/sdss/target_selection/tree/1.0.37/}{1.0.37}

\noindent\textbf{crossmatch plan:} 1.0.0

\noindent\textbf{Summary:} This small supplementary carton provides a sample of
point-like (unresolved) hard-band (2.3-5~keV) X-ray sources detected in
the first 6~months of eROSITA all sky survey data (eRASS:1). The
selection of the eROSITA hard band sample is described by
\citealt{Waddell2026}. The X-ray sources have been cross-matched by the
eROSITA-DE team to
\href{https://www.legacysurvey.org/dr10/}{LegacySurvey DR10} optical/IR
counterparts. All targets are located in the sky hemisphere where MPE
controls the data rights (approx.
180\textless{}\emph{l}\textless360~deg).

\noindent\textbf{Simplified description of selection criteria:} Starting from a
parent catalog of multi-band eRASS:1 X-ray point source $\rightarrow$ LegacySurvey
DR10 associations (method: NWAY assisted by optical/IR priors computed
via a pre-trained Random Forest, building on
\citealt{Salvato2022}), select targets which meet all of the following criteria:
i) have eROSITA 2.3-5keV band detection likelihood\textgreater12.0, ii)
have an X-ray $\rightarrow$ optical/IR cross-match probability (NWAY) of
\texttt{p\_any}\textgreater0.01, iii) have
13.5\textless{}\texttt{fibermag\_r}\textless22.5 or
13.5\textless{}\texttt{fibermag\_i}\textless22.3 or
13.5\textless{}\texttt{fibermag\_z}\textless21.5, iv) are not saturated
in LegacySurvey imaging, v) if detected by Gaia DR2/DR3 then have
\emph{G}\textgreater13.5 and \emph{RP}\textgreater13.5~Vega. We
\emph{deprioritise} targets if any of the following criteria are met: a)
the target already has existing good quality SDSS spectroscopy, b) the
target is a secondary X-ray$\rightarrow$optical/IR association, c) the target falls
outside the core magnitude range (i.e. doesn't satisfy one of
16.5\textless{}\texttt{fibermag\_r}\textless22.0 or
16.5\textless{}\texttt{fibermag\_i}\textless21.8 or
16.5\textless{}\texttt{fibermag\_z}\textless21.0), d) the target lies at
\textbar{}\emph{b}\textbar\textless10~deg or at Dec\textless-75~deg.
Cadence choices (exposure time requests per target) are based on optical
brightness.

\noindent\textbf{Target priority options:} 1510-1512, 1710, 3510-3512, 3710

\noindent\textbf{Cadence options:}
\texttt{bright\_flexible\_2x1,\ dark\_flexible\_2x1,\ dark\_flexible\_2x2}

\noindent\textbf{Implementation:}
\href{https://github.com/sdss/target_selection/blob/1.0.37/python/target_selection/cartons/bhm_spiders_agn.py}{bhm\_spiders\_agn.py}

\noindent\textbf{Number of targets:} 2603

\begin{center}\rule{0.5\linewidth}{0.5pt}\end{center}

\hypertarget{bhm_spiders_agn_gaiadr3_plan1.0.37}{%
\subsection{bhm\_spiders\_agn\_gaiadr3}\label{bhm_spiders_agn_gaiadr3_plan1.0.37}}

\noindent\textbf{target\_selection plan:} 1.0.37

\noindent\textbf{target\_selection tag:}
\href{https://github.com/sdss/target_selection/tree/1.0.37/}{1.0.37}

\noindent\textbf{crossmatch plan:} 1.0.0

\noindent\textbf{Summary:} This is the second highest priority carton for SPIDERS
AGN wide area follow up, it is included to expand the survey footprint
beyond the LegacySurvey footprint. This carton provides optical
counterparts to point-like (unresolved) X-ray sources detected in early
reductions of the first 18~months of eROSITA all sky survey data
(eRASS:3). The sample is expected to contain a mixture of QSOs, AGN,
stars and compact objects. The X-ray sources have been cross-matched by
the eROSITA-DE team to Gaia DR3 optical counterparts. All targets are
located in the sky hemisphere where MPE controls the data rights
(approx. 180\textless{}\emph{l}\textless360~deg). The targets in this
carton are distributed over a wide range of Galactic latitudes, but
targets at low Galactic latitudes
\textbar{}\emph{b}\textbar\textless15~deg do not drive survey strategy.

\noindent\textbf{Simplified description of selection criteria:} Starting from a
parent catalog of eRASS:3 point source $\rightarrow$ Gaia DR3 associations (method:
NWAY assisted by Gaia photometric+astrometric priors computed via a
pre-trained Random Forest, building on
\citealt{Salvato2022}), select targets which meet all of the following criteria:
i) have eROSITA detection likelihood\textgreater6.0, ii) have an
X-ray$\rightarrow$Gaia cross-match probability of \texttt{p\_any}\textgreater0.1,
iii) have \emph{G}\textgreater13.5 and \emph{RP}\textgreater13.5~Vega.
We deprioritise targets if any of the following criteria are met: a) the
target already has existing good quality SDSS spectroscopy, b) the X-ray
detection likelihood is \textless8.0, c) the target is a secondary
X-ray$\rightarrow$Gaia association, d) the target falls outside the core magnitude
range (i.e. doesn't satisfy 16.0\textless{}\emph{G}\textless21.0), e)
the target lies at \textbar{}\emph{b}\textbar\textless15~deg or at
Dec\textless-75~deg, f) the X-ray flux is
\textless2x10\textsuperscript{-14}~erg~s\textsuperscript{-1}~cm\textsuperscript{-2}.
We slightly boost the priorities of targets which have hard X-ray
detections. Cadence choices (exposure time requests per target) are
based on optical brightness.

\noindent\textbf{Target priority options:} 1540-1547, 1740-1747, 3540-3547,
3740-3747

\noindent\textbf{Cadence options:}
\texttt{bright\_flexible\_2x1,\ dark\_flexible\_2x1,\ dark\_flexible\_2x2}

\noindent\textbf{Implementation:}
\href{https://github.com/sdss/target_selection/blob/1.0.37/python/target_selection/cartons/bhm_spiders_agn.py}{bhm\_spiders\_agn.py}

\noindent\textbf{Number of targets:} 717805

\begin{center}\rule{0.5\linewidth}{0.5pt}\end{center}

\hypertarget{bhm_spiders_agn_tda_plan1.0.37}{%
\subsection{bhm\_spiders\_agn\_tda}\label{bhm_spiders_agn_tda_plan1.0.37}}

\noindent\textbf{target\_selection plan:} 1.0.37

\noindent\textbf{target\_selection tag:}
\href{https://github.com/sdss/target_selection/tree/1.0.37/}{1.0.37}

\noindent\textbf{crossmatch plan:} 1.0.0

\noindent\textbf{Summary:} This is a small supplementary carton of
transient/variable X-ray sources selected by comparing X-ray
measurements between several independent 6-month eROSITA sky surveys.
All targets are located in the sky hemisphere where MPE controls the
data rights (approx. 180\textless{}\emph{l}\textless360~deg).

\noindent\textbf{Simplified description of selection criteria:} We start from a
parent catalog of transient/variable X-ray sources selected from the
first three eROSITA All Sky surveys. These X-ray sources have been
associated with LegacySurvey DR10 optical counterparts (method: NWAY
assisted by optical/IR priors computed via a pre-trained Random Forest,
building on
\citealt{Salvato2022}). We then select targets which meet all of the following
criteria: i) have 13.5\textless{}\texttt{fibermag\_r}\textless22.5 or
13.5\textless{}\texttt{fibermag\_i}\textless22.3 or
13.5\textless{}\texttt{fibermag\_z}\textless21.5, ii) if detected by
Gaia DR2/DR3 then have \emph{G}\textgreater13.5 and
\emph{RP}\textgreater13.5~Vega. Cadence choices (exposure time requests
per target) are based on optical brightness.

\noindent\textbf{Target priority options:} 1690

\noindent\textbf{Cadence options:}
\texttt{bright\_flexible\_2x1,\ dark\_flexible\_2x1,\ dark\_flexible\_2x2}

\noindent\textbf{Implementation:}
\href{https://github.com/sdss/target_selection/blob/1.0.37/python/target_selection/cartons/bhm_spiders_agn.py}{bhm\_spiders\_agn.py}

\noindent\textbf{Number of targets:} 9627

\begin{center}\rule{0.5\linewidth}{0.5pt}\end{center}

\hypertarget{bhm_spiders_agn_sep_plan1.0.37}{%
\subsection{bhm\_spiders\_agn\_sep}\label{bhm_spiders_agn_sep_plan1.0.37}}

\noindent\textbf{target\_selection plan:} 1.0.37

\noindent\textbf{target\_selection tag:}
\href{https://github.com/sdss/target_selection/tree/1.0.37/}{1.0.37}

\noindent\textbf{crossmatch plan:} 1.0.0

\noindent\textbf{Summary:} This is a small supplementary catalog of eROSITA
point-like (unresolved) X-ray sources detected in the Continuous Viewing
Zone South/South Ecliptic Pole (CVZ-S/SEP) field. Owing to the deepest
exposure of the eROSITA All-Sky Survey (eRASS) in this region, the
CVZ-S/SEP field suffers from severe source confusion, requiring a
specialized source-detection procedure (Liu et al., in prep.). The X-ray
catalog used here was derived from an early data reduction of
0.2-2.3~keV band data from the 2.3-year eRASS:5 dataset, covering an
approximately 6~deg~x~6~deg sky region centered on the CVZ-S/SEP. The
X-ray sources have been cross-matched by the eROSITA-DE team with Gaia
DR3.

\noindent\textbf{Simplified description of selection criteria:} Starting from a
parent catalog of eRASS:5 point source $\rightarrow$ Gaia DR3 associations from the
SEP/CVZ-S region, select targets which meet all of the following
criteria: i) have eROSITA detection likelihood\textgreater6.0, ii) have
an X-ray$\rightarrow$Gaia cross-match probability of \texttt{p\_any}\textgreater0.1,
iii) have \emph{G}\textgreater13.5 and \emph{RP}\textgreater13.5~Vega.
We deprioritise targets if any of the following criteria are met: a) the
target already has existing good quality SDSS spectroscopy, b) the X-ray
detection likelihood is \textless8.0, c) the target is a secondary
X-ray$\rightarrow$Gaia association, d) the target falls outside the core magnitude
range (i.e. doesn't satisfy 16.0\textless{}\emph{G}\textless21.5), e)
the X-ray flux is
\textless1x10\textsuperscript{-14}~erg~s\textsuperscript{-1}~cm\textsuperscript{-2}.
Cadence choices (exposure time requests per target) are based on optical
brightness.

\noindent\textbf{Target priority options:} 1501-1503, 3501-3507

\noindent\textbf{Cadence options:}
\texttt{bright\_flexible\_2x1,\ dark\_flexible\_2x1,\ dark\_flexible\_2x2}

\noindent\textbf{Implementation:}
\href{https://github.com/sdss/target_selection/blob/1.0.37/python/target_selection/cartons/bhm_spiders_agn.py}{bhm\_spiders\_agn.py}

\noindent\textbf{Number of targets:} 1681

\begin{center}\rule{0.5\linewidth}{0.5pt}\end{center}

\hypertarget{bhm_aqmes_med_plan1.0.37}{%
\subsection{bhm\_aqmes\_med}\label{bhm_aqmes_med_plan1.0.37}}

\noindent\textbf{target\_selection plan:} 1.0.37

\noindent\textbf{target\_selection tag:}
\href{https://github.com/sdss/target_selection/tree/1.0.37/}{1.0.37}

\noindent\textbf{crossmatch plan:} 1.0.0

\noindent\textbf{Summary:} Spectroscopically confirmed optically bright SDSS
QSOs, selected from the SDSS QSO catalog (DR16Q,
\citealt{Lyke2020}). Located in 36 mostly disjoint fields within the SDSS QSO
footprint that were pre-selected to contain higher than average numbers
of bright QSOs and CSC targets. The list of field centres can be found
within
\href{https://github.com/sdss/target_selection/blob/0.3.0/python/target_selection/masks/candidate_target_fields_bhm_aqmes_med_v0.3.1.fits}{the
target\_selection repository}.

\noindent\textbf{Simplified description of selection criteria:} Select all
objects from SDSS DR16 QSO catalog that have
16.0\textless{}\texttt{sdss\_psfmag\_i}\textless19.1~AB, that lie within
1.49~degrees of at least one AQMES-medium field location.

\noindent\textbf{Target priority options:} 1100

\noindent\textbf{Cadence options:} \texttt{dark\_10x4\_4yr}

\noindent\textbf{Implementation:}
\href{https://github.com/sdss/target_selection/blob/1.0.37/python/target_selection/cartons/bhm_aqmes.py}{bhm\_aqmes.py}

\noindent\textbf{Number of targets:} 2663

\begin{center}\rule{0.5\linewidth}{0.5pt}\end{center}

\hypertarget{bhm_aqmes_med_faint_plan1.0.37}{%
\subsection{bhm\_aqmes\_med\_faint}\label{bhm_aqmes_med_faint_plan1.0.37}}

\noindent\textbf{target\_selection plan:} 1.0.37

\noindent\textbf{target\_selection tag:}
\href{https://github.com/sdss/target_selection/tree/1.0.37/}{1.0.37}

\noindent\textbf{crossmatch plan:} 1.0.0

\noindent\textbf{Summary:} Spectroscopically confirmed optically faint SDSS QSOs,
selected from the SDSS QSO catalog (DR16Q,
\citealt{Lyke2020}). Located in 36 mostly disjoint fields within the SDSS QSO
footprint that were pre-selected to contain higher than average numbers
of bright QSOs and CSC targets. The list of field centres can be found
within
\href{https://github.com/sdss/target_selection/blob/0.3.0/python/target_selection/masks/candidate_target_fields_bhm_aqmes_med_v0.3.1.fits}{the
target\_selection repository}.

\noindent\textbf{Simplified description of selection criteria:} Select all
objects from SDSS DR16 QSO catalog that have
19.1\textless{}\texttt{sdss\_psfmag\_i}\textless21.0~AB, that lie within
1.49~degrees of at least one AQMES-medium field location.

\noindent\textbf{Target priority options:} 3100

\noindent\textbf{Cadence options:} \texttt{dark\_10x4\_4yr}

\noindent\textbf{Implementation:}
\href{https://github.com/sdss/target_selection/blob/1.0.37/python/target_selection/cartons/bhm_aqmes.py}{bhm\_aqmes.py}

\noindent\textbf{Number of targets:} 16853

\begin{center}\rule{0.5\linewidth}{0.5pt}\end{center}

\hypertarget{bhm_aqmes_wide2_plan1.0.37}{%
\subsection{bhm\_aqmes\_wide2}\label{bhm_aqmes_wide2_plan1.0.37}}

\noindent\textbf{target\_selection plan:} 1.0.37

\noindent\textbf{target\_selection tag:}
\href{https://github.com/sdss/target_selection/tree/1.0.37/}{1.0.37}

\noindent\textbf{crossmatch plan:} 1.0.0

\noindent\textbf{Summary:} Spectroscopically confirmed optically bright SDSS
QSOs, selected from the SDSS QSO catalog (DR16Q,
\citealt{Lyke2020}). Located in 425 fields within the SDSS QSO footprint,
where the choice of survey area prioritized field that overlapped with
the SPIDERS footprint (approx 180\textless{}\emph{b}\textless360~deg),
and/or had higher than average numbers of bright QSOs and CSC targets.
The list of field centres can be found
\href{https://github.com/sdss/target_selection/blob/0.3.0/python/target_selection/masks/candidate_target_fields_bhm_aqmes_wide_v0.3.1.fits}{within
the target\_selection repository}. The targets in this carton request
two epochs of SDSS-V spectroscopy.

\noindent\textbf{Simplified description of selection criteria:} Select all
objects from SDSS DR16 QSO catalog that have
16.0\textless{}\texttt{sdss\_psfmag\_i}\textless19.1~AB, and that lie
within 1.49~degrees of at least one AQMES-wide field location.

\noindent\textbf{Target priority options:} 1210

\noindent\textbf{Cadence options:} \texttt{dark\_2x4}

\noindent\textbf{Implementation:}
\href{https://github.com/sdss/target_selection/blob/1.0.37/python/target_selection/cartons/bhm_aqmes.py}{bhm\_aqmes.py}

\noindent\textbf{Number of targets:} 24142

\begin{center}\rule{0.5\linewidth}{0.5pt}\end{center}

\hypertarget{bhm_aqmes_wide2_faint_plan1.0.37}{%
\subsection{bhm\_aqmes\_wide2\_faint}\label{bhm_aqmes_wide2_faint_plan1.0.37}}

\noindent\textbf{target\_selection plan:} 1.0.37

\noindent\textbf{target\_selection tag:}
\href{https://github.com/sdss/target_selection/tree/1.0.37/}{1.0.37}

\noindent\textbf{crossmatch plan:} 1.0.0

\noindent\textbf{Summary:} Spectroscopically confirmed optically faint SDSS QSOs,
selected from the SDSS QSO catalog (DR16Q,
\citealt{Lyke2020}). Located in 425 fields within the SDSS QSO footprint,
where the choice of survey area prioritized field that overlapped with
the SPIDERS footprint (approx 180\textless{}\emph{b}\textless360~deg),
and/or had higher than average numbers of bright QSOs and CSC targets.
The list of field centres can be found
\href{https://github.com/sdss/target_selection/blob/0.3.0/python/target_selection/masks/candidate_target_fields_bhm_aqmes_wide_v0.3.1.fits}{within
the target\_selection repository}. The targets in this carton request
two epochs of SDSS-V spectroscopy.

\noindent\textbf{Simplified description of selection criteria:} Select all
objects from SDSS DR16 QSO catalog that have
19.1\textless{}\texttt{sdss\_psfmag\_i}\textless21.0~AB, and that lie
within 1.49~degrees of at least one AQMES-wide field location.

\noindent\textbf{Target priority options:} 3210

\noindent\textbf{Cadence options:} \texttt{dark\_2x4}

\noindent\textbf{Implementation:}
\href{https://github.com/sdss/target_selection/blob/1.0.37/python/target_selection/cartons/bhm_aqmes.py}{bhm\_aqmes.py}

\noindent\textbf{Number of targets:} 99586

\begin{center}\rule{0.5\linewidth}{0.5pt}\end{center}

\hypertarget{bhm_aqmes_bonus_core_plan1.0.37}{%
\subsection{bhm\_aqmes\_bonus\_core}\label{bhm_aqmes_bonus_core_plan1.0.37}}

\noindent\textbf{target\_selection plan:} 1.0.37

\noindent\textbf{target\_selection tag:}
\href{https://github.com/sdss/target_selection/tree/1.0.37/}{1.0.37}

\noindent\textbf{crossmatch plan:} 1.0.0

\noindent\textbf{Summary:} Spectroscopically confirmed optically bright SDSS
QSOs, selected from the SDSS QSO catalog (DR16Q,
\citealt{Lyke2020}). Located anywhere within the SDSS DR16Q footprint.

\noindent\textbf{Simplified description of selection criteria:} Select all
objects from SDSS DR16 QSO catalog that have
16.0\textless{}\texttt{sdss\_psfmag\_i}\textless19.1~AB

\noindent\textbf{Target priority options:} 3300

\noindent\textbf{Cadence options:} \texttt{dark\_flexible\_2x2}

\noindent\textbf{Implementation:}
\href{https://github.com/sdss/target_selection/blob/1.0.37/python/target_selection/cartons/bhm_aqmes.py}{bhm\_aqmes.py}

\noindent\textbf{Number of targets:} 83163

\begin{center}\rule{0.5\linewidth}{0.5pt}\end{center}

\hypertarget{bhm_aqmes_bonus_bright_plan1.0.37}{%
\subsection{bhm\_aqmes\_bonus\_bright}\label{bhm_aqmes_bonus_bright_plan1.0.37}}

\noindent\textbf{target\_selection plan:} 1.0.37

\noindent\textbf{target\_selection tag:}
\href{https://github.com/sdss/target_selection/tree/1.0.37/}{1.0.37}

\noindent\textbf{crossmatch plan:} 1.0.0

\noindent\textbf{Summary:} Spectroscopically confirmed, extremely optically
bright SDSS QSOs, selected from the SDSS QSO catalog (DR16Q,
\citealt{Lyke2020}). Located anywhere within the SDSS DR16Q footprint.

\noindent\textbf{Simplified description of selection criteria:} Select all
objects from SDSS DR16 QSO catalog that have
14.0\textless{}\texttt{sdss\_psfmag\_i}\textless18.0~AB

\noindent\textbf{Target priority options:} 4040

\noindent\textbf{Cadence options:} \texttt{bright\_flexible\_2x1}

\noindent\textbf{Implementation:}
\href{https://github.com/sdss/target_selection/blob/1.0.37/python/target_selection/cartons/bhm_aqmes.py}{bhm\_aqmes.py}

\noindent\textbf{Number of targets:} 10848

\begin{center}\rule{0.5\linewidth}{0.5pt}\end{center}

\hypertarget{bhm_aqmes_bonus_faint_plan1.0.37}{%
\subsection{bhm\_aqmes\_bonus\_faint}\label{bhm_aqmes_bonus_faint_plan1.0.37}}

\noindent\textbf{target\_selection plan:} 1.0.37

\noindent\textbf{target\_selection tag:}
\href{https://github.com/sdss/target_selection/tree/1.0.37/}{1.0.37}

\noindent\textbf{crossmatch plan:} 1.0.0

\noindent\textbf{Summary:} Spectroscopically confirmed optically faint SDSS QSOs,
selected from the SDSS QSO catalog (DR16Q,
\citealt{Lyke2020}). Located anywhere within the SDSS DR16Q footprint.

\noindent\textbf{Simplified description of selection criteria:} Select all
objects from SDSS DR16 QSO catalog that have
19.1\textless{}\texttt{sdss\_psfmag\_i}\textless21.0~AB

\noindent\textbf{Target priority options:} 3302

\noindent\textbf{Cadence options:} \texttt{dark\_flexible\_2x2}

\noindent\textbf{Implementation:}
\href{https://github.com/sdss/target_selection/blob/1.0.37/python/target_selection/cartons/bhm_aqmes.py}{bhm\_aqmes.py}

\noindent\textbf{Number of targets:} 424163

\begin{center}\rule{0.5\linewidth}{0.5pt}\end{center}

\hypertarget{bhm_colr_galaxies_lsdr10_plan1.0.38}{%
\subsection{bhm\_colr\_galaxies\_lsdr10}\label{bhm_colr_galaxies_lsdr10_plan1.0.38}}

\noindent\textbf{target\_selection plan:} 1.0.38

\noindent\textbf{target\_selection tag:}
\href{https://github.com/sdss/target_selection/tree/1.0.38/}{1.0.38}

\noindent\textbf{crossmatch plan:} 1.0.0

\noindent\textbf{Summary:} A supplementary magnitude limited sample of optically
bright galaxies selected from the LegacySurvey DR10 optical/IR imaging
catalog (supplemented with the LegacySurvey DR9 catalog at
Dec\textgreater+32.375~deg). Selection is based on optical morphology,
lack of Gaia parallax, and several magnitude cuts.

\noindent\textbf{Simplified description of selection criteria:} Starting from the
\texttt{legacy\_survey\_dr10} catalog (lsdr10), select entries
satisfying all of the following criteria: i) lsdr10 morphological
\texttt{type} != 'PSF', ii) lsdr10
\texttt{shape\_r}\textgreater1.0~arcsec, iii) no detection of parallax
by Gaia, iv) dereddened \emph{z}-band model mag\textless19.0~AB, and
dereddened \emph{z}-band fiber mag\textless19.5~AB, v) within all the
following apparent fiber mag limits: 16\textless{}\emph{g}\textless22.5~
and 16\textless{}\emph{r}\textless21.5~ and
16\textless{}\emph{z}\textless20.0~AB, vi) if detected in Gaia, then
\emph{G}\textgreater15.0~ and \emph{RP}\textgreater15.0~Vega, vii) at
moderate to high Galactic latitude
\textbar{}\emph{b}\textbar\textgreater15~deg.

\noindent\textbf{Target priority options:} 7100

\noindent\textbf{Cadence options:}
\texttt{bright\_1x1,\ dark\_1x1,\ dark\_flexible\_2x2}

\noindent\textbf{Implementation:}
\href{https://github.com/sdss/target_selection/blob/1.0.38/python/target_selection/cartons/bhm_galaxies.py}{bhm\_galaxies.py}

\noindent\textbf{Number of targets:} 6246703

\begin{center}\rule{0.5\linewidth}{0.5pt}\end{center}

\hypertarget{openfibertargets_bhm_racsradio_boss_plan1.2.0}{%
\subsection{openfibertargets\_bhm\_racsradio\_boss}\label{openfibertargets_bhm_racsradio_boss_plan1.2.0}}

\noindent\textbf{target\_selection plan:} 1.2.0

\noindent\textbf{target\_selection tag:}
\href{https://github.com/sdss/target_selection/tree/1.2.7/}{1.2.7}

\noindent\textbf{crossmatch plan:} 1.0.0

\noindent\textbf{Summary:} Optically bright counterparts to bright 888~MHz radio
sources identified by in the ASKAP/RACS-low survey
(\citealt{Hale2021}).

\noindent\textbf{Simplified description of selection criteria:} Subset of radio
sources from the RACS-low ASKAP survey (entire sky region between
declination -80 deg and +30~deg), with radio (888~MHz) flux exceeding
3~mJy and optical counterparts (with
14.5\textless{}\emph{r}\textless20.5) identified in the footprints of
LegacySurvey DR10-South using NWAY (adopting separation, positional
errors, number density and a prior based on various features).

\noindent\textbf{Target priority options:} 6085

\noindent\textbf{Cadence options:} \texttt{bright\_1x1,\ dark\_1x1}

\noindent\textbf{Implementation:} n/a

\noindent\textbf{Number of targets:} 149989

\begin{center}\rule{0.5\linewidth}{0.5pt}\end{center}

\hypertarget{openfibertargets_bhm_varagn_boss_plan1.2.0}{%
\subsection{openfibertargets\_bhm\_varagn\_boss}\label{openfibertargets_bhm_varagn_boss_plan1.2.0}}

\noindent\textbf{target\_selection plan:} 1.2.0

\noindent\textbf{target\_selection tag:}
\href{https://github.com/sdss/target_selection/tree/1.2.7/}{1.2.7}

\noindent\textbf{crossmatch plan:} 1.0.0

\noindent\textbf{Summary:} Optical variability selected AGN candidates. We have
used custom made forced photometry in ZTF difference images to build
accurate light curves of complicated objects such as low redshift AGN,
where the variable AGN point source is contaminated by the host galaxy
in a PSF dependent way. With these light curves and complementary color
information we built a random forest algorithm to classify all objects,
obtaining 350k AGN candidates, some of which fall outside the usual
locus of AGN in color-color diagrams.

\noindent\textbf{Simplified description of selection criteria:} All ZTF
forced-photometry based AGN candidates that are not included in the
Gaia-unWISE color-selected AGN candidate catalog of
\citet{Shu2019} and that: a) have -30\textless Dec\textless15 and
\textbar{}\emph{b}\textbar\textgreater20~deg, have zero proper motion
according to Gaia (at 2 sigma), are not in the labeled set of
well-characterized objects used in
\citealt{SanchezSaez2023}, and are classified as either low-z AGN, mid-z AGN or
Blazar by our classifier. For candidates classified as high-z AGN the
same limits apply except that in this case
\textbar{}\emph{b}\textbar\textgreater30~deg and the Gaia proper motion
must be within 1 sigma of zero.

\noindent\textbf{Target priority options:} 6085

\noindent\textbf{Cadence options:} \texttt{bright\_1x1,\ dark\_1x1}

\noindent\textbf{Implementation:} n/a

\noindent\textbf{Number of targets:} 6966

\begin{center}\rule{0.5\linewidth}{0.5pt}\end{center}

\hypertarget{openfibertargets_bhm_morexray_boss_plan1.2.0}{%
\subsection{openfibertargets\_bhm\_morexray\_boss}\label{openfibertargets_bhm_morexray_boss_plan1.2.0}}

\noindent\textbf{target\_selection plan:} 1.2.0

\noindent\textbf{target\_selection tag:}
\href{https://github.com/sdss/target_selection/tree/1.2.7/}{1.2.7}

\noindent\textbf{crossmatch plan:} 1.0.0

\noindent\textbf{Summary:} Counterparts of X-ray sources identified by Swift-XRT
and XMM-Newton

\noindent\textbf{Simplified description of selection criteria:} Select sources
from ExSeSS (Swift,
\citealt{Delaney2023}) and
\href{http://xmmssc.irap.omp.eu/Catalogue/4XMM-DR13/4XMM_DR13.html}{4XMM-DR13}
(\citealt{Webb2020}), in any X-ray band. Cross-match to LegacySurvey DR10
counterparts (using NWAY informed by WISE \emph{W1-W2} colors, with
\texttt{p\_any}\textgreater0.3), and then apply magnitude cuts
(14.5\textless{}\emph{r}\textless20~AB for bright time and
20\textless{}\emph{r}\textless21~AB for dark time)

\noindent\textbf{Target priority options:} 6085

\noindent\textbf{Cadence options:} \texttt{bright\_1x1,\ dark\_1x1}

\noindent\textbf{Implementation:} n/a

\noindent\textbf{Number of targets:} 103105

\begin{center}\rule{0.5\linewidth}{0.5pt}\end{center}

\hypertarget{openfibertargets_bhm_hightderates_psb_boss_plan1.2.0}{%
\subsection{openfibertargets\_bhm\_hightderates\_psb\_boss}\label{openfibertargets_bhm_hightderates_psb_boss_plan1.2.0}}

\noindent\textbf{target\_selection plan:} 1.2.0

\noindent\textbf{target\_selection tag:}
\href{https://github.com/sdss/target_selection/tree/1.2.7/}{1.2.7}

\noindent\textbf{crossmatch plan:} 1.0.0

\noindent\textbf{Summary:} Galaxies that have been photometrically identified as
or post-starburst (E+A). Such galaxies are known to have an elevated
rate of tidal disruption events of stars by their supermassive black
holes.

\noindent\textbf{Simplified description of selection criteria:} All or
post-starburst galaxies identified by the machine learning algorithm in
\citealt{French2018}

\noindent\textbf{Target priority options:} 6085

\noindent\textbf{Cadence options:} \texttt{dark\_1x1}

\noindent\textbf{Implementation:} n/a

\noindent\textbf{Number of targets:} 8934

\begin{center}\rule{0.5\linewidth}{0.5pt}\end{center}

\hypertarget{openfibertargets_bhm_hightderates_qbs_boss_plan1.2.0}{%
\subsection{openfibertargets\_bhm\_hightderates\_qbs\_boss}\label{openfibertargets_bhm_hightderates_qbs_boss_plan1.2.0}}

\noindent\textbf{target\_selection plan:} 1.2.0

\noindent\textbf{target\_selection tag:}
\href{https://github.com/sdss/target_selection/tree/1.2.7/}{1.2.7}

\noindent\textbf{crossmatch plan:} 1.0.0

\noindent\textbf{Summary:} Galaxies that have been photometrically identified as
quiescent Balmer-strong (QBS). Such galaxies are known to have an
elevated rate of tidal disruption events of stars by their supermassive
black holes.

\noindent\textbf{Simplified description of selection criteria:} All quiescent
Balmer-strong galaxies identified by the machine learning algorithm in
\citealt{French2018}

\noindent\textbf{Target priority options:} 6085

\noindent\textbf{Cadence options:} \texttt{dark\_1x1}

\noindent\textbf{Implementation:} n/a

\noindent\textbf{Number of targets:} 29659

\begin{center}\rule{0.5\linewidth}{0.5pt}\end{center}

\hypertarget{openfibertargets_bhm_quaia_boss_plan1.2.0}{%
\subsection{openfibertargets\_bhm\_quaia\_boss}\label{openfibertargets_bhm_quaia_boss_plan1.2.0}}

\noindent\textbf{target\_selection plan:} 1.2.0

\noindent\textbf{target\_selection tag:}
\href{https://github.com/sdss/target_selection/tree/1.2.7/}{1.2.7}

\noindent\textbf{crossmatch plan:} 1.0.0

\noindent\textbf{Summary:} Quasar candidates from the Quaia Catalog
(\citealt{StoreyFisher2024_Quaia}) that have not previously been spectroscopically observed
by any generation of SDSS

\noindent\textbf{Simplified description of selection criteria:} All Quaia
G\textless20.5 quasars not previously spectroscopically observed by SDSS

\noindent\textbf{Target priority options:} 6085

\noindent\textbf{Cadence options:} \texttt{bright\_1x1}

\noindent\textbf{Implementation:} n/a

\noindent\textbf{Number of targets:} 999906

\begin{center}\rule{0.5\linewidth}{0.5pt}\end{center}

\hypertarget{openfibertargets_bhm_lofarradio_boss_plan1.2.1}{%
\subsection{openfibertargets\_bhm\_lofarradio\_boss}\label{openfibertargets_bhm_lofarradio_boss_plan1.2.1}}

\noindent\textbf{target\_selection plan:} 1.2.1

\noindent\textbf{target\_selection tag:}
\href{https://github.com/sdss/target_selection/tree/1.3.0/}{1.3.0}

\noindent\textbf{crossmatch plan:} 1.0.0

\noindent\textbf{Summary:} Optically bright counterparts of low-frequency radio
sources identified by LOFAR Two Metre Sky Survey (LOTSS DR2,
\citealt{Hardcastle2023})

\noindent\textbf{Simplified description of selection criteria:} Extract LOFAR
radio detected sources from the LOTSS survey with published matches to
LegacySurvey DR8 counterparts
(\citealt{Hardcastle2023}). Apply a radio flux cut
(\texttt{Total\ flux}\textgreater2~mJy), and apply magnitude limits
(14.5\textless{}\emph{r}\textless20.0)

\noindent\textbf{Target priority options:} 6085

\noindent\textbf{Cadence options:} \texttt{bright\_1x1}

\noindent\textbf{Implementation:} n/a

\noindent\textbf{Number of targets:} 210483

\begin{center}\rule{0.5\linewidth}{0.5pt}\end{center}

\hypertarget{bhm_csc_boss_d3_plan1.2.4}{%
\subsection{bhm\_csc\_boss\_d3}\label{bhm_csc_boss_d3_plan1.2.4}}

\noindent\textbf{target\_selection plan:} 1.2.4

\noindent\textbf{target\_selection tag:}
\href{https://github.com/sdss/target_selection/tree/1.3.3/}{1.3.3}

\noindent\textbf{crossmatch plan:} 1.0.0

\noindent\textbf{Summary:} X-ray sources from the CSC2.1 source catalog with
counterparts in LegacySurvey DR10, Panstarrs1-DR1 or Gaia DR3.

\noindent\textbf{Simplified description of selection criteria:} Starting from the
parent catalog of CSC sources with optical/IR counterparts
(\texttt{bhm\_csc\_v3}). Select entries satisfying the following
criteria: i) optical counterpart is from the LegacySurvey DR10,
PanSTARRS1 or Gaia DR3 catalogs, ii) optical flux/magnitude is in the
accepted range for SDSS-V: all \texttt{fibermag\_{[}g,r,i,z{]}}
\textgreater14.5 and at least one \texttt{fibermag\_{[}g,r,i,z{]}}
\textless21.0~AB (objects with LegacySurvey DR10 counterparts); all
\texttt{psfmag\_{[}g,r,i,z{]}} \textgreater{} 14.0 and at least one
\texttt{psfmag\_{[}g,r,i,z{]}} \textless{} 22.0~AB (objects with
PanSTARRS1 counterparts); 14.0 \textless{} \emph{G}\textless20.0~Vega
(Gaia DR3 counterparts). De-prioritize targets which already have good
quality SDSS spectroscopy. The '\_d3' variation of this carton provides
a shallower cadence option for targets toward the optically faint end of
the sample.

\noindent\textbf{Target priority options:} 1921-1930

\noindent\textbf{Cadence options:} \texttt{dark\_flexible\_3x1}

\noindent\textbf{Implementation:}
\href{https://github.com/sdss/target_selection/blob/1.3.3/python/target_selection/cartons/bhm_csc.py}{bhm\_csc.py}

\noindent\textbf{Number of targets:} 97267

\begin{center}\rule{0.5\linewidth}{0.5pt}\end{center}

\hypertarget{bhm_spiders_clusters_lsdr10_d3_plan1.2.4}{%
\subsection{bhm\_spiders\_clusters\_lsdr10\_d3}\label{bhm_spiders_clusters_lsdr10_d3_plan1.2.4}}

\noindent\textbf{target\_selection plan:} 1.2.4

\noindent\textbf{target\_selection tag:}
\href{https://github.com/sdss/target_selection/tree/1.3.3/}{1.3.3}

\noindent\textbf{crossmatch plan:} 1.0.0

\noindent\textbf{Summary:} This is the main carton for SPIDERS galaxy cluster
wide area follow up. The carton provides a catalog of galaxies which are
candidate members of clusters selected from early reductions of the
first 24-months of eROSITA all sky survey data (eRASS:4). The X-ray
clusters have been associated by the eROSITA-DE team to
\href{https://www.legacysurvey.org/dr10/}{legacysurvey.org/dr10}
optical/IR counterparts using the eROMAPPER red-sequence finder
algorithm
(\citealt{Rykoff2014};
\citealt{IderChitham2020}). All targets are located in the sky hemisphere
where MPE controls the data rights (approx.
180\textless{}\emph{l}\textless360~deg). Targets in this carton are
located at high Galactic latitudes
\textbar{}\emph{b}\textbar\textgreater20~deg.

\noindent\textbf{Simplified description of selection criteria:} Starting from a
parent catalog of eRASS:4 $\rightarrow$ legacysurvey.org/dr10 eROMAPPER cluster
associations, select targets which meet all of the following criteria:
i) have 13.5\textless{}\texttt{fibertotmag\_i}\textless21.0 or
13.5\textless{}\texttt{fibertotmag\_z}\textless20.5~AB, ii) if detected
by Gaia DR3 then have \emph{G}\textgreater13.5 and
\emph{RP}\textgreater13.5~Vega, and iii) do not have existing good
quality SDSS spectroscopy. We assign a range of priorities to targets in
this carton, with Brightest Cluster Galaxies (BCGs) top ranked, followed
by candidate member galaxies according their probability of membership
and whether they are within the 'core'
16\textless{}\texttt{fibertotmag\_i}\textless20.5 or
16\textless{}\texttt{fibertotmag\_z}\textless20.0~AB). The '\_d3'
variation of this carton provides a shallower cadence option for targets
toward the optically faint end of the sample.

\noindent\textbf{Target priority options:} 1502, 1632-1635, 3502, 3632-3636

\noindent\textbf{Cadence options:} \texttt{dark\_flexible\_3x1}

\noindent\textbf{Implementation:}
\href{https://github.com/sdss/target_selection/blob/1.3.3/python/target_selection/cartons/bhm_spiders_clusters.py}{bhm\_spiders\_clusters.py}

\noindent\textbf{Number of targets:} 427445

\begin{center}\rule{0.5\linewidth}{0.5pt}\end{center}

\hypertarget{bhm_spiders_agn_gaiadr3_d3_plan1.2.4}{%
\subsection{bhm\_spiders\_agn\_gaiadr3\_d3}\label{bhm_spiders_agn_gaiadr3_d3_plan1.2.4}}

\noindent\textbf{target\_selection plan:} 1.2.4

\noindent\textbf{target\_selection tag:}
\href{https://github.com/sdss/target_selection/tree/1.3.3/}{1.3.3}

\noindent\textbf{crossmatch plan:} 1.0.0

\noindent\textbf{Summary:} This is the second highest priority carton for SPIDERS
AGN wide area follow up, it is included to expand the survey footprint
beyond the LegacySurvey footprint. This carton provides optical
counterparts to point-like (unresolved) X-ray sources detected in early
reductions of the first 18~months of eROSITA all sky survey data
(eRASS:3). The sample is expected to contain a mixture of QSOs, AGN,
stars and compact objects. The X-ray sources have been cross-matched by
the eROSITA-DE team to Gaia DR3 optical counterparts. All targets are
located in the sky hemisphere where MPE controls the data rights
(approx. 180\textless{}\emph{l}\textless360~deg). The targets in this
carton are distributed over a wide range of Galactic latitudes, but
targets at low Galactic latitudes
\textbar{}\emph{b}\textbar\textless15~deg do not drive survey strategy.

\noindent\textbf{Simplified description of selection criteria:} Starting from a
parent catalog of eRASS:3 point source $\rightarrow$ Gaia DR3 associations (method:
NWAY assisted by Gaia photometric+astrometric priors computed via a
pre-trained Random Forest, building on
\citealt{Salvato2022}), select targets which meet all of the following criteria:
i) have eROSITA detection likelihood\textgreater6.0, ii) have an
X-ray$\rightarrow$Gaia cross-match probability of \texttt{p\_any}\textgreater0.1,
iii) have \emph{G}\textgreater13.5 and \emph{RP}\textgreater13.5~Vega.
We deprioritise targets if any of the following criteria are met: a) the
target already has existing good quality SDSS spectroscopy, b) the X-ray
detection likelihood is \textless8.0, c) the target is a secondary
X-ray$\rightarrow$Gaia association, d) the target falls outside the core magnitude
range (i.e. doesn't satisfy 16.0\textless{}\emph{G}\textless21.0), e)
the target lies at \textbar{}\emph{b}\textbar\textless15~deg or at
Dec\textless-75~deg, f) the X-ray flux is
\textless2x10\textsuperscript{-14}~erg~s\textsuperscript{-1}~cm\textsuperscript{-2}.
We slightly boost the priorities of targets which have hard X-ray
detections. The '\_d3' variation of this carton provides a shallower
cadence option for targets toward the optically faint end of the sample.

\noindent\textbf{Target priority options:} 1541-1548, 3541-3548

\noindent\textbf{Cadence options:} \texttt{dark\_flexible\_3x1}

\noindent\textbf{Implementation:}
\href{https://github.com/sdss/target_selection/blob/1.3.3/python/target_selection/cartons/bhm_spiders_agn.py}{bhm\_spiders\_agn.py}

\noindent\textbf{Number of targets:} 398493

\begin{center}\rule{0.5\linewidth}{0.5pt}\end{center}

\hypertarget{bhm_spiders_agn_sep_d3_plan1.2.4}{%
\subsection{bhm\_spiders\_agn\_sep\_d3}\label{bhm_spiders_agn_sep_d3_plan1.2.4}}

\noindent\textbf{target\_selection plan:} 1.2.4

\noindent\textbf{target\_selection tag:}
\href{https://github.com/sdss/target_selection/tree/1.3.3/}{1.3.3}

\noindent\textbf{crossmatch plan:} 1.0.0

\noindent\textbf{Summary:} This is a small supplementary catalog of eROSITA
point-like (unresolved) X-ray sources detected in the Continuous Viewing
Zone South/South Ecliptic Pole (CVZ-S/SEP) field. Owing to the deepest
exposure of the eROSITA All-Sky Survey (eRASS) in this region, the
CVZ-S/SEP field suffers from severe source confusion, requiring a
specialized source-detection procedure (Liu et al., in prep.). The X-ray
catalog used here was derived from an early data reduction of
0.2-2.3~keV band data from the 2.3-year eRASS:5 dataset, covering an
approximately 6~deg~x~6~deg sky region centered on the CVZ-S/SEP. The
X-ray sources have been cross-matched by the eROSITA-DE team with Gaia
DR3.

\noindent\textbf{Simplified description of selection criteria:} Starting from a
parent catalog of eRASS:5 point source $\rightarrow$ Gaia DR3 associations from the
SEP/CVZ-S region, select targets which meet all of the following
criteria: i) have eROSITA detection likelihood\textgreater6.0, ii) have
an X-ray$\rightarrow$Gaia cross-match probability of \texttt{p\_any}\textgreater0.1,
iii) have \emph{G}\textgreater13.5 and \emph{RP}\textgreater13.5~Vega.
We deprioritise targets if any of the following criteria are met: a) the
target already has existing good quality SDSS spectroscopy, b) the X-ray
detection likelihood is \textless8.0, c) the target is a secondary
X-ray$\rightarrow$Gaia association, d) the target falls outside the core magnitude
range (i.e. doesn't satisfy 16.0\textless{}\emph{G}\textless21.5), e)
the X-ray flux is
\textless1x10\textsuperscript{-14}~erg~s\textsuperscript{-1}~cm\textsuperscript{-2}.
Cadence choices (exposure time requests per target) are based on optical
brightness. The '\_d3' variation of this carton provides a shallower
cadence option for targets toward the optically faint end of the sample.

\noindent\textbf{Target priority options:} 1502-1504, 3502-3508

\noindent\textbf{Cadence options:} \texttt{dark\_flexible\_3x1}

\noindent\textbf{Implementation:}
\href{https://github.com/sdss/target_selection/blob/1.3.3/python/target_selection/cartons/bhm_spiders_agn.py}{bhm\_spiders\_agn.py}

\noindent\textbf{Number of targets:} 863

\begin{center}\rule{0.5\linewidth}{0.5pt}\end{center}

\hypertarget{bhm_aqmes_wide1_plan1.2.4}{%
\subsection{bhm\_aqmes\_wide1}\label{bhm_aqmes_wide1_plan1.2.4}}

\noindent\textbf{target\_selection plan:} 1.2.4

\noindent\textbf{target\_selection tag:}
\href{https://github.com/sdss/target_selection/tree/1.3.3/}{1.3.3}

\noindent\textbf{crossmatch plan:} 1.0.0

\noindent\textbf{Summary:} Spectroscopically confirmed optically bright SDSS
QSOs, selected from the SDSS QSO catalog (DR16Q,
\citealt{Lyke2020}). Located in 425 fields within the SDSS QSO footprint,
where the choice of survey area prioritized field that overlapped with
the SPIDERS footprint (approx 180\textless{}\emph{b}\textless360~deg),
and/or had higher than average numbers of bright QSOs and CSC targets.
The list of field centres can be found
\href{https://github.com/sdss/target_selection/blob/0.3.0/python/target_selection/masks/candidate_target_fields_bhm_aqmes_wide_v0.3.1.fits}{within
the target\_selection repository}. The targets in this carton request a
single epoch of SDSS-V spectroscopy.

\noindent\textbf{Simplified description of selection criteria:} Select all
objects from SDSS DR16 QSO catalog that have
16.0\textless{}\texttt{sdss\_psfmag\_i}\textless19.1~AB, and that lie
within 1.49~degrees of at least one AQMES-wide field location.

\noindent\textbf{Target priority options:} 1211

\noindent\textbf{Cadence options:} \texttt{dark\_1x4}

\noindent\textbf{Implementation:}
\href{https://github.com/sdss/target_selection/blob/1.3.3/python/target_selection/cartons/bhm_aqmes.py}{bhm\_aqmes.py}

\noindent\textbf{Number of targets:} 24142

\begin{center}\rule{0.5\linewidth}{0.5pt}\end{center}

\hypertarget{bhm_aqmes_wide1_faint_plan1.2.4}{%
\subsection{bhm\_aqmes\_wide1\_faint}\label{bhm_aqmes_wide1_faint_plan1.2.4}}

\noindent\textbf{target\_selection plan:} 1.2.4

\noindent\textbf{target\_selection tag:}
\href{https://github.com/sdss/target_selection/tree/1.3.3/}{1.3.3}

\noindent\textbf{crossmatch plan:} 1.0.0

\noindent\textbf{Summary:} Spectroscopically confirmed optically faint SDSS QSOs,
selected from the SDSS QSO catalog (DR16Q,
\citealt{Lyke2020}). Located in 425 fields within the SDSS QSO footprint,
where the choice of survey area prioritized field that overlapped with
the SPIDERS footprint (approx 180\textless{}\emph{b}\textless360~deg),
and/or had higher than average numbers of bright QSOs and CSC targets.
The list of field centres can be found
\href{https://github.com/sdss/target_selection/blob/0.3.0/python/target_selection/masks/candidate_target_fields_bhm_aqmes_wide_v0.3.1.fits}{within
the target\_selection repository}. The targets in this carton request a
single epoch of SDSS-V spectroscopy.

\noindent\textbf{Simplified description of selection criteria:} Select all
objects from SDSS DR16 QSO catalog that have
19.1\textless{}\texttt{sdss\_psfmag\_i}\textless21.0~AB, and that lie
within 1.49~degrees of at least one AQMES-wide field location.

\noindent\textbf{Target priority options:} 3211

\noindent\textbf{Cadence options:} \texttt{dark\_1x4}

\noindent\textbf{Implementation:}
\href{https://github.com/sdss/target_selection/blob/1.3.3/python/target_selection/cartons/bhm_aqmes.py}{bhm\_aqmes.py}

\noindent\textbf{Number of targets:} 99586

\begin{center}\rule{0.5\linewidth}{0.5pt}\end{center}

\hypertarget{bhm_gua_dark_d3_plan1.2.4}{%
\subsection{bhm\_gua\_dark\_d3}\label{bhm_gua_dark_d3_plan1.2.4}}

\noindent\textbf{target\_selection plan:} 1.2.4

\noindent\textbf{target\_selection tag:}
\href{https://github.com/sdss/target_selection/tree/1.3.3/}{1.3.3}

\noindent\textbf{crossmatch plan:} 1.0.0

\noindent\textbf{Summary:} A sample of optically faint candidate AGN lacking
spectroscopic confirmations, derived from the parent sample presented by
\citet{Shu2019}, who applied a machine-learning approach to select QSO
candidates from a combination of the Gaia DR2 and unWISE catalogs.

\noindent\textbf{Simplified description of selection criteria:} Starting with the
\citet{Shu2019} catalog, select targets which satisfy the following
criteria: i) have a Random Forest probability of being a QSO
of\textgreater0.8, ii) are in the magnitude range suitable for BOSS
spectroscopy in dark time (\emph{G}\textgreater16.5 and
\emph{RP}\textgreater16.5, as well as \emph{G}\textless21.2 or
\emph{RP}\textless21.0,~Vega mag), iii) do not yet have good SDSS
optical spectroscopic measurements. Note that the selection criteria are
based on apparent magnitudes, rather than the dereddened magnitudes that
were used in an earlier iteration of this carton. The '\_d3' variation
of this carton provides a shallower cadence option for targets in this
sample.

\noindent\textbf{Target priority options:} 3401

\noindent\textbf{Cadence options:} \texttt{dark\_flexible\_3x1}

\noindent\textbf{Implementation:}
\href{https://github.com/sdss/target_selection/blob/1.3.3/python/target_selection/cartons/bhm_gua.py}{bhm\_gua.py}

\noindent\textbf{Number of targets:} 1936784

\begin{center}\rule{0.5\linewidth}{0.5pt}\end{center}

\hypertarget{bhm_spiders_agn_hard_d3_plan1.2.4}{%
\subsection{bhm\_spiders\_agn\_hard\_d3}\label{bhm_spiders_agn_hard_d3_plan1.2.4}}

\noindent\textbf{target\_selection plan:} 1.2.4

\noindent\textbf{target\_selection tag:}
\href{https://github.com/sdss/target_selection/tree/1.3.3/}{1.3.3}

\noindent\textbf{crossmatch plan:} 1.0.0

\noindent\textbf{Summary:} This small supplementary carton provides a sample of
point-like (unresolved) hard-band (2.3-5~keV) X-ray sources detected in
the first 6~months of eROSITA all sky survey data (eRASS:1). The
selection of the eROSITA hard band sample is described by
\citealt{Waddell2026}. The X-ray sources have been cross-matched by the
eROSITA-DE team to
\href{https://www.legacysurvey.org/dr10/}{LegacySurvey DR10} optical/IR
counterparts. All targets are located in the sky hemisphere where MPE
controls the data rights (approx.
180\textless{}\emph{l}\textless360~deg).

\noindent\textbf{Simplified description of selection criteria:} Starting from a
parent catalog of multi-band eRASS:1 X-ray point source $\rightarrow$ LegacySurvey
DR10 associations (method: NWAY assisted by optical/IR priors computed
via a pre-trained Random Forest, building on
\citealt{Salvato2022}), select targets which meet all of the following criteria:
i) have eROSITA 2.3-5keV band detection likelihood\textgreater12.0, ii)
have an X-ray $\rightarrow$ optical/IR cross-match probability (NWAY) of
\texttt{p\_any}\textgreater0.01, iii) have
13.5\textless{}\texttt{fibermag\_r}\textless22.5 or
13.5\textless{}\texttt{fibermag\_i}\textless22.3 or
13.5\textless{}\texttt{fibermag\_z}\textless21.5, iv) are not saturated
in LegacySurvey imaging, v) if detected by Gaia DR2/DR3 then have
\emph{G}\textgreater13.5 and \emph{RP}\textgreater13.5~Vega. We
\emph{deprioritise} targets if any of the following criteria are met: a)
the target already has existing good quality SDSS spectroscopy, b) the
target is a secondary X-ray$\rightarrow$optical/IR association, c) the target falls
outside the core magnitude range (i.e. doesn't satisfy one of
16.5\textless{}\texttt{fibermag\_r}\textless22.0 or
16.5\textless{}\texttt{fibermag\_i}\textless21.8 or
16.5\textless{}\texttt{fibermag\_z}\textless21.0), d) the target lies at
\textbar{}\emph{b}\textbar\textless10~deg or at Dec\textless-75~deg.
Cadence choices (exposure time requests per target) are based on optical
brightness. The '\_d3' variation of this carton provides a shallower
cadence option for targets in this sample.

\noindent\textbf{Target priority options:} 1511-1513, 1711, 3511-3513, 3711

\noindent\textbf{Cadence options:} \texttt{dark\_flexible\_3x1}

\noindent\textbf{Implementation:}
\href{https://github.com/sdss/target_selection/blob/1.3.3/python/target_selection/cartons/bhm_spiders_agn.py}{bhm\_spiders\_agn.py}

\noindent\textbf{Number of targets:} 798

\begin{center}\rule{0.5\linewidth}{0.5pt}\end{center}

\hypertarget{bhm_spiders_agn_tda_d3_plan1.2.4}{%
\subsection{bhm\_spiders\_agn\_tda\_d3}\label{bhm_spiders_agn_tda_d3_plan1.2.4}}

\noindent\textbf{target\_selection plan:} 1.2.4

\noindent\textbf{target\_selection tag:}
\href{https://github.com/sdss/target_selection/tree/1.3.3/}{1.3.3}

\noindent\textbf{crossmatch plan:} 1.0.0

\noindent\textbf{Summary:} This is a small supplementary carton of
transient/variable X-ray sources selected by comparing X-ray
measurements between several independent 6-month eROSITA sky surveys.
All targets are located in the sky hemisphere where MPE controls the
data rights (approx. 180\textless{}\emph{l}\textless360~deg).

\noindent\textbf{Simplified description of selection criteria:} We start from a
parent catalog of transient/variable X-ray sources selected from the
first three eROSITA All Sky surveys. These X-ray sources have been
associated with LegacySurvey DR10 optical counterparts (method: NWAY
assisted by optical/IR priors computed via a pre-trained Random Forest,
building on
\citealt{Salvato2022}). We then select targets which meet all of the following
criteria: i) have 13.5\textless{}\texttt{fibermag\_r}\textless22.5 or
13.5\textless{}\texttt{fibermag\_i}\textless22.3 or
13.5\textless{}\texttt{fibermag\_z}\textless21.5, ii) if detected by
Gaia DR2/DR3 then have \emph{G}\textgreater13.5 and
\emph{RP}\textgreater13.5~Vega. Cadence choices (exposure time requests
per target) are based on optical brightness. The '\_d3' variation of
this carton provides a shallower cadence option for targets in this
sample.

\noindent\textbf{Target priority options:} 1691

\noindent\textbf{Cadence options:} \texttt{dark\_flexible\_3x1}

\noindent\textbf{Implementation:}
\href{https://github.com/sdss/target_selection/blob/1.3.3/python/target_selection/cartons/bhm_spiders_agn.py}{bhm\_spiders\_agn.py}

\noindent\textbf{Number of targets:} 8141

\begin{center}\rule{0.5\linewidth}{0.5pt}\end{center}

\hypertarget{bhm_spiders_agn_lsdr10_d3_plan1.2.4}{%
\subsection{bhm\_spiders\_agn\_lsdr10\_d3}\label{bhm_spiders_agn_lsdr10_d3_plan1.2.4}}

\noindent\textbf{target\_selection plan:} 1.2.4

\noindent\textbf{target\_selection tag:}
\href{https://github.com/sdss/target_selection/tree/1.3.3/}{1.3.3}

\noindent\textbf{crossmatch plan:} 1.0.0

\noindent\textbf{Summary:} This is the highest priority carton for SPIDERS AGN
wide area follow up. The carton provides optical counterparts to
point-like (unresolved) X-ray sources detected in early reductions of
the first 18~months of eROSITA all sky survey data (eRASS:3). The sample
is expected to contain a mixture of QSOs, AGN, stars and compact
objects. The X-ray sources have been cross-matched by the eROSITA-DE
team to \href{https://www.legacysurvey.org/dr10/}{legacysurvey.org/DR10}
optical/IR counterparts (supplemented with the DR9 catalog at
Dec\textgreater+32.375~deg). All targets are located in the sky
hemisphere where MPE controls the data rights (approx.
180\textless{}\emph{l}\textless360~deg). Due to the LegacySurvey
footprint, nearly all targets in this carton are located at high
Galactic latitudes \textbar{}\emph{b}\textbar\textgreater15~deg.

\noindent\textbf{Simplified description of selection criteria:} Starting from a
parent catalog of eRASS:3 point source $\rightarrow$ legacysurvey.org/DR10(+DR9)
associations (method: NWAY assisted by optical/IR priors computed via a
pre-trained Random Forest, building on
\citealt{Salvato2022}), select targets which meet all of the following criteria:
i) have eROSITA detection likelihood\textgreater6.0, ii) have an X-ray $\rightarrow$
optical/IR cross-match probability (NWAY) of
\texttt{p\_any}\textgreater0.1, iii) have
13.5\textless{}\texttt{fibermag\_r}\textless22.5 or
13.5\textless{}\texttt{fibermag\_i}\textless22.3 or
13.5\textless{}\texttt{fibermag\_z}\textless21.5, iv) are not saturated
in LegacySurvey imaging, v) if detected by Gaia DR2/DR3 then have
\emph{G}\textgreater13.5 and \emph{RP}\textgreater13.5~Vega. We
\emph{deprioritise} targets if any of the following criteria are met: a)
the target already has existing good quality SDSS spectroscopy, b) the
X-ray detection likelihood is \textless8.0, c) the target is a secondary
X-ray$\rightarrow$optical/IR association, d) the target falls outside the core
magnitude range (i.e. doesn't satisfy one of
16.5\textless{}\texttt{fibermag\_r}\textless22.0 or
16.5\textless{}\texttt{fibermag\_i}\textless21.8 or
16.5\textless{}\texttt{fibermag\_z}\textless21.0), e) the target lies at
\textbar{}\emph{b}\textbar\textless15~deg or at Dec\textless-75~deg, f)
the X-ray flux is
\textless2x10\textsuperscript{-14}~erg~s\textsuperscript{-1}~cm\textsuperscript{-2}.
We slightly boost the priorities of targets which have hard X-ray
detections. The '\_d3' variation of this carton provides a shallower
cadence option for targets toward the optically faint end of the sample.

\noindent\textbf{Target priority options:} 1521-1528, 1721-1728, 3521-3528,
3721-3728

\noindent\textbf{Cadence options:} \texttt{dark\_flexible\_3x1}

\noindent\textbf{Implementation:}
\href{https://github.com/sdss/target_selection/blob/1.3.3/python/target_selection/cartons/bhm_spiders_agn.py}{bhm\_spiders\_agn.py}

\noindent\textbf{Number of targets:} 851160

\begin{center}\rule{0.5\linewidth}{0.5pt}\end{center}

\hypertarget{bhm_colr_galaxies_lsdr10_d3_plan1.2.16}{%
\subsection{bhm\_colr\_galaxies\_lsdr10\_d3}\label{bhm_colr_galaxies_lsdr10_d3_plan1.2.16}}

\noindent\textbf{target\_selection plan:} 1.2.16

\noindent\textbf{target\_selection tag:}
\href{https://github.com/sdss/target_selection/tree/1.3.15/}{1.3.15}

\noindent\textbf{crossmatch plan:} 1.0.0

\noindent\textbf{Summary:} A supplementary magnitude limited sample of optically
bright galaxies selected from the LegacySurvey DR10 optical/IR imaging
catalog (supplemented with the LegacySurvey DR9 catalog at
Dec\textgreater+32.375~deg). Selection is based on optical morphology,
lack of Gaia parallax, and several magnitude cuts.

\noindent\textbf{Simplified description of selection criteria:} Starting from the
\texttt{legacy\_survey\_dr10} catalog (lsdr10), select entries
satisfying all of the following criteria: i) lsdr10 morphological
\texttt{type} != 'PSF', ii) lsdr10
\texttt{shape\_r}\textgreater1.0~arcsec, iii) no detection of parallax
by Gaia, iv) dereddened \emph{z}-band model mag\textless19.0~AB, and
dereddened \emph{z}-band fiber mag\textless19.5~AB, v) within all the
following apparent fiber mag limits: 16\textless{}\emph{g}\textless22.5~
and 16\textless{}\emph{r}\textless21.5~ and
16\textless{}\emph{z}\textless20.0~AB, vi) if detected in Gaia, then
\emph{G}\textgreater15.0~ and \emph{RP}\textgreater15.0~Vega, vii) at
moderate to high Galactic latitude
\textbar{}\emph{b}\textbar\textgreater15~deg. The '\_d3' variation of
this carton provides a shallower cadence option for targets toward the
optically faint end of the sample.

\noindent\textbf{Target priority options:} 7101

\noindent\textbf{Cadence options:} \texttt{dark\_flexible\_3x1}

\noindent\textbf{Implementation:}
\href{https://github.com/sdss/target_selection/blob/1.3.15/python/target_selection/cartons/bhm_galaxies.py}{bhm\_galaxies.py}

\noindent\textbf{Number of targets:} 6091854

\begin{center}\rule{0.5\linewidth}{0.5pt}\end{center}

\hypertarget{bhm_rm_core_plan1.0.48}{%
\subsection{bhm\_rm\_core}\label{bhm_rm_core_plan1.0.48}}

\noindent\textbf{target\_selection plan:} 1.0.48

\noindent\textbf{target\_selection tag:}
\href{https://github.com/sdss/target_selection/tree/1.0.48/}{1.0.48}

\noindent\textbf{crossmatch plan:} 1.0.0

\noindent\textbf{Summary:} A sample of candidate QSOs selected via the methods
presented by
\citet{Yang2022}. These targets are located within five (+1 backup) well
known survey fields (SDSS-RM, COSMOS, XMM-LSS, ECDFS, CVZ-S/SEP, and
ELIAS-S1).

\noindent\textbf{Simplified description of selection criteria:} Starting from a
parent catalog of optically selected objects in the RM fields (as
presented by
\citealt{Yang2022}), select candidate QSOs that satisfy all of the
following: i) are identified via the Skew-T algorithm
(\texttt{skewt\_qso\ ==\ 1}); ii) have
17\textless{}\texttt{psfmag\_i}\textless21.5~AB
(16\textless{}\emph{G}\textless21.7~AB in the CVZ-S/SEP field); iii) do
not have significant detections (\textgreater3$\sigma$) of parallax and/or
proper motion in Gaia DR2; iv) are not vetoed due to results of visual
inspections of recent spectroscopy; vi) have detections in all of the
gri bands (a Gaia detection is sufficient in the CVZ-S/SEP field);and
vii) do not lie in the SDSS-RM field. This carton version starts with an
updated parent catalog, and fixes an issue which could have caused a few
targets to be unintentionally excluded.

\noindent\textbf{Target priority options:} 1000-1100

\noindent\textbf{Cadence options:} \texttt{dark\_100x8,\ dark\_174x8}

\noindent\textbf{Implementation:}
\href{https://github.com/sdss/target_selection/blob/1.0.48/python/target_selection/cartons/bhm_rm.py}{bhm\_rm.py}

\noindent\textbf{Number of targets:} 3757

\begin{center}\rule{0.5\linewidth}{0.5pt}\end{center}

\hypertarget{bhm_rm_known_spec_plan1.0.48}{%
\subsection{bhm\_rm\_known\_spec}\label{bhm_rm_known_spec_plan1.0.48}}

\noindent\textbf{target\_selection plan:} 1.0.48

\noindent\textbf{target\_selection tag:}
\href{https://github.com/sdss/target_selection/tree/1.0.48/}{1.0.48}

\noindent\textbf{crossmatch plan:} 1.0.0

\noindent\textbf{Summary:} A sample of known QSOs identified through optical
spectroscopy from various projects, as collated by
\citet{Yang2022}. These targets are located within five (+1 backup) well
known survey fields (SDSS-RM, COSMOS, XMM-LSS, ECDFS, CVZ-S/SEP, and
ELIAS-S1).

\noindent\textbf{Simplified description of selection criteria:} Starting from a
parent catalog of optically selected objects in the RM fields (as
presented by
\citealt{Yang2022}), select targets which satisfy all of the following: i)
are flagged as having a spectroscopic identification (in the parent
catalog); ii) have 15\textless{}\texttt{psfmag\_i}\textless21.7~AB
(SDSS-RM, CDFS, ELIAS-S1 field), 16\textless{}\emph{G}\textless21.7~Vega
(CVZ-S/SEP field), 15\textless{}\texttt{psfmag\_i}\textless21.5~AB
(COSMOS and XMM-LSS fields); iii) have a spectroscopic redshift in the
range 0.005\textless{}\emph{z}\textless7; iv) are not vetoed due to
results of visual inspections of recent spectroscopy. This carton
version starts with an updated parent catalog, and fixes an issue which
could have caused a few targets to be unintentionally excluded.

\noindent\textbf{Target priority options:} 1000-1100

\noindent\textbf{Cadence options:} \texttt{dark\_100x8,\ dark\_174x8}

\noindent\textbf{Implementation:}
\href{https://github.com/sdss/target_selection/blob/1.0.48/python/target_selection/cartons/bhm_rm.py}{bhm\_rm.py}

\noindent\textbf{Number of targets:} 3212

\begin{center}\rule{0.5\linewidth}{0.5pt}\end{center}

\hypertarget{bhm_rm_var_plan1.0.48}{%
\subsection{bhm\_rm\_var}\label{bhm_rm_var_plan1.0.48}}

\noindent\textbf{target\_selection plan:} 1.0.48

\noindent\textbf{target\_selection tag:}
\href{https://github.com/sdss/target_selection/tree/1.0.48/}{1.0.48}

\noindent\textbf{crossmatch plan:} 1.0.0

\noindent\textbf{Summary:} A sample of candidate QSOs selected via their optical
variability properties, as presented by
\citet{Yang2022}. These targets are located within five (+1 backup) well
known survey fields (SDSS-RM, COSMOS, XMM-LSS, ECDFS, CVZ-S/SEP, and
ELIAS-S1).

\noindent\textbf{Simplified description of selection criteria:} Starting from a
parent catalog of optically selected objects in the RM fields (as
presented by
\citealt{Yang2022}), select candidate QSOs that satisfy all of the
following: i) have significant variability in the DES or PanSTARRS1
multi-epoch photometry (\texttt{var\_sn{[}g{]}}\textgreater3 and
\texttt{var\_rms{[}g{]}}\textgreater0.05); ii) have
17\textless{}\texttt{psfmag\_i}\textless20.5~AB
(16\textless{}\emph{G}\textless21.7~AB in the CVZ-S/SEP field); iii) do
not have significant detections (\textgreater3$\sigma$) of parallax and/or
proper motion in Gaia DR2; iv) are not vetoed due to results of visual
inspections of recent spectroscopy; and vi) do not lie in the SDSS-RM
field. This carton version starts with an updated parent catalog, and
fixes an issue which could have caused a few targets to be
unintentionally excluded.

\noindent\textbf{Target priority options:} 1000-1100

\noindent\textbf{Cadence options:} \texttt{dark\_174x8}

\noindent\textbf{Implementation:}
\href{https://github.com/sdss/target_selection/blob/1.0.48/python/target_selection/cartons/bhm_rm.py}{bhm\_rm.py}

\noindent\textbf{Number of targets:} 871

\begin{center}\rule{0.5\linewidth}{0.5pt}\end{center}

\hypertarget{bhm_rm_ancillary_plan1.0.48}{%
\subsection{bhm\_rm\_ancillary}\label{bhm_rm_ancillary_plan1.0.48}}

\noindent\textbf{target\_selection plan:} 1.0.48

\noindent\textbf{target\_selection tag:}
\href{https://github.com/sdss/target_selection/tree/1.0.48/}{1.0.48}

\noindent\textbf{crossmatch plan:} 1.0.0

\noindent\textbf{Summary:} A supporting sample of candidate QSOs which have been
selected by the Gaia-unWISE AGN catalog
(\citealt{Shu2019}) and/or the SDSS XDQSO catalog
(\citealt{Bovy2011}). These targets are located within five (+1 backup) well
known survey fields (SDSS-RM, COSMOS, XMM-LSS, ECDFS, CVZ-S/SEP, and
ELIAS-S1).

\noindent\textbf{Simplified description of selection criteria:} Starting from a
parent catalog of optically selected objects in the RM fields (as
presented by
\citealt{Yang2022}), select candidate QSOs that satisfy all of the
following: i) are identified via external ancillary methods
(\texttt{photo\_bitmask\ \&\ 3\ !=\ 0}); ii) have
15\textless{}\texttt{psfmag\_i}\textless21.5~AB
(16\textless{}\emph{G}\textless21.7~AB in the CVZ-S/SEP field); iii) do
not have significant detections (\textgreater3$\sigma$) of parallax and/or
proper motion in Gaia DR2; iv) are not vetoed due to results of visual
inspections of recent spectroscopy; and v) do not lie in the SDSS-RM
field. This carton version starts with an updated parent catalog, and
fixes an issue which could have caused a few targets to be
unintentionally excluded.

\noindent\textbf{Target priority options:} 1000-1100

\noindent\textbf{Cadence options:} \texttt{dark\_100x8,\ dark\_174x8}

\noindent\textbf{Implementation:}
\href{https://github.com/sdss/target_selection/blob/1.0.48/python/target_selection/cartons/bhm_rm.py}{bhm\_rm.py}

\noindent\textbf{Number of targets:} 1114

\begin{center}\rule{0.5\linewidth}{0.5pt}\end{center}

\hypertarget{manual_bhm_spiders_comm_lco_plan0.5.23}{%
\subsection{manual\_bhm\_spiders\_comm\_lco}\label{manual_bhm_spiders_comm_lco_plan0.5.23}}

\noindent\textbf{target\_selection plan:} 0.5.23

\noindent\textbf{target\_selection tag:}
\href{https://github.com/sdss/target_selection/tree/0.3.22/}{0.3.22}

\noindent\textbf{crossmatch plan:} 0.5.0

\noindent\textbf{Summary:} A utility sample of X-ray selected targets located in
the XMM-XXL-N and Stripe-82X fields having reliable spectroscopic
redshifts. This carton was designed to support SDSS-V commissioning (of
FPS at LCO), as a reference sample to test the accuracy and completeness
of pipeline-derived redshifts extracted from new SDSS-V spectroscopic
data.

\noindent\textbf{Simplified description of selection criteria:} Starting from a
compilation of X-ray selected sources in the XMM-XXL-N and Stripe-82X
fields with optical counterparts in the LegacySurvey DR8 catalog, select
those having i) reliable spectroscopic redshifts (confirmed visually)
from previous SDSS observations, and ii) which lie in the magnitude
range 16\textless{}\texttt{mag\_r}\textless23.0~AB.

\noindent\textbf{Target priority options:} 1000

\noindent\textbf{Cadence options:} \texttt{dark\_1x4}

\noindent\textbf{Implementation:} n/a

\noindent\textbf{Number of targets:} 3160

\bibliography{main}{}
\bibliographystyle{aasjournalv7}

\end{document}